\documentclass[11pt,a4paper,oneside,openright]{book}
\let\counterwithout\relax

\usepackage{listings}
\usepackage{tcolorbox}

\usepackage{mystyle}
\usepackage{braket}
\usepackage{graphicx}
\usepackage{caption}
\usepackage{subcaption}
\usepackage{bm}

\usepackage{empheq}
\usepackage{wrapfig, framed}
\usepackage{color}
\usepackage{textcomp}
\usepackage{amssymb,amsmath,amsfonts,mathrsfs}

\usepackage[utf8]{inputenc}
\usepackage[compat=1.1.0]{tikz-feynman}
\usetikzlibrary{arrows, decorations.markings,shapes.geometric}
\usepackage{tocloft}
\usepackage{contour}
\definecolor{light-gray}{gray}{0.95}

\usepackage{stmaryrd}
\usepackage{slashed}

\usepackage{booktabs}
\usepackage{graphicx}
\usepackage{dcolumn}
\usepackage{bm}
\usepackage{xcolor}
\usepackage{hyperref}
\hypersetup{
    colorlinks=false,
    linktoc=page,
    citecolor=blue,
    linkcolor=cyan,
    urlcolor=blue
}

\makeatletter
\@addtoreset{section}{chapter}
\makeatother
\newlength\mylen

\renewcommand\cftpartpresnum{Part~}
\newcommand{\bs}{\boldsymbol}

\begin{document}

\begin{titlepage}

\begin{centering}

\begin{figure}[h]
\centering
\includegraphics[scale=0.8]{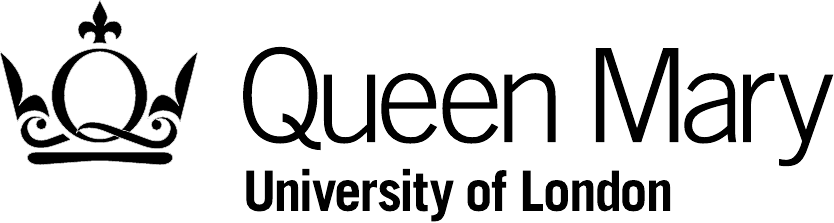}
\end{figure}\ \\

{\LARGE
\textsc{Thesis submitted for the degree of \\[0.1cm] Doctor of Philosophy}
}\\[1.1cm]

\HRule\\[0.5cm]
{\huge
\bfseries{The Eikonal Approximation and the Gravitational Dynamics of Binary Systems}
}\\[0.5cm]
\HRule\\[1.5cm]

{\huge
\textsc{Arnau Robert Koemans Collado}
}\\[2cm]

{\Large
Supervisors \\[0.3cm]
\textsc{Dr Rodolfo Russo \& Prof Steven Thomas}\\[1.5cm]
\today}\\[0.5cm]

{\large
\textsf{
Centre for Research in String Theory \\[0.1cm]
School of Physics and Astronomy \\[0.1cm]
Queen Mary University of London}}

\end{centering}

\end{titlepage}

\thispagestyle{empty}
\topskip0pt

\vspace*{3cm}
\begin{center}
\textit{This thesis is dedicated to my family and friends,}\\
\textit{without their unconditional love and support this would not have been possible.}
\end{center}
\vspace*{\fill}

\newpage
\thispagestyle{empty}
\topskip0pt
\vspace*{7cm}
\begin{center}
\begin{minipage}[][][c]{0.8\textwidth}

\textit{
\hspace{-5pt}
It doesn't matter how beautiful your theory is, it doesn't matter how smart you are. If it doesn't agree with experiment, it's wrong.}
\begin{flushright}
-- Richard P. Feynman
\end{flushright}

\textit{
\hspace{-5pt}
If you don't know, ask. You will be a fool for the moment, but a wise man for the rest of your life.}
\begin{flushright}
-- Seneca the Younger
\end{flushright}

\textit{
\hspace{-5pt}
When you are courting a nice girl an hour seems like a second. When you sit on a red-hot cinder a second seems like an hour. That's relativity.}
\begin{flushright}
-- Albert Einstein
\end{flushright}

\end{minipage}
\end{center}

\vspace*{\fill}

\chapter*{Abstract}

In this thesis we study the conservative gravitational dynamics of binary systems using the eikonal approximation; allowing us to use scattering amplitude techniques to calculate dynamical quantities in classical gravity. This has implications for the study of binary black hole systems and their resulting gravitational waves.

In the first three chapters we introduce some of the basic concepts and results that we will use in the rest of the thesis. In the first chapter an overview of the topic is discussed and the academic context is introduced. The second chapter includes a basic discussion of gravity as a quantum field theory, the post-Newtonian (PN) and post-Minkowskian (PM) expansions and the eikonal approximation. In the third chapter we consider various Feynman integrals that are used extensively in subsequent chapters. Specifically, we give a recipe for expanding the relevant integrands in a so-called high energy expansion and then calculating the resulting integrals. 

The fourth chapter involves the study of massless states scattering off of a stack of D$p$-branes in $N=8$ supergravity. The setup we consider provides an ideal scenario to study inelastic contributions to the scattering process and their impact on the formulation of the eikonal approximation. These results will give us a better understanding of the eikonal approximation presented in the second chapter. The fifth chapter involves studying the eikonal and corresponding dynamical quantities in a Kaluza-Klein theory of gravity providing further interesting insight into the eikonal approximation and allowing us to compare with various known results.

The sixth and seventh chapter apply the concepts developed in this thesis to the problem of binary Schwarzschild black holes in $D$ spacetime dimensions. This allows us to apply the framework exposed in previous chapters to a physically realistic scenario giving us a better understanding of how to extract the relevant dynamical information from scattering amplitudes. The results derived in chapter six also have an impact on our understanding at higher orders in the PM expansion beyond the ones considered in this text. In the seventh chapter we present the Hamiltonian for a system of binary Schwarzschild black holes and show how to extract the Hamiltonian from other dynamical quantities calculated using the eikonal.

In the last chapter we provide some concluding remarks and a brief outlook.

\chapter*{Acknowledgements}

It is my pleasure to thank my supervisors, Dr Rodolfo Russo and Prof Steve Thomas. Their pastoral and academic guidance has been a crucial component in getting me to where I am academically. The writing of this thesis would not have been possible without them. From good humour to complicated equations they have always been there to support me. I couldn't have asked for a better pair of supervisors. I would also like to give a very special thank you to our collaborator throughout my PhD, Paolo Di Vecchia, for his unparalleled wisdom and mathematical accuracy. Additionally I would like to thank my examiners Andreas Brandhuber and Poul H. Damgaard for hosting a fantastic viva and providing insightful corrections.

I would also like to thank my fellow PhD students at CRST. Through fantastic discussions ranging from physics to philosophy as well as great jokes and laughs, they have added the spark that has made my PhD experience wonderful. In particular Edward Hughes, Zac Kenton, Joe Hayling, Martyna Jones, Emanuele Moscato, Rodolfo Panerai, Zolt\'an Laczk\'o, Christopher Lewis--Brown, Ray Otsuki, Luigi Alfonsi, Nadia Bahjat--Abbas, Nejc $\check{\mathrm{C}}$eplak, Linfeng Li, Ricardo Stark--Muchão, Manuel Accettulli Huber, Rashid Alawadhi, Enrico Andriolo, Stefano De Angelis, Marcel Hughes, Gergely Kantor, David Peinador Veiga, Rajath Radhakrishnan, and Shun-Qing Zhang. 

It has also been a great pleasure to form part of CRST and I would also like to thank the faculty members that comprise this group. Their wisdom and guidance, always on offer, has been an important part of my experience. In particular David Berman, Matt Buican, Constantinos Papageorgakis, Sanjaye Ramgoolam, Gabriele Travaglini and Christopher White. 

Lastly, my family has been an incredibly important support mechanism and I would like to thank my parents and siblings for providing the environment necessary for me to thrive. My friends outside of academia have helped in making sure I didn't get too intertwined with the fabric of the cosmos and I would especially like to thank Jacob Swambo for putting up with my bad humour and incessant questioning of everything. I would also like to thank all my friends from Puente de Monta\~{n}ana for helping me disconnect and enjoy the Spanish countryside sunshine.

This work was supported by an STFC research studentship.

\chapter*{Declaration}
\vspace{-0.5cm}
I, Arnau Robert Koemans Collado, confirm that the research included within this thesis is my own work or that where it has been carried out in collaboration with, or supported by others, that this is duly acknowledged below and my contribution indicated. Previously published material is also acknowledged below. \\

\noindent I attest that I have exercised reasonable care to ensure that the work is original, and does not to the best of my knowledge break any UK law, infringe any third party’s copyright or other Intellectual Property Right, or contain any confidential material. \\

\noindent I accept that the College has the right to use plagiarism detection software to check the electronic version of the thesis. \\

\noindent I confirm that this thesis has not been previously submitted for the award of a degree by this or any other university. \\

\noindent The copyright of this thesis rests with the author and no quotation from it or information derived from it may be published without the prior written consent of the author. \\

\noindent Signature: 
\begin{figure}[h]
\hspace{0.1cm}
\includegraphics[scale=0.05]{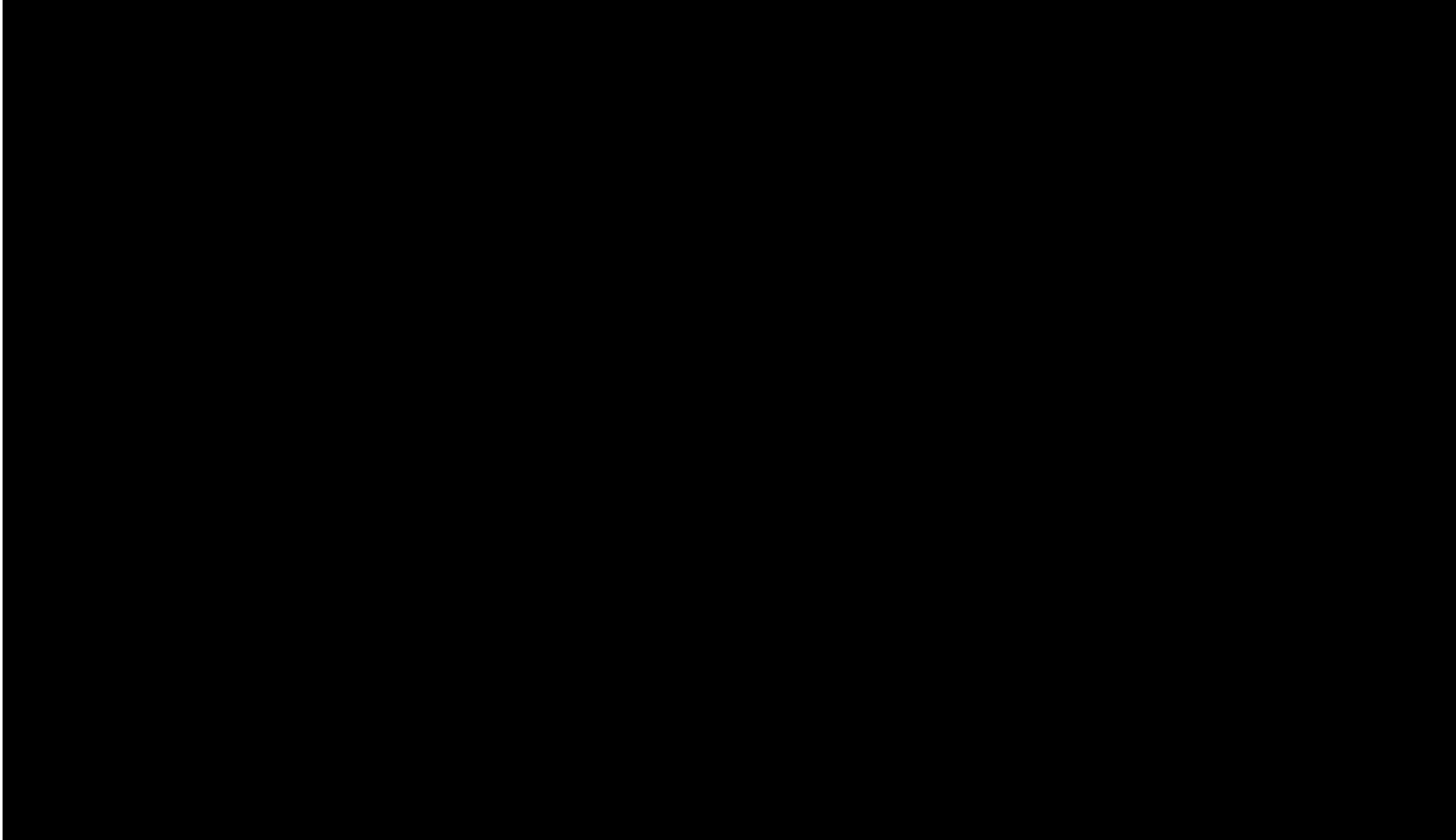}
\end{figure}

\noindent Date: \today \\
 
\noindent  Details of collaboration and publications:\\

\noindent This thesis describes research carried out with my supervisors Rodolfo Russo and Steve Thomas, which was published in \cite{Collado:2018isu, KoemansCollado:2019lnh, KoemansCollado:2019ggb}. It also contains some unpublished material. We collaborated with Paolo Di Vecchia in \cite{Collado:2018isu, KoemansCollado:2019ggb}. Where other sources have been used, they are cited in the bibliography.

\newpage

\tableofcontents

\chapter{Introduction}\label{chap:intro}

Modern theoretical physics contains two incredible yet seemingly unreconcilable theories with which it describes reality. We have Quantum Field Theory (QFT) \cite{Gordon1926, Klein:1926tv, Dirac:1928hu, Dirac:1930ek, Feynman:1949zx, Yang:1954ek} which is used to describe physics at small scales; at colliders and in certain condensed matter setups. We also have General Relativity (GR) \cite{Einstein:1916vd} which details the behaviour of matter at very large scales; for black holes as well as the entire cosmos. Although the Standard Model \cite{Glashow:1961tr, Higgs:1964pj, Guralnik:1964eu, Weinberg:1967tq, Salam:1968rm} is a fantastically accurate theory at the scales relevant for QFT it nonetheless lacks a description and inclusion of gravity, one of the forces most well known to everyday people.

The Standard Model and QFT are not without their problems. The chief of these issues, as mentioned above, is its lack of a complementary theory of quantum gravity. Furthermore the Standard Model does not naturally incorporate neutrino masses which have been experimentally found to have non-zero mass \cite{Schechter:1980gr, Fukuda:1998mi, Ahmad:2001an}. The Standard Model also has no coherent description of dark matter or dark energy as well as other cosmological theories such as inflation.

General Relativity was developed by Einstein as a classical description of gravitation replacing the older Newtonian gravity. Although its successes are unmatched by any other theory of gravity it too suffers from a variety of issues. There is some overlap with the issues in QFT in the sense that a coherent understanding of quantum gravity could potentially lead to a better understanding of dark energy. GR also suffers from technical issues if one tries to naively quantize gravity in the same way as electrodynamics. This is discussed in more detail in section \ref{sec:QGUVissues} but fundamentally the naive quantum version of Einstein's theory of gravity turns out to be non-renormalizable due to the dimensionful coupling constant.

In this thesis we will overcome these shortcomings and use both QFT and GR to describe the dynamics of binary systems with the objective of describing the dynamics of binary black holes. We know that the quantization of quantum gravity leads to a perturbatively non-renormalizable quantum field theory. This inherently limits the predictive power and usefulness that canonical quantum gravity fundamentally provides. However, as we will see throughout this thesis, in certain limits, such as two particle high energy forward scattering, it can produce useful and calculable results.

In 2015 the first observation of gravitational waves was made via the LIGO experiment \cite{Abbott:2016blz, castelvecchi2016einstein, Abbott:2016nmj, Abbott:2017vtc}. This ushered in a new paradigm for precision measurements with which to test both GR and other theories of gravity. Originally predicted by Einstein \cite{Einstein:1916cc,Einstein:1918btx} these observations give theoreticians a new problem to study in depth. Through this new and exciting experimental tool we expect to be able to uncover and test our understanding of cosmology, black holes and gravitation. All of which are areas of study relevant to our understanding of quantum gravity.

\begin{figure}[h]
  \centering
  \includegraphics[scale=0.3]{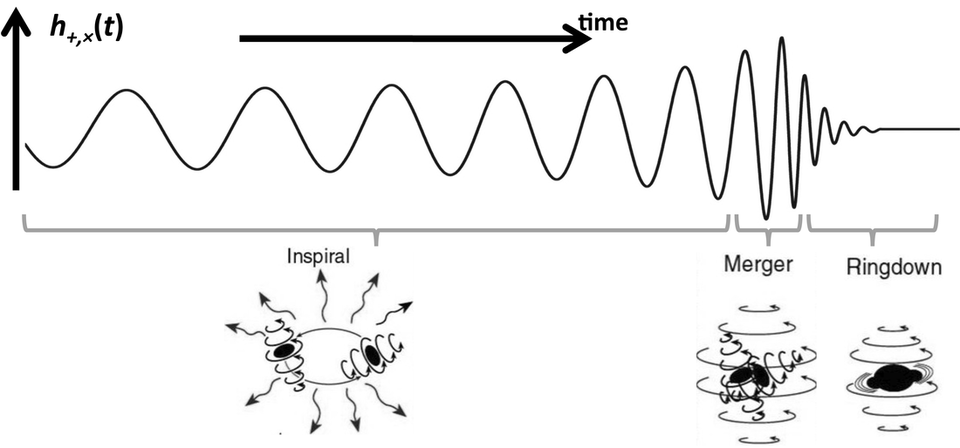}
  \caption{A figure illustrating the various phases of a black hole merger. The inspiral and ringdown phases can be aided via analytical calculations whereas the merger phase is simulated using purely numerical calculations. Image taken from \cite{soundsofspacetime}.}
  \label{fig:bhwave}
\end{figure}

Remarkably, using the raw data of these observations, experimentalists can deduce a lot about the properties of the objects that are colliding; whether they be black holes, neutron stars or mixed binary systems. In order to do this they use gravitational wave templates which are usually generated from a mixture of numerical and analytical techniques. Although there has been a lot of success numerically \cite{Campanelli:2005dd, Pretorius:2005gq, Baker:2005vv, Lange:2017wki, Blackman:2017pcm} it has been observed that combining numerical with analytic results provides more accurate templates for experimentalists to use \cite{Buonanno:1998gg, Buonanno:2000ef, Damour:2001tu, Antonelli:2019ytb}. Figure \ref{fig:bhwave} illustrates the various phases of a black hole merger. Analytical calculations can assist numerical computations in the inspiral and ringdown phases. The objective of this thesis will be to study the analytical component related to the inspiral phase of this problem.

There are two analytical perturbative regimes in which the inspiral phase of a black hole merger is usually studied; the post-Newtonian (PN) \cite{droste1917field, Einstein:1938yz, Bini:2017wfr, Foffa:2019hrb} and post-Minkowskian (PM) regime \cite{bertotti1956gravitational, Kerr1959, BERTOTTI1960169, Portilla:1979xx, Westpfahl:1979gu, Portilla:1980uz, Bel:1981be, Ledvinka:2008tk, Damour:2016gwp,Damour:2017zjx,Bjerrum-Bohr:2018xdl, Bern:2019nnu, Bern:2019crd, Kalin:2019rwq, Cristofoli:2019neg, Bjerrum-Bohr:2019kec, Kalin:2019inp, Damour:2019lcq, Bini:2020flp, Bern:2020gjj}. As is described in more detail in section \ref{sec:pmexpansionbg}, these regimes use different parameters in their perturbative expansion. The PN regime is effectively an expansion in the relative velocity of the two black holes\footnote{Note that this includes varying powers of the gravitational constant $G_N$ due to the virial theorem.} and is therefore non-relativistic. Whilst the PM regime is an expansion in the gravitational constant $G_N$ at all orders in velocity and is therefore relativistic. Although the PN regime has been studied since the time of Einstein \cite{droste1917field,Einstein:1938yz} and is therefore more well established, the PM regime has found renewed interest due to new developments and techniques in QFT. We will focus on the PM regime in this thesis\footnote{We will be considering the conservative dynamics of the inspiral process, the non-conservative component can also be studied \cite{Ademollo:1990sd, Amati:1990xe, Gruzinov:2014moa, Ciafaloni:2015xsr, Ciafaloni:2018uwe, Addazi:2019mjh} but will not be considered in this thesis.}.

The PM expansion, being an expansion in the gravitational coupling $G_N$, at all orders in velocity, naturally leads to a fully relativistic QFT-based description. One can quite quickly come to this realization by considering the fact that at tree-level one has an amplitude proportional to $G_N$, at one-loop it is proportional to $G_N^2$ and so on. The main caveat to this is that we are trying to describe a classical process, the inspiraling phase of two black holes, but the gravitational amplitudes naturally include quantum contributions. There are various techniques that can be used to extract the relevant classical information; such as by calculating the potential using an EFT-matching based approach \cite{Cheung:2018wkq, Bern:2019nnu, Bern:2019crd} or by using various other approaches which relate the amplitudes to classical observables such as the scattering angle \cite{Kabat:1992tb,Akhoury:2013yua, Melville:2013qca, Luna:2016idw, Bjerrum-Bohr:2018xdl, Kosower:2018adc, DiVecchia:2019myk, DiVecchia:2019kta, Kalin:2019rwq, Bjerrum-Bohr:2019kec, Kalin:2019inp, Damour:2019lcq, Bini:2020flp, Bern:2020gjj}. Of this last set of techniques there is the so called eikonal approximation which will be the main tool used here.

The eikonal approximation \cite{Blankenbecler:1962ez,Levy:1969cr,Abarbanel:1969ek,Wallace:1977ae,Kabat:1992tb} is a technique in quantum field theory for calculating the behaviour of high-energy scattering. This technique can be further used to describe the behaviour of high-energy classical scattering which is what we will study in order to relate the eikonal approximation to quantities relevant for the dynamics of binary systems. As we will see in section \ref{sec:eikonalapproxbg} the eikonal approximation is closely related to the Regge limit of the scattering amplitudes. 

In the Regge high energy limit the $2\to 2$ scattering process is dominated by the contributions of the highest spin states in the theory~\cite{tHooft:1987vrq, Amati:1987wq, Muzinich:1987in}. So, in a gravitational theory, this scattering is dominated at large values of the impact parameter by ladder diagrams involving the exchange of gravitons between the external states. The leading energy contributions of this class of diagrams resums into an exponential phase \cite{Kabat:1992tb,Akhoury:2013yua}; this is the so-called eikonal phase. This can also be done for high-energy potential scattering in a non-relativistic scenario \cite{Schiff:1956zz,Sugar:1969rn,Wallace:1973iu}. The eikonal phase can then be directly related to classical observables such as the scattering angle or Shapiro time delay as we will see in more detail in section \ref{sec:eikonalapproxbg}.

The high energy limit of scattering amplitudes in gravitational theories has been thoroughly studied as a gedanken-experiment that provides a non-trivial test of the consistency of the gravitational theory. A particularly tractable regime is the Regge limit, where both the energies and the impact parameter are large and unitarity is preserved due to a resummation of Feynman diagrams which reproduces the effect of a classical geometry~\cite{tHooft:1987vrq, Amati:1987wq,Muzinich:1987in,Sundborg:1988tb, Verlinde:1991iu, Amati:1993tb}. These early studies focused on the case of external massless states whose high energy Regge scattering matches the gravitational interaction of two well-separated shock-waves. However it is possible to generalise the same approach to the scattering of massive states~\cite{Kabat:1992tb} where the large centre of mass energy is due to both the kinetic and rest mass energy. It is then possible to interpolate between the ultra-relativistic Regge scattering mentioned above and the study of the non-relativistic large distance interaction between massive objects. This can be done both for pure General Relativity (GR) as well as for string theory, see for instance \cite{D'Appollonio:2010ae} for the analysis of the scattering of a perturbative massless state off a D-brane which we recall is a massive object\footnote{See~\cite{Bjerrum-Bohr:2014zsa,Bjerrum-Bohr:2016hpa} for the study of light/heavy scattering in standard GR including the derivation of quantum correction to the gravitational potential.}. The technique of deriving the relativistic interaction of two massive objects from an amplitude approach has recently attracted renewed attention~\cite{Neill:2013wsa,Akhoury:2013yua,Luna:2016idw,Cachazo:2017jef,Bjerrum-Bohr:2018xdl,Cheung:2018wkq,Kosower:2018adc,Bern:2019nnu, Bern:2019crd, Kalin:2019rwq, Bjerrum-Bohr:2019kec, Kalin:2019inp, Damour:2019lcq, Bini:2020flp, Bern:2019crd} since it links directly to the post-Minkowskian approximation of the classical gravitational dynamics relevant for the inspiraling phase of binary black hole systems~\cite{Buonanno:1998gg,Damour:2016gwp,Damour:2017zjx,Antonelli:2019ytb}.

The amplitude approach to the relativistic two-body problem can be stated in the following conceptually simple way. Consider $2 \to 2$ scattering where the external states have the quantum numbers necessary to describe the classical objects one is interested in (massless states describe shock-waves, massive scalars can describe Schwarzschild black holes, then spin and charge can be added to describe Kerr \cite{Bini:2017xzy,Vines:2017hyw, Vines:2018gqi, Guevara:2018wpp,Chung:2018kqs,Bautista:2019tdr, Guevara:2019fsj, Maybee:2019jus, Damgaard:2019lfh, Siemonsen:2019dsu} and Reissner-Nordstr\"om black holes). Then the limit is taken where Newton's gravitational constant $G_N$ is small, but all classical parameters, such as the Schwarzschild radius or the classical angular momentum, are kept finite. Since in this thesis we will generally be interested in studying the scattering of scalar states, the only classical parameter in the problem is the effective Schwarzschild radius, $R_s^{D-3} \sim G_N M^*$, where $M^*$ is the largest mass scale in the process. We can have $M^*= \sqrt{s}$ in the ultra-relativistic/massless case or $M^*=m_1$ in the probe-limit with $m_1^2 \gg (s-m_1^2), m_2^2$. In either case the relevant kinematic regime is the Regge limit, since the centre of mass energy $\sqrt{s}$ has to be much larger than the momentum transferred $\sqrt{|t|}$. Since $G_N$ is small, one might think that the perturbative diagrams with graviton exchanges yield directly the effective two-body potential, but one must be careful in performing this step. In the limit mentioned above the perturbative amplitude at a fixed order in $G_N$ is divergent thus creating tension with unitarity. These divergent terms should exponentiate when resumming the leading contributions at large energy at different orders in $G_N$. This exponential, called the eikonal phase\footnote{In more general gravitation theories the eikonal phase can become an operator; this already happens~at~leading~order~in string theory~\cite{Amati:1987wq,Amati:1987uf,Amati:1988tn,D'Appollonio:2010ae} and also in an effective theory of gravity including higher derivative corrections \cite{Camanho:2014apa,DAppollonio:2015fly}.}, is the observable that we wish to calculate and that, as we will see, contains the relevant information for the two-body potential.

While the picture described in this chapter applies to any weakly coupled gravitational theory, new features arise when one goes beyond two derivative gravity. For instance, in string theory the eikonal phase is promoted to an eikonal operator; since we are now dealing with objects that have a characteristic length, in certain regimes tidal forces~\cite{Giddings:2006vu,D'Appollonio:2013hja} can become important and excite the incoming state to different final states so as to produce an inelastic transition. At the leading order in the high energy, large impact parameter expansion, this stringy eikonal operator is obtained~\cite{Amati:1987wq,Amati:1987uf,Amati:1988tn,D'Appollonio:2010ae} from the standard eikonal phase, written in terms of the impact parameter $b$, simply via a shift $b\to b+ \hat{X}$, where $\hat{X}$ contains the bosonic string oscillation modes. A non-trivial eikonal operator also appears in the context of a gravitational effective field theory with higher derivative terms that modify the onshell 3-graviton vertex~\cite{Camanho:2014apa}. If the scale $\ell_{hd}$ at which the higher derivative corrections become important is much bigger than the Planck scale $\ell_P$, then,  by resumming the leading energy behaviour of the ladder diagrams as mentioned above, it is possible to use the effective field theory description to derive an eikonal operator also valid at scales $b \sim \ell_{hd}\gg \ell_P$. Again from this result it is possible to derive classical quantities, such as the time delay, that are now obtained from the eigenvalues of the eikonal operator. Generically when $b \sim \ell_{hd}$ the time delay for some scattering processes calculated in the effective field theory becomes negative. This causality violation most likely signals a breakdown of the effective field theory approach and in fact is absent when the same process is studied in a full string theory setup~\cite{Camanho:2014apa,DAppollonio:2015fly}.

Since the appearance of inelastic processes in the leading eikonal approximation is the signal of novel physical phenomena such as those mentioned above, it is interesting to see whether there are new features of this type in the {\em subleading} eikonal, which captures the first corrections in the large impact parameter expansion for the same $2 \to 2$ scattering. One of the aims of chapter \ref{chap:background} is to provide an explicit algorithm that allows us to derive this subleading eikonal from the knowledge of the amplitudes contributing to the scattering process under consideration. In the literature there are several explicit calculations of the subleading eikonal in various gravitational field theories in the two derivative approximation, see for instance~\cite{D'Appollonio:2010ae, Giddings:2010pp,Akhoury:2013yua,Melville:2013qca,Bjerrum-Bohr:2014zsa,Bjerrum-Bohr:2016hpa,Luna:2016idw}. However in these studies the process was {\em assumed} to be elastic to start with, while in chapter \ref{chap:sugraeik} we wish to spell out the conditions under which this is the case. Hopefully this will also provide a step towards a full understanding of the subleading eikonal operator at the string level~\cite{Amati:1990xe, Amati:1992zb}. Another goal of our analysis is to highlight that both the leading and the subleading eikonal depend on onshell data. The leading eikonal follows from the spectrum of the highest spin states and the onshell three-point functions, while in the case of the subleading eikonal some further information is necessary as new states in the spectrum may become relevant and the onshell four-point functions provide a non-trivial contribution. 

For the sake of concreteness, in chapter \ref{chap:sugraeik} we cast our analysis in the setup of type II supergravities focusing on the scattering of massless states off a stack of $N$ D$p$-branes~\cite{D'Appollonio:2010ae}, but the same approach can be applied in general to capture the subleading contributions of the large impact parameter scattering in any gravitational theory. In the limit where the mass (density) $N T_p / \kappa_D$ of the target D$p$-branes is large and the gravitational constant  $\kappa_D$ is small, with $N T_p \kappa_D$ fixed,  the process describes the scattering in a classical potential given by the gravitational backreaction of the target. In this case the eikonal phase is directly related (by taking its derivative with respect to the impact parameter) to the deflection angle of a geodesic in a known background. When considering the scattering of a dilaton in the maximally supersymmetric case, there is perfect agreement for the deflection angle between the classical geodesic and the amplitude calculations including the first subleading order~\cite{D'Appollonio:2010ae}. However, in the Feynman diagram approach there are inelastic processes, where a dilaton is transformed into a Ramond-Ramond (RR) field, at the same order in energy as the elastic terms contributing to the subleading eikonal (see section~\ref{ineldilRRallE}). Thus it is natural to ask what the role of these inelastic contributions is and why they should not contribute to the classical eikonal even if they grow with the energy of the scattering process. We will see in section~\ref{HElimit} that these contributions arise from the interplay of the {\em leading} eikonal and the inelastic part of the tree-level S-matrix. One should subtract these types of contributions from the expression for the amplitude in order to isolate the terms that exponentiate to provide the classical eikonal. In the example of the dilaton scattering off a stack of D$p$-branes analysed in detail in chapter \ref{chap:sugraeik}, this subtraction cancels completely the contribution of the inelastic processes and one recovers for the subleading eikonal the result found in~\cite{D'Appollonio:2010ae}. At further subleading orders this procedure may be relevant for isolating the terms that are exponentiated even in the elastic channel and thus providing a precise algorithm for extracting the classical contribution (the eikonal) from a Feynman diagram calculation may assist in analysing them.

In chapter \ref{chap:kkeik} we focus on the case when one of the spatial dimensions is compactified on a circle of radius $R$, i.e. with a background  manifold,  ${\mathbb R}^{1,D-2} \times S^1$. There is a choice of where to orient the $S^1$ with respect to the plane of scattering. Here we assume it is along one of the transverse direction, so that  the transverse momentum exchange $q = (q', n/R) $ where $q $ and $ q' $  are continuous momenta  in $ {\mathbb R}^{1,D-1}  $ and ${\mathbb R}^{1,D-2}$ respectively and $n/R$, $n  \in {\mathbb Z } $,  is the quantized momenta along the $S^1 $.  This simple  Kaluza-Klein compactification  produces a surprisingly rich theory in  ${\mathbb R}^{1,D-2}$  where we find an infinite Kaluza-Klein tower of charged states emerging from the scalar and graviton in ${\mathbb R}^{1,D-1}$. In particular from this lower dimensional viewpoint, 2 $ \rightarrow $ 2 scattering of scalar particles now involves generally massive and charged (with respect to the $U(1) $ gauge field emerging from the metric in higher dimension) Kaluza-Klein particles, involving  both elastic and inelastic processes. The latter involve massive Kaluza-Klein scalars which change their species via exchange of a massive spin-2, spin-1 and spin-0 states in the Kaluza-Klein tower. By contrast elastic scattering of scalars involves only the exchange of a massless graviton, photon and dilaton.  

In chapter \ref{chap:kkeik} we specifically focus on the elastic scattering of Kaluza-Klein scalars and analyse the eikonal in various kinematic limits, focussing mainly on the case where we compactify from $D=5$ to $D=4$. In the ultra-relativistic limit, for fixed Kaluza-Klein masses $m_1$, $m_2$  namely,  $s' \gg |t'|$, $s' \gg m_1^2 , m_2^2$, we find the eikonal phase is related to a compactified version of the Aichelburg-Sexl shock wave metric, which has appeared in the previous literature in the study of shock waves in brane world scenarios \cite{Emparan:2001ce}. In the second kinematic limit we consider elastic scattering of a massless Kaluza-Klein scalar off a heavy Kaluza-Klein scalar of mass $ m_2 $ with $m_2^2 \gg (s' - m_2^2)$. Here we find that the leading eikonal is related to the leading order contribution (in inverse powers of the impact parameter $b$) to the deflection angle in the background of a Schwarzschild black hole of mass $m_2$. However this is not the whole story. In fact at leading order in $1/b$,  the same result would hold for a charged dilatonic black hole background which is a solution of Einstein-Maxwell-dilaton (EMd) theory \cite{Horne:1992zy}. These are precisely the black hole backgrounds we should expect to be relevant in our model because the heavy Kaluza-Klein scalars are electrically charged and also couple to the dilaton field. This becomes clear when we move on to consider the subleading contributions to the eikonal which a priori contribute terms at order $1/b^2 $ to the deflection angle. These corrections  involve  the exchange of massless  $U(1)$ gauge fields and dilatons as well as gravitons between the two scalars. We find that because of the precise charge mass relation, $Q=2M$ (in appropriate units), for the Kaluza-Klein states, these subleading corrections to the eikonal vanish. In terms of the corresponding deflection angle we understand the vanishing of the subleading $1/b^2 $ terms as a consequence of the extremal $Q=2M$ limit of the deflection angle in the background of a EMd black hole.

In chapter \ref{chap:graveik} we will focus on the $2 \to 2$ scattering of massive scalar particles\cite{Kabat:1992tb,Akhoury:2013yua,Luna:2016idw,Bjerrum-Bohr:2018xdl} up to order $G_N^2$ (i.e. 2PM level) in pure gravity. Here we keep the spacetime dimension $D$ general, which serves as an infrared regulator, and also consider the subleading ${\cal O}(G_N^2)$ contributions that do not directly enter in the 2PM classical interaction but that could be relevant for the 3PM result~\cite{Bern:2019nnu, Bern:2019crd, Damour:2019lcq}. Since our analysis is $D$-dimensional we cannot apply the standard 4$D$ spinor-helicity description, but we construct the relevant parts of the amplitudes with one and two graviton exchanges by using an approach similar in spirit where tree-level amplitudes are glued together. Further discussion on applying unitarity methods in $D>4$ can be found in \cite{Bern:2011qt, Bern:2019crd, Huber:2019fea} and the references therein. This is the same technique that is also used in chapter \ref{chap:sugraeik}.

In chapter \ref{chap:graveik} we will carry out the approach discussed in section \ref{sec:eikonalamplitudebg} explicitly up to the 2PM order. The $D$-dimensional case is slightly more intricate than the 4$D$ one as we find that the contribution from the scalar box integral not only contributes to the exponentiation of the 1PM result, but also yields non-trivial subleading terms that have to be combined with the triangle contributions to obtain the full 2PM eikonal. We also see that our result smoothly interpolates between the general, the light-bending (when $m_1^2 \gg (s-m_1^2) \gg m_2^2$) and  the ultra-relativistic cases (when $s\gg m^2_1,m^2_2$); this holds not just for the classical part of the 2PM eikonal phase, which is trivially zero in the massless case, but also for the quantum part~\cite{Amati:1990xe,Amati:1992zb, Ciafaloni:2018uwe}. This feature does not seem to be realised in the recent 3PM result~\cite{Bern:2019nnu, Bern:2019crd} and it would be interesting to understand this issue better\footnote{However this interpolation seems to hold in the more recent 3PM result found in \cite{Damour:2019lcq}. This is discussed in more detail in chapter \ref{chap:conclusion}.}. Finally in chapter \ref{chap:potential} we use the results in chapter \ref{chap:graveik} to derive and write down an explicit expression for the corresponding $D$-dimensional two-body potential.

This thesis is organised as follows. In chapter \ref{chap:background} we will give a brief overview of the various tools necessary for the rest of the thesis. Chapter \ref{chap:integrals} focuses on taking the various high-energy limits for the propagator integrals we will need in order to calculate the various scattering processes. In chapters \ref{chap:sugraeik},  \ref{chap:kkeik} and \ref{chap:graveik} we will study the eikonal and classical dynamics in different scenarios including a supergravity scenario, scattering in a Kaluza-Klein background and scattering in pure Einstein gravity respectively. In chapter \ref{chap:potential} we will use the results in chapter \ref{chap:graveik} to derive the two-body Hamiltonian in Einstein gravity for arbitrary spacetime dimensions. Concluding remarks will be given in chapter \ref{chap:conclusion}.

\chapter{Background}\label{chap:background}

In this chapter we will review some of the basic concepts that we will need to develop the content in subsequent chapters. We will start with an overview of the quantum field theory produced by a massive scalar coupled to pure Einstein gravity; including a derivation of the Feynman rules and some discussion on the various problems which arise when trying to quantize gravity. This will also include a brief discussion of the setup for $N=8$ supergravity which we will need for chapter \ref{chap:sugraeik}. We will then review the eikonal approximation and apply it to the context of gravity. In doing so we will derive the full first post-Minkowskian expression for the two-body eikonal including a discussion of how to derive the two-body deflection angle. A brief overview of the post-Newtonian and post-Minkowskian expansion regimes will also be given.

\section{Conventions \& Kinematics}\label{eq:convandkin}

Throughout this thesis we will be working in $D$ spacetime dimensions unless otherwise specified and we will use the mostly-plus metric signature,
\begin{equation}
\eta_{\mu \nu} = \text{diag}(-,+,+,\ldots) \;.
\end{equation}
We will also be working in natural units where $c=\hbar=1$ throughout unless otherwise specified.

For the $2 \to 2$ scattering problems that we will be considering, the incoming momenta will be taken to be $k_1, k_2$ and the outgoing momenta will be taken as $k_3, k_4$ for the massive states with masses $m_1,m_2$ respectively, unless otherwise indicated. The Mandelstam variables we will be using are defined as,
\begin{equation}\label{eq:mandelstamdefn}
s=-(k_1+k_2)^2 ~~;~~ u=-(k_1+k_4)^2 ~~;~~ t=-(k_1+k_3)^2 \;.
\end{equation}
We also have the usual relation between Mandelstam variables,
\begin{equation}
s+t+u = 2 (m_1^2+m_2^2) \;.
\end{equation}
By choosing the center of mass frame we can parametrise the momenta by,
\begin{equation}
k_1 = (E_1,\mathbf{p})~,~k_2 = (E_2,-\mathbf{p})~,~k_3=(-E_3,\mathbf{p}')~,~k_4=(-E_4,-\mathbf{p}') \;.
\end{equation}
From this we can find relations between the Mandelstam variables, the energy, the center of mass momenta and the scattering angle $\theta$ between two momentum variables $\mathbf{p}$ and $\mathbf{p}'$,
\begin{subequations}\label{eq:mandeltoangleandp}
\begin{align}
s &= (E_1+E_2)^2 = E^2 \;, \\
t &= -2p^2 (1-\cos \theta) \;, \\
u &= -2p^2 (1+\cos \theta) \;,
\end{align}
\end{subequations}
where $p=|\mathbf{p}|$. We can also calculate,
\begin{equation}\label{eq:1geEP}
E p = \sqrt{(k_1 k_2)^2 - k_1^2 k_2^2} = \frac{1}{2} \sqrt{(s-m_1^2 -m_2^2)^2 - 4 m_1^2 m_2^2} \;,
\end{equation}
where we have denoted $k_{1 \mu} k_{2}^{\mu} = k_1 k_2$. We will define an explicit exchanged momentum between the two bodies which will generally be defined as, 
\begin{equation}
q=k_1+k_3,
\end{equation}
such that $t=-q^2$. However in chapter \ref{chap:sugraeik} we have one high-energy state scattering off of a very heavy static state and so we only have two external momenta. As such the exchanged momenta between the high-energy state and the very heavy static state in chapter \ref{chap:sugraeik} will be given by,
\begin{equation}
q = k_1 + k_2,
\end{equation}
such that $s=-q^2$. In chapters \ref{chap:integrals} and \ref{chap:sugraeik} we will usually refer to the momentum exchange vector, $q$, as $q_\perp$ when discussing scattering off of a stack of D-branes where $q_\perp$ refers to the $D-p-2$ spatial components representing the momentum transferred transverse to the stack of branes. Further details are discussed in section \ref{kinematics}.

\section{Quantum Gravity} \label{sec:quantumgrav}

In this section we will discuss the basic building blocks we will require in the rest of this chapter as well as this thesis. We will start by discussing a massive scalar theory weakly coupled to pure gravity and follow this on with a parallel discussion in the case of the bosonic sector of $N=8$ supergravity. The various modifications required for the Kaluza-Klein gravity discussion in chapter \ref{chap:kkeik} can be found in section \ref{kktheory}.

\subsection{Pure Gravity} \label{sec:puregravbackground}

We start by writing the action for a massive scalar, $\phi$, of mass $m$ weakly coupled to Einstein gravity in $D$ spacetime dimensions,
\begin{equation} \label{eq:gravmatteraction}
S= \int d^{D} x \sqrt{-g} \left( \frac{1}{2 \kappa_D^2} R + \mathcal{L}_m \right) \;,
\end{equation}
where $R$ is the Ricci scalar, $g=\det(g_{\mu \nu})$ and we have defined the gravitational coupling as,
\begin{equation}
\kappa_D = \sqrt{8 \pi G_N} \;,
\end{equation}
where $G_N$ is the $D$-dimensional Newton's gravitational constant. The Lagrangian for the matter sector is given by,
\begin{equation}
\mathcal{L}_m = - \frac{1}{2}\nabla_{\mu}\phi\, \nabla^{\mu}\phi - \frac{1}{2} m^2 \phi^2  \;. 
\end{equation} 
We can perform a quick check that these expressions can be used to derive the Einstein field equations. The variation of the action with respect to the metric, $g_{\mu \nu}$, is given by,
\begin{equation}
\frac{\delta S}{\delta g^{\mu \nu}} = \int d^{D} x  \frac{1}{2 \kappa_D^2}\left(\sqrt{-g} \frac{\delta R}{\delta g^{\mu \nu}} + R \frac{\delta \sqrt{-g}}{\delta g^{\mu \nu}} + 2 \kappa_D^2 \frac{\delta (\sqrt{-g} \mathcal{L}_m)}{\delta g^{\mu \nu}} \right) \;.
\end{equation} 
We can calculate the variation of the Ricci scalar and the determinant of the metric \cite{misner2017gravitation},
\begin{equation}
\frac{\delta R}{\delta g^{\mu \nu}} = R_{\mu \nu}~~;~~\frac{\delta \sqrt{-g}}{\delta g^{\mu \nu}}=-\frac{1}{2}\sqrt{-g} g_{\mu \nu} \;.
\end{equation}
The variation of the matter sector by definition gives us the stress-energy tensor so we have,
\begin{equation}
\frac{\delta (\sqrt{-g}\mathcal{L}_m)}{\delta g^{\mu \nu}} = - \frac{1}{2} \sqrt{-g} T_{\mu \nu} \;.
\end{equation}
Putting this all together we therefore find,
\begin{eqnarray}
\frac{\delta S}{\delta g^{\mu \nu}} &=& \int d^{D} x \sqrt{-g}  \frac{1}{2 \kappa_D^2}  \left( R_{\mu \nu} - \frac{1}{2} g_{\mu \nu} - \kappa_D^2 T_{\mu \nu} \right) = 0 \nonumber \\
& \implies & R_{\mu \nu} - \frac{1}{2} g_{\mu \nu} = 8 \pi G_N T_{\mu \nu} \;.
\end{eqnarray}
These are the well-known Einstein field equations \cite{Einstein:1916vd}, as required. 

In order to linearise this action and be able to discuss and derive the Feynman rules we will study the action in the weak-field limit \cite{hamber2008quantum, Donoghue:2017pgk} where we introduce the graviton field, $h_{\mu \nu}$, via,
\begin{equation}
g_{\mu \nu} = \eta_{\mu \nu} + 2 \kappa_D h_{\mu \nu} \;, \label{eq:bgfield}
\end{equation}
where $\eta_{\mu \nu}$ is the usual flat Minkowski metric. From now on we will use $\eta_{\mu \nu}$ to raise and lower indices. In order to derive the graviton propagator we need to substitute \eqref{eq:bgfield} in \eqref{eq:gravmatteraction} and collect all the terms up to $\mathcal{O}(h^2)$. Note that we have,
\begin{eqnarray}\label{eq:gravhsquareint1}
\frac{1}{2 \kappa_D^2} \sqrt{-g}R &=& \frac{1}{2 \kappa_D^2} \left( \partial_{\mu}\partial_{\nu} h^{\mu \nu} - \square h \right)  - \frac{1}{2} \left( \partial_{\alpha} h_{\mu \nu} \partial^{\alpha} h^{\mu \nu} - \frac{1}{2} \partial_{\alpha} h \partial^{\alpha} h \right) \nonumber \\
&& \quad -  \partial^{\alpha} \left( h_{\mu \alpha} - \frac{1}{2} \eta_{\mu \alpha} h \right) \partial_{\beta} \left( h^{\mu \beta} - \frac{1}{2} \eta^{\mu \beta} h \right)  \;,
\end{eqnarray}
where we have defined $h=h^{\mu}_{\, \mu}$. We also need to add a gauge fixing term in order to fix the gauge since $h_{\mu \nu}$ in \eqref{eq:bgfield} is not uniquely defined. We will choose the de Donder gauge (harmonic gauge) given by the condition,
\begin{equation}
\partial_{\mu} h^{\mu}{}_{\nu} - \frac{1}{2} \partial_{\nu} h = 0 \;.
\end{equation}
Note that in de Donder gauge we have,
\begin{eqnarray}\label{eq:gravhsquareint2}
\square = \frac{1}{\sqrt{-g}} \partial_{\mu} ( \sqrt{-g} g^{\mu \nu} \partial_{\nu}) &=& g^{\mu \nu} \partial_{\mu}\partial_{\nu} + \frac{1}{\sqrt{-g}} \partial_{\mu}( \sqrt{-g} g^{\mu \nu}) \partial_{\nu} \nonumber \\
& = & g^{\mu \nu} \partial_{\mu} \partial_{\nu} - \frac{2 \kappa_D}{\sqrt{-g}} \left( \partial_\mu h^{\mu \nu} - \frac{1}{2} \eta^{\mu \nu} \partial_\mu h + \mathcal{O}(h^2) \right) \partial_{\nu} \nonumber \\
& \simeq & g^{\mu \nu} \partial_{\mu} \partial_{\nu} \;.
\end{eqnarray}
The term that we need to add to the Lagrangian to fix this gauge is given by,
\begin{equation}\label{eq:gravhsquareint3}
\mathcal{L}_f = \partial_{\mu} \left( h^{\mu \nu} - \frac{1}{2} \eta^{\mu \nu} h \right) \partial^{\rho} \left( h_{\rho \nu} - \frac{1}{2} \eta_{\rho \nu} h \right)  \;.
\end{equation}
Note that we will not be giving the full details of how to implement the gauge fixing condition at the level of the partition function but as with other gauge theories one would have to use the Faddeev-Popov method, details can be found here \cite{Donoghue:1994dn, hamber2008quantum, Donoghue:2017pgk}. Putting together \eqref{eq:gravhsquareint1}, \eqref{eq:gravhsquareint2} and \eqref{eq:gravhsquareint3} with the linear in $h$ interaction part of the matter sector we have,
\begin{eqnarray} \label{eq:Laglinear}
\mathcal{L}_{\mathcal{O}(h^2)} &=& - \frac{1}{2} \left( \partial_{\alpha} h_{\mu \nu} \partial^{\alpha} h^{\mu \nu} - \frac{1}{2} \partial_{\alpha} h \partial^{\alpha} h \right) - \kappa_D h^{\mu \nu} T_{\mu \nu} \;.
\end{eqnarray}
We can integrate the term in the round brackets by parts yielding,
\begin{eqnarray}
\mathcal{L}_{\mathcal{O}(h^2)} &=& \frac{1}{2} \left(  h_{\mu \nu} \partial_{\alpha} \partial^{\alpha} h^{\mu \nu} - \frac{1}{2}  h \partial_{\alpha} \partial^{\alpha} h \right) - \kappa_D h^{\mu \nu} T_{\mu \nu} \nonumber \\
&=& \frac{1}{2} h^{\alpha \beta} \left( \frac{1}{2} \eta_{\alpha \mu} \eta_{\beta \nu} + \frac{1}{2} \eta_{\alpha \nu} \eta_{\beta \mu} - \frac{1}{2} \eta_{\alpha \beta} \eta_{\mu \nu} \right) \square h^{\mu \nu} - \kappa_D h^{\mu \nu} T_{\mu \nu} \;.
\end{eqnarray}
To find the propagator for the above linearised Lagrangian we can schematically write the partition function as,
\begin{equation}
\mathcal{Z}_h[T] \sim \int [\mathcal{D}h] \exp\left( i \int d^D x  \left[ \frac{1}{2} h^{\alpha \beta} \Delta^{-1}_{\alpha \beta \mu \nu} h^{\mu \nu} - \kappa_D h^{\mu \nu} T_{\mu \nu} \right] \right) \;,
\end{equation}
where we have the matter sector acting as a source for the graviton field and where the Green's function is given by $\Delta_{\alpha \beta \mu \nu}$. In order to find the propagator we need to invert the differential operator, $\Delta^{-1}_{\alpha \beta \mu \nu}$, that appears above. The tensor structure can be found by multiplying by an ansatz,
\begin{equation}
a \, \eta_{\alpha \mu} \eta_{\beta \nu} + b \, \eta_{\alpha \beta} \eta_{\mu \nu} \;,
\end{equation}
and requiring the product be equal to the identity. For the undetermined coefficients, $a,b$, we find,
\begin{equation}
a = 1~~;~~b = - \frac{1}{D-2} \;.
\end{equation}
The inverse of the differential operator, $\square$, can be found easily by first Fourier transforming into momentum space,
\begin{eqnarray}
&& \square G(x,y) = \delta(x-y) \nonumber \\
&\implies & - p^2 G(p) = 1 \;,
\end{eqnarray}
we therefore have,
\begin{equation}
G(x,y) = - \int \frac{d^D k}{(2\pi)^D} e^{-i k (x-y)} \frac{1}{p^2 + i \epsilon} \;.
\end{equation}
Putting all this together we then find for the graviton propagator,
\begin{equation}
i\Delta_{\alpha \beta \mu \nu}(x-y) = - \int \frac{d^D k}{(2\pi)^D} \frac{i}{k^2 + i \epsilon} e^{-i k (x-y)} \frac{1}{2} \left( \eta_{\alpha \mu} \eta_{\beta \nu} + \eta_{\alpha \nu} \eta_{\beta \mu} - \frac{2}{D-2} \eta_{\alpha \beta} \eta_{\mu \nu} \right) \;.
\end{equation}

In order to find the vertices describing the interactions between the various states in our theory we need to continue expanding the action in powers of $h$ and collect the various terms corresponding to the various interactions. We will explicitly calculate the $\phi \phi h$ vertex in momentum space but the method carries through easily (albeit cumbersomely) for the various other vertices that we will list below. 

The stress-energy tensor is given by,
\begin{equation}
T_{\mu \nu} = - \partial_{\mu} \phi \partial_{\nu} \phi + \frac{1}{2} \eta_{\mu \nu} \left( \partial_{\alpha} \phi \partial^{\alpha} \phi + m^2 \phi^2 \right) \;,
\end{equation}
and so from the $\phi \phi h$-interaction component of \eqref{eq:Laglinear} we have,
\begin{equation}
\mathcal{L}_{\phi \phi h} = \kappa_D \left( h^{\mu \nu} \partial_{\mu} \phi \partial_{\nu} \phi - \frac{1}{2} h \left( \partial_{\alpha} \phi \partial^{\alpha} \phi + m^2 \phi^2 \right) \right) \;.
\end{equation}
By Fourier transforming into momentum space we have that, $\partial_{\mu} \to -i k_{\mu}$, and doing so in the equation above as well as taking the appropriate functional derivatives we have,
\begin{eqnarray}
 \frac{\delta^3 \tilde{\mathcal{L}}_{\phi \phi h}}{\delta h_{\mu \nu} \delta \phi_1 \delta \phi_2} &=& - \kappa_D \left( k_1^{\mu} k_2^{\nu} + k_1^{\nu} k_2^{\mu} - \eta^{\mu \nu} \left( k_1 k_2 - m^2 \right)  \right) \nonumber \\
\implies  [V_{\phi_1 \phi_2 h}]^{\mu \nu} &=& - i \kappa_D \left( k_1^{\mu} k_2^{\nu} + k_1^{\nu} k_2^{\mu} - \eta^{\mu \nu} \left( k_1 k_2 - m^2 \right)  \right) \;,
\end{eqnarray}
where in the last line we have included the factor of $i$ coming from the path integral. Note that we have chosen not to write the momentum conserving delta function $(2 \pi)^{D} \delta^{D}\left( \sum_i k_i \right)$.

Using this method any other interaction between the various fields can be derived systematically. However one needs to include higher powers in $h$ in the expansion of \eqref{eq:gravmatteraction} to capture interactions such as $hhh$ or $\phi \phi h h$ and other higher order interactions. Since the method is similar to the one shown above we will simply list these expressions below.

\subsubsection{Feynman Rules}

In this subsection we will list all the Feynman rules in momentum space for a massive scalar weakly coupled to pure gravity that we will require in the rest of the thesis. The derivation of the propagator as well as the method to derive the vertices has been discussed above. We will not be including the various momentum conserving delta functions for each vertex. These results have been checked against known results in the literature \cite{DeWitt:1967yk, DeWitt:1967ub, DeWitt:1967uc}.
\begin{itemize}
\item For the graviton propagator we have,
\begin{equation}\label{eq:dedop}
\begin{tikzpicture}[baseline=(eq)]
    \begin{feynman}[inline=(eq)]
      \vertex[dot] (m) at ( 0, 0) {};
      \vertex [dot] (c) at (-1.5, 0) {};
      \vertex (eq) at (0,0) {};
      \diagram* {
      (c) -- [boson, edge label=$q$] (m),
      };
    \end{feynman}
\end{tikzpicture}
 = [G_{h}]^{\mu \nu ; \rho \sigma} = \frac{-i}{2q^2} \left ( \eta^{\mu \rho} \eta^{\nu \sigma} + \eta^{\mu \sigma} \eta^{\nu \rho} - \frac{2}{D-2} \eta^{\mu \nu} \eta^{\rho \sigma} \right ) \;.
\end{equation}
\item For the massive scalar propagator we have,
\begin{equation}
\begin{tikzpicture}[baseline=(eq)]
    \begin{feynman}[inline=(eq)]
      \vertex[dot] (m) at ( 0, 0) {};
      \vertex [dot] (c) at (-1.5, 0) {};
      \vertex (eq) at (0,0) {};
      \diagram* {
      (c) -- [edge label=$q$] (m),
      };
    \end{feynman}
\end{tikzpicture}
 = [G_{\phi}] = \frac{-i}{q^2+m^2} \;.
\end{equation}
\item For the $\phi \phi h$ vertex we have,
\begin{equation}\label{ddgV}
\begin{tikzpicture}[baseline=(eq)]
    \begin{feynman}[inline=(eq)]
      \vertex[circle,inner sep=0pt,minimum size=0pt] (m) at ( 0, 0) {\contour{white}{}};
      \vertex (a) at (0,-2) {};
      \vertex (c) at (-2, 0.2) {};
      \vertex (d) at ( 2, 0.2) {};
      \vertex (eq) at (0,-1) {};
      \diagram* {
      (a) -- [boson,edge label=${\scriptstyle(\mu \nu)}$] (m),
      (c) -- [fermion,edge label=$k_1$] (m),
      (d) -- [fermion,edge label=$k_2$, swap] (m),
      };
    \end{feynman}
\end{tikzpicture}
 =  [V_{\phi_1 \phi_2 h}]^{\mu \nu} = - i \kappa_D \left( k_1^{\mu} k_2^{\nu} + k_1^{\nu} k_2^{\mu} - \eta^{\mu \nu} \left( k_1 k_2 - m^2 \right)  \right)
\end{equation}
\item For the $\phi \phi h h$ vertex we have,
\begin{eqnarray}
&&
\begin{tikzpicture}[baseline=(eq)]
    \begin{feynman}[inline=(eq)]
      \vertex[circle,inner sep=0pt,minimum size=0pt] (m) at ( 0, 0) {\contour{white}{}};
      \vertex (a) at (-1,-2) {};
      \vertex (b) at ( 1,-2) {};
      \vertex (c) at (-2, 0) {};
      \vertex (d) at ( 2, 0) {};
      \vertex (eq) at (0,-1) {};
      \diagram* {
      (a) -- [boson,edge label=${\scriptstyle(\rho \sigma)}$] (m),
      (b) -- [boson,edge label'=${\scriptstyle(\lambda \tau)}$] (m),
      (c) -- [fermion,edge label=$k_1$] (m),
      (d) -- [fermion,edge label=$k_2$, swap] (m),
      };
    \end{feynman}
\end{tikzpicture}
= [V_{\phi_1 \phi_2 h h}]_{\rho \sigma ; \lambda \tau}  \nonumber \\
&& = (i \kappa_D^2) \left[ (k_1 k_2 - m^2) \left(\frac{1}{2} \eta_{\rho \sigma}\eta_{\lambda \tau} - \eta_{\rho \tau} \eta_{\sigma \lambda} \right) + 4 k_{1\rho } k_{2\tau} \eta_{\sigma \lambda} \right. \nonumber \\
&&  - 2 k_{1\rho} k_{2 \sigma}\eta_{\lambda \tau} \biggr]  + \ldots \;,
\end{eqnarray}
where the dots stand for the symmetrisation of the various permutations of the two massive scalars and the two gravitons.
\item For the three graviton $h h h$ vertex we find,
\begin{eqnarray}\label{3graV}
&&
\begin{tikzpicture}[baseline=(eq)]
    \begin{feynman}[inline=(eq)]
      \vertex[circle,inner sep=0pt,minimum size=0pt] (m) at ( 0, 0) {\contour{white}{}};
      \vertex (a) at (0,-2) {};
      \vertex (c) at (-2, 0.2) {};
      \vertex (d) at ( 2, 0.2) {};
      \vertex (eq) at (0,-1) {};
      \diagram* {
      (a) -- [boson,edge label=$k \text{,} \, {\scriptstyle(\mu \nu)}$] (m),
      (c) -- [boson,edge label'=$p \text{,} \, {\scriptstyle(\lambda \rho)}$] (m),
      (d) -- [boson,edge label=$q \text{,} \, {\scriptstyle(\tau \sigma)}$] (m),
      };
    \end{feynman}
\end{tikzpicture}
 = [V_{h_{1k} h_{2p} h_{3q}}]_{\mu \nu; \lambda \rho; \tau \sigma}\nonumber \\
&& = - 2 i \kappa_D \Biggl( - \frac{1}{2} p_{\mu} q_{\nu} \eta_{\lambda \rho} 
\eta_{\tau \sigma} + 2 p_{\mu} q_{\sigma } \eta_{\lambda \nu} \eta_{\tau \rho} - \eta_{\rho \sigma} p_{\mu} q_{\lambda} \eta_{\tau \nu} 
\nonumber \\
&& - \frac{1}{2} \eta_{\mu \nu} p_{\tau} \eta_{\lambda \rho} q_{\sigma} + \frac{1}{4} \eta_{\mu \nu} \eta_{\lambda \rho} \eta_{\tau \sigma } p\cdot q  - \eta_{\rho \sigma}
p_{\lambda} \eta_{\mu \tau} q_{\nu} + \frac{1}{2} \eta_{\rho \sigma} p_{\mu} 
\eta_{\lambda \tau} q_{\nu}   \nonumber  \\ 
&& - \eta_{\mu \rho} \eta_{\nu \sigma} p_{\lambda} q_{\tau} + \frac{1}{2} \eta_{\mu \nu} \eta_{\rho \sigma} p_{\lambda} q_{\tau} + \eta_{\mu \rho} \eta_{\nu \sigma} \eta_{\lambda \tau} p \cdot q  - \frac{1}{4} \eta_{\mu \nu} \eta_{\rho \sigma}  \eta_{\lambda \tau} p \cdot q  \nonumber \\
&& - \eta_{\mu \sigma} \eta_{\nu \lambda} \eta_{\rho \tau} p \cdot q + \eta_{\mu \sigma} p_{\lambda} \eta_{\nu \tau} q_{\rho}  \Biggr)   + \ldots \;, 
\end{eqnarray}
where the dots stand for the sum over the various permutations of the gravitons.
\end{itemize}

\subsection{Supergravity} \label{sec:sugrabackground}

In order to complete the materials needed for chapter \ref{chap:sugraeik} we need to consider gravity extended to include the dilaton and RR gauge fields present in $N=8$ supergravity. As we will consider scattering off of a stack of D$p$-branes we will also need the boundary action sourced by the branes. We will be denoting the dilaton as above with $\phi$ and use the relevant equations for the massive scalar in the previous section with $m=0$. The RR gauge fields will be denoted by $C^{(n-1)}$, where $n=p+2$ and $p$ is the dimension of the D$p$-brane world-volume and so we can also define the field strength $F_n = \textrm{d}C^{(n-1)}$.

The bulk action we need to consider is given by,
\begin{eqnarray}\label{eq:bulkaction}
S= \int d^{D} x \sqrt{-g} \left[ \frac{1}{2 \kappa_D^2} R- \frac{1}{2} 
\partial_{\mu}\phi\, \partial^{\mu}\phi -
\frac{1}{2n!}e^{-a(D) \sqrt{2}\kappa_D \phi }F^2_{n} \right] + S_{\text{D-Brane}} \;,
\end{eqnarray}
where the expression for $a(D)$ can be found in \cite{Duff:1994an} and is given by\footnote{In 10D type II supergravity, $a(D=10)=\frac{p-3}{2}$.},
\begin{equation}\label{eq:aDforSUGRA}
a^2(D)=4-\frac{2(p+1)(D-p-3)}{D-2} \;.
\end{equation}
Note that by convention we will use the positive root of the expression above. We will use the following boundary action which is being sourced by the D$p$-branes \cite{Garousi:1998fg},
\begin{eqnarray}
S_{\text{D-Brane}} = \int d^D x \,\,\delta^{D-p-1} (x) \left( J_h h_{\alpha}^{\,\,\, \alpha} (x)+
J_{\phi} \phi (x) + \mu_p C_{01 \dots p} (x) \right) \;,
\label{branecoup}
\end{eqnarray}
where the quantities $J_h, J_{\phi}, \mu_p$ are the couplings of the graviton, dilaton and RR to the D$p$-brane respectively. These are given by,
\begin{eqnarray}
J_h = - T_p~~~;~~~~
\mu_p = \sqrt{2} T_p  \,\,;\,\,\,\, J_\phi= - \frac{a(D)}{\sqrt{2}}T_p \;,
\label{JJmu}
\end{eqnarray}
where $T_p$ is the D$p$-brane tension. As in the previous section discussing pure gravity we will be expanding the metric around the flat metric using \eqref{eq:bgfield}.

\subsubsection{Feynman Rules} 

Following the techniques used in section \ref{sec:puregravbackground} we can derive the additional Feynman rules needed to calculate the various quantities in chapter \ref{chap:sugraeik}. These are listed below and have been checked with various known results in the literature \cite{Garousi:2012yr,Bakhtiarizadeh:2013zia}. Note that the Feynman rules for the interactions between the dilaton and graviton are equivalent to the Feynman rules for the massive scalar in \ref{sec:puregravbackground} with $m=0$. As before we are not writing momentum conserving delta functions at vertices.

\begin{itemize}
\item The propagator for the RR fields is given by,
\begin{equation}\label{eq:dedop}
\begin{tikzpicture}[baseline=(eq)]
    \begin{feynman}[inline=(eq)]
      \vertex[dot] (m) at ( 0, 0) {};
      \vertex [dot] (c) at (-1.5, 0) {};
      \vertex (eq) at (0,0) {};
      \diagram* {
      (c) -- [scalar, edge label=$q$] (m),
      };
    \end{feynman}
\end{tikzpicture}
 = [G_{C^{(n-1)}}]^{\mu_1\ldots\mu_{n-1}}_{\nu_1\ldots\nu_{n-1}} = \frac{-i}{q^2}\,(n-1)! \delta^{\mu_1}_{[\nu_1} \ldots \delta^{\mu_{n-1}}_{\nu_{n-1}]} \;.
\end{equation}
\item The $\phi \; C^{(n-1)} C^{(n-1)}$ vertex is given by,
\begin{eqnarray}
&&
\begin{tikzpicture}[baseline=(eq)]
    \begin{feynman}[inline=(eq)]
      \vertex[circle,inner sep=0pt,minimum size=0pt] (m) at ( 0, 0) {\contour{white}{}};
      \vertex (a) at (0,-2) {};
      \vertex (c) at (-2, 0.2) {};
      \vertex (d) at ( 2, 0.2) {};
      \vertex (eq) at (0,-1) {};
      \diagram* {
      (a) -- [fermion] (m),
      (c) -- [charged scalar,edge label=$k_1$] (m),
      (d) -- [charged scalar,edge label=$k_2$, swap] (m),
      };
    \end{feynman}
\end{tikzpicture}
 =  [V_{\phi F^{(n)}_{1} C_2^{(n-1)}}] \nonumber \\
&=& - \frac{i a(D) \sqrt{2} \kappa_D}{(n-1)!} (F_{1\; \mu_1 \mu_2 \ldots \mu_n} k^{\mu_1}_{2} C_2^{\mu_2 \ldots \mu_n}) \;,
\end{eqnarray}
where (1) and (2) are labels of the two RR fields and $F_{i\; \mu_1 \ldots \mu_n}$ is the field strength associated with the $(p+1)$-form gauge field $C_{i \; \mu_2 \ldots \mu_n}$, $F_{i\; \mu_1 \ldots \mu_n}=(\text{d}C_{i})_{\mu_1 \ldots \mu_n}$.
\item The $h \; C^{(n-1)} C^{(n-1)}$ vertex is given by,
\begin{equation}
\begin{tikzpicture}[baseline=(eq)]
    \begin{feynman}[inline=(eq)]
      \vertex[circle,inner sep=0pt,minimum size=0pt] (m) at ( 0, 0) {\contour{white}{}};
      \vertex (a) at (0,-2) {};
      \vertex (c) at (-2, 0.2) {};
      \vertex (d) at ( 2, 0.2) {};
      \vertex (eq) at (0,-1) {};
      \diagram* {
      (a) -- [boson,edge label=${\scriptstyle(\mu \nu)}$] (m),
      (c) -- [charged scalar,edge label=$k_1$] (m),
      (d) -- [charged scalar,edge label=$k_2$, swap] (m),
      };
    \end{feynman}
\end{tikzpicture}
 =  [V_{F^{(n)}_{1} F^{(n)}_{2} h}] = \frac{i \kappa_D}{n!} (2n F_{12}^{\mu \nu} - \eta^{\mu \nu} F_{12}) h^{\mu \nu} \;,
\end{equation}
where $F_{12}^{\mu \nu} = F^{\mu}_{1 \; \mu_2 \ldots \mu_n} F_{2}^{\nu \mu_2 \ldots \mu_n}$ and we also have $F_{12} = F_{12}^{\mu \nu} \eta_{\mu \nu}$.
\item The $\phi \phi \; C^{(n-1)} C^{(n-1)}$ vertex is given by,
\begin{eqnarray}
&&
\begin{tikzpicture}[baseline=(eq)]
    \begin{feynman}[inline=(eq)]
      \vertex[circle,inner sep=0pt,minimum size=0pt] (m) at ( 0, 0) {\contour{white}{}};
      \vertex (a) at (-1,-2) {};
      \vertex (b) at ( 1,-2) {};
      \vertex (c) at (-2, 0) {};
      \vertex (d) at ( 2, 0) {};
      \vertex (eq) at (0,-1) {};
      \diagram* {
      (a) -- [charged scalar,edge label=$k_1$] (m),
      (b) -- [charged scalar,edge label=$k_2$, swap] (m),
      (c) -- [fermion] (m),
      (d) -- [fermion] (m),
      };
    \end{feynman}
\end{tikzpicture}
= [V_{\phi \phi' F^{(n)}_{1} F^{(n)}_{2}}] = - \frac{2 i \kappa_D^2}{n!} 
a^2(D) 
F_{12} \;.
\end{eqnarray}
\item From \eqref{branecoup} we read the D$p$-brane graviton coupling,
\begin{equation}
\begin{tikzpicture}[baseline=(eq)]
    \begin{feynman}[inline=(eq)]
      \vertex[circle,inner sep=0pt,minimum size=0pt] (m) at ( 0, 0) {\contour{white}{}};
      \draw[fill=light-gray] (0,-1.5) ellipse (0.5cm and 0.25cm);
      \vertex (a) at (0,-1.5) {};
      \diagram* {
      (a) -- [boson,edge label=${\scriptstyle(\mu \nu)}$] (m),
      };
    \end{feynman}
\end{tikzpicture}
 =  [B_{h}] = -i T_p \eta^{\mu \nu}_{\parallel} h_{\mu \nu} \int \frac{d^{\perp}k}{(2 \pi)^{\perp}} \;, \label{branecouplinggrav}
\end{equation}
where $\parallel$ denotes the $p+1$ directions along the D$p$-brane and $\perp$ denotes the $D-p-1$ directions transverse to the D$p$-brane.
\item The boundary coupling with the dilaton is given by
\begin{equation}
\begin{tikzpicture}[baseline=(eq)]
    \begin{feynman}[inline=(eq)]
      \vertex[circle,inner sep=0pt,minimum size=0pt] (m) at ( 0, 0) {\contour{white}{}};
      \draw[fill=light-gray] (0,-1.5) ellipse (0.5cm and 0.25cm);
      \vertex (a) at (0,-1.5) {};
      \diagram* {
      (a) -- (m),
      };
    \end{feynman}
\end{tikzpicture}
 =  [B_{\phi}] = - i T_p \frac{a(D)}{\sqrt{2}} \int \frac{d^{\perp}k}{(2 \pi)^{\perp}} \;. 
\end{equation}
\item The boundary coupling with the RR gauge potential is given by,
\begin{equation}
\begin{tikzpicture}[baseline=(eq)]
    \begin{feynman}[inline=(eq)]
      \vertex[circle,inner sep=0pt,minimum size=0pt] (m) at ( 0, 0) {\contour{white}{}};
      \draw[fill=light-gray] (0,-1.5) ellipse (0.5cm and 0.25cm);
      \vertex (a) at (0,-1.5) {};
      \diagram* {
      (a) -- [scalar] (m),
      };
    \end{feynman}
\end{tikzpicture}
 =  [B_{C^{(n-1)}}] = i \mu_p C_{01\ldots p} \int \frac{d^{\perp}k}{(2 \pi)^{\perp}} \;. 
\end{equation}
\end{itemize}

\subsection{Sourcing Schwarzschild Black Holes}

In this section we will show schematically that the massive scalar states we have been discussing are appropriate states with which to model Schwarzschild black holes. This will be particularly important in chapter \ref{chap:graveik} when we consider the scattering of massive scalars to describe binary black holes in pure gravity. It will also be relevant in chapter \ref{chap:sugraeik}. We will be working in $D=4$ here for simplicity but the general $D$ analogue is discussed in section \ref{Einstein}.
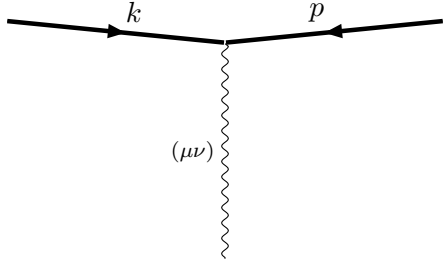
\begin{figure}[h]
\centering
\begin{tikzpicture}[scale=1.5]
    \begin{feynman}
      \vertex[circle,inner sep=0pt,minimum size=0pt] (m) at ( 0, 0) {\contour{white}{}};
      \vertex (a) at (0,-2) {};
      \vertex (c) at (-2, 0.2) {};
      \vertex (d) at ( 2, 0.2) {};
      
      \diagram* {
      (a) -- [boson,edge label=${\scriptstyle(\mu \nu)}$] (m),
      (c) -- [fermion,edge label=$k$] (m),
      (d) -- [fermion,edge label=$p$, swap] (m),
      };
      \draw [line width=0.6mm] (c) -- (m);
      \draw [line width=0.6mm] (d) -- (m);
    \end{feynman}
\end{tikzpicture}
\caption{This figure illustrates the sourcing of a Schwarzschild black hole by a massive scalar state. The thick line represents a static Schwarzschild black hole of mass $M$ and the wavy line represents a graviton.} \label{fig:oneptforSchw}
\end{figure}

In order to show that we can source a Schwarzschild black hole at large distances by using the field theory described in section \ref{sec:puregravbackground} we want to calculate the effective one-point function as illustrated in figure \ref{fig:oneptforSchw}. This will give us the first contribution, in the large distance limit, to the Schwarzschild solution. Further contributions can be calculated by going to higher effective loop orders, further details can be found in section \ref{lntoeikonal}. The one-point function can be readily calculated using the Feynman rules derived above and in $D=4$ we find,
\begin{equation}
\langle h^{\mu \nu} \rangle = - \frac{\kappa}{q^2} \left( k^{\mu} p^{\nu} + k^{\nu}  p^{\mu} - M^2 \eta^{\mu \nu} \right) \;,
\end{equation}
where $q=k+p$. Since we want an effective description of a massive static heavy object with no spin we can write the external momenta as,
\begin{equation}
k^{\mu} = (E,\mathbf{0}) = (M,\mathbf{0})~~~;~~~p^{\mu} = (-E,-\mathbf{q}) \sim (-M,\mathbf{0}) \;,
\end{equation}
where for $M \gg |\mathbf{q}|$ we have taken $E \sim M$. Implementing this we find,
\begin{eqnarray}
\langle h^{00} \rangle &=& - \frac{M \kappa}{2q^2} \eta^{00} \\
\langle h^{ij} \rangle &=& \frac{M \kappa}{2q^2} \eta^{i j}
\end{eqnarray}
where we have also normalised by a factor of $1/\sqrt{2E} \sim 1/\sqrt{2M}$ per scalar leg and we have split the metric into its time-like and space-like components. Note that $\eta^{00}=-1$ represents the time-like direction and $\eta^{i j}=\text{diag}(+1,+1,+1)$ represents the space-like directions.

To be able to compare this with the appropriate expansion of the Schwarzschild metric in harmonic gauge we need to convert the above from momentum into position space. Using \eqref{eq:impoformu} we can schematically write,
\begin{eqnarray}
\langle 2 \kappa \tilde{h}^{\mu \nu} \rangle & \sim & - \frac{2 G_N M}{r} \left(\eta^{00} - \eta^{i j} \right) \label{eq:d4schwsource}\\
\implies  ds^2 & \supset & \frac{2 G_N M}{r} \left(dt^2 + d\mathbf{x}^2 \right) \;,
\end{eqnarray}
where we have introduced the factor of $2\kappa$ as per the definition of the graviton metric in \eqref{eq:bgfield}. Comparing with the first terms in the $R_s/r$ expansion of the Schwarzschild metric in harmonic gauge found in \cite{Aguirregabiria:2001vk} we find agreement. Note that \eqref{eq:d4schwsource} also agrees with \eqref{gra13} for $D=4$ and $p=0$.

\subsection{High-Energy Issues in Quantum Gravity} \label{sec:QGUVissues}

Although we have been discussing gravity as a quantum theory in order to be able to derive the various Feynman rules in section \ref{sec:quantumgrav}, we know that such a construction of quantum gravity suffers from UV divergences \cite{tHooft:1974toh,Deser:1974cz,Deser:1974xq}. We also know that it is non-renormalisable \cite{Deser:1974cy,Goroff:1985th,vandeVen:1991gw} and due to this we find that we need to include an entire series of higher order curvature terms that are allowed by the diffeomorphism invariance and we are not allowed, a-priori, to just use the Einstein-Hilbert action \cite{Donoghue:1995cz}. 

We can construct an effective field theory (EFT) with the allowed higher order curvature terms and appropriate couplings as,
\begin{equation}\label{eq:qgeftaction}
S_{\text{EFT}} = \frac{1}{2 \kappa_D^2} \int d^D x \sqrt{-g} \left( R + c_1 R^2 + c_2 R_{\mu \nu}R^{\mu \nu} + c_3 R_{\mu \nu \rho \sigma}R^{\mu \nu \rho \sigma} + \ldots \right) \;,
\end{equation}
where we have not included the matter sector as we will be restricting discussion to the gravity sector here for simplicity. From the first term in the equation above we can deduce the mass dimension of the gravitational coupling,
\begin{equation}
\kappa_D^2 \sim G_N = [M]^{2-D} \;,
\end{equation}
and so in terms of the Plank scale we have,
\begin{equation}
\frac{1}{\kappa_D^2} \sim \frac{1}{G_N} = (M_p)^{D-2} \;,
\end{equation}
where $M_p$ is the Plank mass. This tells us that for $D > 2$ we will have a non-renormalisable theory and we notice that for energies $E > M_p$ we will get results which will not be useful, for example non-unitary results for scattering amplitudes. 

In general the higher order curvature terms would begin to contribute at some secondary mass scale which we can call $M_c$ with $M_c < M_p$ and so we would have,
\begin{equation}
c_{1,2,3} \sim \frac{1}{M_c^2} \;.
\end{equation}
For example in string theory this scale would be $M_c \sim 1/l_s$ where $l_s$ is the length of the string and so we would have $c_{1,2,3} \sim l_s^2 = \alpha'$, these curvature corrections in string theory are usually referred to as stringy corrections \cite{green1988superstring,Polchinski:1998rq,Polchinski:1998rr,Tong:2009np}. In principle we would then proceed as we did in section \ref{sec:puregravbackground} and calculate all the Feynman rules with the action defined in \eqref{eq:qgeftaction}.

Since we are ultimately interested in the classical dynamics of binary systems the quantum corrections to various quantities, such as scattering amplitudes, will not be particularly important and will be dropped. However for consistency with the discussion above we would, in principle, need to include the contributions from the curvature corrections. We can do a quick schematic analysis following \cite{Donoghue:1994dn,Donoghue:1995cz} to show that this is unnecessary. For simplicity we will set $c_n=0$ with $n>2$ in \eqref{eq:qgeftaction}. The corresponding equations of motion is schematically given by,
\begin{equation}
\square h + c_1 \square \square h = 8 \pi G_N T \;.
\end{equation}
The Green's function for this wave equation has the form,
\begin{eqnarray}
G(x) & \sim & \int \frac{d^D k}{(2\pi)^D} \frac{e^{i k x}}{k^2 + c_1 k^4} \nonumber \\
&=& \int \frac{d^D k}{(2\pi)^D} e^{i k x} \left( \frac{1}{k^2} - c_1 + \ldots \right) \;,
\end{eqnarray}
where in the last step we have used $c_1 \ll 1$. Calculating the static potential between two masses using this we find,
\begin{equation}
V(r) \sim \int d^{D-1} \mathbf{k} \; e^{i \mathbf{k} \mathbf{r}} \left( \frac{1}{\mathbf{k}^2} - c_1 + \ldots \right) \sim \frac{1}{r^{D-3}} - c_1 \delta^{D-1}(r) + \ldots \;.
\end{equation}
From this we can see that the classical effects of the higher order curvature corrections will be local.

As we will study in more detail in section \ref{sec:eikonalapproxbg} we are interested in the high-energy classical dynamics when trying to describe heavy objects such as black holes. This can be intuited from the fact that heavy objects are inherently high-energy objects since they are very massive. Classical physics starts to become relevant when the particle separation is larger then the de-Broglie wavelength. The separation for a scattering problem can be taken to be the impact parameter and so we have \cite{Bern:2019crd}, 
\begin{equation}\label{eq:brogliereltob}
b \gg \lambda \sim \frac{1}{p} \;,
\end{equation}
where $b$ is the impact parameter and $p$ is some momentum scale of the scattering problem. We therefore realise that the classical dynamics is described by physics at large distance scales. As such we can ignore the effects of curvature corrections as they will be suppressed by the fact that we are interested in large distances. This means that the framework built up in section \ref{sec:puregravbackground} can be used without concerning ourselves with the UV complete description of the full quantum gravity theory.

\section{The Post-Newtonian and Post-Minkowskian Expansion}\label{sec:pmexpansionbg}

There are two well-known perturbative regimes in which general relativity is studied; the post-Newtonian (PN) regime \cite{droste1917field, Einstein:1938yz, Bini:2017wfr, Foffa:2019hrb} and the post-Minkowskian (PM) regime \cite{bertotti1956gravitational, Kerr1959, BERTOTTI1960169, Portilla:1979xx, Westpfahl:1979gu, Portilla:1980uz, Bel:1981be, Ledvinka:2008tk, Damour:2016gwp,Damour:2017zjx,Bjerrum-Bohr:2018xdl}. In this subsection we will describe these two perturbative regimes and specifically discuss what the perturbation parameters are.

In the post-Newtonian regime we expand the two-body interaction potential or other quantities describing the dynamics of the two bodies for small velocities and hence in powers of,
\begin{equation}
\left( \frac{v}{c} \right)^2 \sim \frac{G_N m}{|\mathbf{r}| c^2} \ll 1 \;,
\end{equation}
where the second expression is due to the virial theorem. In the expression above, $v$ is the relative velocity between the two bodies, $m$ is the total mass and $|\mathbf{r}|$ is their separation. This regime has been well studied \cite{Damour:1999cr, Bini:2017wfr, Levi:2018nxp, Foffa:2019hrb} but will not generally be used in this thesis except to compare results calculated here with known results in the literature.

\begin{figure}[h]
  \centering
  \includegraphics[scale=1.5]{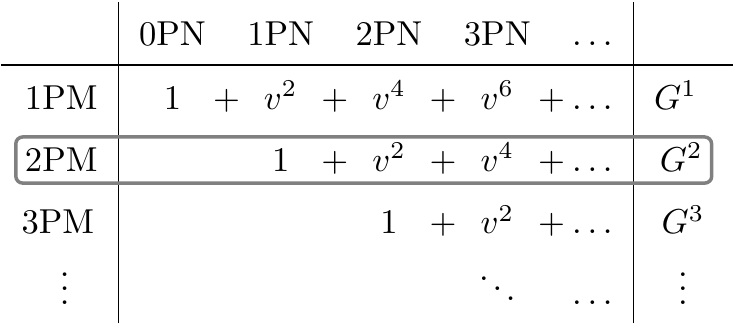}
  \caption{A figure illustrating the connection between the post-Newtonian and post-Minkowskian regimes.}
  \label{fig:PnPmConnection}
\end{figure}

In the post-Minkowskian regime the two-body interaction potential is instead expanded in terms of Newton's gravitational constant $G_N$ but includes all orders in velocities. Since the PM regime includes all orders in velocities it is effectively the relativistic analogue to the PN non-relativistic (i.e. small velocities) expansion. Since the PM result is valid to all orders in velocity it naturally includes the relevant PN result at each matching order in $G_N$, this is illustrated in figure \ref{fig:PnPmConnection}.

As we will see in more detail in section \ref{sec:eikonalamplitudebg} and in chapters \ref{chap:graveik} and \ref{chap:potential} we can relate the classical pieces of quantum scattering amplitudes directly to the PM expansion and it is therefore a particularly tractable and interesting regime to study. In this thesis we will mainly focus on PM results.

In order to compare our PM results with known PN results in the literature we need to define a method to get from PM results to PN results. From equations (28) and (32) in \cite{Bini:2017wfr} we can write,
\begin{equation}
v_{\infty}^2 = 2 \frac{E - (m_1+m_2)}{\mu} \implies \gamma = - \frac{k_1 k_2}{m_1 m_2} = (1+v_{\infty}^2)^{1/2}
\end{equation}
where $v_{\infty}$ is the relative velocity between the two bodies at infinity for unbounded motion as discussed in \cite{Bini:2017wfr}. Notice that this velocity variable is different to the standard definition given by $\gamma = (1-v^2)^{-1/2}$, although they coincide at lowest order in the non-relativistic limit. This can be related directly to the energy via,
\begin{equation}\label{eq:EinPNexp}
\gamma = \frac{E^2-m_1^2-m_2^2}{2m_1m_2} \implies E^2 = 2m_1m_2 \sqrt{1+v_{\infty}^2} + m_1^2 + m_2^2 \;.
\end{equation}
In subsequent sections we will sometimes want to perform a velocity expansion (PN expansion) of the various PM results. As we will see in section \ref{sec:leadingeikbg}, our PM results are usually written in terms of the total angular momentum $J$. However in order to perform the PN expansion it is useful to write our results in terms of a rescaled angular momentum variable defined as,
\begin{equation}
j = \frac{J}{G_N m_1 m_2} \;.
\label{PN6}
\end{equation}
We can do dimensional analysis on this quantity and find,
\begin{equation}
j = \frac{[L^2 T^{-1} M]}{[L^3 M^{-1} T^{-2} M^2]} = [L^{-1} T] \;,
\label{PN7}
\end{equation}
and so we note that $j$ has the dimensions of inverse velocity and so to make this dimensionless (reinstating $c$ and noting that quantities such as the angle should be dimensionless) we must treat $j \sim jc$. Similarly in order to make $v_{\infty}$ dimensionless we must treat it as $v_{\infty} \sim v_{\infty} / c$. The order of the PN expansion can then be thought of as an expansion in powers of $1/c^2$ \cite{Bini:2017wfr}. In general we can write,
\begin{equation}\label{eq:idforPNterms}
\frac{v_{\infty}^{m}}{j^{n}} \to \frac{v_{\infty}^{m}}{j^{n} c^{(m+n)}} \implies \frac{1}{2}(m+n)\text{-PN} \;. 
\end{equation}
This expression is used in section \ref{sec:leadeikPN} and chapter \ref{chap:potential} to identify the PN order of various terms.

\section{The Eikonal Approximation}\label{sec:eikonalapproxbg}

Following on from the arguments presented in section \ref{sec:QGUVissues} we know that the non-renormalisability of quantum gravity as described in section \ref{sec:quantumgrav} does not prevent us from using the theory to describe the high-energy classical dynamics of a binary system. In this section we will describe how to extract the classical dynamics, in the form of a two-body deflection angle, from scattering amplitudes.

As we have previously discussed, we are interested in the high-energy dynamics because ultimately we want to describe the dynamics of heavy massive objects such as black holes. For a $2 \to 2$ scattering problem with incoming momenta $k_1,k_2$ and outgoing momenta $k_3,k_4$ we can use the Mandelstam variable $s$, defined in \eqref{eq:mandelstamdefn}, as a proxy for the energy and $t$ as a proxy for the exchanged momenta, notice that $s=E^2$ and $t=-q^2$ where $q$ is the momentum exchanged between the two scattering objects. As we have stated in section \ref{sec:QGUVissues} classical dynamics is associated with large distances where, $b \gg G_N M^*$. Since the momentum exchanged is the conjugate variable to the impact parameter for a scattering problem we can similarly state that,
\begin{equation}
b \sim \frac{1}{q} \gg G_N M^* \;,
\end{equation}
which implies that the exchanged momenta are small. Since we are interested in the high-energy limit we are also want to take $E$ to be large. Putting all this together we see that what we are describing is a scattering process in the Regge limit,
\begin{equation}\label{eq:reggelimitdefn}
s \to \infty~~\text{with}~~\frac{t}{s}~~\text{small}\;.
\end{equation}
It is worth mentioning a few more scaling observations that we can draw. Since we are taking $b \gg G_N M^*$ as well as a high-energy limit we notice that the center of mass momentum, $p$, is therefore also large. From this we can therefore also write,
\begin{equation}
J \sim p b \gg 1\;.
\end{equation}
We can also notice from \eqref{eq:reggelimitdefn} that we are taking $s \to \infty$ with $t=-q^2 \to 0$ to keep the ratio small but fixed and this implies that,
\begin{equation}
p, m_1, m_2 \gg q \;.
\end{equation} 

A well known high-energy approximation in potential scattering called the eikonal approximation \cite{Schiff:1956zz,Sugar:1969rn,Wallace:1973iu} can be used in this context to describe the dynamics that we are interested in. It has been shown that this approximation can be used to describe the high-energy (Regge) limit of scattering amplitudes in a relativistic field theory context \cite{Blankenbecler:1962ez,Levy:1969cr,Abarbanel:1969ek,Wallace:1977ae} and was used in \cite{Kabat:1992tb} to show equivalence with high-energy semi-classical calculations in quantum gravity \cite{tHooft:1987vrq}. This will be the framework we will use to derive and discuss the dynamics of high-energy $2 \to 2$ scattering. Furthermore the calculation of the eikonal has been shown to be related to the sum of an infinite class of Feynman diagrams \cite{Cheng:1969eh,Levy:1969cr}, in the case of gravity dominated by graviton exchanges \cite{tHooft:1987vrq,Muzinich:1987in,Amati:1987wq}. We will use these facts extensively in the discussion below.

\subsection{The Eikonal Phase \& Amplitude}\label{sec:eikonalamplitudebg}

In general any $2 \to 2$ scattering amplitude with incoming momenta $k_1, k_2$ and outgoing momenta $k_3,k_4$ can be written as,
\begin{equation}
\langle k_1 k_2 | T | k_3 k_4 \rangle = (2 \pi)^{D} \delta^{D}( k_1+k_2+k_3+k_4 )  \mathcal{A}( k_1 k_2 \to k_3 k_4 ) \;,
\end{equation}
where $T$ represents the non-trivial component of the scattering S-matrix, i.e. $S=1+iT$, and the $|k_i k_j \rangle$ are two-particle states. To simplify the notation and using the redundancy offered by the conservation of momentum we will drop the momentum conserving $\delta$ functions and write the amplitudes as,
\begin{equation}
\mathcal{A}( k_1 k_2 \to k_3 k_4 ) = \mathcal{A}(s,m_i,q) \;,
\end{equation}
where the $m_i$ represent the masses of the two scattering particles. We can further represent scattering amplitudes of this sort, with $n$ graviton exchanges between the two scattering particles, by dressing the RHS above with a subscript $n$.

As is standard in the discussion of the eikonal approximation, we introduce an auxiliary $(D-2)$-dimensional vector $\mathbf{q}$ such that $\mathbf{q}^2=-t$ and then take the Fourier transform to rewrite the result in terms of the conjugate variable $\mathbf{b}$ (the impact parameter). We can now define the corresponding amplitudes in impact parameter space which are defined via,
\begin{equation}
\tilde{\mathcal{A}}_{n}(s,m_i,\mathbf{b}) = \frac{1}{4 E p} \int \frac{{\textrm{d}}^{D-2}\mathbf{q}}{(2\pi)^{D-2}}e^{i\mathbf{q} \mathbf{b}}  \mathcal{A}_{n}(s,m_i,q) \label{eq:eikips}\;,
\end{equation}
where $E=E_1+E_2$ is the total energy, $p=|\mathbf{p}|$ is the absolute value of the space-like momentum in the center of mass frame of the two scattering particles and we have written $q=|\mathbf{q}|$. The normalisation is a non-relativistic normalisation factor that we have introduced to make the definition of the eikonal phase below more straightforward. We recall that the $Ep$ factor can be written in terms of the Mandelstam variables or incoming momenta as per \eqref{eq:1geEP}.

We can generally write the gravitational S-matrix in impact parameter space as \cite{Giddings:2009gj,Giddings:2011xs},
\begin{equation}\label{eq:smatrixasampsum}
S(s,m_i,\mathbf{b}) = 1 + i \sum_{n=1}^{\infty} \tilde{\mathcal{A}}_{n}(s,m_i,\mathbf{b}) \;,
\end{equation} 
where $\tilde{\mathcal{A}}_{n}(s,m_i,\mathbf{b})$ is the full amplitude with $n$ graviton exchanges in impact parameter space including the appropriate normalisation as defined in \eqref{eq:eikips}. 

In general, amplitudes in gravity describing graviton exchange between two bodies are divergent in the Regge limit. This will be made more explicit in the following subsections. However since we need to preserve the unitarity of the S-matrix, we notice that this can only be done if the resummation of these divergent diagrams resums into a phase. It has been extensively shown that this resummation and exponentiation does indeed happen \cite{Cheng:1969eh, Levy:1969cr, Abarbanel:1969ek, Kabat:1992tb, Akhoury:2013yua, Ciafaloni:2014esa, Bjerrum-Bohr:2018xdl} and is made particularly explicit in impact parameter space. In order to preserve the unitarity we can express the gravitational S-matrix in terms of phases, 
\begin{eqnarray}
S(s,m_i,\mathbf{b})  &=&  \left(1 + i T(s,m_i,\mathbf{b}) \right) \exp \left[{i (\delta^{(1)}(s,m_i,\mathbf{b}) +  \delta^{(2)}(s,m_i,\mathbf{b}) + \ldots)}\right] \;, \nonumber \\ \label{eq:eiksmatrix} 
\end{eqnarray}
where $\delta^{(1)}(s,m_i,\mathbf{b})$ and $\delta^{(2)}(s,m_i,\mathbf{b})$ are the leading eikonal and subleading eikonal respectively and will be defined further below. The parameter used to define the expansion in~\eqref{eq:eiksmatrix} is $(R_s/\mathbf{b})^{D-3}$ (see~\eqref{S14} for the numerical factors in the definition of $R_s$). The symbol $T(s,m_i,\mathbf{b})$ corresponds to all the non-divergent (in energy or mass) contributions to the amplitudes with any number of graviton exchanges. We have implicitly assumed that the eikonals behave as phases instead of operators since we are dealing with a purely elastic scenario in this chapter. For a more general form of the equation above see \eqref{fullInelEik}.

By using \eqref{eq:smatrixasampsum} and \eqref{eq:eiksmatrix} as well as observations from amplitude calculations \cite{Giddings:2009gj,Akhoury:2013yua} we note that we can write the sum of leading divergent contributions to the amplitudes with $n$ graviton exchanges as,
\begin{eqnarray}\label{eq:sumofleadingcont}
i \sum_{n=1}^{\infty} \tilde{\mathcal{A}}^{(1)}_{n}(s,m_i,\mathbf{b}) &=& i \tilde{\mathcal{A}}^{(1)}_{1}(s,m_i,\mathbf{b}) + i\tilde{\mathcal{A}}^{(1)}_{2}(s,m_i,\mathbf{b}) + \ldots \nonumber \\
&=& i \tilde{\mathcal{A}}^{(1)}_{1}(s,m_i,\mathbf{b}) + \frac{1}{2} \left(i\tilde{\mathcal{A}}^{(1)}_{1}(s,m_i,\mathbf{b})\right)^2 + \ldots \nonumber \\
&=& e^{i \delta^{(1)}(s,m_i,\mathbf{b})} - 1  \;,
\end{eqnarray}
where $\delta^{(1)}(s,m_i,\mathbf{b}) = \tilde{\mathcal{A}}^{(1)}_{1}(s,m_i,\mathbf{b})$ is the leading eikonal. Note that we refer to the leading contribution to each amplitude, with different numbers of graviton exchanges, via the superscript label $(1)$, each subleading order is then referred to by increasing the number in the superscript. As per the description of \eqref{eq:eiksmatrix} we note that only the contributions which are divergent in energy or mass exponentiate. 

Expanding \eqref{eq:eiksmatrix} and collecting all the possible contributions at 2PM (i.e. at $\mathcal{O}(G_N^2)$) we can write an explicit expression for the first subleading eikonal,
\begin{eqnarray}
i \delta^{(2)}(s,m_i,\mathbf{b}) = i \tilde{\mathcal{A}}^{(2)}_{2} - i \tilde{\mathcal{A}}^{(1)}_{1} i \tilde{\mathcal{A}}^{(2)}_{1} = i \tilde{\mathcal{A}}^{(2)}_{2} \;, \label{eq:eiksubleading}
\end{eqnarray}
where in the second step we have used that in Einstein gravity we have $\mathcal{A}^{(2)}_{1}=0$ as we will see from the results in the next subsection. Note that in other theories this may not be the case. For example, in the supergravity case we will discuss in chapter \ref{chap:sugraeik} we find a contribution of the form $\mathcal{A}^{(2)}_{1}$ coming from the tree-level diagram with one RR field exchange between the scalar field and the stack of D-branes as you can see from \eqref{inelE0ips}.

Notice that for each eikonal we have,
\begin{equation}
i \delta^{(k)}(s,m_i,\mathbf{b}) \sim i \tilde{\mathcal{A}}^{(k)}_{k} \sim \mathcal{O}(G_N^{k}) \;,
\end{equation}
where $G_N$ is the usual Newton's constant. This relates the discussion presented here with the so called post-Minkowskian approximation discussed in section \ref{sec:pmexpansionbg}. We will see in more detail in chapter \ref{chap:graveik} that the leading eikonal corresponds to the 1PM order in the post-Minkowskian expansion, the subleading eikonal corresponds to the 2PM order and so on.

For sake of completeness, we can collect the various contributions of the form \eqref{eq:sumofleadingcont} at each order in $G_N$ and write the complete eikonal amplitude in momentum space,
\begin{equation}\label{eq:momspaceeikamp}
\mathcal{A}_{\text{eik}}(s, m_i, q) = 4 E p \int d^2 \mathbf{b} \, e^{-i \mathbf{q} \cdot \mathbf{b}} \left( \exp \left(i \sum_{i} \delta^{(i)}(s, m_i, \mathbf{b}) \right) - 1 \right) \;.
\end{equation}

From this discussion we can see that to calculate the eikonal we need to be able to categorise and collect the various leading, subleading and further contributions of each of the amplitudes with $n=1,2,3,\ldots$ number of graviton exchanges. In general the amplitudes we will calculate will include pieces which are either not relevant classically or are superclassical\footnote{These are contributions which are already taken into account by the exponential series of a process at a lower order in $G_N$.} and we will need an algorithm to be able to distinguish these. The scaling limit required to categorise the various contributions and identify the relevant new information at each order in $G_N$ can be spelled out as follows:
\begin{itemize}
\item We take $G_N$ small by keeping $G_N M^*$ fixed and we are interested in the non-analytic contributions as $t\to 0$ since they determine the large distance interaction
\item The ratios $m_i^2/s$, where $m_{1,2}$ are the masses of the external scalars, can be arbitrary; when they are fixed, one is describing the scattering of two Schwarzschild black holes, but it is possible to smoothly take them to be small or large and make contact with different relativistic regimes
\item At each order in $G_N^n$ the terms that grow faster than $E_1$ or $E_2$ (at large $E_i$ and fixed $G_N M^*$) should not provide new data, but just exponentiate the energy divergent contributions at lower perturbative orders
\item The terms that grow as $E_i$ provide a new contribution to the eikonal phase at order $G_N^n$ from which one can derive the contribution to the classical two-body deflection angle and from it the relevant information on the $n$PM effective two-body potential
\end{itemize}
We can summarise this mathematically as,
\begin{equation}
  \label{eq:heml}
  16 \pi G_N=2\kappa_D^2 \to 0~,~~~s\gg q^2=|t|~,~~~\mbox{with}~\, \frac{G_N M^*}{b} \ll 1 \;.
\end{equation}
Recall that $M^*$ was defined in chapter \ref{chap:intro} as the largest mass scale in the process under consideration. 

\subsection{The Leading Two-Body Eikonal in Pure Gravity}\label{sec:leadingeikbg}

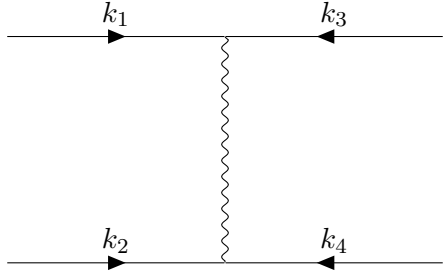
\begin{figure}[h]
  \centering
  \begin{tikzpicture}[scale=1.5]
	    \begin{feynman}
			\vertex (a) at (-2,-2) {};
			\vertex (b) at ( 2,-2) {};
			\vertex (c) at (-2, 0) {};
			\vertex (d) at ( 2, 0) {};
			\vertex[circle,inner sep=0pt,minimum size=0pt] (e) at (0, 0) {};
			\vertex[circle,inner sep=0pt,minimum size=0pt] (g) at (0, -2) {};
			\diagram* {
			(c) -- [fermion,edge label=$k_1$] (e) -- [anti fermion,edge label=$k_3$] (d),
			(g) -- [boson] (e),
			(a) -- [fermion,edge label=$k_2$] (g) -- [anti fermion,edge label=$k_4$] (b),
			};
	    \end{feynman}
    \end{tikzpicture}
  \caption{A figure illustrating the scattering between two massive scalars exchange a single graviton. The solid lines represent massive scalars and the wavy lines represent gravitons.}
  \label{fig:treelevelscatteringgrav}
\end{figure}

In this and subsequent subsections we will give a thorough analysis for the leading two-body eikonal in pure gravity. As the discussion in section \ref{sec:eikonalamplitudebg} implies, to calculate the leading eikonal we first need to calculate the tree-level amplitude between two scalars exchanging a single graviton. As specified before, we will take the two massive scalars with masses $m_1,m_2$ to have incoming momenta $k_1, k_2$ and outgoing momenta $k_3,k_4$ respectively. This scattering process is illustrated in figure \ref{fig:treelevelscatteringgrav}.
Using the Feynman rules derived in section \ref{sec:puregravbackground} we easily find the result,
\begin{equation}
i \mathcal{A}_{1} = 2i \kappa_D^2 \frac{1}{q^2} \left ( \frac{1}{2}(s- m_1^2 - m_2^2)^2 - \frac{2}{D-2} m_1^2 m_2^2 + \frac{1}{2}(s- m_1^2 - m_2^2)t  \right) \;.
\end{equation}
The procedure at the end of section \ref{sec:eikonalamplitudebg} specifies that we ignore the non-analytic terms in the amplitudes as these will only yield local contributions in impact parameter space and can be safely ignored in the large impact parameter limit. This means that we can ignore the last term in the big round brackets above.

For the rest of this section as well as the subsequent subsections we will explicitly work in $D=4$ spacetime dimensions. We do this for two reasons; to compare with widely known results in the literature and to avoid needing to take small-angle approximations when performing the partial wave decomposition in section \ref{sec:introeikpwd}. As such we can write the above amplitude in a more compact form as,
\begin{equation}\label{eq:introtreelevelamp}
i \mathcal{A}_{1} = i \frac{16 \pi G_N}{q^2} \left ( 2 (k_1k_2)^2 - m_1^2 m_2^2 \right) \;.
\end{equation}
We can now also calculate the impact-parameter analogue of this expression which following the observations in \eqref{eq:sumofleadingcont} will serve as the expression for the leading two-body eikonal,
\begin{eqnarray}\label{eq:introleadingeik}
\delta^{(1)}(s,m_i,b) = \tilde{\mathcal{A}}^{(1)}_{1}(s,m_i,b) &=& - 4 G_N \frac{2 (k_1k_2)^2 - m_1^2 m_2^2}{\sqrt{(k_1 k_2)^2 - m_1^2 m_2^2}} \log(b) \nonumber \\
&=& - \delta_1(s,m_i) \log(b) \;,
\end{eqnarray}
where $b=|\mathbf{b}|$ and in the second line we have defined the quantity,
\begin{equation}\label{eq:delta1kin}
\delta_1(s,m_i) = 4 G_N \frac{2 (k_1k_2)^2 - m_1^2 m_2^2}{\sqrt{(k_1 k_2)^2 - m_1^2 m_2^2}} \;.
\end{equation}
We will revisit the results found in this subsection in chapter \ref{chap:graveik} where we will also take various limits of the resulting expressions to compare with a wide variety of known results.

\subsection{Partial Wave Decomposition and the Two-Body Deflection Angle}\label{sec:introeikpwd}

In this subsection we will perform a standard partial wave decomposition to the leading eikonal amplitude given by \eqref{eq:momspaceeikamp} with only $\delta^{(1)}$ non-zero. We can write the leading eikonal amplitude as,
\begin{eqnarray} \label{eq:startingleadeikamp}
\mathcal{A}_{\text{eik}}^{(1)}(s, m_i, q) &=& 4 E p \int d^2 \mathbf{b} \, e^{-i \mathbf{q} \cdot \mathbf{b}} \left( \exp \left(i \delta^{(1)}(s, m_i, b) \right) - 1 \right) \nonumber \\
&=& 4 E p \int d^2 \mathbf{b} \, e^{-i \mathbf{q} \cdot \mathbf{b}} \left( b^{- i \delta_1(s,m_i)} - 1 \right) \;.
\end{eqnarray}
where we have used the expression for $\delta^{(1)}(s, m_i, b)$ calculated in \eqref{eq:introleadingeik}.

The scattering amplitude can be expanded in partial waves as follows,
\begin{eqnarray}
\mathcal{A}_{\text{eik}}^{(1)}(s, m_i, q) = \sum_{J=0}^\infty (2J+1) a_J(s,m_i) P_J(x)~~;~~~x \equiv \cos \theta \;,
\label{ACV28}
\end{eqnarray}
where $P_J$ are Legendre polynomials and the partial wave coefficients are given by,
\begin{eqnarray}
a_J (s,m_i) = \frac{1}{2} \int_{-1}^{+1} dx \, \mathcal{A}_{\text{eik}}^{(1)}(s, m_i, q) P_J(x) \;.
\label{ACV29}
\end{eqnarray}
The Legendre polynomials are normalized as,
\begin{eqnarray}
\int_{-1}^{+1} dx \,  P_J (x) P_L (x) = \frac{2}{2J+1} \delta_{JL}~;~P_J (1) =1~;~ P_J (x) = \frac{(-1)^J}{2^J J!} \frac{d^J}{\partial x^J} (1-x^2)^J \;.
\label{ACV30}
\end{eqnarray}
Notice that the variable $x \equiv \cos \theta$ is related to the Mandelstam variable $t$ as described by the equations in \eqref{eq:mandeltoangleandp}, from these we also have the relation,
\begin{equation}\label{eq:relqtoscatangle}
q=2p \sin \frac{\theta}{2} \;.
\end{equation}
By using the expansion,
\begin{equation}
e^{-iqb \cos\phi}= \sum_{n=-\infty}^{\infty} i^n J_n(qb) e^{-i n\phi} \;,
\end{equation}
where $J_n(x)$ are Bessel functions, we can simplify \eqref{eq:startingleadeikamp} and write it as,
\begin{eqnarray} \label{eq:startingleadeikampSimplified}
\mathcal{A}_{\text{eik}}^{(1)}(s, m_i, q) &=& 8 \pi E p \int_{0}^{\infty} db \, b J_0 (q b) \left( b^{- i \delta_1(s,m_i)} - 1 \right) \;.
\end{eqnarray}
Using \eqref{ACV29} we can write the partial wave coefficients as,
\begin{equation}
\frac{a_J(s,m_i)}{4Ep} = \pi \int_0^{\pi} d \theta \, \sin(\theta) P_J(\cos(\theta)) \left\{ \int_{0}^{\infty} db \, b J_0 (q b) \left( b^{- i \delta_1(s,m_i)} - 1 \right) \right\} \;.
\end{equation}
Using the equation between the exchanged momenta and scattering angle, \eqref{eq:relqtoscatangle}, and using the equation (7.253) in \cite{gradshteyn1996table}, which reads,
\begin{equation}
\int_0^{\pi/2} dx \sin (2x) P_n (\cos(2x)) J_0 (a \sin(x)) = a^{-1} J_{2n+1}(a) \;,
\end{equation}
we find that we can express $a_J$ as,
\begin{equation}
\frac{a_J(s,m_i)}{4Ep} = \frac{\pi}{p} \int_{0}^{\infty} db \, J_{2J+1} (2pb) \left( b^{- i \delta_1(s,m_i)} - 1 \right) \label{eq:aJintegralb} \;.
\end{equation}
This integral can be performed and we finally find,
\begin{equation}\label{eq:leadingPWDinJ}
a_J(s,m_i)  = \frac{2 E \pi}{p} \left( p^{i \delta_1} \frac{\Gamma\left( J - \frac{i \delta_1(s,m_i)}{2} + 1 \right)}{\Gamma\left( J + \frac{i \delta_1(s,m_i)}{2} + 1 \right)} - 1 \right) \;.
\end{equation}
This expression is equivalent to known results in the literature \cite{Amati:1992zb}.

Using the expressions in the discussion above we want to be able to derive the deflection angle as this will effectively translate the amplitude into the dynamics describing the interaction between the two heavy objects\footnote{In this case we are describing two Schwarzschild black holes.}. In chapters \ref{chap:graveik} and \ref{chap:potential} we will revisit this idea in greater detail and include the subleading corrections to the leading eikonal. Following Landau and Lifshitz \cite{landau2013quantum} we can relate the partial waves to the scattering angle by noticing that the asymptotic form of the Legendre polynomial for large $J$ is, 
\begin{equation}
P_J(\cos \theta) \approx \frac{-i}{\sqrt{2\pi J \sin \theta}}\left( e^{i(J+1/2)\theta + i \pi/4} - e^{-i(J+1/2)\theta - i \pi/4}  \right) \;.
\end{equation}
Substituting this into \eqref{ACV28} we find,
\begin{equation}
\mathcal{A}_{\text{eik}}^{(1)}(s, m_i, q) \simeq - i \sum_{J=0}^\infty a_J(s,m_i) \sqrt{\frac{J}{2\pi \sin \theta}}\left( e^{i(J+1/2)\theta + i \pi/4} - e^{-i(J+1/2)\theta - i \pi/4}  \right) \;.
\end{equation}
Since the exponential factors are dominated by large $J$ the value of the sum in the equation above is determined by the appropriate saddle point. The saddle point is given by,
\begin{equation}\label{eq:anglefromaJ}
i \theta_J = - \frac{\partial}{\partial J}\log(a_J(s,m_i)) \;,
\end{equation}
where we have chosen the positive $\theta$ by convention and written $\theta$ as $\theta_J$. We can now use this equation to find the deflection angle from \eqref{eq:leadingPWDinJ},
\begin{eqnarray}\label{eq:leadinglargeJangexp}
\theta_J &=& i \frac{\partial}{\partial J}\log(a_J(s,m_i)) \nonumber \\
&=& i \frac{\partial}{\partial J}\log \left[ \frac{2 E \pi}{p} \left( p^{i \delta_1} \frac{\Gamma\left( J - \frac{i \delta_1(s,m_i)}{2} + 1 \right)}{\Gamma\left( J + \frac{i \delta_1(s,m_i)}{2} + 1 \right)} - 1 \right) \right] \nonumber \\
& \approx & \frac{\delta_1(s,m_i)}{J} - \frac{\delta_1^3(s,m_i)}{12 J^3} + \ldots \;,
\end{eqnarray}
where in the last line we have expanded for large $J$ and kept only terms that contribute in the high-energy limit. In the next subsection we will give an explicit closed form expression for this series.

\subsubsection{Saddle Point Approximation}

In this subsection we will perform a saddle point approximation in order to evaluate the integral given in \eqref{eq:aJintegralb}. We start by recognising that we can write the Bessel function which appears, in an integral form by using \cite{temme2011special},
\begin{equation}
J_n(x) = \frac{1}{2\pi} \int_{-\pi}^{\pi} dt \, e^{i (x \sin t-nt)} \;,
\end{equation}
and using this allows us to write,
\begin{equation}
\frac{a_J(s,m_i)}{4Ep} = \frac{1}{2p} \int_{0}^{\infty} db \int_{-\pi}^{\pi} d\phi \, \left( b^{-i \delta_1(s,m_i)} - 1 \right) e^{i (2pb \sin \phi - (2J+1)\phi)} \;.
\end{equation}
Defining the variable $x=pb/J$ we can rewrite this as,
\begin{eqnarray}
\frac{a_J(s,m_i)}{4Ep} &=& \frac{J}{2p^2} \int_{0}^{\infty} dx \int_{-\pi}^{\pi} d\phi \, \left( \left( \frac{Jx}{p} \right)^{-i \delta_1(s,m_i)}  - 1 \right) e^{i (2xJ \sin \phi - (2J+1)\phi)} \nonumber \\
&\approx & \frac{J}{2p^2} \left( \frac{J}{p} \right)^{-i \delta_1(s,m_i)} \int_{0}^{\infty} dx \int_{-\pi}^{\pi} d\phi \,  x^{-i \tilde{\delta}_1(s,m_i) J}  e^{i (2xJ \sin \phi - (2J+1)\phi)} \;, \nonumber \\ \label{eq:fullaJ}
\end{eqnarray}
where in the second line we have dropped the $-1$ as this is not kinematically relevant in the large $J$ limit and we have defined the quantity,
\begin{equation}
\tilde{\delta}_1(s,m_i) = \frac{\delta_1(s,m_i)}{J} \;. \label{eq:tildtonorm}
\end{equation} 
Collecting all the terms in the exponent that are large in the large $J$ limit we see that we need to perform a saddle-point approximation on the two-variable function given by,
\begin{equation}
f(x,\phi) = -\tilde{\delta}_1(s,m_i) \log(x) + 2x \sin \phi - 2 \phi \;, \label{eq:saddlef}
\end{equation}
where we note the exponent is then $i J f(x, \phi)$. When performing a saddle-point approximation in 2 variables we know that,
\begin{equation}
\int dx \, d\phi \, e^{i J f(x,\phi)} \approx \frac{2\pi}{\sqrt{-\det(i J S''(x_0,\phi_0))}} e^{i J f(x_0, \phi_0)} \;, \label{eq:saddleapprox}
\end{equation}
where $S''(x_0,y_0)$ is the Hessian evaluated at the saddle-point and we have taken into account relevant factors of $i$ and $J$. 

To find the saddle-point we need to solve the following set of simultaneous equations,
\begin{eqnarray}\label{eq:fder}
&&\frac{\partial f(x,\phi)}{\partial x} = -\frac{\tilde{\delta}_1(s,m_i)}{x}  + 2\sin \phi = 0 \;, \label{eq:fder1} \\
&&\frac{\partial f(x,\phi)}{\partial \phi} = 2x \cos \phi - 2 = 0 \;. \label{eq:fder2}
\end{eqnarray}
We can easily express $x$ in terms of $\cos \phi$ from the second equation and substituting this back into the first equation gives us an expression for $\phi$ in terms of $\tilde{\delta}_1(s,m_i)$ and so we find,
\begin{equation}
x_0 = \frac{1}{\cos \phi_0}~~;~~\phi_0 = \arctan \left( \frac{\tilde{\delta}_1(s,m_i)}{2} \right) \;.
\end{equation}
Substituting these results back into our original function \eqref{eq:saddlef} we get,
\begin{equation}\label{eq:saddlefforleadeik}
\frac{-i}{J} f(x_0,\phi_0) = -\tilde{\delta}_1 \log \left( \sqrt{\frac{\tilde{\delta}_1^2}{4}+1} \right) + \tilde{\delta}_1 - 2 \arctan \left( \frac{\tilde{\delta}_1}{2} \right) \;.
\end{equation}
Putting \eqref{eq:fullaJ}, \eqref{eq:saddleapprox} and \eqref{eq:saddlefforleadeik} together allows us to write the partial wave coefficients in a saddle approximation, 
\begin{eqnarray}\label{eq:fullleadingsaddleaJexpr}
&&\frac{a_J(s,m_i)}{4Ep} \approx \frac{J}{2p^2} \left( \frac{J}{p} \right)^{-i \delta_1} \frac{2\pi}{\sqrt{-\det(i J S''(x_0,\phi_0))}} \nonumber \\
&& \qquad \times \exp \left[i J \left( -\tilde{\delta}_1 \log \left( \sqrt{\frac{\tilde{\delta}_1^2}{4}+1} \right) + \tilde{\delta}_1 - 2 \arctan \left(\frac{\tilde{\delta}_1}{2} \right) \right) \right] \;,
\end{eqnarray}
where we have not explicitly written the Hessian as it is not relevant when calculating the deflection angle as we will see below. 

Using \eqref{eq:anglefromaJ} we can now calculate the deflection angle. Note that we need to use $\tilde{\delta}_1=\delta_1/J$ in equation \eqref{eq:fullleadingsaddleaJexpr} above. We will be explicit here as it is instructive,
\begin{eqnarray}\label{eq:fullresumleadangle}
\theta_J &=& i \frac{\partial}{\partial J} \log(a_J) \nonumber \\
&\simeq & i \frac{\partial}{\partial J} \left[ -i \delta_1 \log J - i \delta_1 \log \left( \sqrt{\frac{\delta_1^2}{4J^2}+1} \right) + i \delta_1 - 2 i J \arctan \left( \frac{\delta_1}{2J} \right) \right] \nonumber \\
&=& - \left[ - \frac{\delta_1}{J} + \frac{\delta_1^3}{J (\delta_1^2 + 4J^2)} - 2 \arctan \left( \frac{\delta_1}{2J} \right) + \frac{4 \delta_1 J}{\delta_1^2 +4J^2} \right] \nonumber \\
& \approx & 2 \arctan \left(\frac{\delta_1(s,m_i)}{2J} \right) \;,
\end{eqnarray}
where in the second line we have only kept terms in the exponent of \eqref{eq:fullleadingsaddleaJexpr} as well as the $J^{-i\delta_1}$ term because the rest are not relevant in the high-energy, large $J$ limit and in the third line we note that the combination of the first, second and fourth terms sum to 0. We note that the large $J$ expansion of the last line is equivalent to \eqref{eq:leadinglargeJangexp} as required.

\subsubsection{Saddle Point Approximation to Derive Linear in Eikonal Contributions to the Deflection Angle}

In this subsection we will be performing a saddle point approximation as in the previous subsection but we will include all the corrections to the eikonal, not just the leading eikonal. In order to be able to compute this we will have to make certain approximations which will give a partially complete result. This will prove to be sufficient for the analysis up to 2PM discussed in chapters \ref{chap:sugraeik}, \ref{chap:kkeik} and \ref{chap:graveik}.

To study this we modify \eqref{eq:startingleadeikamp} and write,
\begin{equation}
\mathcal{A}_{\text{eik}}^{(1)}(s, m_i, q) = 4 E p \int d^2 \mathbf{b} \, e^{-i \mathbf{q} \cdot \mathbf{b}} \left( \exp \left(i \sum_k \delta^{(k)}(s, m_i, b) \right) - 1 \right) \;,
\end{equation}
where we have included all the corrections to the eikonal. From this we can then modify the expression for $a_J$ in \eqref{eq:fullaJ} to read, 
\begin{eqnarray}
\frac{a_J(s,m_i)}{4Ep} & \approx & \frac{J}{2p^2} \left( \frac{J}{p} \right)^{-i \delta_1} \int_{0}^{\infty} dx \int_{-\pi}^{\pi} d\phi \nonumber \\
&& \quad \times \left( x^{-i \tilde{\delta}_1 J} \exp\left( \sum_{k=1}^{\infty} \frac{i \tilde{\delta}_{k+1}}{x^k} J \right) \right) e^{i (2xJ \sin \phi - (2J+1)\phi)} \;,
\end{eqnarray}
where we have used \eqref{eq:introleadingeik} for $\delta^{(1)}$ and we have defined,
\begin{equation}
\delta^{(k)}(s, m_i, b) = \frac{\delta_k(s,m_i)}{b^k}~~\text{for}~~k>1 \;,
\end{equation}
as well as,
\begin{equation}\label{eq:tildedeltak}
\tilde{\delta}_{k} = \frac{\delta_{k}}{J^{k}} p^{k-1}~~\text{for}~~k > 1 \;.
\end{equation}
Note that $\delta_k(s,m_i)$ is of order $\mathcal{O}(G_N^k)$. We then see, by collecting all the terms that are large in the large $J$ limit as we have done before, that the function for which we need to find the saddle point is given by,
\begin{equation}
f(x,\phi) = -\tilde{\delta}_1 \log(x) + \sum_{k=1}^{\infty} \frac{ \tilde{\delta}_{k+1}}{x^k} + 2x \sin \phi - 2 \phi \;. \label{eq:saddlefFULL}
\end{equation}
We can now write the equations which we must solve to find the saddle point,
\begin{eqnarray}\label{eq:fderFULL}
&&\frac{\partial f(x,\phi)}{\partial x} = - \frac{\tilde{\delta}_1}{x} - \sum_{k=1}^{\infty} \frac{k\tilde{\delta}_{k+1}}{x^{k+1}} + 2\sin \phi = 0 \label{eq:fder1FULL} \\
&&\frac{\partial f(x,\phi)}{\partial \phi} = 2x \cos \phi - 2 = 0 \;. \label{eq:fder2FULL}
\end{eqnarray}
Since we are now including all the corrections to the eikonal we have significantly complicated the set of equations we must solve. In order to make this problem tractable we will  linearise these equations in $\phi$. Doing so we find the saddle point,
\begin{equation}
x_0 = 1~~;~~\phi_0 = \frac{1}{2} \left(\tilde{\delta}_1 + \sum_{k=1}^{\infty} k \tilde{\delta}_{k+1} \right) \;.
\end{equation}
Employing the exact same method as before we then find for the partial wave coefficients,
\begin{equation}\label{eq:fullleadingsaddleaJexprLINEAR}
\frac{a_J(s,m_i)}{4Ep} \approx \frac{J}{2p^2} \left( \frac{J}{p} \right)^{-i \delta_1} \frac{2\pi}{\sqrt{-\det(i J S''(x_0,\phi_0))}} \exp \left[i J \sum_{k=1}^{\infty} k \tilde{\delta}_{k+1} \right] \;.
\end{equation}
Using this result along with equations \eqref{eq:anglefromaJ}, \eqref{eq:tildedeltak} we can then find the resulting expression for the angle,
\begin{equation}\label{eq:deflanglinearJ}
\theta_J \approx \frac{\delta_1}{J} + \sum_{k=1}^{\infty} \frac{k p^k \delta_{k+1} }{J^{k+1}} \;.
\end{equation}
The equivalent expression in impact parameter space can be found by using the fact that at the saddle $x_0=1$ we have $J=pb$,
\begin{equation}\label{eq:deflanglinearb}
\theta_b \approx \frac{1}{p} \left( \frac{\delta_1}{b} + \sum_{k=1}^{\infty} \frac{k \delta_{k+1} }{b^{k+1}} \right) = - \frac{1}{p} \frac{\partial}{\partial b} \sum_{k=1}^{\infty} \delta^{(k)}(s,m_i,b) \;,
\end{equation}
which we notice is the usual expression used to derive the angle from the eikonal \cite{Amati:1990xe, Akhoury:2013yua, Bjerrum-Bohr:2018xdl}. In the second step above we are using the full eikonal expressions and not just the coefficients. For sake of completeness we can take the massless probe limit of this equation which will be used in chapter \ref{chap:sugraeik}. Using \eqref{eq:1geEP} we see that the momentum factor in the massless probe limit is given to leading order by, $p \approx E$ and so we have,
\begin{equation}\label{eq:deflanglinearbmassless}
\theta_b \approx - \frac{1}{E} \frac{\partial}{\partial b} \sum_{k=1}^{\infty} \delta^{(k)}(E,b) \;,
\end{equation}

\subsection{Post-Newtonian Expansion}\label{sec:leadeikPN}

As a final remark in this section we will look at the post-Newtonian (PN) expansion of the leading two-body deflection angle we have calculated above and compare this with known results. We will briefly revisit the PN expansion when discussing the two-body Hamiltonian in chapter \ref{chap:potential}.

Using \eqref{eq:delta1kin} we can write the leading contribution to the deflection angle \eqref{eq:fullresumleadangle} fully as,
\begin{equation}
\theta_J^{(1)} = \frac{4 G_N}{J} \frac{2 (k_1k_2)^2 - m_1^2 m_2^2}{\sqrt{(k_1 k_2)^2 - m_1^2 m_2^2}} \;.
\end{equation}
Using \eqref{PN5} we can expand this in terms of the relative velocity between the two bodies,
\begin{equation}
\theta_J^{(1)} = \frac{1}{j v_{\infty}} + \frac{2 v_{\infty}}{j} \;,
\end{equation}
where $j$ is defined in \eqref{PN6}. We notice, using \eqref{eq:idforPNterms}, that the first term in this equation is a 0PN contribution since it goes as $\left( 1/c^2 \right)^0$ and the second term is a 1PN contribution since it goes as $\left( 1/c^2 \right)^1$. These are equivalent to the first and second terms of the 0PN and 1PN results for the deflection angle respectively which we quote below \cite{Bini:2017wfr},
\begin{eqnarray}
\theta_{\text{0PN}} = \arctan\left(\frac{1}{j v_{\infty}}\right) &=& \frac{1}{j v_{\infty}} - \frac{1}{3 j^3 v_{\infty}^3} + \ldots \;, \\
\theta_{\text{1PN}} &=& \frac{2 v_{\infty}}{j}+\frac{3 \pi }{2 j^2}+\frac{4}{j^3 v_{\infty}} + \ldots \;.
\end{eqnarray}

\chapter{High-Energy Expansion of Various Integrals}\label{chap:integrals}

In this chapter we are going to discuss the solutions to various integrals that we will need in subsequent chapters. It is based off of the papers \cite{Collado:2018isu} and \cite{KoemansCollado:2019ggb}. We will discuss the integrals needed for massive scalar scattering as well as the integrals needed for scattering massless states off of D-branes. Since we are interested in the high-energy limits of these integrals we will be using various techniques which allow us to expand the integrands in the high-energy limit and find perturbative solutions.

\section{Integrals for Massive Scalar Scattering} \label{app:integrals}

In the classical regime the centre of mass energy $\sqrt{s}$ and the masses $m_i^2$ are much larger than the momentum exchanged $q$. In this limit the integrals appearing in the amplitudes discussed in detail in chapter \ref{chap:graveik} can be performed so as to extract the leading and the subleading contributions (we also calculate the subsubleading contribution to the scalar box integral). Our approach is the following: we first write our starting point in terms of Schwinger parameters $t_i$, then we perform the integrals over the $t$'s parametrising the scalar propagators by using a saddle point approximation, finally the integrals over the graviton propagators reduce to those of an effective two-point function. In the following subsection we give a detailed analysis of the so called scalar box integral, showing that for a general spacetime dimension $D$ it provides a classical contribution proportional to $(D-4)$. We will also give results for the triangle integrals which are necessary to evaluate the full classical contribution at one-loop.

\subsection{Scalar Box Integral} \label{app:boxintegrals}

In this subsection we will discuss the scalar box integral. To start we will be evaluating,
\begin{equation}
  \label{eq:1lbmi}
  \mathcal{I}_4(s,t) = \int \!\frac{d^Dk}{(2\pi)^D}\, \frac{1}{k^2} \,\frac{1}{(k+k_1)^2+m_1^2} \,\frac{1}{(q+k)^2} \, \frac{1}{(k_2-k)^2+m_2^2} \;.
\end{equation}
After a Wick rotation and introducing Schwinger parameters the integral over the loop momentum $k$ is Gaussian and can be readily performed. After evaluating this we find,
\begin{equation}
\mathcal{I}_4(s,t) = i \int_{0}^{\infty} \prod_{i=1}^{4}  \, d t_i \frac{T^{-\frac{D}{2}}}{(4\pi)^{\frac{D}{2}}} \exp \left[-\frac{2 k_1 k_2 t_2 t_4 + q^2 t_1 t_3 + t_2^2 m_1^2 + t_4^2 m_2^2}{T} \right]\;,
\end{equation}
where $q\equiv k_1+k_3$ is the momentum exchanged and we have defined $T=\sum_{i} t_i$. The form of the equation above is suggestive because we are interested in the limit where $|k_1 k_2|, m_i^2 \gg q^2$, this means that the integral over $t_2,t_4$ can be performed with a saddle point approximation around $t_2=t_4=0$.

What makes this integral awkward is that its region of integration is just the positive quadrant in $t_2,t_4$. In order to circumvent this problem it is convenient to sum the contribution of the crossed box integral. In terms of Schwinger parameters this is given by
\begin{equation}
\mathcal{I}_4(u,t) = i \int_{0}^{\infty} \prod_{i=1}^{4}  \, d t_i \frac{T^{-\frac{D}{2}}}{(4\pi)^{\frac{D}{2}}} \exp \left[-\frac{2 k_2 k_3 t_2 t_4 + q^2 t_1 t_3 + t_2^2 m_1^2 + t_4^2 m_2^2}{T} \right] \;.
\end{equation}
Notice that $\mathcal{I}_4(u,t)$ can be obtained from $\mathcal{I}_4(s,t)$ by swapping $k_1 \leftrightarrow k_3$. In order to combine $\mathcal{I}_4(s,t)$ and $\mathcal{I}_4(u,t)$, it is convenient to define,
\begin{equation}
  \tilde{k}^2 = k_1 k_2 + \frac{q^2}{4} = - k_2 k_3 - \frac{q^2}{4} \;.
\end{equation}
Then we can rewrite,
\begin{equation}
  \label{eq:intco}
  \mathcal{I}_4(s,t) = i \int_{0}^{\infty} \prod_{i=1}^{4}  \, d t_i f(\tilde{k}^2,t_i)\;,~~~ \mathcal{I}_4(u,t) = i \int_{0}^{\infty} \prod_{i=1}^{4}  \, d t_i f(-\tilde{k}^2,t_i)\;,~~~
\end{equation}
where,
\begin{equation}\label{eq:fuf}
  f(\tilde{k}^2,t_i) =  \frac{e^{-q^2 \frac{t_1 t_3}{T}}}{(4\pi)^{\frac{D}{2}}\,T^{\frac{D}{2}}} \exp \left[  -\frac{(t_2\; t_4)}{T} \left(\begin{matrix} m_1^2 & \tilde{k}^2 \\ \tilde{k}^2 & m_2^2 \end{matrix} \right) \left(\begin{matrix} t_2 \\ t_4 \end{matrix} \right)  \right]
   e^{\frac{q^2}{2T} |t_2 t_4|}\;. 
\end{equation}
As previously mentioned we are interested in performing these integrals in the limit where $|k_1 k_2|, m_i^2,1/t_2,1/t_4$ are all of the same order and much bigger than $q^2$. We can therefore Taylor expand the integrands for small $t_2$ and $t_4$ and at the leading order we simply obtain the function~\eqref{eq:fuf} where $T$ reduces to $t_1+t_3$ and the last exponential can be neglected. It is therefore convenient to define,
\begin{equation}
T_0 = t_1 + t_3\;.
\end{equation}
By expressing the two integrals in this way, we can see that they are equivalent under the change $t_2 \rightarrow - t_2$ or $t_4 \rightarrow - t_4$. We note that, $I_4(t_2, -t_4) = J_4(t_2, t_4)$, $I_4(-t_2, t_4) = J_4(t_2, t_4)$, $I_4(-t_2, -t_4) = I_4(t_2, t_4)$ where $I_4$ and $J_4$ are the integrands of $\mathcal{I}_4(s,t)$ and $\mathcal{I}_4(u,t)$ respectively. We can therefore write the combination of box and crossed box integrals as,
\begin{eqnarray} \label{eq:i4j4full}
\mathcal{I}_4(s,t) + \mathcal{I}_4(u,t) &=& \frac{i}{2} \int_{0}^{\infty} \, d t_1 d t_3 \int_{-\infty}^{\infty} \, d t_2 d t_4 \frac{e^{-q^2 \frac{t_1 t_3}{T}}}{(4\pi)^{\frac{D}{2}}\,T^{\frac{D}{2}}} \nonumber \\
&& \qquad \times \exp \left[  -\frac{(t_2\; t_4)}{T} \left(\begin{matrix} m_1^2 & \tilde{k}^2 \\ \tilde{k}^2 & m_2^2 \end{matrix} \right) \left(\begin{matrix} t_2 \\ t_4 \end{matrix} \right)  \right] e^{\frac{q^2}{2T} |t_2 t_4|}\;,
\end{eqnarray}
where $T$ is now explicitly defined as $T=T_0+|t_2|+|t_4|$. Note that we have written some quantities as $|t_2|, |t_4|$ since the original domain of integration is for $t_{2,4} \geq 0$.

Expanding \eqref{eq:i4j4full} around $(t_2,t_4)=(0,0)$ the leading contribution (i.e. the one that eikonalises the tree-level amplitude, see comments below \eqref{S12}), which we denote as $\mathcal{I}_4^{(1)}(s,t) + \mathcal{I}_4^{(1)}(u,t)$, can be written as a Gaussian integral,
\begin{eqnarray}\label{eq:i4j4leadingint}
 \mathcal{I}_4^{(1)}(s,t) + \mathcal{I}_4^{(1)}(u,t) &=& i \int_{0}^{\infty}\!\! dT_0\, \frac{T_0^{1-\frac{D}{2}}}{(4\pi)^{\frac{D}{2}}} \int_{0}^{1} d x_1 \, \exp \left[- q^2 x_1 (1-x_1) T_0 \right] \nonumber \\
&&\times  \frac{1}{2} \int_{-\infty}^\infty dt_2 \, dt_4 \, \exp \left[  -\frac{(t_2\; t_4)}{T_0} \left(\begin{matrix} m_1^2 & \tilde{k}^2 \\ \tilde{k}^2 & m_2^2 \end{matrix} \right) \left(\begin{matrix} t_2 \\ t_4 \end{matrix} \right) \right]\;,  
\end{eqnarray}
where $x_1=t_1/T_0$. The quadrants with $t_2, t_4>0$ and $t_2,t_4<0$ yield the same contribution corresponding $\mathcal{I}_4^{(1)}(s,t)$, while those with $t_2>0>t_4$ and $t_4>0>t_2$ are again identical and correspond to $\mathcal{I}_4^{(1)}(u,t)$. 

We should also recall that $\tilde{k}^2$ is not a kinematic variable directly relevant to the amplitude calculations. In the expression resulting from performing the Gaussian integral over $t_2, t_4$ in \eqref{eq:i4j4leadingint} above we need to be careful and also substitute for,
\begin{equation}\label{eq:tildekexp}
\tilde{k}^2 = k_1 k_2 + \frac{q^2}{4} = k_1 k_2 \left(1 + \frac{q^2}{4 k_1 k_2} \right) \;, 
\end{equation}
whilst taking into account that we are interested in the limit where $|k_1 k_2|, m_i^2 \gg q^2$. The remaining two integrals over $T_0$ and $x_1$ can be decoupled and, by collecting everything together, we find
\begin{equation}
\mathcal{I}_4^{(1)}(s,t)  + \mathcal{I}_4^{(1)}(u,t) = \frac{1}{2} \frac{\pi}{(4\pi)^{\frac{D}{2}}} \frac{-1}{\sqrt{(k_1 k_2)^2 - m_1^2 m_2^2}}\,\Gamma \left(\frac{6-D}{2}\right)\, \frac{\Gamma^2\!\left(\frac{D-4}{2}\right)}{\Gamma(D-4)} (q^2)^{\frac{D-6}{2}} \;.
\label{eq:i4leading}
\end{equation}
Since we will look at the resulting amplitudes in impact parameter space it is worth calculating the expressions above in impact parameter space. Using \eqref{eq:eikips} and \eqref{eq:impoformu} we find that \eqref{eq:i4leading} becomes\footnote{Note that in this chapter we are not including the normalisation factor, $1/4Ep$, found in \eqref{eq:eikips}.},
\begin{equation}
\tilde{\mathcal{I}}_{4}^{(1)}(s,t) + \tilde{\mathcal{I}}_{4}^{(1)}(u,t) = \frac{-1}{128 \pi^{D-2}} \frac{1}{\sqrt{(k_1 k_2)^2 - m_1^2 m_2^2}} \Gamma^2\!\left(\frac{D}{2}-2\right) \frac{1}{\mathbf{b}^{2D-8}} \;.
\end{equation}
We can see that in $D=4$ the result is IR divergent and dimensional regularisation can be used to extract the $\log \mathbf{b}^2$ term we are interested in. In order to implement this we would use the substitution,
\begin{eqnarray}
\frac{\Gamma ( \frac{D-4}{2})}{4 \pi^{\frac{D-2}{2}}\mathbf{b}^{D-4}} \Longrightarrow - \frac{1}{2\pi} \log \mathbf{b} \;.
\end{eqnarray}
Note that if we were to directly integrate over one of the quadrants in order to calculate the result for just one of the scalar box diagrams we find at leading order,
\begin{eqnarray}\label{eq:1lb1q}
\mathcal{I}_4^{(1)}(u,t) &=& \frac{i}{(4\pi)^{\frac{D}{2}}} \frac{\ln\left[\left(\frac{\sqrt{k_2 k_3 + m_1 m_2}+\sqrt{k_2 k_3 - m_1 m_2}}{\sqrt{2 m_1 m_2}}\right)^2\right]}{2 \sqrt{(k_2k_3)^2 - m_1^2 m_2^2}}\,\Gamma \left(\frac{6-D}{2}\right)\, \frac{\Gamma^2\!\left(\frac{D-4}{2}\right)}{\Gamma(D-4)} (q^2)^{\frac{D-6}{2}} \nonumber \\
& \approx & \frac{i}{(4\pi)^{\frac{D}{2}}} \frac{\text{arcsinh}\left(\sqrt{\frac{\sigma-1}{2}}\right)}{ \sqrt{(k_1k_2)^2 - m_1^2 m_2^2}}\,\Gamma \left(\frac{6-D}{2}\right)\, \frac{\Gamma^2\!\left(\frac{D-4}{2}\right)}{\Gamma(D-4)} (q^2)^{\frac{D-6}{2}} \;,
\end{eqnarray}
where we have used $-k_2 k_3 = k_1 k_2 + q^2/2$ in the second line to express this in terms of $k_1 k_2$ and we have defined the quantity $\sigma = \frac{- k_1 k_2}{m_1 m_2}$. Note that to find the equivalent expression for $\mathcal{I}_4^{(1)}(s,t)$ we switch $k_3 \leftrightarrow k_1$ in the first line above.

Continuing with the saddle point approximation and expanding \eqref{eq:i4j4full} further we find the following subleading contribution,
\begin{eqnarray} \label{eq:ij4subleading}
\mathcal{I}_4^{(2)}(s,t) + \mathcal{I}_4^{(2)}(u,t) &=& i \int_{0}^{\infty}\!\! dT_0\, \frac{T_0^{1-\frac{D}{2}}}{(4\pi)^{\frac{D}{2}}} \int_{0}^{1} d x_1 \, \exp \left[- q^2 x_1 (1-x_1) T_0 \right] \nonumber \\
&&\times  \frac{1}{2} \int_{-\infty}^\infty dt_2 \, dt_4 \, \exp \left[  -\frac{(t_2\; t_4)}{T_0} \left(\begin{matrix} m_1^2 & \tilde{k}^2 \\ \tilde{k}^2 & m_2^2 \end{matrix} \right) \left(\begin{matrix} t_2 \\ t_4 \end{matrix} \right) \right] \nonumber \\
&&\times \frac{|t_2|+|t_4|}{T_0^2} \biggl(2 k_1 k_2 t_2 t_4+m_1^2 t_2^2+m_2^2 t_4^2  \nonumber \\
&&  +q^2 T_0^2 x_1 (1-x_1) - \frac{D}{2} T_0 \biggr) \nonumber \\
&=& \frac{i\sqrt{\pi}}{2 (4\pi)^{\frac{D}{2}}} \frac{m_1+m_2}{(k_1 k_2)^2 - m_1^2 m_2^2} \,\Gamma \left(\frac{5-D}{2}\right)\,\frac{\Gamma^2\left(\frac{D-3}{2}\right)}{\Gamma(D-4)} (q^2)^{\frac{D-5}{2}}\;, \nonumber \\ 
\end{eqnarray}
where we have also taken into account the fact that we need to substitute for $\tilde{k}^2$ using \eqref{eq:tildekexp} and find the leading contribution in $|k_1 k_2|, m_i^2 \gg q^2$ after performing the substitution. It is worth looking at the impact parameter space expression of the above result. In impact parameter space we have,
\begin{equation}
\tilde{\mathcal{I}}_4^{(2)}(s,t) + \tilde{\mathcal{I}}_4^{(2)}(u,t) = \frac{i}{32 \pi^{D-\frac{3}{2}}} \frac{m_1+m_2}{(k_1 k_2)^2 - m_1^2 m_2^2} \,\Gamma \left(\frac{2D-7}{2}\right)\,\frac{\Gamma^2\left(\frac{D-3}{2}\right)}{\Gamma(D-4)} \frac{1}{(\mathbf{b}^2)^{D-\frac{7}{2}}} \;,
\end{equation}
which we see vanishes for $D=4$.

The subsubleading integral is slightly more nuanced. In this case not only do we need to resolve the third term in the expansion of \eqref{eq:i4j4full} but we also need to take into account the contribution coming from the expansion of $\tilde{k}^2$ in the result for \eqref{eq:i4j4leadingint}. The extra contribution from \eqref{eq:i4j4leadingint} is given by,
\begin{equation} \label{eq:xtraconti4ltoi4ssl}
\frac{i}{2} \frac{\pi}{(4\pi)^{\frac{D}{2}}} \frac{k_1 k_2}{4[m_1^2 m_2^2-(k_1 k_2)^2]^{3/2}}\,\Gamma \left(\frac{6-D}{2}\right)\, \frac{\Gamma^2\!\left(\frac{D-4}{2}\right)}{\Gamma(D-4)} (q^2)^{\frac{D-4}{2}} \;.
\end{equation}
From the expansion of \eqref{eq:i4j4full} to subsubleading order we find,
\begin{eqnarray}
\mathcal{I}_4^{(3)}(s,t) + \mathcal{I}_4^{(3)}(u,t) &=& i \int_{0}^{\infty}\!\! dT_0\, \frac{T_0^{1-\frac{D}{2}}}{(4\pi)^{\frac{D}{2}}} \int_{0}^{1} d x_1 \, \exp \left[- q^2 x_1 (1-x_1) T_0 \right] \nonumber \\
&&\times  \frac{1}{2} \int_{-\infty}^\infty dt_2 \, dt_4 \, \exp \left[  -\frac{(t_2\; t_4)}{T_0} \left(\begin{matrix} m_1^2 & \tilde{k}^2 \\ \tilde{k}^2 & m_2^2 \end{matrix} \right) \left(\begin{matrix} t_2 \\ t_4 \end{matrix} \right) \right] \nonumber \\
&&\times \frac{1}{8 T_0^4} \biggl\{ (|t_2|+|t_4|)^2 \biggl[-4 D T_0 \left(t_4 \left(2 \tilde{k}^2 t_2+m_2^2 t_4\right) \right. \nonumber \\
&&\left. +m_1^2 t_2^2+q^2 t_1 t_3\right)  +(D+2) D T_0^2+4 \left(t_4 \left(2 \tilde{k}^2 t_2+m_2^2 t_4\right) \right. \nonumber \\ 
&& \left. +m_1^2 t_2^2+q^2 t_1 t_3\right) \left(2 \tilde{k}^2 t_2 t_4+m_1^2 t_2^2+m_2^2 t_4^2+q^2 t_1 t_3-2 T_0 \right) \biggr] \nonumber \\
&& + 4 T_0^3 q^2 |t_2| |t_4| \biggr\} \;,
\end{eqnarray}
where to solve the various resulting integrals in the Gaussian integration over $t_2,t_4$ we refer to the building blocks computed in appendix \ref{app:aux}. Taking into account the contribution \eqref{eq:xtraconti4ltoi4ssl} and again substituting as per \eqref{eq:tildekexp} we find for the final result,
\begin{eqnarray}
&& \mathcal{I}_4^{(3)}(s,t) + \mathcal{I}_4^{(3)}(u,t) = \frac{i}{8(4\pi)^{\frac{D}{2}}} \Gamma \left(\frac{4-D}{2}\right)\, \frac{\Gamma^2\!\left(\frac{D-2}{2}\right)}{\Gamma(D-4)}  (q^2)^{\frac{D-4}{2}} \frac{1}{D-4} \nonumber \\
&& \times \biggl[ \frac{4(5-D)}{(k_1 k_2)^2 - m_1^2 m_2^2} \left( 1 + \frac{2k_1 k_2}{[(k_1k_2)^2 - m_1^2 m_2^2]^{1/2}} \text{arcsinh} \left( \sqrt{\frac{\sigma-1}{2}} \right) \right) \nonumber \\
&& + i \frac{\pi (D-4) (k_1+k_2)^2}{[(k_1k_2)^2 - m_1^2 m_2^2]^{3/2}} \biggr]  \;,
\end{eqnarray}
where as before we have used the definition, $\sigma = \frac{- k_1 k_2}{m_1 m_2}$. A curious feature of the expression above is that it is purely imaginary for $D=4$ since the last line representing the real component vanishes. This expression has been checked against known result in $D=4$ found in \cite{Ellis:2007qk}. In impact parameter space the above expression reads,
\begin{eqnarray}
&& \tilde{\mathcal{I}}_4^{(3)}(s,t) + \tilde{\mathcal{I}}_4^{(3)}(u,t) = \frac{i}{128 \pi^{D-1}} \Gamma^2\!\left(\frac{D-2}{2}\right) \frac{1}{(\mathbf{b}^2)^{D-3}} \nonumber \\
&& \times \biggl[ \frac{4(5-D)}{(k_1 k_2)^2 - m_1^2 m_2^2} \left( 1 + \frac{2k_1 k_2}{[(k_1k_2)^2 - m_1^2 m_2^2]^{1/2}} \text{arcsinh} \left( \sqrt{\frac{\sigma-1}{2}} \right) \right) \nonumber \\
&& + i \frac{\pi (D-4) (k_1+k_2)^2}{[(k_1k_2)^2 - m_1^2 m_2^2]^{3/2}} \biggr] \;.
\end{eqnarray}

\subsection{Triangle Integrals} \label{app:triintegrals}

In this subsection we will derive results for the integrals relevant for the triangle amplitudes. The first integral we need to calculate is,
\begin{eqnarray}
\mathcal{I}_{3}(m_i) &=& \int \!\frac{d^Dk}{(2\pi)^D}\, \frac{1}{k^2} \,\frac{1}{(q+k)^2} \,\frac{1}{(k+k_i)^2+m_i^2}  \;.
\end{eqnarray}
As in the previous subsection we can write this integral in terms of Schwinger parameters and perform the Gaussian integral over the loop momenta. This yields,
\begin{equation}
\mathcal{I}_3 = i \int_{0}^{\infty} \prod_{i=1}^{3}  \, d t_i \frac{T^{-\frac{D}{2}}}{(4\pi)^{\frac{D}{2}}} \exp \left[-\frac{ m_i^2 t_3^2 + q^2 t_1 t_2}{T} \right]\;, \label{eq:i3full}
\end{equation}
where $q\equiv k_1+k_3$ is the momentum exchanged and $T=\sum_{i} t_i$. We have written this integral in a suggestive way because we are interested in the limit where $m_i^2 \gg q^2$, this means that the integral over $t_3$ can be performed with a saddle point approximation around $t_3=0$. To make it easier to perform the relevant expansion we write $T=T_0+| t_3 |$ where $T_0=t_1+t_2$. Performing the saddle point approximation, we find at leading order,
\begin{eqnarray}
\mathcal{I}_{3}^{(1)}(m_i) &=& i \int_{0}^{\infty} d T_0 \frac{T_0^{-\frac{D}{2}+1}}{(4\pi)^{\frac{D}{2}}} \int_{0}^{1} dx_2 \exp \left[- q^2 T_0 x_2(1-x_2) \right] \int_{0}^{\infty} dt_3 \exp \left[- \frac{m_i^2 t_3^2}{T_0} \right] \nonumber \\
&=& \frac{i}{(4 \pi)^{\frac{D}{2}}} \frac{\sqrt{\pi}}{2 m_i} \Gamma \left(\frac{5-D}{2} \right)\, \frac{\Gamma^2\!\left(\frac{D-3}{2}\right)}{\Gamma(D-3)} (q^2)^{\frac{D-5}{2}} \;, \label{eq:i3m1}
\end{eqnarray}
where we have written $x_2=t_2/T_0$. Expanding \eqref{eq:i3full} further we find for the subleading contribution,
\begin{eqnarray}
\mathcal{I}_{3}^{(2)}(m_i) &=& i \int_{0}^{\infty} d T_0 \frac{T_0^{-\frac{D}{2}+1}}{(4\pi)^{\frac{D}{2}}} \int_{0}^{1} dx_2 \exp \left[- q^2 T_0 x_2(1-x_2) \right] \int_{0}^{\infty} dt_3 \exp \left[- \frac{m_i^2 t_3^2}{T_0} \right] \nonumber \\
&& \quad \times \frac{1}{2 T_0^2} |t_3| \left( 2 q^2 t_1 t_2 - D T_0 + 2 m_i^2 t_3^2 \right) \nonumber \\
&=& -\frac{i}{(4 \pi)^{\frac{D}{2}}} \frac{1}{2 m_i^2} \Gamma \left(\frac{4-D}{2} \right) \, \frac{\Gamma^2\!\left(\frac{D-2}{2}\right)}{\Gamma(D-3)} (q^2)^{\frac{D-4}{2}} \;. \label{eq:i3m1sub}
\end{eqnarray}
The equations above in impact parameter space read,
\begin{eqnarray}
\tilde{\mathcal{I}}_{3}^{(1)}(m_i) &=& \frac{i}{(\pi)^{D-\frac{3}{2}}} \frac{\sqrt{\pi}}{64 m_i} \Gamma \left(\frac{2D-7}{2} \right)\, \frac{\Gamma^2\!\left(\frac{D-3}{2}\right)}{\Gamma(D-3)} \frac{1}{(\mathbf{b}^2)^{D-\frac{7}{2}}} \;, \label{eq:i3m1ips}
\end{eqnarray}
and,
\begin{eqnarray}
\tilde{\mathcal{I}}_{3}^{(2)}(m_i) &=& -\frac{i}{\pi^{D-1}} \frac{1}{32 m_i^2} \, \Gamma^2\!\left(\frac{D-2}{2}\right) \frac{1}{(\mathbf{b}^2)^{D-3}} \;. \label{eq:i3m1subips}
\end{eqnarray}

We also want to consider the tensor triangle integral given by,
\begin{equation}
\mathcal{I}_{3}^{\mu \nu}(m_i) = \int \!\frac{d^Dk}{(2\pi)^D}\, \frac{1}{k^2}  \,\frac{1}{(q+k)^2} \, \frac{1}{(k_i+k)^2 + m_i^2} k^{\mu} k^{\nu} \;.
\end{equation}
Employing Schwinger parameters as before we find that,
\begin{eqnarray}
\mathcal{I}_3 &=& i \int_{0}^{\infty} \prod_{i=1}^{3}  \, d t_i \frac{T^{-\frac{D}{2}}}{(4\pi)^{\frac{D}{2}}} \exp \left[-\frac{ m_i^2 t_3^2 + q^2 t_1 t_2}{T} \right] \nonumber \\
&& \quad \times \left( \frac{1}{2T} \eta^{\mu \nu} + \frac{1}{T^2}(q t_2 + k_i t_3)^{\mu}(q t_2 + k_i t_3)^{\nu} \right) \;, \label{eq:imunu3full}
\end{eqnarray}
where the various symbols have been defined previously. Employing the same method as for the previous two integrals we find the following result at leading order,
\begin{eqnarray}
\mathcal{I}_{3}^{(1)\,\mu \nu}(m_i) &=& \frac{i}{4m_i} \frac{1}{(4\pi)^{\frac{D}{2}}}\biggl[ (q^2)^{\frac{D-3}{2}} \sqrt{\pi} \frac{\Gamma{\left( \frac{3-D}{2} \right)} \Gamma^2{\left( \frac{D-1}{2} \right)}}{\Gamma{\left( D-1 \right)}} \left( \eta^{\mu \nu} + \frac{k_i^{\mu} k_i^{\nu}}{m_i^2} \right. \nonumber \\
&& \left. - (D-1)\frac{q^{\mu}q^{\nu}}{q^2} \right) + 2(q^2)^{\frac{D-4}{2}} \frac{\Gamma{\left( \frac{4-D}{2} \right)} \Gamma{\left( \frac{D-2}{2} \right)}\Gamma{\left( \frac{D}{2} \right)}}{\Gamma{\left( D-1 \right)}} \frac{q^{(\mu}k_i^{\nu)}}{m_i} \biggr] \;. \nonumber \\ \label{eq:i3munuleading}
\end{eqnarray}
Although this is the result at leading order in the expansion around the saddle point $t_3=0$ we can identify the second line as subleading contributions to the integral in the limit given by \eqref{eq:heml}. We can see this by looking at how a contraction between $q$ and an external momentum behaves,
\begin{equation}
k_1^{\mu} q_{\mu} = k_1^{\mu} (k_{1 \mu}+k_{3 \mu}) = \frac{1}{2} (k_1+k_3)^2 = \frac{1}{2}q^2
\end{equation}
where we have used the fact that $k_1^2=k_3^2$. Power counting with the above relation identifies the last line of \eqref{eq:i3munuleading} as subleading. This type of argument extends to contractions between any external momenta and $q$ since we can always write $q=k_1+k_3=-k_2-k_4$.

For the next order in the expansion around the saddle point we have,
\begin{eqnarray}
\mathcal{I}_{3}^{(2)\,\mu \nu}(m_i) &=& \frac{-i}{(4\pi)^{\frac{D}{2}}} \frac{1}{4m_i^2} \biggl[ (q^2)^{\frac{D-2}{2}} \frac{\Gamma{\left( \frac{2-D}{2} \right)} \Gamma^2{\left( \frac{D}{2} \right)}}{\Gamma{\left( D-1 \right)}} \left( \eta^{\mu \nu} + \frac{2 k_i^{\mu} k_i^{\nu}}{m_i^2} \right. \nonumber \\
&& \left. - D \frac{q^{\mu}q^{\nu}}{q^2} \right) + \frac{\sqrt{\pi}(D-1)}{2}(q^2)^{\frac{D-3}{2}} \frac{\Gamma{\left( \frac{3-D}{2} \right)} \Gamma^2{\left( \frac{D-1}{2} \right)}}{\Gamma{\left( D-1 \right)}} \frac{q^{(\mu} k_i^{\nu)}}{m_i} \biggr] \;. \nonumber \\ \label{eq:i3munusubleading}
\end{eqnarray}
We note that as with \eqref{eq:i3munuleading}, the second line above is kinematically subleading with respect to the first line. Note that although we will not be evaluating the equivalent vector triangle integral $\mathcal{I}_3^{\mu}$, since we will not need it in the subsequent chapters, the calculation follows straightforwardly from the method outlined above. A parallel discussion in the case of bubble integrals is given in section \ref{integralref2prop}.

We can write the above expressions in impact parameter space as we have done with previous results. To make these expressions clear we will write them after we have contracted with external momenta. So we have,
\begin{eqnarray}\label{eq:i3munuleadingips}
k_{j \, \mu} k_{j \, \nu} \tilde{\mathcal{I}}_{3}^{(1)\,\mu \nu}(m_i) &=& \frac{i}{32m_i} \frac{1}{\pi^{D-\frac{3}{2}}} \biggl[ \frac{1}{(\mathbf{b}^2)^{D-\frac{5}{2}}} \frac{\Gamma{\left( \frac{2D-5}{2} \right)} \Gamma^2{\left( \frac{D-1}{2} \right)}}{\Gamma{\left( D-1 \right)}} \nonumber \\
&& \times \left( -m_j^2 + \frac{(k_i k_j)^2}{m_i^2} - (D-1) \frac{-2D^2 +7D +5}{4\mathbf{b}^2} \right) \nonumber \\
&& \quad + (-1)^{j+1} \frac{k_i k_j}{m_i} \frac{2}{(\mathbf{b}^2)^{D-2}} \frac{\csc \left( \frac{\pi D}{2} \right)  \Gamma{\left( \frac{D}{2} \right)}}{\Gamma{\left( \frac{4-D}{2} \right)}} \biggr] \;,
\end{eqnarray}
and for the subleading expression,
\begin{eqnarray}
k_{j \, \mu} k_{j \, \nu} \tilde{\mathcal{I}}_{3}^{(2)\,\mu \nu}(m_i) &=& \frac{-i}{16 m_i^2} \frac{1}{\pi^{D-1}} \biggl[ \frac{1}{(\mathbf{b}^2)^{D-2}} \frac{\Gamma^2{\left( \frac{D}{2} \right)}}{D-2} \left( -m_j^2 + \frac{2(k_i k_j)^2}{m_i^2} + \frac{D^2(D-2)}{2\mathbf{b}^2} \right) \nonumber \\
&& - (-1)^{j+1} \frac{k_i k_j}{m_i} \frac{2 \pi^{\frac{3}{2}}}{(\mathbf{b}^2)^{D-\frac{3}{2}}} \frac{\Gamma{\left( \frac{2D-3}{2} \right)} \Gamma{\left( \frac{D+1}{2} \right)} \sec \left( \frac{\pi D}{2} \right)}{\Gamma{\left( D-1 \right)} \Gamma{\left( \frac{1-D}{2} \right)} } \biggr] \;. \label{eq:i3munusubleadingips}
\end{eqnarray}

\section{Integrals for D-brane Scattering} \label{sec:dbraneintegrals}

In this section we will work out the high-energy limits of the integrals needed in chapter \ref{chap:sugraeik}. The discussion will heavily parallel the discussion in section \ref{app:integrals} and so some of the details are neglected. Note that we define the symbol, $\perp=D-p-1$, which will be used extensively below. 

\subsection{Scalar Triangle Integral}

For the scalar triangle integral, relevant when discussing the scattering of a massless state off of a stack of D-branes, we have,
\begin{equation}
\mathcal{I}_{3}(E, q_{\perp}) = \int \frac{d^{\perp} k}{(2\pi)^{\perp}}\frac{1}{k^2 (k_1-k)_{\perp}^2 (k+k_2)_{\perp}^2}  \;, \label{I3dbranefull}
\end{equation}
where the $\perp$ subscript refers to taking only the components in the space perpendicular to the D-branes and we have $k^2= k_{\perp}^2 - E^2$. As in the previous subsection we can write this integral in terms of Schwinger parameters and perform the Gaussian integral over the loop momenta. This yields,
\begin{equation}
\mathcal{I}_3(E, q_{\perp}) = \int_{0}^{\infty} \prod_{i=1}^{3}  \, d t_i \frac{T^{-\frac{D-p-1}{2}}}{(4\pi)^{\frac{D-p-1}{2}}} \exp \left[\frac{ E^2 t_1^2 - q_{\perp}^2 t_2 t_3}{T} \right]\;, \label{eq:i3dbranefullschw}
\end{equation}
where $q\equiv k_1+k_2$ is the momentum exchanged and $T=\sum_{i} t_i$. Note that we have used some of the kinematic relations given in section \ref{kinematics}. We have again written this integral in a suggestive way because we are interested in the limit where $E^2 \gg q^2$, this means that the integral over $t_1$ can be performed with a saddle point approximation around $t_1=0$. To make it easier to perform the relevant expansion we write $T=T_0+| t_1 |$ where $T_0=t_2+t_3$. Doing so we find at leading order,
\begin{eqnarray}
\mathcal{I}_{3}^{(1)}(E,q_{\perp}) &=& \int_{0}^{\infty} d T_0 \frac{T_0^{-\frac{D-p-1}{2}+1}}{(4\pi)^{\frac{D-p-1}{2}}} \int_{0}^{1} dx_3 \exp \left[- q_{\perp}^2 T_0 x_3(1-x_3) \right]  \nonumber \\
&& \quad \times \, \int_{0}^{\infty} dt_1 \exp \left[\frac{E^2 t_1^2}{T_0} \right] \nonumber \\
&=& \frac{\sqrt{\pi}}{(4\pi)^{\frac{D-p-1}{2}}} (q^2_{\perp})^{\frac{D-p-6}{2}}  \frac{i}{2 E} \frac{\Gamma{\left(\frac{6-D+p}{2} \right)} \Gamma^2 \left(\frac{D-p-4}{2} \right)}{\Gamma(D-p-4)} \;,
\end{eqnarray}
where we have written $x_3=t_3/T_0$. Expanding \eqref{eq:i3dbranefullschw} further we find for the subleading contribution,
\begin{eqnarray}
\mathcal{I}_{3}^{(2)}(E,q_{\perp}) &=& \int_{0}^{\infty} d T_0 \frac{T_0^{-\frac{D-p-1}{2}+1}}{(4\pi)^{\frac{D-p-1}{2}}} \int_{0}^{1} dx_2 \exp \left[- q_{\perp}^2 T_0 x_2(1-x_2) \right]  \nonumber \\
&& \, \times \, \int_{0}^{\infty} dt_1 \exp \left[\frac{E^2 t_3^2}{T_0} \right] \frac{1}{2 T_0^2} |t_1| \left( 2 q_{\perp}^2 t_2 t_3 - (D-p-1) T_0 - 2 E^2 t_1^2 \right) \nonumber \\
&=& - \frac{1}{(4\pi)^{\frac{D-p-1}{2}}} (q^2_{\perp})^{\frac{D-p-5}{2}}  \frac{1}{2 E^2} \frac{\Gamma{\left(\frac{5-D+p}{2} \right)} \Gamma^2 \left(\frac{D-p-3}{2} \right)}{\Gamma(D-p-4)} \;.
\end{eqnarray}
As we have done for the integrals relevant to the massive scalar process, we can find the corresponding expressions in impact parameter space. Using \eqref{ttoips} and \eqref{eq:impoformu} we find,
\begin{eqnarray}
\tilde{\mathcal{I}}_{3}^{(1)}(E,\mathbf{b}) &=& \frac{i}{64 \pi^{D-p-2} E} \Gamma^2 \left(\frac{D-p-4}{2} \right) \frac{1}{\mathbf{b}^{2D-2p-8}}\\
\tilde{\mathcal{I}}_{3}^{(2)}(E,\mathbf{b}) &=& - \frac{1}{32 \pi^{D-p-\frac{3}{2}} E^2} \frac{\Gamma^2 \left(\frac{D-p-3}{2} \right) \Gamma \left(\frac{2D-2p-7}{2} \right)}{\Gamma \left(D-p-4 \right)} \frac{1}{\mathbf{b}^{2D-2p-7}}  \;.
\end{eqnarray}

\subsubsection{Removing a Propagator}

We briefly review what happens to $\mathcal{I}_3$ when we remove the $k+k_2$ propagator in \eqref{I3dbranefull} to explicitly recognise that these contributions are localised on the D-branes when working in impact parameter space. We find after introducing Schwinger parameters,
\begin{equation}
\mathcal{I}_{3,3} = \int \frac{d^{\perp} k}{(2\pi)^{\perp}}\frac{1}{k^2 (k_1-k)_{\perp}^2} =  \int dt_1 dt_2 \left( \frac{\pi}{T} \right)^{\frac{\perp}{2}} \text{exp} \left[ \frac{t_2^2 k_{1 \perp}^2}{T} +t_1^2 E^2 - t_2 k_{1 \perp}^2 \right] \;,
\end{equation}
where $\mathcal{I}_{3,3}$ refers to the fact that we've killed the third propagator in the integral given by \eqref{I3dbranefull} and $T=t_1+t_2$ where $t_1$, $t_2$ are the Schwinger parameters. From this we can see that $\mathcal{I}_{3,3}=f(E)$ which means that the result is not a function of the momentum exchanged, $q_{\perp}$. If we calculate the impact parameter space expression for this integral we find that $\mathcal{\tilde{I}}_{3,3}= f(E) \delta^{\perp-1}(\mathbf{b})$, which as described before suggests that these types of terms can only produce contributions which are localised on the D-branes. Note that the same happens when you remove the second propagator $k_1-k$. However if one removes the propagator $k$ we find that $\mathcal{I}_{3,1}=\mathcal{I}_{2}$ as expected.

\subsection{Bubble Integrals} \label{integralref2prop}

For the scalar bubble integrals relevant when discussing the scattering of a massless state off of a stack of D-branes, we have,
\begin{eqnarray}
\mathcal{I}_{2}(q_{\perp}) &=& \int \frac{d^{\perp} k}{(2\pi)^{\perp}} \frac{1}{k_{\perp}^2 (k-q)_{\perp}^2} \;,
\end{eqnarray}
where as in the previous subsection the $\perp$ subscript refer to the components in the directions perpendicular to the D-branes. Continuing in the same way as we have done before we can rewrite this in terms of Schwinger parameters and perform the Gaussian integral over the loop momenta, yielding,
\begin{eqnarray}
\mathcal{I}_{2}(q_{\perp}) &=& \int_{0}^{\infty} \prod_{i=1}^{2}  \, d t_i \frac{T^{-\frac{D-p-1}{2}}}{(4\pi)^{\frac{D-p-1}{2}}} \exp \left[q_{\perp}^2 t_2 \left( \frac{t_2}{T} - 1 \right) \right]\;,
\end{eqnarray}
This can be solved exactly and we find,
\begin{eqnarray}\label{eq:bubblescalarint}
\mathcal{I}_{2}(q_{\perp}) &=& \frac{1}{(4\pi)^{\frac{D-p-1}{2}}} (q^2_{\perp})^{\frac{D-p-5}{2}}  \frac{\Gamma{\left( \frac{3-D+p}{2} \right)} \Gamma^2{\left( \frac{D-p-1}{2} \right)}}{ \Gamma{\left( D-p-1 \right)}} (-2(D-p-2))  \;.
\end{eqnarray}
We can perform an equivalent procedure for the vector bubble integral,
\begin{eqnarray}
\mathcal{I}_{2}^{\mu}(q_{\perp}) &=& \int \frac{d^{\perp} k}{(2\pi)^{\perp}} \frac{k^{\mu}}{k_{\perp}^2 (k-q)_{\perp}^2} \nonumber \\
&=&  \int_{0}^{\infty} \prod_{i=1}^{2}  \, d t_i \frac{T^{-\frac{D-p-1}{2}}}{(4\pi)^{\frac{D-p-1}{2}}} \, \frac{q_{\perp}^{\mu} t_2}{T} \exp \left[q_{\perp}^2 t_2 \left( \frac{t_2}{T} - 1 \right) \right] \nonumber \\
&=& \frac{1}{(4\pi)^{\frac{D-p-1}{2}}} (q^2_{\perp})^{\frac{D-p-5}{2}} q_{\perp}^{\mu} \frac{\Gamma{\left( \frac{5-D+p}{2} \right)} \Gamma{\left( \frac{D-p-1}{2} \right)} \Gamma{\left( \frac{D-p-3}{2} \right)}}{\Gamma{\left( D-p-2 \right)}} \;.
\end{eqnarray}
Similarly for the tensor bubble integral,
\begin{eqnarray}
\mathcal{I}_{2}^{\mu \nu}(q_{\perp}) &=& \int \frac{d^{\perp} k}{(2\pi)^{\perp}} \frac{k^{\mu}k^{\nu}}{k_{\perp}^2 (k-q)_{\perp}^2}  \nonumber \\
&=&  \int_{0}^{\infty} \prod_{i=1}^{2}  \, d t_i \frac{T^{-\frac{D-p-1}{2}}}{(4\pi)^{\frac{D-p-1}{2}}}  \left( \frac{1}{2T}  \eta^{\mu \nu}_{\perp} + \frac{t_2^2}{T^2} q_{\perp}^{\mu} q_{\perp}^{\nu} \right) \exp \left[q_{\perp}^2 t_2 \left( \frac{t_2}{T} - 1 \right) \right] \nonumber \\
&=&\frac{1}{(4\pi)^{\frac{D-p-1}{2}}} (q^2_{\perp})^{\frac{D-p-3}{2}}  \frac{\Gamma{\left( \frac{3-D+p}{2} \right)} \Gamma^2{\left( \frac{D-p-1}{2} \right)}}{2 \Gamma{\left( D-p-1 \right)}}  \nonumber \\
&& \quad \times \, \bigl( \eta^{\mu \nu}_{\perp} - (D-p-1) \frac{q_{\perp}^{\mu}q_{\perp}^{\nu}}{q_{\perp}^2} \bigl) \;.
\end{eqnarray}
The corresponding impact parameter space expressions are found to be,
\begin{eqnarray}
\tilde{\mathcal{I}}_{2}(\mathbf{b}) &=& \frac{1}{8 \pi^{D-p-\frac{3}{2}}}  \frac{\Gamma{\left( \frac{2D-2p-7}{2} \right)} \Gamma^2{\left( \frac{D-p-1}{2} \right)}}{ \Gamma{\left( D-p-1 \right)}}\frac{2(p-D+2)}{3-D+p} \frac{1}{\mathbf{b}^{2D-2p-7}} \;.
\end{eqnarray}
We neglect to write the impact parameter expressions for the vector and tensor bubble integrals because they are only meaningful when their exact form is calculated and this depends on what object the integrals are contracting with.

\section{Impact Parameter Space Integral}\label{sec:ipsintegrals}

In this section we will discuss the result for the integral needed to calculate the impact parameter space expressions discussed above and in chapters \ref{chap:sugraeik}, \ref{chap:kkeik} and \ref{chap:graveik}. We are interested in solving the integral,
\begin{eqnarray}
\mathcal{I}_{\text{IPS}} = \int \frac{d^d \mathbf{q}}{(2\pi)^d} {\rm e}^{i  \mathbf{q}  \mathbf{b}} (q^2)^\nu \;.
\end{eqnarray}
To evaluate this integral we introduce hyper spherical coordinates \cite{wolf2013integral} where we let the angle between $\mathbf{q}$ and $\mathbf{b}$ be denoted by $\theta_{d-1}$ and the magnitude of $\bf{q}$ be denoted by $q$. We can then write,
\begin{eqnarray}
\mathcal{I}_{\text{IPS}} &=& \int \frac{d^d \mathbf{q}}{(2\pi)^d} e^{iq b \cos{\theta_{d-1}}} (q^2)^\nu \nonumber \\
& = & \frac{1}{(2\pi)^{d}} \frac{2 \pi^{\frac{d-1}{2}}}{\Gamma(\frac{d-1}{2})} \int_0^{\infty} dq \, q^{2\nu} q^{d-1} \int_0^{\pi} d \theta_{d-1} \sin^{d-2}(\theta_{d-1}) e^{iq b\cos{\theta_{d-1}}} \nonumber \\
&=& \frac{1}{(2\pi)^{\frac{d}{2}}}  b^{1-\frac{d}{2}} \int_0^{\infty} dq \, q^{2\nu} q^{\frac{d}{2}} J_{\frac{d-2}{2}}(q b) \;,
\end{eqnarray}
where in the second line we have evaluated the angular integral excluding the $\theta_{d-1}$ integral and in the last line we have evaluated the $\theta_{d-1}$ integral. The $J_n(x)$ is a Bessel function of the first kind. Evaluating the final line above yields,
\begin{eqnarray}
\mathcal{I}_{\text{IPS}} = \int \frac{d^d \mathbf{q}}{(2\pi)^d} {\rm e}^{i  \mathbf{q}  \mathbf{b}} ( q^2)^\nu = \frac{2^{2\nu}}{\pi^{d/2}}
\frac{\Gamma ( \nu + \frac{d}{2})}{\Gamma (-\nu)} 
\frac{1}{ ( \mathbf{b}^2)^{\nu + \frac{d}{2}}} \;. \label{eq:impoformu}
\end{eqnarray}

\chapter{The Eikonal in Supergravity}\label{chap:sugraeik}

This chapter is based on the paper \cite{Collado:2018isu} where we discuss the probe-limit eikonal in supergravity. The specific scenario we study is the scattering of massless scalar states off of a heavy stack of D$p$-branes. Through studying this we are able to derive some properties of the inelastic contributions to the eikonal which generalises some of the ideas presented in chapter \ref{chap:background}.

The chapter is structured as follows. In section~\ref{1braneamps} we briefly review the kinematics of the process under study and provide the results for the tree-level amplitudes describing the elastic dilaton to dilaton and the inelastic dilaton to RR scatterings. In section~\ref{2braneamps} we study the one-loop diagrams that contribute to the same processes. We perform the calculation in two ways; one is the traditional approach of using Feynman rules, while in a second approach we provide a prescription for treating onshell bulk amplitudes as effective vertices and gluing them to the D$p$-branes. We check that these two approaches provide the same classical eikonal since they agree at the level of the amplitudes except for possible contributions that are localised on the D$p$-branes (i.e. terms that are proportional to a delta function in the impact parameter space). In section~\ref{HElimit} we study the Regge high energy limit of the amplitudes we derived and, as mentioned above, provide a prescription to extract the classical eikonal at subleading orders from the amplitude. In section~\ref{lntoeikonal} we rederive the same diagrams analysed in section~\ref{2braneamps} in a slightly different way, which allows us to extract the classical solution representing the gravitational backreaction of the target D$p$-branes. In this section we also compare the eikonal with the appropriate classical deflection angle. In all our calculations the contributions of the different fields are separated, thus it is straightforward to focus just on the graviton exchanges and obtain both the metric and the deflection angle for pure Einstein gravity which agrees with the results in the literature \cite{Bjerrum-Bohr:2016hpa,Akhoury:2013yua,Luna:2016idw} as well as the more general discussion for pure Einstein gravity presented in chapter \ref{chap:graveik}.

\section{Scattering in the Born Approximation} \label{1braneamps}

In the Born approximation the interaction between a perturbative state and a stack of D$p$-branes is described by a tree-level diagram with two external states~\cite{Klebanov:1995ni,Garousi:1996ad,Hashimoto:1996bf}. In the limit where the distance between the D$p$-branes and the external states is large, this interaction is captured by a tree-level Feynman diagram with the exchange of a single massless state between the D$p$-branes and a bulk three-point vertex. In this section, we briefly summarise the kinematics of this interaction and then discuss its large energy behaviour. The leading term in this limit is dominated by the exchange of the particles with the highest spin; here we focus mainly on the field-theory limit of the full string setup and so the highest spin state is the graviton. This leading term is elastic and universal, {\em i.e.} the polarisation of the in and the out states are identical and, the result depends only on the momentum exchanged and the energy density of the D$p$-brane target. 

In this section we are also interested in the first subleading correction in the large energy limit. As expected, this contribution depends on the exchange of lower spin states, such as the Ramond-Ramond forms in supergravity. This means that the result depends on other features (besides the energy density) of the target D$p$-brane, such as its charge density or its angular momentum. In general at this order, the transition is not elastic and so displays a non-trivial Lorentz structure. As mentioned in the introduction, this result will be important in defining the eikonal limit beyond the leading order in the large distance limit.

\subsection{Kinematics} \label{kinematics}

We can write the momenta of the two massless external particles scattering off a stack of D$p$-branes as follows, 
\begin{equation}
k_1=(E,\ldots,E) \qquad k_2=(-E,\ldots,\mathbf{q}, -E + q_{D-1}) \text{ ,}
\label{k1k2}
\end{equation}
where the dots are over the $p$ spatial components along the D$p$-brane in $k_2$, $\mathbf{q}$ denotes the $D-p-2$ spatial components transverse to the direction of the incoming particle of the momentum exchange vector $q=k_1+k_2$ and $q_{D-1}$ is the last component of $q$. Note that in the Regge limit 
\begin{equation}
  \label{eq:ReggeLd}
  q^2 \ll E^2,
\end{equation}
and $q_{D-1}$ is of order $E^{-1}$. Writing out the explicit kinematics as above we can see that $(k_1)_{\parallel}^2=(k_2)_{\parallel}^2=-E^2$ and $(k_1 \cdot k_2)_{\parallel}=E^2$.

\subsection{Elastic and Inelastic Diagrams} \label{inelex1}

The elastic scattering of a dilaton with a graviton being exchanged with the D-branes can easily be calculated in supergravity by using the Feynman rules in the relevant subsections of~\ref{sec:quantumgrav} 
\begin{equation}
  \label{eq:dil-dil0}
  A^{\rm dd}_1 = i (2\pi)^{p+1} \delta^{p+1}(k_1+k_2)\; {\cal A}^{\rm dd}_1\,,~~\mbox{where}~~ {\cal A}^{\rm dd}_1= \frac{2 NT_p \kappa_D E^2}{q^2}~,
\end{equation}
where $N$ is the number of D-branes in the stack. Notice that the result does not depend on the dimensionality $p$ of the D-branes. In the limit~\eqref{eq:ReggeLd}, the leading energy contribution of any elastic scattering is still described by~\eqref{eq:dil-dil0} multiplied by a kinematic factor forcing the polarisation of the ingoing and outgoing polarisation to be the same (for instance, $\epsilon_{1 \mu}^{~~\nu} \epsilon_{2 \nu}^{~~\mu}$ in the graviton-graviton case). For general states there are subleading energy corrections to this formula, but they start at order $E^0$. 

In the inelastic case, in contrast, it is possible to have order $E$ contributions. As an example, let us start from the amplitude where the incoming particle is a dilaton and the outgoing one is an RR state. Again the first amplitude contributing to this process can be derived by using the Feynman rules in the relevant subsections of~\ref{sec:quantumgrav}
\begin{equation}
  \label{eq:dil-RR0}
   A^{\rm dR}_1 = i (2\pi)^{p+1} \delta^{p+1}(k_1+k_2)\ \; {\cal A}^{\rm dR}_1\,,~~\mbox{where}~~{\cal A}^{\rm dR}_1= \frac{2 a(D) N T_p \kappa_D E \, q^\mu C_{\mu 1\ldots p}}{q^2}~,
\end{equation}
where $C_{\mu_1 \ldots \mu_{p+1}}$ is the polarisation of the RR potential describing the second external state and $a(D)$ is defined in \eqref{eq:aDforSUGRA}; in 10D type II supergravity we find $a(D=10)=\frac{p-3}{2}$.

Notice that it is possible to derive the same results by using a different approach that uses only on onshell data. The idea is simply to start from an onshell 3-particle vertex in the bulk\footnote{Strictly speaking the onshell vertices between three massless states often vanish in Minkowski space; as usual, one can define a non-trivial three-point vertex by analytic continuation on the momenta or equivalently by thinking of changing the spacetime signature.} instead of using the full Feynman rules. As an example, consider the vertex with two dilatons and one graviton~\eqref{ddgV}: on shell we can ignore the term proportional to $k_1 \cdot k_2 = (k_1+k_2)^2/2=q^2/2$, where $q$ is the momentum of the graviton. When this effective vertex is used in a diagram, we exploit the condition $q^2=0$ to simplify the numerator of the momentum space amplitude. Terms proportional to $q^2$ appearing in the standard Feynman diagram calculation would produce contributions localised on the D-branes as they cancel the pole of the massless propagator $1/q^2$, so we can ignore them for our purposes. Indeed, by using the onshell two dilatons and one graviton vertex, the standard propagator~\eqref{eq:dedop}, multiplying by $-T_p \eta_{\parallel}^{\rho \sigma}$ for the boundary coupling and imposing the onshell conditions in the numerator obtained in this way, one can easily reproduce~\eqref{eq:dil-dil0} up to terms that do not depend on $q$ and so are localised on the D-branes in the impact parameter space (after performing the Fourier transform~\eqref{ttoips}). 

We conclude this section by mentioning that it is possible to write the amplitudes above including all string theory corrections simply by implementing the following change to the expression ${\cal A}_1$ above
\begin{eqnarray}
T_p \kappa_D & \to & T_p \kappa_D\; \frac{\Gamma\left(1-\alpha' E^2\right) \Gamma\left(1+ \frac{\alpha' q^2}{4}\right)}{\Gamma\left(1-\alpha' E^2 + \frac{\alpha' q^2}{4} \right)}\nonumber \\
&\sim &  T_p \kappa_D\; \Gamma\left(1 + \frac{\alpha'q^2}{4}\right) e^{i \pi \frac{\alpha' q^2}{4}} (\alpha' E^2)^{1-\frac{ \alpha' q^2}{4}}\;,   \label{eq:scD}
\end{eqnarray}
where in the final step we have written the result explicitly in the Regge limit.

\section{Double Exchange Scattering} \label{2braneamps}

\begin{figure}[h]
  \centering
  \begin{tikzpicture}[scale=1.5]
    \begin{feynman}
      \vertex[blob, minimum size=1.5cm] (m) at ( 0, 0) {\contour{white}{}};
      \vertex (a) at (-1,-2) {};
      \vertex (b) at ( 1,-2) {};
      \vertex (c) at (-2, 0) {};
      \vertex (d) at ( 2, 0) {};
	  \draw[fill=light-gray] (-1,-2) ellipse (0.5cm and 0.25cm);
	  \draw[fill=light-gray] (1,-2) ellipse (0.5cm and 0.25cm);       
      \diagram* {
      (a) -- [fermion,edge label=$k_3$] (m),
      (b) -- [fermion,edge label=$k_4$,swap] (m),
      (c) -- [fermion,edge label=$k_1$] (m),
      (d) -- [fermion,edge label=$k_2$, swap] (m),
      };
    \end{feynman}
  \end{tikzpicture}
  \caption{A schematic diagram showing our procedure for calculating effective one-loop amplitudes. The circular blob represents the four-point effective vertex and the two oval blobs represent the D-branes. The four-point vertex is sewed with the D-branes by using the appropriate propagator and boundary coupling. \label{fig:1lc}}
\end{figure}
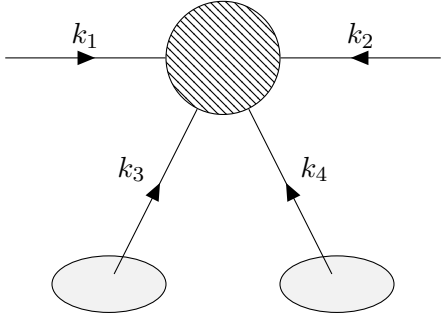

In this section we use the onshell approach mentioned in the previous section to calculate the  amplitudes with a double exchange of particles between the probe and the D-branes. As before we are interested in the classical limit where the gravitational constant is small; $\kappa_D \to 0$, with $N T_p \kappa_D$ fixed. The general idea is that we can use the bulk four-point amplitudes $\mathcal{A}_{bulk}$ as effective vertices and sew them with the relevant propagator to the D-branes, so as to construct diagrams such as the one sketched schematically  in figure~\ref{fig:1lc}.

Again this procedure requires an offshell extension of the bulk four-point vertex, but, as we argue below, the ambiguity related to this step is irrelevant for the large distance (small $q$) scattering. Thus after sewing the D-brane boundary couplings to the relevant external legs of the onshell effective vertex, schematically, we can write the double exchange amplitude as,
\begin{equation}
\mathcal{A}_{2} = \int \frac{d^{\perp}k_i}{(2 \pi)^{\perp}} \frac{d^{\perp}k_j}{(2 \pi)^{\perp}} \left( \frac{1}{2} [B_i] \frac{[G_i]}{k_i^2}\; [B_j] \frac{[G_j]}{k_j^2} \; \delta^{\perp}{(k_i + k_j - q)} \mathcal{A}_{bulk}(k_1,\ldots,k_4)\right) \text{ ,}\label{attachbtoamp}
\end{equation}
where $\perp=D-p-1$ is the number of directions transverse to the D-branes and the overall factor of $1/2$ is a symmetry factor due to the two identical sources. Here we have ``attached" the $i^{th}$ and $j^{th}$ external leg by using the boundary couplings $[B_i]$ and the standard propagators $[G_i]$, see the relevant subsections of~\ref{sec:quantumgrav}. Terms proportional to $k_i^2$ or $k_j^2$ in $\mathcal{A}_{bulk}$ are absent in the onshell result and would kill one of the propagators attached to the D-branes. We then have a variation of the cancelled propagator argument discussed after~\eqref{eq:dil-RR0}; terms without one of the propagators attached to the D-branes either yield integrals without scale and so can be set to zero in dimensional regularisation or can only produce contributions that are independent of $q$ and so are localised on the D-branes.

\begin{figure}[h]
  \begin{subfigure}[t]{0.29\textwidth}
    \centering
    \begin{tikzpicture}
	    \begin{feynman}
			\vertex (a) at (-1,-2) {};
			\vertex (b) at ( 1,-2) {};
			\vertex (c) at (-2.4, 0) {};
			\vertex (d) at ( 2.4, 0) {};
			\vertex[circle,inner sep=0pt,minimum size=0pt] (e) at (-1, 0) {};
			\vertex[circle,inner sep=0pt,minimum size=0pt] (f) at (1, 0) {};
			\draw[fill=light-gray] (-1,-2) ellipse (0.5cm and 0.25cm);
			\draw[fill=light-gray] (1,-2) ellipse (0.5cm and 0.25cm);       
			\diagram* {
			(c) -- [fermion,edge label=$k_1$] (e) -- [scalar] (f) -- [anti fermion,edge label=$k_2$] (d),
			(a) -- [scalar] (e),
			(b) -- [scalar] (f),
			};
	    \end{feynman}
    \end{tikzpicture}
    \caption{}
    \label{fig:2a}
  \end{subfigure}
  \quad
  \begin{subfigure}[t]{0.29\textwidth}
    \centering
    \begin{tikzpicture}
		\begin{feynman}
			\vertex (a) at (-1,-2) {};
			\vertex (b) at ( 1,-2) {};
			\vertex (c) at (-2.4, 0) {};
			\vertex (d) at ( 2.4, 0) {};
			\vertex[circle,inner sep=0pt,minimum size=0pt] (e) at (0, -1) {};
			\vertex[circle,inner sep=0pt,minimum size=0pt] (m) at (0, 0) {};
			\draw[fill=light-gray] (-1,-2) ellipse (0.5cm and 0.25cm);
			\draw[fill=light-gray] (1,-2) ellipse (0.5cm and 0.25cm);       
			\diagram* {
			(c) -- [fermion,edge label=$k_1$] (m) -- [anti fermion,edge label=$k_2$] (d),
			(m) -- [boson] (e),
			(a) -- [scalar] (e),
			(b) -- [scalar] (e),
			};
		\end{feynman}
    \end{tikzpicture}
    \caption{}
    \label{fig:2b}
  \end{subfigure}
  \quad
  \begin{subfigure}[t]{0.29\textwidth}
    \centering
    \begin{tikzpicture}
    	\begin{feynman}
			\vertex (a) at (-1,-2) {};
			\vertex (b) at ( 1,-2) {};
			\vertex (c) at (-2.4, 0) {};
			\vertex (d) at ( 2.4, 0) {};
			\vertex[circle,inner sep=0pt,minimum size=0pt] (e) at (0, -1) {};
			\vertex[circle,inner sep=0pt,minimum size=0pt] (m) at (0, 0) {};
			\draw[fill=light-gray] (-1,-2) ellipse (0.5cm and 0.25cm);
			\draw[fill=light-gray] (1,-2) ellipse (0.5cm and 0.25cm);       
			\diagram* {
			(c) -- [fermion,edge label=$k_1$] (m) -- [anti fermion,edge label=$k_2$] (d),
			(a) -- [scalar] (m),
			(b) -- [scalar] (m),
			};
		\end{feynman}
    \end{tikzpicture}
    \caption{}
    \label{fig:2c}
  \end{subfigure}
  \caption{The various topologies of diagrams that contribute to $\mathcal{A}^{\rm ddRR}$. In \ref{fig:2a} we have the t- and u-channels, in \ref{fig:2b} we have the s-channel diagram and finally in \ref{fig:2c} we have the contact diagram. The solid lines represent dilatons, wavy lines represent gravitons and the dashed lines represent RR fields.}
  \label{fig:2}
\end{figure}
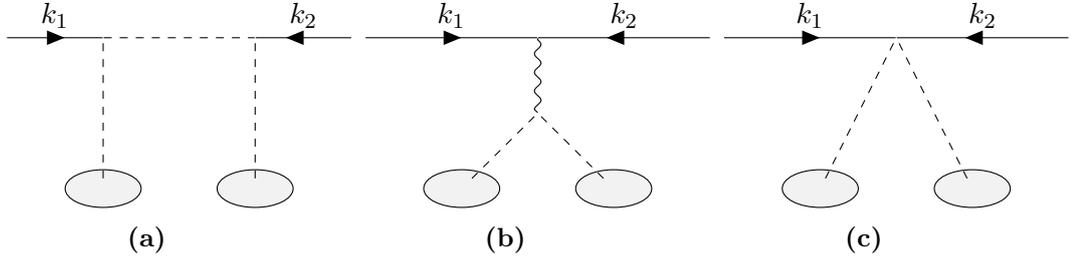

In section \ref{sec:dbraneintegrals} we have calculated the integrals that are relevant for the amplitude in figure~\ref{fig:1lc}. For instance the 3-propagator integral~\eqref{I3dbranefull}, relevant for the diagrams in figure~\ref{fig:2a} and figure~\ref{fig:3a}, is
\begin{equation}
  \mathcal{I}_{3}(q_{\perp}) = \int \frac{d^{\perp} k}{(2\pi)^{\perp}}\frac{1}{k^2 (k_1-k)_{\perp}^2 (k+k_2)_{\perp}^2}~.
 \end{equation}
We can see that, when one of the perpendicular propagators is cancelled, the integral can only depend on the quantities $k_1$ (or $k_2$), $\eta_{\mu \nu}$ and $\eta_{\parallel \mu \nu}$. Since these terms do not depend on the exchanged momentum $q=k_1+k_2$, we find that in impact parameter space their contributions are delta functions. The presence of factors proportional to $q$ in the numerator (arising from the vertices) does not spoil the argument, as in this case the impact parameter result is proportional to derivatives of the delta function and so is still localised on the D-branes. Notice that this cancelled propagator argument generalises to the ladder type diagrams with any number of propagators. Another type of integral appearing in the explicit evaluation of~\eqref{attachbtoamp} is (see section~\ref{integralref2prop})
\begin{equation}
  \mathcal{I}_{2}(q_{\perp}) = \int \frac{d^{\perp} k}{(2\pi)^{\perp}} \frac{1}{k_{\perp}^2 (k-q)_{\perp}^2}\;.
\end{equation}
In this case, if one of the two propagators in the integrand is cancelled, one obtains an integral without a scale and so again the ambiguities related to the offshell extension of the four-point bulk amplitudes are irrelevant for the calculation we are interested in.

In order to complete the argument and show that using the bulk four-point amplitude in~\eqref{attachbtoamp} is sufficient for our purposes, one should consider also the transverse conditions that are enforced on the onshell vector and graviton fields, such as $k_i^\mu \epsilon^{(i)}_{\mu\nu}=0$ for the case of a graviton. This same issue does not arise when attaching RR fields as we will see in subsection \ref{AttachRR}. We will discuss this point in more detail in the following subsections where we calculate~\eqref{attachbtoamp} explicitly for the elastic dilaton-brane scattering amplitudes when the interaction is mediated by RR fields, gravitons and dilatons.

\subsection{Dilaton to Dilaton Elastic Scattering}

We first apply the approach sketched above to the elastic dilaton-brane scattering deriving the full subleading amplitude. From a diagrammatic point of view there are three types of contributions due to the exchange of RR, graviton and dilaton fields between the external particles and the D-branes. We will also compare these results with those obtained from using the supergravity Feynman rules outlined in the relevant subsections of~\ref{sec:quantumgrav}.

\subsubsection{RR Sources} \label{AttachRR}

We start by analysing the RR exchange. By using the four-point two NS-NS (with these states taken to be dilatons), two RR closed string amplitude found in \cite{Bakhtiarizadeh:2013zia} we obtain, in the field theory limit, the following onshell vertex
\begin{eqnarray}
i \mathcal{A}_{bulk}^{\rm ddRR} &=& \frac{i \kappa_D^2}{2}\frac{1}{n!} \frac{2}{s t u} \Bigl[a(D) \, s t u F_{34} + n F_{34}^{\alpha \mu} \Bigl(a(D) \, st k_{2 \alpha} k_{3 \mu}   \nonumber \\
&&  + a(D) \, su k_{2 \mu} k_{4 \alpha} + (2 a^2(D) \, s^2 -8tu)k_{2 \alpha}k_{2 \mu} \Bigr) \Bigr] \nonumber \\
&=& \frac{i \kappa_D^2}{n!} \Bigl[a(D) \, F_{34} + n F_{34}^{\alpha \mu} \Bigl(a(D) \, \frac{1}{u} k_{2 \alpha} k_{3 \mu}   \nonumber \\
&&  + a(D) \, \frac{1}{t} k_{2 \mu}k_{4 \alpha} + \left(2 a^2(D) \, \left(-\frac{1}{t} - \frac{1}{u} \right) - \frac{8}{s} \right) k_{2 \alpha}k_{2 \mu} \Bigr) \Bigr] \text{ ,} \label{STampRR}
\end{eqnarray}
where the various symbols are defined in the relevant subsections of~\ref{sec:quantumgrav} and $n=p+2$. In order to properly attach the D-branes to $\mathcal{A}_{bulk}^{\rm ddRR}$ we need to express $F_{34}^{\alpha \mu}$ as,
\begin{eqnarray}
F_{34}^{\mu \nu} &=& F_3^{\mu \mu_1 \ldots \mu_{n-1}} F_{4 \mu_1 \ldots \mu_{n-1}}^{\nu} \nonumber \\
&=& \left( k_3^{\mu} C^{(3) \mu_1 \ldots \mu_{n-1}} + (-1)^{n-1} k_3^{\mu_1} C^{(3) \mu_2 \ldots \mu_{n-1} \mu} + \ldots \right) \nonumber \\
&& \times \left( k_4^{\nu} C^{(4)}_{\mu_1 \ldots \mu_{n-1}} + (-1)^{n-1} k_{4 \mu_1} C^{(4)}_{\mu_2 \ldots \mu_{n-1}}{}^{\nu} + \ldots \right) \nonumber \\
&=&  k_3^{\mu} k_4^{\nu} C^{(3) \mu_1 \ldots \mu_{n-1}} C^{(4)}_{\mu_1 \ldots \mu_{n-1}} + (n-1) k_3 \cdot k_4 C^{(3) \mu \mu_2 \ldots \mu_{n-1}} C^{(4) \nu}_{\mu_2 \ldots \mu_{n-1}}  \;, \nonumber \\ 
\end{eqnarray}
where we have used the facts that $k_3$ and $k_4$ only have components perpendicular to the D-branes and $C^{(3)}$ and $C^{(4)}$ only have components parallel to the D-branes, which therefore implies that $k_i \cdot C^{(j)} = 0$. We now take derivatives with respect to the gauge fields to make this an effective vertex to use when we attach the D-branes to the RR fields. We need to also take into account the different sets of labels that the $C$ fields can carry, i.e. $\mu_1\mu_2 \ldots \mu_{n-1} = 01\ldots p$ or $\mu_1\mu_2 \ldots \mu_{n-1} = 12 \ldots p0$,  etc., for which we note there are $(n-1)!$ sets of possible labels for $C^{(3) \mu_1 \ldots \mu_{n-1}} C^{(4)}_{\mu_1 \ldots \mu_{n-1}}$, as there are $(n-1)$ contracted indices, and $(n-2)!$ for $C^{(3) \mu \mu_2 \ldots \mu_{n-1}} C^{(4) \nu}_{\mu_2 \ldots \mu_{n-1}}$. Putting this together allows us to write,

\begin{eqnarray}
F_{34}^{\mu \nu} &=& (n-1)! k_3^{\mu} k_4^{\nu}  + (n-1)(n-2)! k_3 \cdot k_4 \eta_{\parallel}^{\mu \nu} \nonumber \\
&=& (n-1)!(k_3^{\mu} k_4^{\nu}  + k_3 \cdot k_4 \eta_{\parallel}^{\mu \nu}) \text{ .} \label{F12braneattach}
\end{eqnarray}
From the last line above we can also deduce that $F_{34} = n! k_3 \cdot k_4$. Note that these expressions only hold when both RR fields are attached to the D-branes. We can now use \eqref{F12braneattach} as well as $(k_2)_{\parallel}^2=-E^2$ and $(k_2 \cdot k_3)_{\parallel}=(k_2 \cdot k_4)_{\parallel} = 0$ to rewrite the contribution to \eqref{attachbtoamp} due to the onshell vertex \eqref{STampRR} when the RR fields are attached to the D-branes. In this case the integrand of \eqref{attachbtoamp} then reads
\begin{eqnarray}
 \frac{i (N \mu_p\kappa_D)^2}{2} \frac{1}{n!} \frac{(n-1)!}{k_3^2 k_4^2} \Bigl[ 2 a^2(D) n \left( \frac{s}{4} + \frac{s^2}{2 t u}E^2 \right) + n \left( -\frac{2 t u}{s} - 4 E^2 \right) \Bigr] \text{ ,} \label{FullRRSTAmp1}
\end{eqnarray}
where $N$ has been inserted to take into account the $N$ D-branes in the stack. We can write the full answer in terms of the momentum integrals defined in section \ref{sec:dbraneintegrals},
\begin{equation}
i \mathcal{A}_{2}^{\rm ddRR} = i (N T_p\kappa_D)^2 \Bigl[ 2 a^2(D) \, \left( \frac{s}{4} \mathcal{I}_2 + s E^2 \mathcal{I}_3 \right) - \left( \frac{8}{s} k_{1 \mu} k_{2 \nu} \mathcal{I}_{2}^{\mu \nu} + 4 E^2 \mathcal{I}_2 \right) \Bigr] \text{ .} \label{FullRRSTAmp2}
\end{equation}

We want to compare \eqref{FullRRSTAmp2} with the equivalent result arising from performing the same calculation using Feynman diagrams. We can calculate all the relevant onshell Feynman diagrams for this process. The four contributions to the full amplitude are given by,
\begin{eqnarray}
i \mathcal{A}^{\rm ddRR}_{{\rm FT}, t} &=&  [V_{\phi_1 F^{(n)}_{3} C^{(n-1)}}]_{\mu_2 \ldots \mu_n} [V_{\phi_2 F^{(n)}_{4} C^{(n-1)}}]^{\mu_2 \ldots \mu_n} [G_{C^{(n-1)}}] \nonumber \\
&=& \frac{i \kappa_D^2}{(n-1)!} \frac{2 a^2(D)}{(k_1+k_3)^2} F_{34}^{\mu \nu} (k_1+k_3)_{\mu} (k_1+k_3)_{\nu} \label{FTampRR1} \\ 
&& \nonumber \\
i \mathcal{A}^{\rm ddRR}_{{\rm FT}, u} &=&  [V_{\phi_2 F^{(n)}_{3} C^{(n-1)}}]_{\mu_2 \ldots \mu_n} [V_{\phi_1 F^{(n)}_{4} C^{(n-1)}}]^{\mu_2 \ldots \mu_n} [G_{C^{(n-1)}}] \nonumber \\
&=& \frac{i \kappa_D^2}{(n-1)!} \frac{2 a^2(D)}{(k_1+k_4)^2} F_{34}^{\mu \nu} (k_1+k_4)_{\mu} (k_1+k_4)_{\nu}  \label{FTampRR2} \\
&& \nonumber \\
i \mathcal{A}^{\rm ddRR}_{{\rm FT}, s} &=&  [V_{\phi_1 \phi_2 h}]_{\mu \nu} [G_{h}]^{\mu \nu ; \rho \sigma} [V_{F^{(n)}_{3} F^{(n)}_{4} h}]_{\rho \sigma} \nonumber \\
&=& - \frac{2i \kappa_D^2}{n!} \frac{1}{(k_1+k_2)^2} (n F_{34}^{\mu \nu} (k_{1 \mu} k_{2 \nu} + k_{2 \mu} k_{1 \nu}) - k_1 \cdot k_2 F_{34}) \label{FTampRR3} \\
&& \nonumber \\
i \mathcal{A}^{\rm ddRR}_{{\rm FT}, c} &=& [V_{\phi_1 \phi_2 F^{(n)}_{3} F^{(n)}_{4}}] \nonumber \\
&=& - \frac{2 i \kappa_D^2}{n!} a^2(D) F_{34} \text{ ,} \label{FTampRR4}
\end{eqnarray}
where we have neglected to write the various momentum conserving delta functions. For simplicity we have written the expressions above without including the boundary vertex corresponding to the D-branes. In order to obtain the amplitudes with the D-branes attached one needs to multiply the amplitudes above by $[G_{C^{(n-1)}}] [B_{C^{(n-1)}}]$ for every D-brane that is attached.

We now need to sum \eqref{FTampRR1}-\eqref{FTampRR4} and include the factors of $[G_{C^{(n-1)}}] [B_{C^{(n-1)}}]$ we excluded earlier. We also need to use expressions such as, $F_{34}^{\mu \nu} (k_{1 \mu} k_{2 \nu} + k_{2 \mu} k_{1 \nu}) = -2 F_{34}^{\mu \nu} k_{2 \mu}k_{2 \nu} + \frac{k_3 \cdot k_4}{n}F_{34}$, $F^{ \alpha \mu}_{34} k_{2 \alpha} k_{3 \mu} = - \frac{u}{2n} F_{34}$ and $F^{ \alpha \mu}_{34} k_{4 \alpha} k_{2 \mu} = - \frac{t}{2n} F_{34}$, which are straightforward to derive using \eqref{F12braneattach} as a reference. We find that the full amplitude is given by,

\begin{eqnarray}
i \mathcal{A}^{\rm ddRR}_{{\rm FT}} &=& -(i N \mu_p)^2 \frac{1}{2} \int \frac{d^{\perp}k_3}{(2 \pi)^{\perp}} \frac{d^{\perp}k_4}{(2 \pi)^{\perp}} \frac{1}{k_3^2} \frac{1}{k_4^2} \delta^{\perp}{(k_3 + k_4 - q)} \nonumber \\ &&\times \frac{i \kappa_D^2}{2} \frac{2}{(n-1)!} \left( 2 a^2(D) \, \frac{s}{tu} - \frac{8}{s} \right) F_{34}^{\mu \nu} k_{2 \mu}k_{2 \nu} \text{ .}
\end{eqnarray}
Using equation \eqref{F12braneattach} we find,

\begin{eqnarray}
i \mathcal{A}^{\rm ddRR}_{{\rm FT}} = i (N T_p\kappa_D)^2 \Bigl[ 2 a^2(D) \, \left( \frac{s}{4} \mathcal{I}_2 + s E^2 \mathcal{I}_3 \right) - \left( \frac{8}{s} k_{1 \mu} k_{2 \nu} \mathcal{I}_{2}^{\mu \nu} + 4 E^2 \mathcal{I}_2 \right) \Bigr] \text{ .}
\label{FullRRFTAmp}
\end{eqnarray}
Comparing \eqref{FullRRFTAmp} with \eqref{FullRRSTAmp2} we find that we have been able to reproduce the same results we produced using our ``effective bulk vertex" prescription by using traditional supergravity Feynman rules.

\subsubsection{Graviton Sources} \label{AttachGrav}

As we have done in the previous subsection for RR fields, we want to derive the full field theory amplitude for graviton exchange by using the four-point NS-NS closed string amplitude (with two external states taken to be dilatons and two taken to be gravitons) as the effective four-point vertex. When attaching a D-brane sourcing a graviton one replaces the polarisation of the relevant external graviton in $\mathcal{A}_{bulk}^{\rm ddgg}$ as follows,
\begin{equation}
\epsilon^{\mu \nu} \to [G_{h}]^{\mu \nu ; \rho \sigma} [B_{h}]_{\rho \sigma} = - N T_p \left(\eta^{\mu \nu}_{\parallel} - \frac{p+1}{D-2} \eta^{\mu \nu} \right) \text{ ,} \label{ebraneattach}
\end{equation}
which is effectively the combination one needs to use in \eqref{attachbtoamp} alongside the bulk vertex in order to obtain the amplitude with the D-branes attached. 

In the case when we sew D-branes that are sourcing gravitons we have the added complication that, as one can see from \eqref{ebraneattach}, the polarisations of the legs we attach the D-branes are neither transverse nor traceless. However the bulk four-point amplitudes we will use as effective four-point vertices in this subsection assume that the external graviton polarisations are traceless and transverse. This implies that by using momentum conservation and the onshell conditions, it is easy to write equivalent onshell vertices that in general yield different results\footnote{For instance by using directly the expression~\eqref{eq:ddggv} one does not obtain~\eqref{STampGG}, as discussed below.} when sewn to the D-branes. Thus we need to add a prescription on what additional properties the effective vertex should have before sewing it to the D-branes. The onshell vertex vanishes for any longitudinal polarisation of any massless particle, i.e. in the case of gravitons it is zero when we substitute $\epsilon_i^{\mu \nu} = \zeta_i^{\mu} k_i^{\nu} + \zeta_i^{\nu} k_i^{\mu}$. Of course when checking this property one needs in general to use momentum conservation and the onshell properties of the remaining external states. However, the momenta of the gravitons glued to the D-branes will appear as integrated variables in the final expression and at that stage it is not always possible to use momentum conservation to write them in terms of the external momenta. Thus we require a further constraint on the onshell bulk effective vertex that can be used to derive a loop diagram: when one of the gravitons that will be glued to the D-branes is longitudinal, the bulk amplitude must vanish whilst not explicitly using momentum conservation in the products $k_i\epsilon_j$ and $\zeta_i k_j$, but only doing so on products between momenta $k_i k_j$ (i.e. only using $s+t+u=0$ in our analysis). In the case of a four-point bulk onshell amplitude, as long as it includes both momenta of the external legs that will be attached to the D-branes in its ``momentum set'' (i.e. the three independent momenta with which the amplitude is expressed), then the condition mentioned above is met.

We start by recalling the field theory limit of the four-point two dilaton, two graviton amplitude \cite{green1988superstring} which we can write as,
\begin{eqnarray}\label{eq:ddggv}
&& i \mathcal{A}_{bulk}^{\rm ddgg} = \frac{i \kappa_D^2}{2} \frac{2}{stu} \left( u^2 t^2 \, \epsilon_3^{\mu \nu}\epsilon_{4 \mu \nu} + 4 t^2 \, k_1^{\mu} k_1^{\nu} k_2^{\rho} k_2^{\sigma} \epsilon_{3 \rho \sigma} \epsilon_{4 \mu \nu} - 4 u t^2 \, k_1^{\mu} k_2^{\nu} \epsilon_{3 \nu}{}^{\rho} \epsilon_{4 \mu \rho}  \right. \nonumber \\
&& \left. - 4 u^2 t \, k_1^{\mu} k_2^{\nu} \epsilon_{3 \mu}{}^{\rho} \epsilon_{4 \nu \rho} + 8 u t \, k_1^{\mu} k_1^{\nu} k_2^{\rho} k_2^{\sigma} \epsilon_{3 \mu \rho} \epsilon_{4 \nu \sigma} + 4 u^2 \, k_1^{\mu} k_1^{\nu} k_2^{\rho} k_2^{\sigma} \epsilon_{3 \mu \nu} \epsilon_{4 \rho \sigma} \right) \,.
\end{eqnarray}
In this form the bulk vertex does not satisfy the requirement mentioned above for the two gravitons, but if we take this equation and use momentum conservation to express it using $(k_1, k_3, k_4)$ or $(k_2, k_3, k_4)$, we obtain an expression that can be glued to the D-branes simply by replacing the graviton polarisations with \eqref{ebraneattach}. Then we find for the integrand of \eqref{attachbtoamp},
\begin{eqnarray} \label{STampGG}
\frac{i (N T_p \kappa_D)^2}{4} \frac{1}{k_3^2 k_4^2} \frac{2}{stu} \left(4 E^4 s^2 + 4 E^2 stu + \frac{(D-p-3)(1+p)}{D-2} u^2 t^2 \right) 
\text{ ,} 
\end{eqnarray}
where we have also used the relevant kinematics mentioned in section \ref{kinematics}. By including the appropriate integrals one obtains,
\begin{equation}
i \mathcal{A}_{2}^{\rm ddgg} = i (N T_p \kappa_D)^2 \left( 4E^4 \mathcal{I}_{3} + \frac{(D-p-3)(1+p)}{D-2}\frac{2}{s} k_{1 \mu} k_{2 \nu} \mathcal{I}_{2}^{\mu \nu} + 2 E^2 \mathcal{I}_{2}  \right) \text{ .} \label{FullGGSTAmp}
\end{equation}

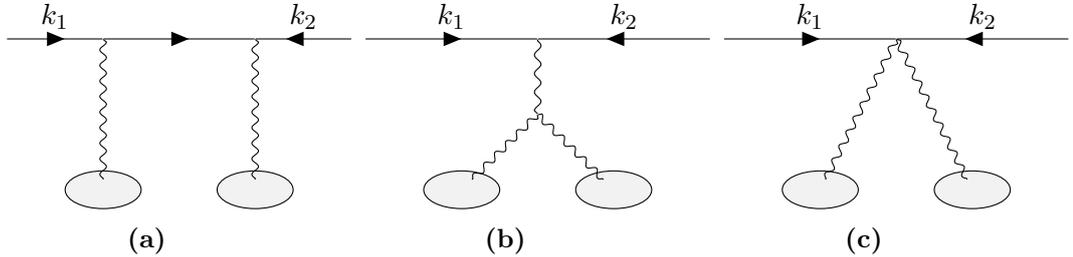
\begin{figure}[h]
  \begin{subfigure}[t]{0.29\textwidth}
    \centering
    \begin{tikzpicture}
	    \begin{feynman}
			\vertex (a) at (-1,-2) {};
			\vertex (b) at ( 1,-2) {};
			\vertex (c) at (-2.4, 0) {};
			\vertex (d) at ( 2.4, 0) {};
			\vertex[circle,inner sep=0pt,minimum size=0pt] (e) at (-1, 0) {};
			\vertex[circle,inner sep=0pt,minimum size=0pt] (f) at (1, 0) {};
			\draw[fill=light-gray] (-1,-2) ellipse (0.5cm and 0.25cm);
			\draw[fill=light-gray] (1,-2) ellipse (0.5cm and 0.25cm);       
			\diagram* {
			(c) -- [fermion,edge label=$k_1$] (e) -- [fermion] (f) -- [anti fermion,edge label=$k_2$] (d),
			(a) -- [boson] (e),
			(b) -- [boson] (f),
			};
	    \end{feynman}
    \end{tikzpicture}
    \caption{}
    \label{fig:3a}
  \end{subfigure}
  \quad
  \begin{subfigure}[t]{0.29\textwidth}
    \centering
    \begin{tikzpicture}
		\begin{feynman}
			\vertex (a) at (-1,-2) {};
			\vertex (b) at ( 1,-2) {};
			\vertex (c) at (-2.4, 0) {};
			\vertex (d) at ( 2.4, 0) {};
			\vertex[circle,inner sep=0pt,minimum size=0pt] (e) at (0, -1) {};
			\vertex[circle,inner sep=0pt,minimum size=0pt] (m) at (0, 0) {};
			\draw[fill=light-gray] (-1,-2) ellipse (0.5cm and 0.25cm);
			\draw[fill=light-gray] (1,-2) ellipse (0.5cm and 0.25cm);       
			\diagram* {
			(c) -- [fermion,edge label=$k_1$] (m) -- [anti fermion,edge label=$k_2$] (d),
			(m) -- [boson] (e),
			(a) -- [boson] (e),
			(b) -- [boson] (e),
			};
		\end{feynman}
    \end{tikzpicture}
    \caption{}
    \label{fig:3b}
  \end{subfigure}
  \quad
  \begin{subfigure}[t]{0.29\textwidth}
    \centering
    \begin{tikzpicture}
    	\begin{feynman}
			\vertex (a) at (-1,-2) {};
			\vertex (b) at ( 1,-2) {};
			\vertex (c) at (-2.4, 0) {};
			\vertex (d) at ( 2.4, 0) {};
			\vertex[circle,inner sep=0pt,minimum size=0pt] (e) at (0, -1) {};
			\vertex[circle,inner sep=0pt,minimum size=0pt] (m) at (0, 0) {};
			\draw[fill=light-gray] (-1,-2) ellipse (0.5cm and 0.25cm);
			\draw[fill=light-gray] (1,-2) ellipse (0.5cm and 0.25cm);       
			\diagram* {
			(c) -- [fermion,edge label=$k_1$] (m) -- [anti fermion,edge label=$k_2$] (d),
			(a) -- [boson] (m),
			(b) -- [boson] (m),
			};
		\end{feynman}
    \end{tikzpicture}
    \caption{}
    \label{fig:3c}
  \end{subfigure}
  \caption{The various topologies of diagrams that contribute to $\mathcal{A}^{\rm ddgg}$. In \ref{fig:3a} we have the t- and u-channels, in \ref{fig:3b} we have the s-channel diagram and finally in \ref{fig:3c} we have the contact diagrams. The solid lines represent dilatons and the wavy lines represent gravitons.}
  \label{fig:3}
\end{figure}

We want to compare \eqref{FullGGSTAmp} with the equivalent result arising from using Feynman diagrams as we have done in the RR case. We first calculate the relevant Feynman diagrams for this process. Note that since we will be attaching the D-branes to the graviton external legs we will not be imposing $k_i \cdot \epsilon_i = 0$ or $\text{Tr}(\epsilon_i)= 0$, i.e. we will keep the gravitons offshell. The four contributions to the full amplitude are given by the following diagrams,
\begin{eqnarray}
i\mathcal{A}^{\rm ddgg}_{{\rm FT}, u} &=& \epsilon_{3 \mu \nu} \epsilon_{4 \rho \sigma} [V_{\phi_1 \phi_2 h}]^{\mu \nu} [V_{\phi_1 \phi_2 h}]^{\rho \sigma} [G_{\phi}] \\
&& \nonumber \\
i\mathcal{A}^{\rm ddgg}_{{\rm FT}, t} &=&  \epsilon_{3 \rho \sigma} \epsilon_{4 \mu \nu} [V_{\phi_1 \phi_2 h}]^{\mu \nu} [V_{\phi_1 \phi_2 h}]^{\rho \sigma} [G_{\phi}] \\
&& \nonumber \\
i\mathcal{A}^{\rm ddgg}_{{\rm FT}, s} &=&  \epsilon_{3 \rho \sigma} \epsilon_{4 \lambda \tau} [V_{\phi_1 \phi_2 h}]^{\gamma \delta} [G_{h}]_{\mu \nu ; \gamma \delta} \left( T^{\mu \nu;\rho\sigma;\lambda \tau} (q,k_3,k_4)  \right. \\ \nonumber
&+&  T^{\rho\sigma;\mu \nu;\lambda \tau} (k_3,q,k_4) + \left. T^{\mu \nu;\lambda \tau;\rho\sigma} (q,k_4,k_3) + T^{\rho\sigma;\lambda \tau;\mu \nu} (k_3,k_4,q)  \right. \\ \nonumber
&+& \left. T^{\lambda \tau;\mu \nu;\rho\sigma} (k_4,q,k_3) + T^{\lambda \tau;\rho\sigma;\mu \nu} (k_4,k_3,q) \right)  \\
&& \nonumber \\
i\mathcal{A}^{\rm ddgg}_{{\rm FT}, c} &=& \epsilon_{3 \rho \sigma} \epsilon_{4 \lambda \tau} [V_{\phi \phi' h h'}]^{\rho \sigma \lambda \tau} \text{ ,}
\end{eqnarray}
where we have not explicitly written the resulting Lorentz structure for brevity and $\epsilon_3$, $\epsilon_4$ are the graviton polarisations which need to be replaced with $[G_{h}]^{\mu \nu ; \rho \sigma} [B_{h}]_{\rho \sigma}$. Doing so and using the kinematics outlined in section \ref{kinematics} we have for the t-channel,
\begin{eqnarray}
i \mathcal{A}^{\rm ddgg}_{{\rm FT}, t} &=& -(-i)(-i \kappa_D)^2 (N T_p)^2 4 E^4 \frac{1}{2} \int \frac{d^{\perp}k_3}{(2 \pi)^{\perp}} \frac{d^{\perp}k_4}{(2 \pi)^{\perp}} \frac{1}{k_3^2} \frac{1}{k_4^2} \delta^{\perp}{(k_3 + k_4 - q)} \frac{1}{t} \nonumber \\
&=& i (N T_p \kappa_D)^2 2 E^4 \mathcal{I}_3 \;, \label{GGuFTAmp} 
\end{eqnarray}
with an equivalent contribution for the u-channel. These two diagrams are the only ones that contribute to leading order in energy, $\mathcal{O}(E^3)$, in the full amplitude as we have seen with the result derived from using our effective bulk vertex method. We can now look at the two remaining diagrams which contribute to subleading order in energy, $\mathcal{O}(E^2)$. The s-channel diagram gives,
\begin{eqnarray}
&& i \mathcal{A}^{\rm ddgg}_{{\rm FT}, s} = - (-i \kappa_D)(-2 i \kappa_D) \left( - \frac{i}{2} \right) (N T_p)^2 \frac{1}{2}  \nonumber \\
&& \times \int \frac{d^{\perp}k_3}{(2 \pi)^{\perp}} \frac{d^{\perp}k_4}{(2 \pi)^{\perp}} \frac{1}{k_3^2} \frac{1}{k_4^2} \delta^{\perp}{(k_3 + k_4 - q)} \left[ 4 E^2 \frac{(D-2p-4)}{D-2} \right. \nonumber \\
&& \quad \, \left. + \frac{(D-p-3)(1+p)}{D-2} \left( \frac{4}{s}(k_2 \cdot k_3)(k_2 \cdot k_3) + 2 (k_1 \cdot k_3) \right) \right] \;.  \label{GGsFTAmp}
\end{eqnarray}
We also have for the contact diagram,
\begin{eqnarray}
i \mathcal{A}^{\rm ddgg}_{{\rm FT}, c} &=& \left( \frac{i \kappa_D^2}{2} \right) (N T_p)^2 \frac{1}{2} \int \frac{d^{\perp}k_3}{(2 \pi)^{\perp}} \frac{d^{\perp}k_4}{(2 \pi)^{\perp}} \frac{1}{k_3^2} \frac{1}{k_4^2} \delta^{\perp}{(k_3 + k_4 - q)} \biggl( \frac{16 E^2 (D-p-3)}{D-2} \nonumber \\
&& + \frac{4(1+p)(D-p-3)}{D-2} (k_1 \cdot k_2) \biggr) \text{ .} \label{GGcFTAmp} 
\end{eqnarray}
Summing the above two contributions yields
\begin{eqnarray}
i \mathcal{A}^{\rm ddgg}_{{\rm FT}, c} + i \mathcal{A}^{\rm ddgg}_{{\rm FT}, s} &=& i (N T_p \kappa_D)^2 \int \frac{d^{\perp}k_3}{(2 \pi)^{\perp}} \frac{d^{\perp}k_4}{(2 \pi)^{\perp}} \frac{1}{k_3^2} \frac{1}{k_4^2} \delta^{\perp}{(k_3 + k_4 - q)}  \nonumber \\
&& \times \biggl( 4 E^2 +  \frac{(D-p-3)(1+p)}{D-2} (k_1 \cdot k_3) (k_2 \cdot k_3) \biggr) \text{ .} \label{GGcsFTAmp}
\end{eqnarray}
We can easily see by summing \eqref{GGuFTAmp} and \eqref{GGcsFTAmp} that we are able to reproduce \eqref{FullGGSTAmp} using the supergravity Feynman rules.

\subsubsection{Dilaton Sources}\label{AttachDil}

Here we calculate the amplitude for elastic dilaton-brane scattering with dilaton exchange by using the four-point dilaton string amplitude as the effective vertex. We have in the field theory limit of the string theory amplitude \cite{Garousi:2012yr},
\begin{eqnarray}
i \mathcal{A}_{bulk}^{\rm dddd} = i \kappa_D^2 \left( \frac{st}{u} + \frac{su}{t} + \frac{ut}{s} \right) \text{ .}
\end{eqnarray}
As before, using our prescription, we include the relevant integrals arising from \eqref{attachbtoamp}. When attaching dilatons the relevant factor, arising from $[G_{\phi}] [B_{\phi}]$, is $-(i N T_p \frac{a(D)}{\sqrt{2}})^2$. Using this we have,
\begin{eqnarray}
i \mathcal{A}_{2}^{\rm dddd} &=& - i \left( N T_p \kappa_D \frac{a(D)}{\sqrt{2}} \right)^2 \left( s k_{2 \mu} \mathcal{I}_{2}^{\mu} + s k_{1 \mu} \mathcal{I}_{2}^{\mu} + \frac{2}{s} k_{1 \mu} k_{2 \nu} \mathcal{I}_{2}^{\mu \nu}  \right) \text{ .}
\end{eqnarray}
It is trivial to compare these results with those found using supergravity Feynman rules and so we will not be comparing them explicitly here.

\subsection{Dilaton to RR Inelastic Scattering} \label{ineldilRRallE}

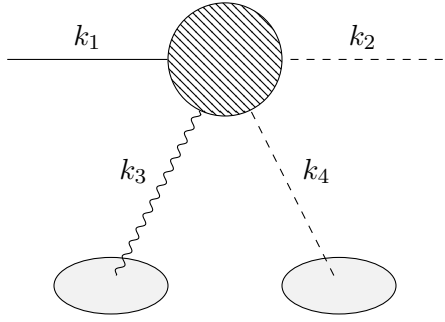
\begin{figure}[h]
  \centering
  \begin{tikzpicture}[scale=1.5]
    \begin{feynman}
      \vertex[blob, minimum size=1.5cm] (m) at ( 0, 0) {\contour{white}{}};
      \vertex (a) at (-1,-2) {};
      \vertex (b) at ( 1,-2) {};
      \vertex (c) at (-2, 0) {};
      \vertex (d) at ( 2, 0) {};
	  \draw[fill=light-gray] (-1,-2) ellipse (0.5cm and 0.25cm);
	  \draw[fill=light-gray] (1,-2) ellipse (0.5cm and 0.25cm);       
      \diagram* {
      (a) -- [boson,edge label=$k_3$] (m),
      (b) -- [scalar,edge label=$k_4$,swap] (m),
      (c) -- [edge label=$k_1$] (m),
      (d) -- [scalar,edge label=$k_2$, swap] (m),
      };
    \end{feynman}
  \end{tikzpicture}
  \caption{A schematic diagram showing our procedure for calculating the effective one-loop amplitude for dilaton to RR inelastic scattering. The circular blob represents the four-point effective vertex $\mathcal{A}_{bulk}^{\rm dRgR}$ and the two oval blobs represent the D-branes. As before the solid lines correspond to dilatons, the wavy lines correspond to gravitons and the dashed lines correspond to RR fields. \label{fig:4}}
\end{figure}

As with the elastic dilaton scattering case that we have considered in the previous subsection we can use the relevant equation in \cite{Bakhtiarizadeh:2013zia} for the four-point two NS-NS (with one state taken to be a dilaton and the other a graviton), two RR closed string amplitude as an effective vertex for calculating the amplitude for an inelastic transition from a dilaton to an RR field via the exchange of a graviton and an RR field with the D-branes. The bulk vertex needed is given by,
\begin{eqnarray}
&& i \mathcal{A}_{bulk}^{\rm dRgR} = - \frac{i \kappa_D^2}{\sqrt{2}}\frac{16 a(D)}{n!} \Bigl[ F_{24} \epsilon_3^{\mu \nu} \left( s^2 k_{2 \mu} k_{2 \nu} + u k_{4 \mu} \left( u k_{4 \nu} - 2 s k_{2 \nu} \right) \right) + nt   \nonumber \\
&&\times \left( (n-1) t F_{24}^{\alpha \beta \mu \nu} \epsilon_{3 \beta \nu} k_{3 \alpha} k_{3 \mu} - F_{24}^{\alpha \mu} \left(u k_4^{\nu} - s k_2^{\nu} \right)\left( \epsilon_{3 \mu \nu} k_{3 \alpha} - \epsilon_{3 \alpha \nu} k_{3 \mu} \right)      \right) \Bigr] \;,  \nonumber \\ \label{stringdRgRvertex}
\end{eqnarray}
where the labels here correspond as follows; label 1 is associated with the external dilaton, label 2 is associated with the external RR field, label 3 is associated with the internal graviton and label 4 is associated with the internal RR field as shown in figure \ref{fig:4}. After using our prescription and expressing \eqref{stringdRgRvertex} with momentum set $(k_1, k_3, k_4)$ we find that the expression satisfies the condition described at the beginning of section \ref{2braneamps} as required.

By following the same approach discussed in the previous subsection, we find that after attaching the graviton to the D-branes, the integrand of \eqref{attachbtoamp} reads
\begin{eqnarray}
&& \frac{16 i N^2 T_p \mu_p \kappa_D^2}{\sqrt{2}} \frac{1}{k_3^2}  \frac{1}{k_4^2} \frac{a(D)}{n!} \biggl[\frac{s}{ut} F_{24} E^2 + F_{24} \frac{1+p}{D-2} \frac{s}{u} + n F_{24}^{\alpha \beta} \left\lbrace \frac{1}{u} \left(\eta_{\parallel \beta \nu}k_1^{\nu}k_{3 \alpha} \right. \right. \nonumber \\
&& \left. \left. - \eta_{\parallel \alpha \nu}k_1^{\nu}k_{3 \beta} \right)  + \frac{1+p}{D-2} \left( \frac{1}{u} \left(k_{1 \alpha} k_{3 \beta} - k_{1 \beta} k_{3 \alpha} \right) + \frac{t}{us} \left((n-1)k_{3 \alpha} k_{3 \beta} - k_{4 \alpha}  k_{3 \beta} \right) \right) \right\rbrace \nonumber \\
&&  - \frac{t}{us} F_{24}^{\alpha \beta \mu \nu} \eta_{\parallel \beta \nu}k_{3 \alpha} k_{3 \mu} \biggr] \;,
\label{RRRRdgrawamp}
\end{eqnarray}
where the first term has been identified in advance as the term with the possible features needed to contribute to the leading energy behaviour of the amplitude. 

In order to attach an RR leg to the D-branes, we need to rewrite the field strengths in terms of the potentials. We can explicitly calculate all the necessary combinations of field strengths that arise in \eqref{RRRRdgrawamp} and rewrite them in terms of the potentials. It is important to recall that in these expressions one of the RR fields is attached to the D-branes (with label 4) and one is an external state (with label 2). We will first focus on the term that is relevant for the leading energy contribution and subsequently look at all other combinations. We have,
\begin{eqnarray}
E^2 F_{24} &=& E^2 \left( k_2^{\mu_1} C^{(2) \mu_2 \ldots \mu_{n}} + (-1)^{n-1} k_2^{\mu_2} C^{(2) \mu_3 \ldots \mu_{n} \mu_1} + \ldots \right) \nonumber \\
&& \times \left[ k_{4 \mu_1} C^{(4)}_{\mu_2 \ldots \mu_{n}} + (-1)^{n-1} k_{4 \mu_2} C^{(4)}_{\mu_3 \ldots \mu_{n} \mu_1} + \ldots \right) \\ \nonumber
&=& E^2 \left( n (k_2 \cdot k_4) C^{(2) \mu_2 \ldots \mu_{n}} C^{(4)}_{\mu_2 \ldots \mu_{n}} \right. \nonumber \\
&& \left. + (-1)^{n-1} n(n-1) k_2^{\mu_1} C^{(2) \mu_2 \ldots \mu_{n}} k_{4 \mu_2} C^{(4)}_{\mu_3 \ldots \mu_{n} \mu_1} \right] \;,
\end{eqnarray}
where we have identified the two terms in the last line above to be the only two distinct types of terms that arise, with the associated counting taken into account. The indices on $C^{(4)}$ have to lie in the space parallel to the D-brane world volume so $\mu_3 \ldots \mu_{n} = 1 \ldots p$ and we note that there are $(n-2)!$ ways to do this. Furthermore $(k_2^{\mu_1})_{\parallel}$ only has a non-zero component for $\mu_1 = 0$ ($\mu_1$ has to also lie along the D-branes as it is one of the indices on $C^{(4)}$) and so we have to all orders in $E$,
\begin{eqnarray}
E^2 F_{24} &=& E^3 n! (k_4 \cdot C^{(2)})^{1 \ldots p} C^{(4)}_{0 \ldots p} + E^2 n! (k_2 \cdot k_4) C^{(2) 0 \ldots p} C^{(4)}_{0 \ldots p} \;,
\end{eqnarray}
where we have used $(k_{2})^{0}=-E$ and $(-1)^{n-1} C^{(4)}_{1 \ldots p 0} = - C^{(4)}_{0 \ldots p}$. 

We now look at types of terms which contribute to the subleading energy behaviour of \eqref{RRRRdgrawamp},
\begin{eqnarray}
F_{24}^{\alpha \mu} \eta_{\parallel \alpha \nu}k_1^{\nu}k_{3 \mu} &=&  F_{24}^{0 \mu} (k_{1})_{0} k_{3 \mu} \nonumber \\
&=&  \left( (k_2)^{0} C^{(2) \mu_2 \ldots \mu_{n}} + (-1)^{n-1} k_2^{\mu_2} C^{(2) \mu_3 \ldots \mu_{n} 0} + \ldots \right) \nonumber \\
&& \times \left( k_{4}^{\mu} C^{(4)}_{\mu_2 \ldots \mu_{n}} + (-1)^{n-1} k_{4 \mu_2} C^{(4)}_{\mu_3 \ldots \mu_{n}}{}^{\mu} + \ldots \right) (k_{1})_{0} k_{3 \mu} \nonumber \\
&=& \left( (k_2)^{0} C^{(2) \mu_2 \ldots \mu_{n}} + (-1)^{n-1} k_2^{\mu_2} C^{(2) \mu_3 \ldots \mu_{n} 0} + \ldots \right) \nonumber \\
&& \qquad \times \left( k_{4}^{\mu} C^{(4)}_{\mu_2 \ldots \mu_{n}} \right) (k_{1})_{0} k_{3 \mu} \nonumber \\
&=&  (n-1)! \left( (k_2)^{0} C^{(2) 0 \ldots p} + (-1)^{n-1} k_2^{0} C^{(2) 1 \ldots p 0} \right)\nonumber \\
&& \qquad \times \left( (k_3 \cdot k_{4}) C^{(4)}_{0 \ldots p} \right) (k_{1})_{0}    \nonumber \\
&=& 0 \;,
\end{eqnarray}
where we have used the fact that the $\mu$ index has to be along the D-branes but $(k_3)_{\parallel}=0$ in the third line and in the last line we have again used the fact that $(-1)^{n-1} C^{(2)}_{1 \ldots p 0} = - C^{(2)}_{0 \ldots p}$. We also need,
\begin{eqnarray}
F_{24}^{\alpha \beta} \eta_{\parallel \beta \mu} k_1^{\mu} &=& F_{2}^{\alpha \mu_2 \ldots \mu_n} \left( k_{4}^{\beta} C^{(4)}_{\mu_2 \ldots \mu_{n}} + (-1)^{n-1} k_{4 \mu_2} C^{(4)}_{\mu_3 \ldots \mu_{n}}{}^{\beta} + \ldots \right) \eta_{\parallel \beta \mu} k_1^{\mu} \nonumber \\
&=& - (n-1)!  F_{2}^{\alpha \mu_2 1 \ldots p} k_{4 \mu_2} (k_1)^{0} C^{(4)}_{0 \ldots p} \nonumber \\
&=& - (n-1)! E \left[( k_4 \cdot C^{(2)})^{1 \ldots p} k_2^{\alpha} - (k_2 \cdot k_4) C^{(2) \alpha 1 \ldots p} \right] C^{(4)}_{0 \ldots p} \;,
\end{eqnarray}
where in the second line we have used the fact that $(k_1 \cdot k_4)_{\parallel} = 0$. We also require,
\begin{eqnarray}
F_{24}^{\alpha \beta} k_{1 \beta} &=& F_{2}^{\alpha \mu_2 \ldots \mu_n} \left( k_{4}^{\beta} C^{(4)}_{\mu_2 \ldots \mu_{n}} + (-1)^{n-1} k_{4 \mu_2} C^{(4)}_{\mu_3 \ldots \mu_{n}}{}^{\beta} + \ldots \right) k_{1 \beta} \nonumber \\
&=& (n-1)! \left( (k_1 \cdot k_{4}) F_{2}^{\alpha 0 \ldots p} C^{(4)}_{0 \ldots p} - k_{4 \mu_2} (k_1)^{0} F_{2}^{\alpha \mu_2 1 \ldots p} C^{(4)}_{0 \ldots p} + \ldots \right)\nonumber \\
&=& (n-1)! \bigl[ (k_1 \cdot k_4) \left(k_2^{\alpha} C^{(2) 0 \ldots p} + E C^{(2) \alpha 1 \ldots p} \right) \nonumber \\
&& - E ( k_4 \cdot C^{(2)})^{1 \ldots p} k_2^{\alpha} + E (k_2 \cdot k_4) C^{(2) \alpha 1 \ldots p} \bigr] C^{(4)}_{0 \ldots p} \;,
\end{eqnarray}
where we have used the properties of the gauge potentials outlined previously. We can also have,
\begin{eqnarray}
F_{24}^{\alpha \beta} k_{3 \beta} &=& F_{2}^{\alpha \mu_2 \ldots \mu_n} \left( k_{4}^{\beta} C^{(4)}_{\mu_2 \ldots \mu_{n}} + (-1)^{n-1} k_{4 \mu_2} C^{(4)}_{\mu_3 \ldots \mu_{n}}{}^{\beta} + \ldots \right) k_{3 \beta} \nonumber \\
&=& (n-1)! \left( (k_3 \cdot k_{4}) F_{2}^{\alpha 0 \ldots p} C^{(4)}_{0 \ldots p} \right)\nonumber \\
&=& (n-1)! (k_3 \cdot k_4) \left(k_2^{\alpha} C^{(2) 0 \ldots p} + E C^{(2) \alpha 1 \ldots p} \right) C^{(4)}_{0 \ldots p} \;,
\end{eqnarray}
where in the second line we have again used the fact that the index $\beta$ must lie parallel to the D-branes but $(k_3)_{\parallel}=0$. Finally we have,
\begin{eqnarray}
F_{24}^{\alpha \beta \mu \nu} \eta_{\parallel \beta \nu} k_{3 \alpha} k_{3 \mu} &=& F_{2}^{\alpha \beta \mu_3 \ldots \mu_n} \bigl( k_{4}^{\mu} C^{(4) \nu}{}_{\mu_3 \ldots \mu_{n}} + (-1)^{n-1} k_4^{\nu} C^{(4)}_{\mu_3 \ldots \mu_{n}}{}^{\mu}  \nonumber \\
&& + (-1)^{n-1} k_{4 \mu_3} C^{(4)}_{\mu_4 \ldots \mu_{n}}{}^{\mu \nu} + \ldots \bigr) \eta_{\parallel \beta \nu} k_{3 \alpha} k_{3 \mu} \nonumber \\
&=& (n-2)! \left( F_{2}^{\alpha 0 \ldots p} k_4^{\mu} C^{(4)}_{0 \ldots p} \right. \nonumber \\ 
&& \left. + (-1)^{n-1} F_{2}^{\alpha \beta \mu_3 2 \ldots p} k_{4 \mu_3} C^{(4)}_{2 \ldots p}{}^{\mu \nu}  \eta_{\parallel \beta \nu} \right) k_{3 \alpha} k_{3 \mu} \nonumber \\
&=& (n-2)! (k_3 \cdot k_4) \left((k_2 \cdot k_3) C^{(2) 0 \ldots p} + E (k_3 \cdot C^{(2)} )^{1 \ldots p} \right) C^{(4)}_{0 \ldots p} \;, \nonumber \\
\end{eqnarray}
where in the third line we have again used the fact that the index $\mu$ must lie parallel to the D-branes but $(k_3)_{\parallel}=0$.

Using these expressions we can write the integrand in \eqref{attachbtoamp} relevant to the inelastic dilaton to RR amplitude,
\begin{eqnarray}
&& - i (N^2 T_p \mu_p \kappa_D^2  ) \frac{16 a(D)}{\sqrt{2}} \frac{1}{k_3^2 k_4^2} \biggl[ \frac{s}{ut} \left( -E^3 (k_4 \cdot C^{(2)})^{1 \ldots p} - E^2 (k_2 \cdot k_4) C^{(2) 0 \ldots p} \right)  \nonumber \\
&& - \frac{t}{4} C^{(2) 0 \ldots p} + \frac{E}{2} (k_4 \cdot C^{(2)})^{1 \ldots p} + \frac{1+p}{D-2} \biggl\lbrace C^{(2) 0 \ldots p} \left( \frac{s}{2u}E^2 + \frac{(3+n)t}{4} + \frac{t^2}{4u} \right)  \nonumber \\
&& + \frac{s}{2u}E (k_4 \cdot C^{(2)})^{1 \ldots p} + (k_3 \cdot C^{(2)})^{1 \ldots p} E \left( \frac{1}{2} - \frac{t}{2u}n \right) \biggr\rbrace \biggr] \;.
\end{eqnarray}
Inserting the integrals as per our prescription we obtain,
\begin{eqnarray}
\label{eq:dRgRf}
&& i \mathcal{A}_{2}^{\rm dRgR} = - i (N^2 T_p \mu_p \kappa_D^2) \frac{8 a(D)}{\sqrt{2}} \biggl[ - ( q \cdot C^{(2)})^{1 \ldots p} E^3 \mathcal{I}_3   \nonumber \\
&& - (q \cdot k_2) C^{(2) 0 \ldots p} E^2 \mathcal{I}_3 + \frac{1}{2} k_{1 \mu} \mathcal{I}_2^{\mu} C^{(2) 0 \ldots p} + \frac{1}{2}E C_{\mu}^{(2)}{}^{1 \ldots p} \mathcal{I}_2^{\mu} \nonumber \\
&& + \frac{1+p}{D-2} \biggl\lbrace C^{(2) 0 \ldots p} \left( -\frac{s}{2} E^2 \mathcal{I}_3 - \frac{(2+n)}{2} k_{1 \mu} \mathcal{I}_2^{\mu} + \frac{s}{4}\mathcal{I}_2 - \frac{s^2}{4}\mathcal{I}_3 \right) - \frac{s}{4}E ( q \cdot C^{(2)})^{1 \ldots p} \mathcal{I}_3 \nonumber \\
&&  + \frac{1}{2}E C_{\mu}^{(2)}{}^{1 \ldots p} \mathcal{I}_2^{\mu} + E C_{\mu}^{(2)}{}^{1 \ldots p} \left(\frac{1}{2} \mathcal{I}_2^{\mu} - \frac{s}{4}q^{\mu} \mathcal{I}_3 \right) \biggr\rbrace \biggr] \;,
\end{eqnarray}
where $q=k_1+k_2$ is the momentum exchanged.

\section{The Supergravity Eikonal} \label{HElimit}

In this section we will focus on and derive explicit expressions for the leading and subleading high-energy behaviour of the various amplitudes we considered in section \ref{2braneamps} and analyse their behaviour in the context of the eikonal approximation. 

We can transform any of the amplitudes we have considered into impact parameter space by using,
\begin{equation}\label{ttoips}
\mathcal{A}_{h}(E,\mathbf{b}) = \int \frac{\text{d}^{D-p-2} \mathbf{q}}{(2 \pi)^{D-p-2}} e^{i \mathbf{b} \cdot \mathbf{q}} \mathcal{A}_{h}(E,q) \text{ ,}
\end{equation}
where $h$ refers to the number of boundaries of the amplitude (i.e. the number of exchanges with the D-branes). The notation here is slightly different from chapter \ref{chap:background} to emphasise the difference between the scattering off of D-branes we have here and the scattering of massive scalars we had there. We have also normalised the impact parameter transform differently and moved the normalisation to the definition of the eikonal below. We start by focusing on the elastic case where the leading energy behaviour of the tree-level amplitude, one-loop amplitude and amplitudes with a higher number of boundaries is universal and so does not display any non-trivial Lorentz structure. As we saw in chapter \ref{chap:background} by summing these contributions, we find that the S-matrix approximates to,
\begin{equation} \label{leadingeikgen}
S_l(E,\mathbf{b}) \approx 1+ \sum_{h=1}^{\infty} \frac{\mathcal{A}^{(1)}_{h}(E,\mathbf{b})}{2E} = e^{i \delta^{(1)}(E,\mathbf{b})}  \text{ ,}
\end{equation}
where $\mathcal{A}^{(1)}_{h}(E,\mathbf{b})$ is the leading energy contribution of the amplitude with $h$ boundaries and $\delta^{(1)}(E,\mathbf{b}) = \mathcal{A}^{(1)}_{h=1}(E,\mathbf{b})/(2 E)$ is called the leading eikonal. Note that $S_l(E,\mathbf{b})$ captures all the information in the leading energy term of all amplitudes. We can write something similar for the subleading energy contribution by starting from $h=2$ and summing all the subleading contributions of the amplitudes at each number of boundaries. In this case we have the subleading eikonal given by $\delta^{(2)}(E,\mathbf{b}) = \mathcal{A}^{(2)}_{h=2}(E,\mathbf{b})/(2 E)$ where $\mathcal{A}^{(2)}_{h}(E,\mathbf{b})$ is the subleading contribution to the amplitude with $h$ boundaries.

In the following we want to emphasise the construction of the S-matrix in the eikonal approximation discussed in chapter \ref{chap:background} to include more general situations as for instance the presence of inelastic processes. Traditionally, for elastic processes, we write, including all contributions to all orders,
\begin{equation}
S(E,\mathbf{b}) \approx \text{exp}\left({i \delta^{(1)}(E,\mathbf{b}) + i \delta^{(2)}(E,\mathbf{b}) + \ldots}\right) \text{ ,}
\label{Sela}
\end{equation}
where $\delta^{(1)}(E,\mathbf{b})$ is the leading eikonal and $\delta^{(2)}(E,\mathbf{b})$ is the subleading eikonal mentioned above. In subsection \ref{elasticeik} we show that this statement holds in the case of elastic dilaton scattering that we have already studied. In order to study this let us  write the tree-level ($h=1$) and one-loop ($h=2$) amplitudes as,
\begin{eqnarray}
\frac{i \mathcal{A}_{1}(E,\mathbf{b})}{2E} &=& i(N T_p \kappa_D) (A_{h=1}^{(1)}(\mathbf{b}) E + A_{h=1}^{(2)}(\mathbf{b}) E^{0} + \ldots) \label{diskampinE} \\
\frac{i \mathcal{A}_{2}(E,\mathbf{b})}{2E} &=& i(N T_p \kappa_D)^2 (A_{h=2}^{(1)}(\mathbf{b}) E^2 + A_{h=2}^{(2)}(\mathbf{b}) E + A_{h=2}^{(3)}(\mathbf{b}) E^{0} + \ldots) \label{annampinE} \text{ ,}
\end{eqnarray}
where we have divided by $\frac{1}{\sqrt{2E}}$ for each of the two external particles involved and where the $A$ symbols correspond to $\mathcal{A}/2E$ where the dependence on energy has been factored out\footnote{We also note here that we are using the amplitudes once stripped of factors of $i (2\pi)^{p+1} \delta^{p+1}(k_1+k_2)$.}. Note here that in order to express the leading contributions as an exponential of the leading eikonal we have $i A_{h=2}^{(1)}(\mathbf{b})=-\frac{1}{2}(A_{h=1}^{(1)}(\mathbf{b}))^2$. 

In equations (\ref{diskampinE}) and (\ref{annampinE}) we have also allowed for terms of order $E^0$ that, as we will see, are not present in the elastic dilaton scattering, but appear in the inelastic dilaton to RR scattering. We would like to extend the construction of the S-matrix in the eikonal approximation when these extra terms are present. Our proposal is that, in this more general case, (\ref{Sela}) is written as follows,
\begin{eqnarray}
S(E,\mathbf{b})  &=& \text{exp}\left[{\frac{i}{2} (\delta^{(1)}(E,\mathbf{b}) +  \delta^{(2)}(E,\mathbf{b}) + \ldots)}\right] \left(1 + i T(E,\mathbf{b}) \right) \times \nonumber \\
&& \;\;\; \text{exp}\left[{\frac{i}{2} (\delta^{(1)}(E,\mathbf{b}) +  \delta^{(2)}(E,\mathbf{b}) + \ldots)}\right] \text{ ,} \label{fullInelEik} 
\end{eqnarray}
where $\delta^{(1)}(E,\mathbf{b})$ and $\delta^{(2)}(E,\mathbf{b})$ are the leading eikonal and subleading eikonal respectively. The symbol $T(E,\mathbf{b})$ corresponds to all the non-diverging contributions to the amplitudes with any number of boundaries. For example the first contribution to $T(E,\mathbf{b})$ is $A_{h=1}^{(2)}(\mathbf{b})$; the first contribution to the tree-level dilaton to dilaton scattering process that does not grow with $E$. We have written \eqref{fullInelEik} in this way to account for when the eikonal and subleading eikonal behave as operators instead of phases. As can be seen from \cite{D'Appollonio:2013hja}, in string theory eikonal operators become important and it could therefore be useful for future considerations to be aware of this fact. In the cases considered in this chapter the eikonal operators behave as phases and one can therefore recombine the exponentials. From the definitions above we see that to properly define the subleading eikonal we need,
\begin{equation}
i \frac{\delta^{(2)}(E,\mathbf{b})}{(NT_p \kappa_D)^2} = i A_{h=2}^{(2)}(\mathbf{b}) E - \left( \frac{1}{2}i A_{h=1}^{(2)}(\mathbf{b}) i A_{h=1}^{(1)}(\mathbf{b}) E + \frac{1}{2}i A_{h=1}^{(1)}(\mathbf{b}) i A_{h=1}^{(2)}(\mathbf{b}) E \right) \;, \label{subleadingEik}
\end{equation}
where all the symbols have been defined above and we note that $A_{h=2}^{(2)}(\mathbf{b})$ represents the full subleading energy contribution derived from the one-loop amplitude in \eqref{annampinE}. Note that we have written (\ref{subleadingEik}) in the most general way possible accounting for the possibility that $i A_{h=1}^{(1)}(\mathbf{b})$ is an operator instead of a phase. In the cases we consider in this chapter the eikonal operators become phases and the equation above reads $i \delta^{(2)}(E,\mathbf{b}) = (NT_p \kappa_D)^2 \left(i A_{h=2}^{(2)}(\mathbf{b}) E - i A_{h=1}^{(2)}(\mathbf{b}) i A_{h=1}^{(1)}(\mathbf{b}) E\right) $. \\

\subsection{Elastic Contributions to the Eikonal} \label{elasticeik}

We will calculate and discuss some explicit results in the high-energy limit for the interactions discussed in section \ref{2braneamps} for elastic dilaton scattering and show how this relates to the elastic eikonal scattering amplitude framework discussed at the start of this section. All the results for the integrals used below can be found in chapter \ref{sec:dbraneintegrals}.

\subsubsection{RR Sources} \label{AttachRRhighE}

The first and second terms of \eqref{FullRRSTAmp2} do not contribute to the high-energy limit. The first term is trivially $E^0$ as can be seen from the explicit expression for $\mathcal{I}_2$ in section \ref{integralref2prop}. The second term is more subtle but is also not of $\mathcal{O}(E^2)$ due to the extra propagator present in the integrals (the $1/u$ and $1/t$) which brings down a factor of $1/E$ after performing the integral, $\mathcal{I}_3$. The remaining terms we have are,

\begin{eqnarray}
i \mathcal{A}^{\rm dd \; (2)}_{h=2} &\approx& - i(N T_p \kappa_D)^2 \left( \frac{4}{s} k_{1 \mu} k_{2 \nu} \mathcal{I}_{2}^{\mu \nu} + 2 E^2 \mathcal{I}_{2} \right) \text{ .} 
\end{eqnarray}
Substituting the results for the various integrals,
\begin{eqnarray}
i \mathcal{A}^{\rm dd \; (2)}_{h=2} & \approx & i(N T_p \kappa_D)^2 E^2 \frac{1}{(4 \pi)^{\frac{D-p-1}{2}}} \frac{\Gamma{\left( \frac{3-D+p}{2} \right)} \Gamma^2{\left( \frac{D-p-1}{2} \right)}}{\Gamma{\left( D-p-1 \right)}} \nonumber \\
&& \qquad \times (q^2_{\perp})^{\frac{D-p-5}{2}} (4(D-p-2)-2) \;. \label{finalampST_RR}
\end{eqnarray}
Note that here and throughout this and the following section the $\approx$ signifies that we have dropped some terms that are subleading in energy which arise from performing the integrals. 

\subsubsection{Graviton Sources} \label{AttachGravhighE}

We can now substitute the relevant results for the integrals in \eqref{FullGGSTAmp} and identify which terms contribute to each power of energy. We have for the leading contribution,

\begin{eqnarray}
i \mathcal{A}^{\rm dd \; (1)}_{h=2} &=& i (N T_p \kappa_D)^2 4 E^4 \mathcal{I}_3^{(1)} \nonumber \\
&\approx& - (N T_p \kappa_D)^2 E^3 \frac{2 \sqrt{\pi}}{(4\pi)^{\frac{D-p-1}{2}}} \frac{\Gamma{\left(\frac{6-D+p}{2} \right)}  \Gamma^2\left(\frac{D-p-4}{2}\right)}{\Gamma(D-p-4)}(q^2_{\perp})^{\frac{D-p-6}{2}}  \text{ .} \label{GGAmpHighEu}
\end{eqnarray}
Note that in the last line we have used the solution for the leading energy contribution of $\mathcal{I}_3$ in section \ref{sec:dbraneintegrals} which we have denoted as $\mathcal{I}_3^{(1)}$. This is the only contribution at leading order in energy. The u- and t-channel diagrams which produce this leading contribution also have subleading contributions arising from the subleading term in $\mathcal{I}_3$,
\begin{eqnarray}
i \mathcal{A}^{\rm dd \; (2)}_{h=2} &=& i (N T_p \kappa_D)^2 4 E^4 \mathcal{I}_3^{(2)} \nonumber \\
&=& - i (N T_p \kappa_D)^2 E^2 \frac{2}{(4\pi)^{\frac{D-p-1}{2}}}  \frac{\Gamma{\left(\frac{5-D+p}{2} \right)} \Gamma^2(\frac{D-p-3}{2})}{\Gamma(D-p-4)} (q^2_{\perp})^{\frac{D-p-5}{2}} \;,
\label{GGAmpSubEu}
\end{eqnarray} 
where $\mathcal{I}_3^{(2)}$ the subleading energy contribution to $\mathcal{I}_3$. The other subleading contributions that arise from the second and third terms in \eqref{FullGGSTAmp} are,
\begin{eqnarray}
i \mathcal{A}^{\rm dd \; (2)}_{h=2} &=& i (N T_p \kappa_D)^2 \left(- \frac{2(D-2p-4)}{D-2} E^2 \mathcal{I}_2 + \frac{2(D-p-3)(1+p)}{D-2} \frac{1}{s} k_{1 \mu} k_{2 \nu}\mathcal{I}_{2}^{\mu \nu} \right) \nonumber \\
&\approx & i (N T_p \kappa_D)^2 E^2 \left(\frac{4(D-2p-4)(D-p-2)}{D-2} + \frac{(D-p-3)(1+p)}{D-2} \right) \nonumber \\
&& \times \frac{1}{(4\pi)^{\frac{D-p-1}{2}}}  \frac{\Gamma{\left( \frac{3-D+p}{2} \right)} \Gamma^2{\left( \frac{D-p-1}{2} \right)}}{ \Gamma{\left( D-p-1 \right)}} (q^2_{\perp})^{\frac{D-p-5}{2}} \text{ ,} \label{GGAmpHighEs}
\end{eqnarray}
and,
\begin{eqnarray}
i \mathcal{A}^{\rm dd \; (2)}_{h=2} &=& i (N T_p \kappa_D)^2 E^2 \frac{4 (D-p-3)}{D-2} \mathcal{I}_2 \nonumber \\
&=& - i (N T_p \kappa_D)^2 E^2 \frac{8(D-p-3)(D-p-2)}{D-2} \frac{1}{(4\pi)^{\frac{D-p-1}{2}}} \nonumber \\
&& \times  \frac{\Gamma{\left( \frac{3-D+p}{2} \right)} \Gamma^2{\left( \frac{D-p-1}{2} \right)}}{ \Gamma{\left( D-p-1 \right)}} (q^2_{\perp})^{\frac{D-p-5}{2}} \text{ ,} \label{GGAmpHighEc}
\end{eqnarray}
where we have separated the second and third terms of \eqref{FullGGSTAmp} into \eqref{GGAmpHighEs} and \eqref{GGAmpHighEc} purposefully in order to be able to more easily compare with the results obtained in section  \ref{lntoeikonal}.

\subsubsection{Dilaton Sources}\label{AttachDilhighE}

The only term contributing to the leading energy behaviour in this case is,
\begin{eqnarray}
i\mathcal{A}^{\rm dd \; (2)}_{h=2} &=& - 2 i \kappa_D^2 \left( N T_p \frac{a(D)}{\sqrt{2}} \right)^2  \frac{1}{s} k_{1 \mu} k_{2 \nu}\mathcal{I}_{2}^{\mu \nu} \nonumber \\
&\approx & i (N T_p \kappa_D)^2 \left(\frac{a(D)}{\sqrt{2}} \right)^2 E^2 \frac{1}{(4\pi)^{\frac{D-p-1}{2}}}  \frac{\Gamma{\left( \frac{3-D+p}{2} \right)} \Gamma^2{\left( \frac{D-p-1}{2} \right)}}{ \Gamma{\left( D-p-1 \right)}} (q^2_{\perp})^{\frac{D-p-5}{2}} \;, \nonumber \\ \label{finalampFT_Dil}
\end{eqnarray}
where we have used the kinematics outlined in section \ref{kinematics}.

\subsubsection{Eikonal Scattering} \label{ElasticEikDiscussion}

We can use the results derived above to explicitly show that \eqref{leadingeikgen} holds for the elastic scattering of dilatons from D-branes. Note that we drop the vector notation for the impact parameter as only the magnitude, $b=|\mathbf{b}|$ appears below. Writing the leading energy behaviour of the tree-level and one-loop amplitudes in the form of \eqref{diskampinE} and \eqref{annampinE} respectively and by converting these expressions into impact parameter space using \eqref{eq:impoformu}, we find for the tree-level amplitude,
\begin{eqnarray}
i A_{h=1}^{(1)\,e}(b) &=& \frac{i}{4 \pi^{\frac{D-p-2}{2}}} \frac{\Gamma \left(\frac{D-p-4}{2} \right) }{b^{D-p-4}}\;, \label{elEips} \\
i A_{h=1}^{(2)\,e}(b) &=& 0 \;,
\end{eqnarray}
and for the one-loop amplitude,
\begin{eqnarray}
i A_{h=2}^{(1)\,e}(b) &=& - \frac{1}{32 \pi^{D-p-2}} \frac{\Gamma^2 \left(\frac{D-p-4}{2} \right) }{b^{2D-2p-8}}\;, \\
i A_{h=2}^{(2)\,e}(b) &=& i \frac{1}{16 \pi^{D-p-3/2}} \frac{\Gamma^2 \left(\frac{D-p-3}{2} \right) \Gamma \left(\frac{2D-2p-7}{2} \right)}{\Gamma \left(D-p-4 \right)} \frac{1}{b^{2D-2p-7}} \;,
\end{eqnarray}
where here we are focusing on the elastic component as reminded by the superscript $e$. We note that for the one-loop amplitude the contributions to $A_{h=2}^{(2)\,e}(b)$ arising from \eqref{finalampST_RR}, \eqref{GGAmpHighEs}, \eqref{GGAmpHighEc} and \eqref{finalampFT_Dil} sum to zero. This means that the only contribution to the subleading eikonal arises from the subleading contribution to the leading energy contribution, \eqref{GGAmpSubEu}, where we recall that $\mathcal{I}_3$ has contributions at different powers of $E$.

We can now easily confirm that $i A_{h=2}^{(1)\,e}(b)=-\frac{1}{2}(A_{h=1}^{(1)\,e}(b))^2$ as required in order to see the exponentiation of the leading eikonal $\chi^{(1)}(E,b)$ in the elastic channel. We therefore find that the elastic dilaton scattering process we have considered behaves as predicted by the leading eikonal expression \eqref{leadingeikgen}. 

\subsection{Inelastic Contributions to the Eikonal} \label{inelasticeik}

As we have done in section \ref{elasticeik} for the elastic dilaton scattering process, we can find the leading energy behaviour of the inelastic scattering of a dilaton and an RR field from the stack of D-branes that we studied in section \ref{ineldilRRallE}. Looking at the leading energy contribution of \eqref{eq:dRgRf} we find,
\begin{eqnarray}
&& i \mathcal{A}_{h=2}^{\rm dR \; (2)} =  i (N T_p \kappa_D)^2 8 a(D)( q \cdot C^{(2)})^{1 \ldots p} E^3 \mathcal{I}_3^{(1)} \nonumber \\
&& \approx   - (N T_p \kappa_D)^2 2 a(D) E^2 ( q \cdot C^{(2)})^{1 \ldots p} \frac{2 \sqrt{\pi}}{(4\pi)^{\frac{D-p-1}{2}}} \frac{\Gamma{\left(\frac{6-D+p}{2} \right)}  \Gamma^2\left(\frac{D-p-4}{2}\right)}{\Gamma(D-p-4)}(q^2_{\perp})^{\frac{D-p-6}{2}} \;, \nonumber \\ 
\label{dRRAmpHighE}
\end{eqnarray}
We can now apply the prescription outlined in \eqref{fullInelEik} to \eqref{dRRAmpHighE}. Writing the tree-level amplitude \eqref{eq:dil-RR0} in the form of \eqref{diskampinE} we find that
\begin{eqnarray}
i A_{h=1}^{{\rm dR} (1)}(b) &=& 0 \\
i T(E,b) \approx i A_{h=1}^{{\rm dR} (2)}(b) &=& i \frac{a(D)}{4} ( q \cdot C)^{1 \ldots p} \frac{1}{\pi^{\frac{D-p-2}{2}}} \frac{\Gamma \left(\frac{D-p-4}{2} \right) }{b^{D-p-4}} \;. \label{inelE0ips}
\end{eqnarray}
The other ingredients we need are $A_{h=2}^{(1)}(b)$ and $A_{h=2}^{(2)}(b)$ that we can read by comparing~\eqref{dRRAmpHighE} and \eqref{annampinE},
\begin{eqnarray}
i A_{h=2}^{{\rm dR} (1)}(b) &=& 0 \\
i A_{h=2}^{{\rm dR} (2)}(b) &=& - \frac{a(D)}{16} ( q \cdot C)^{1 \ldots p} \frac{1}{\pi^{D-p-2}} \frac{\Gamma^2 \left(\frac{D-p-4}{2} \right) }{b^{2D-2p-8}} \;.
\end{eqnarray}
We then need to calculate $i A_{h=1}^{(1) \,e}(b) i A_{h=1}^{{\rm dR} (2)}(b)$ as this will show us what to subtract in order to obtain the well defined subleading eikonal $\delta^{(2)}(E,b)$, including the inelastic contributions discussed above. We note here that although this inelastic process does not contribute to the total $A_{h=1}^{(1)}(b)$ we have to take into account the contribution from the elastic processes described in section \ref{ElasticEikDiscussion}.  We can easily verify by using \eqref{elEips} and \eqref{inelE0ips} that,
\begin{equation}
  \label{eq:dRchi0}
  i A_{h=2}^{{\rm dR} (2)}(b)- i A_{h=1}^{(1) \,e}(b) i A_{h=1}^{{\rm dR} (2)}(b) = 0
\end{equation}
and so we see that the inelastic dilaton to RR channel does not contribute to the subleading eikonal~\eqref{subleadingEik}.


\section{Alternative Computation of the Leading and Subleading Eikonal}
\label{lntoeikonal}

In this section we discuss a more conventional way to compute the elastic scattering of a dilaton from a stack of D$p$-branes both in Einstein gravity and in a theory of gravity extended to include the dilaton and RR fields.  We will take the high energy limit and extract the leading and subleading eikonal, which agrees with the ones computed in the previous section. The leading eikonal is obtained from the tree diagram corresponding to the exchange of a graviton, while the subleading eikonal is derived from a number of one-loop  diagrams that depend on which theory of gravity we consider. Since three of the one-loop diagrams are most easily obtained by first computing the one-point graviton amplitude and then attaching the three-point vertex containing two dilatons and one graviton, in the first subsection we compute the one-point amplitudes for the graviton, dilaton and RR field at the tree and one-loop level and we show that they are  directly related to the large distance behaviour of the classical solution describing the 
D$p$-branes to which the graviton, dilaton and RR field are coupled. In the second subsection we compute the contribution of the various field theory diagrams to the elastic dilaton scattering and from them we extract the leading and subleading eikonal. 

\subsection{One-point Amplitudes and the Classical Solution}
\label{1point}

In this subsection we will write the one-point functions for the graviton, dilaton and RR field in the gravity theory described by the bulk action given in \eqref{eq:bulkaction} and the boundary action given by \eqref{branecoup} as we've done in the previous sections. Using these two actions one can compute the contribution to the one-point amplitude of the diagram with the 3-graviton vertex yielding\footnote{In all one-point amplitudes we omit to explicitly write a $\delta$-function that constrains the longitudinal component of the momentum to be vanishing.},

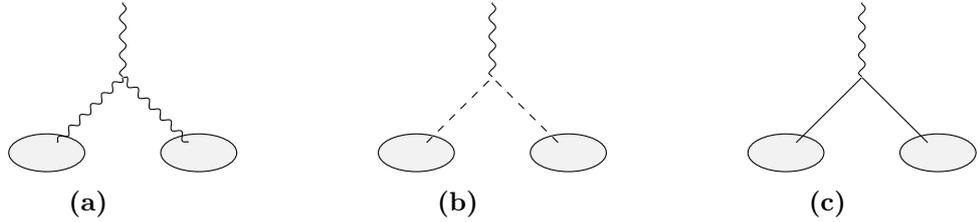
\begin{figure}[h]
  \begin{subfigure}[t]{0.3\textwidth}
    \centering
    \begin{tikzpicture}
		\begin{feynman}
			\vertex (a) at (-1,-2) {};
			\vertex (b) at ( 1,-2) {};
			\vertex (c) at (-2.5, 0) {};
			\vertex (d) at ( 2.5, 0) {};
			\vertex[circle,inner sep=0pt,minimum size=0pt] (e) at (0, -1) {};
			\vertex[circle,inner sep=0pt,minimum size=0pt] (m) at (0, 0) {};
			\draw[fill=light-gray] (-1,-2) ellipse (0.5cm and 0.25cm);
			\draw[fill=light-gray] (1,-2) ellipse (0.5cm and 0.25cm);       
			\diagram* {
			(m) -- [boson] (e),
			(a) -- [boson] (e),
			(b) -- [boson] (e),
			};
		\end{feynman}
    \end{tikzpicture}
    \caption{}
    \label{fig:5a}
  \end{subfigure}
  \quad
  \begin{subfigure}[t]{0.3\textwidth}
    \centering
    \begin{tikzpicture}
		\begin{feynman}
			\vertex (a) at (-1,-2) {};
			\vertex (b) at ( 1,-2) {};
			\vertex (c) at (-2.5, 0) {};
			\vertex (d) at ( 2.5, 0) {};
			\vertex[circle,inner sep=0pt,minimum size=0pt] (e) at (0, -1) {};
			\vertex[circle,inner sep=0pt,minimum size=0pt] (m) at (0, 0) {};
			\draw[fill=light-gray] (-1,-2) ellipse (0.5cm and 0.25cm);
			\draw[fill=light-gray] (1,-2) ellipse (0.5cm and 0.25cm);       
			\diagram* {
			(m) -- [boson] (e),
			(a) -- [scalar] (e),
			(b) -- [scalar] (e),
			};
		\end{feynman}
    \end{tikzpicture}
    \caption{}
    \label{fig:5b}
  \end{subfigure}
  \quad
  \begin{subfigure}[t]{0.3\textwidth}
    \centering
    \begin{tikzpicture}
		\begin{feynman}
			\vertex (a) at (-1,-2) {};
			\vertex (b) at ( 1,-2) {};
			\vertex (c) at (-2.5, 0) {};
			\vertex (d) at ( 2.5, 0) {};
			\vertex[circle,inner sep=0pt,minimum size=0pt] (e) at (0, -1) {};
			\vertex[circle,inner sep=0pt,minimum size=0pt] (m) at (0, 0) {};
			\draw[fill=light-gray] (-1,-2) ellipse (0.5cm and 0.25cm);
			\draw[fill=light-gray] (1,-2) ellipse (0.5cm and 0.25cm);       
			\diagram* {
			(m) -- [boson] (e),
			(a) -- (e),
			(b) -- (e),
			};
		\end{feynman}
    \end{tikzpicture}
    \caption{}
    \label{fig:5c}
  \end{subfigure}
  \caption{The various contributions to the one-point function at subleading order used to construct the classical solution. Figure \ref{fig:5a} is the contribution with the 3-graviton vertex and figures \ref{fig:5b} and \ref{fig:5c} are the contributions with RR fields and dilaton sources respectively. As before the solid lines correspond to dilatons, the wavy lines correspond to gravitons and the dashed lines correspond to RR fields.}
  \label{fig:5}
\end{figure}

\begin{eqnarray}
&&  \langle h_{\lambda \tau} \rangle_{(\ref{fig:5a})} = N^2 \kappa_D
J_h^2 
\left[\frac{ |q_{\perp}|^{D-p-5}  }{
(4\pi)^{ \frac{D-p-1}{2} }  } \Gamma \left(\frac{3-D+p}{2}\right) 
\frac{\Gamma^2 (\frac{D-p-1}{2})}{
\Gamma (D-p-1)} \right] \nonumber \\
&& \times \Big\{ \frac{(D-3-p) (p+1)}{2 (D-2)} 
\Big[ \eta_{\perp \lambda \tau} - (3-D+p) \frac{q_{\perp \lambda} 
q_{\perp \tau}}{q_{\perp}^2}   \Big] \nonumber \\
&& - 2 \frac{(D-p-2) (p+1) (D-p-3)}{D-2} 
\frac{q_{\perp \lambda} 
q_{\perp \tau}}{q_{\perp}^2}   \nonumber \\
&& - \frac{2(D-p-2)(D-p-3)^2}{(D-2)^2} \eta_{\parallel \lambda \tau}
- \frac{2 (D-p-2) (p+1)^2}{(D-2)^2} \eta_{\perp \lambda \tau}
\Big\} \;,
\label{3gra1}
\end{eqnarray}
The sum of the contributions from the diagrams with the dilaton and the RR field is given by
\begin{eqnarray}
&&\langle h \rangle_{(\ref{fig:5b}) + (\ref{fig:5c})} = N^2 \kappa_D 
\left[\frac{ |q_{\perp}|^{D-p-5}  }{
(4\pi)^{ \frac{D-p-1}{2} }  } \Gamma \left(\frac{3-D+p}{2}\right) 
\frac{\Gamma^2 (\frac{D-p-1}{2})}{
\Gamma (D-p-1)} \right]  \nonumber \\
&& \times \Big\{ \frac{J_{\phi}^2}{2}
\Big[ \eta_{\perp \lambda \tau} - (3-D+p) \frac{q_{\perp \lambda} 
q_{\perp \tau}}{q_{\perp}^2}   \Big]  \label{3gra2} \\
&& + \frac{\mu_p^2}{2} \Big[  - \frac{2(D-p-2) (D-p-3)}{D-2} 
 \eta_{\parallel \lambda \tau} + \frac{2 (D-p-2) (p+1)}{D-2} 
\eta_{\perp \lambda \tau}  \nonumber \\ \nonumber
&& - \Big[ \eta_{\perp \lambda \tau} - (3-D+p) \frac{q_{\perp \lambda} 
q_{\perp \tau}}{q_{\perp}^2}   \Big]  \Big]  \Big\} \;.
\end{eqnarray}
Summing the  three contributions  we get
\begin{eqnarray}
&&\langle h \rangle_{(\ref{fig:5a}) + (\ref{fig:5b}) + (\ref{fig:5c})} =N^2 \kappa_D 
\left[\frac{ |q_{\perp}|^{D-p-5}  }{
(4\pi)^{ \frac{D-p-1}{2} }  } \Gamma \left(\frac{3-D+p}{2}\right) 
\frac{\Gamma^2 (\frac{D-p-1}{2})}{
\Gamma (D-p-1)} \right] \nonumber \\   
&& \times \Big\{  - 2 \frac{(D-p-2) (p+1) (D-p-3)}{D-2} 
\frac{q_{\perp \lambda} 
q_{\perp \tau}}{q_{\perp}^2}   \nonumber \\
&& - \frac{2(D-p-2)}{D-2}  (D-p-3)\eta_{\parallel \lambda \tau} 
\Big[ J_h^2 \frac{D-p-3}{D-2} + \frac{\mu_p^2}{2} \Big] 
\nonumber \\
&& - \frac{2 (D-p-2)}{D-2} (p+1) \eta_{\perp \lambda \tau} \Big[ J_h^2 
\frac{p+1}{D-2} - 
\frac{\mu_p^2}{2} \Big] \Big\} \;,
\label{htotal}
\end{eqnarray}
where from the expressions for $J_{\phi}$, $\mu_p$ and $a(D)$ defined in the relevant subsections of~\ref{sec:quantumgrav}, we have used the fact that the following quantity vanishes,
\begin{eqnarray}
 \frac{(D-p-3)(p+1)}{2(D-2)}J_h^2 + 
\frac{J_{\phi}^2}{2}- \frac{\mu_p^2}{2}  = 0 \;.
\label{gra5r}
\end{eqnarray}
We neglect for a moment the term in the second line of \eqref{htotal} that corresponds to a gauge transformation  of the metric as we will discuss it in subsection \ref{Einstein} where we will see that it must be neglected if we want the metric in the harmonic gauge.

Going from momentum to position space, (\ref{htotal}) becomes,
\begin{eqnarray}
&&\langle \tilde{h}_{\mu \nu} \rangle_{(\ref{fig:5a}) + (\ref{fig:5b}) + (\ref{fig:5c})} = \frac{N^2 \kappa_D}{D-2} 
\left( \frac{1}{(D-p-3) \Omega_{D-p-2} r^{D-p-3}}
  \right)^2 \nonumber \\
&& \times
 \left\{ \eta_{\parallel \mu \nu} (D-p-3) \left[  J_h^2 \frac{D-p-3}{D-2}
 +\frac{ \mu_p^2}{2} \right] + (p+1) \eta_{\perp \mu \nu}
 \left[  J_h^2 \frac{p+1}{D-2}
 -\frac{ \mu_p^2}{2} \right]  \right\} \;, \nonumber \\
 \label{gra7r}
 \end{eqnarray}
where we note that the tilde signifies the Fourier transform to position space. The expression in position space can be obtained by using \eqref{eq:impoformu}. Inserting the explicit quantities (\ref{gra7r}) becomes,
 \begin{eqnarray}
 &&\langle 2 \kappa_D \tilde{h}_{\mu \nu} \rangle_{(\ref{fig:5a}) + (\ref{fig:5b}) + (\ref{fig:5c})} = 
\frac{1}{2}\left(\frac{R_p}{r} \right)^{2(D-p-3)} 
\nonumber \\
&& \times
 \left\{ \eta_{\parallel \mu \nu} \frac{D-p-3}{D-2} \left[   \frac{D-p-3}{D-2}
 + 1 \right] + \frac{p+1}{D-2}  \eta_{\perp \mu \nu}
 \left[   \frac{p+1}{D-2}  - 1 \right]  \right\} \;, \nonumber \\
\label{gra7b}
\end{eqnarray}
where we have introduced the following quantity,
\begin{eqnarray}
 \frac{2 N \kappa_D T_p}{(D-p-3) \Omega_{D-p-2} r^{D-p-3}}
\equiv \left(\frac{R_p}{r} \right)^{D-p-3}  ~~;~~
 \Omega_d \equiv \frac{ 2 \pi^{\frac{d+1}{2} }}{\Gamma (
\frac{d+1}{2})} \;.
\label{Rp}
\end{eqnarray}
We note that (\ref{gra7b}) provides the  total one-loop contribution to the one-point graviton amplitude. The tree contribution can
also be easily computed from the bulk and boundary actions yielding,
\begin{eqnarray}
\langle 2 \kappa_D \tilde{h}_{\mu \nu} (x) \rangle _{1}= - 
\left(\frac{R_p}{r} \right)^{D-p-3} 
 \left( \frac{D-p-3}{D-2} \eta_{\mu \nu}^{\parallel} 
- \frac{p+1}{D-2} \eta_{\mu \nu}^{\perp} \right) \;,
\label{gra13}
\end{eqnarray}
which is the Fourier transform of the following amplitude in momentum space,
\begin{eqnarray}
\langle h_{\mu \nu} \rangle_{1} = - \frac{N T_p}{q_\perp^2} \left( \frac{D-p-3}{D-2} \eta_{\parallel \mu \nu} - \frac{p+1}{D-2} \eta_{\perp\mu \nu} \right) \;.
\label{gra13f}
\end{eqnarray}
Note that we are using the same notation as in section \ref{2braneamps} with subscripts $1$ and $2$ representing tree diagrams and one-loop diagrams respectively. In an extended gravity theory also containing the dilaton and the RR field we have to include the one-point amplitude for the dilaton and the RR field. The one-loop one-point amplitude for the dilaton is given  by the sum of two diagrams. One with the vertex containing two dilatons and one graviton and the other with the vertex with one dilaton and two RR fields.
It turns out that the first diagram is vanishing while the second one gives,
\begin{eqnarray}
\langle \phi \rangle_{2} = 
\frac{a(D)  \sqrt{2} N^2 \kappa_D   }{2 }  \mu_p^2   
(2-D+p)
\left[\frac{ |q_{\perp}|^{D-p-5}  }{
(4\pi)^{ \frac{D-p-1}{2} }  } \Gamma \left(\frac{3-D+p}{2}\right) 
\frac{\Gamma^2 (\frac{D-p-1}{2})}{
\Gamma (D-p-1)} \right] 
\label{OD15}
\end{eqnarray} 
where the dilaton field has been canonically normalised. From (\ref{OD15}) we can go to position space,
\begin{equation}
\langle \sqrt{2} \kappa_D \tilde{\phi} \rangle_{2} = 
\frac{ a(D) }{4 }
 \left( \frac{2 N \kappa_D T_p}{(D-p-3) \Omega_{D-p-2} r^{D-p-3}} 
\right)^2 = \frac{ a(D) }{4 }  \left(\frac{R_p}{r} \right)^{2(D-p-3)} \;,
\label{OD18}
\end{equation}
where we have used that $\mu_p = \sqrt{2} T_p$.  We also have the tree diagram that in momentum space gives the following contribution,
\begin{eqnarray}
 \langle \phi \rangle_{1}  = N
J_\phi \frac{1}{q_\perp^2} \;,
\label{OD23}
\end{eqnarray} 
which in position space becomes,
\begin{eqnarray}
\langle  \sqrt{2} \kappa_D \tilde{\phi} \rangle_{1} = \frac{J_\phi}{\sqrt{2}T_p} 
\left(\frac{R_p}{r} \right)^{D-p-3} 
=- \frac{a(D)}{2}   \left(\frac{R_p}{r} \right)^{D-p-3} \;.
\label{OD24}
\end{eqnarray} 
The one-loop one-point amplitude for the RR fields gets a contribution from two diagrams; one with the vertex involving two RR fields and one graviton and the other involving again two RR fields and a dilaton. The sum of the two is equal, in momentum space, to,
\begin{eqnarray}
\langle C_{01 \dots p}\rangle_{2} =  4 N^2 T_p \kappa_D \mu_p (D-p-2)  \left[\frac{ |q_{\perp}|^{D-p-5}  }{
(4\pi)^{ \frac{D-p-1}{2} }  } \Gamma \left(\frac{3-D+p}{2}\right) 
\frac{\Gamma^2 (\frac{D-p-1}{2})}{
\Gamma (D-p-1)} \right] \;, \nonumber \\
\label{RR15}
\end{eqnarray} 
which in position space becomes,
\begin{eqnarray}
&&\langle \tilde{C}_{01 \dots p}\rangle_{2} = - 4 N^2 T_p \kappa_D \mu_p  \frac{1}{2}   
  \left( \frac{1}{(D-p-3) \Omega_{D-p-2} r^{D-p-3}} \right)^2 \;,
\label{RR16}
\end{eqnarray} 
where the field $C_{01 \dots p}$ is canonically normalised. In order to compare this with the classical solution, we need the quantity,
\begin{eqnarray}
\langle \sqrt{2} \kappa_D \tilde{C}_{01 \dots p}\rangle_{2}  = -  
  \left( \frac{2 N \kappa_D T_p}{(D-p-3) \Omega_{D-p-2} r^{D-p-3}} \right)^2  =- \left(\frac{R_p}{r} \right)^{2(D-p-3)}  \;.
\label{RR18}
\end{eqnarray} 
The tree diagram can also be easily computed, we find
\begin{eqnarray}
\langle \sqrt{2} \kappa_D \tilde{C}_{01 \dots p} \rangle_{1}= 
-\frac{2 N T_p \kappa_D}{(D-p-3) \Omega_{D-p-2} r^{D-p-3}} = \left(\frac{R_p}{r} \right)^{D-p-3} \;.
\label{RR22}
\end{eqnarray}

The previous diagrammatic results, obtained for the various one-point amplitudes, can be compared with the large distance expansion of the classical solution. It turns out that  the tree diagrams reproduce the first correction to the flat limit  of the classical solution when $r \rightarrow \infty$, while the one-loop diagrams reproduce the subleading correction to the flat limit. The classical solution is given by \cite{Duff:1994an},
\begin{eqnarray}
&& ds^2 \equiv g_{\mu \nu} dx^\mu d x^\nu =  [H(r)]^{-\frac{D-p-3}{D-2}  } dx_{\parallel}^2 + [H(r)]^{\frac{p+1}{D-2}}
 dx_{\perp}^2   \nonumber \\
 && {\rm e}^{- \sqrt{2} \kappa_D \phi} = ( H (r) )^{a (D)/2}~~;~~~\sqrt{2} \kappa_D C_{01 \dots p} = 1 -
  H^{-1} (r)  \;,
\label{ds2xy}
\end{eqnarray}
where,
\begin{eqnarray}
H(r) = 1 + \left(\frac{R_p}{r} \right)^{D-p-3} \;.
\label{gra8}
\end{eqnarray}
Expanding the two terms appearing in the metric, we get,
\begin{eqnarray}
&&[H(r)]^{-\frac{D-p-3}{D-2}  }  = 1 - \frac{D-p-3}{D-2} 
 \left(\frac{R_p}{r} \right)^{D-p-3} 
 \nonumber \\ 
 &&
 + \frac{1}{2} \frac{D-p-3}{D-2} \left( \frac{D-p-3}{D-2} +1 \right) \left(\frac{R_p}{r} \right)^{2(D-p-3)}   + \dots \;,
\label{gra10x}
\end{eqnarray}
and
\begin{eqnarray}
&& [H(r)]^{\frac{p+1}{D-2}} = 1 + \frac{p+1}{D-2}  \left(\frac{R_p}{r} \right)^{D-p-3} 
 \nonumber \\ 
 && 
 + \frac{1}{2} \frac{p+1}{D-2} \left( \frac{p+1}{D-2} -1 \right)  \left(\frac{R_p}{r} \right)^{2(D-p-3)}  +\dots \;.
\label{gra11x}
\end{eqnarray}
Remembering that in our notation $g_{\mu \nu} =\eta_{\mu \nu} + 2 \kappa_D h_{\mu \nu}$, we see that
for $r \rightarrow \infty$ we get the flat Minkowski metric. Then, comparing (\ref{ds2xy}), (\ref{gra10x}) and (\ref{gra11x}) with (\ref{gra13}) and (\ref{gra7b}), we see that the first correction to the flat space metric is given by the tree diagram of the one-point graviton amplitude, while the next correction is given by the one-loop diagram contribution to the one-point graviton amplitude. Expanding in a similar way the classical solution for the dilaton we get,
\begin{eqnarray}
&& - \sqrt{2} \kappa_D \phi = \frac{a(D)}{2} \log \left(1+ \left(\frac{R_p}{r} \right)^{D-p-3} \right) 
\nonumber \\
&& = \frac{a(D)}{2} \left(  \left(\frac{R_p}{r} \right)^{D-p-3} - \frac{1}{2}  \left(\frac{R_p}{r} \right)^{2(D-p-3)}  +
\dots \right) \;.
\nonumber \\
\label{dilaFF}
\end{eqnarray}
These two terms are  reproduced by the tree diagram in (\ref{OD24}) and the one-loop term in (\ref{OD18}) respectively. Similarly expanding the solution for the RR field yields,
\begin{eqnarray}
&& \sqrt{2} \kappa_D C_{01 \dots p} =  1 - H^{-1}  =   \left(\frac{R_p}{r} \right)^{D-p-3}  -
 \left(\frac{R_p}{r} \right)^{2(D-p-3)}  + \dots \;.
\label{dilRR}
\end{eqnarray}
Again we find that these two terms are equal to those in (\ref{RR22}) and (\ref{RR18}).

In conclusion, we have shown  that the various terms of the expansion of the classical solution
can be reproduced by computing the one-point amplitude of the corresponding fields.

\subsection{Elastic Dilaton Scattering in Extended Gravity}
\label{ElaExtGra}

In this subsection we compute the elastic dilaton scattering amplitude in an extended theory of gravity with a dilaton and an RR field as in section \ref{elasticeik}. It consists of one tree diagram and five one-loop diagrams. The tree diagram and the sum of three one-loop diagrams can be obtained directly from the one-point amplitude computed in the previous subsection by saturating it with the three-point amplitude of two dilatons and one graviton given in (\ref{ddgV}). For the tree diagram we find the following\footnote{In this case we also omit writing the factor $(2\pi)^{p+1} \delta^{(p+1)} (k_1 +k_2)$ of momentum conservation along the directions of the D$p$-brane.},
\begin{eqnarray}
i\mathcal{A}_{1}^{\rm dd} = i \frac{2 N T_p \kappa_D}{(-s)} \left(\frac{D-p-3}{D-2} (k_{1} \cdot k_2)_{\parallel}  - \frac{p+1}{D-2}  (k_{1} \cdot k_2)_{\perp}   \right)  =  i \frac{2 N T_p \kappa_D E^2}{(-s)} \;,
\label{EDS1}
\end{eqnarray}
where we have neglected terms without the pole at $s \sim 0$ as well as terms negligible at high energy (see kinematics in (\ref{k1k2})). We find that this is in agreement with \eqref{eq:dil-dil0}.

The first one-loop diagram corresponds to the separate exchange of two gravitons that  are then attached to the D$p$-branes. One gets,
\begin{eqnarray}
i \mathcal{A}_{2, t}^{\rm ddgg} = i (N \kappa_D T_p)^2 4E^4 \int \frac{d^{D-p-1} k}{(2\pi)^{D-p-1}} \frac{1}{(k_1 - k)_{\perp}^2 k^2 (k_2 +k)^2_{\perp} } \;,
\label{EDS2}
\end{eqnarray}
where $k^2 \equiv -E^2 + k_{\perp}^2$. At high energy we obtain a leading term given by,
\begin{eqnarray}
i \mathcal{A}_{2, t}^{\rm ddgg} \approx  -  \frac{2  (N \kappa_D T_p)^2E^3 \sqrt{\pi} }{  (4 \pi)^{\frac{D-p-1}{2}}} 
\frac{\Gamma{\left(\frac{6-D+p}{2} \right)} \Gamma^2(\frac{D-p-4}{2})}{\Gamma(D-p-4)}  (q^2_{\perp})^{\frac{D-p-6}{2}} \;,
\label{EDS3}
\end{eqnarray}
and a subleading term equal to,
\begin{eqnarray}
i \mathcal{A}_{2, t}^{\rm ddgg} \approx - i \frac{(N \kappa_D T_p)^2  2E^2}{(4\pi)^{\frac{D-p-1}{2}}}
\frac{\Gamma (\frac{5-D+p}{2}) \Gamma^2 ( \frac{D-p-3}{2}) }{ 
\Gamma (D-p-4)} (q^2_{\perp})^{\frac{D-p-5}{2}} \;. 
\label{EDS4}
\end{eqnarray}
Comparing \eqref{EDS3} and \eqref{EDS4} with the equivalent results \eqref{GGAmpHighEu} and \eqref{GGAmpSubEu} derived in section \ref{HElimit} we again find agreement. The second diagram contains a vertex with two dilatons and two gravitons with the gravitons attached to the D-branes.  We find that,
\begin{eqnarray}
&& i \mathcal{A}_{2, c}^{\rm ddgg}  = i   (N \kappa_D T_p)^2     
\int \frac{d^{D-p-1} k}{(2 \pi)^{D-p-1}} 
 \frac{1}{k^{2}_{\perp}  (q-k)^{2}_{\perp}}   \nonumber \\
&& \times \frac{D-3-p}{D-2} \left( -(p+1) (k_1 \cdot k_2)  + 
4 (k_{1} \cdot k_{2} )_{\parallel} \right) \;.
\label{EDS5}
\end{eqnarray}
In the high energy limit we can neglect the first term in the round bracket in the second line and we find,
\begin{eqnarray}
i \mathcal{A}_{2, c}^{\rm ddgg} &\approx & - i \frac{(N \kappa_D T_p)^2 8 E^2}{(4\pi)^{\frac{D-1-p}{2}}}  \frac{(D-p-3)(D-p-2)}{D-2}\,\,
\nonumber \\
&& \qquad \times \frac{ \Gamma( \frac{3-D+p}{2}) \Gamma^2( \frac{D-p-1}{2})}{
\Gamma ( D-p-3)} (q^2_{\perp})^{\frac{D-p-5}{2}}  \;. 
\label{EDS6}
\end{eqnarray}
We can easily see that this is equivalent to \eqref{GGAmpHighEc}. Finally, the last three one-loop diagrams are obtained by saturating the one-point amplitudes in (\ref{3gra1}) and (\ref{3gra2}) with the vertex in (\ref{ddgV}).   Let us start with the one-loop diagram in (\ref{3gra1}). The term with $(k_1 \cdot k_2)$ in (\ref{ddgV})  and the second term in the second line and the term in the third line of (\ref{3gra1}) do not contribute at high energy. The remaining terms give,
\begin{eqnarray}
&&i \mathcal{A}_{2, s}^{\rm ddgg} \approx i \frac{(N \kappa_D T_p E)^2}{D-2} \frac{1}{(4\pi)^{\frac{D-p-1}{2}}}  \frac{ \Gamma (\frac{3-D+p}{2}) \Gamma^2 (\frac{D-p-1}{2} )}{ \Gamma (D-p-1)} (q^2_{\perp})^{\frac{D-p-5}{2}} 
\nonumber \\
&& \times \left( (D-p-3)(p+1) + 4 (D-p-2) (D-2p -4)  \right) \;.
\label{EDS7}
\end{eqnarray}
Once again comparing this with our results from section \ref{HElimit} we see that the equation above is equivalent to \eqref{GGAmpHighEs}. Let us do the same analysis with (\ref{3gra2}). Again the term with $(k_1 \cdot k_2)$ in (\ref{ddgV}) does not contribute at high energy. Also the terms with $q_{\perp \lambda} q_{\perp \tau}$ do not contribute at high energy. We are therefore left with the following expression,
\begin{eqnarray}
&&i \mathcal{A}_{2, s}^{\rm ddRR} + i \mathcal{A}_{2, s}^{\rm dddd}  \approx i N^2 \kappa_D^2 E^2 \frac{1}{(4\pi)^{\frac{D-p-1}{2}}}  \frac{ \Gamma (\frac{3-D+p}{2}) \Gamma^2 (\frac{D-p-1}{2} )}{ \Gamma (D-p-1)} (q^2_{\perp})^{\frac{D-p-5}{2}} 
\nonumber \\
&& \times  \left( J_\phi^2 + \mu_p^2 \left( 2(D-p-2) -1 \right) \right)  \;.
\label{EDS8}
\end{eqnarray}
Inserting the relevant expression for $J_\phi$ and using $\mu_p = \sqrt{2} T_p$,
\begin{eqnarray}
&& i \mathcal{A}_{2, s}^{\rm ddRR} + i \mathcal{A}_{2, s}^{\rm dddd}  \approx i (N \kappa_D T_p E)^2   \frac{1}{(4\pi)^{\frac{D-p-1}{2}}}  \frac{ \Gamma (\frac{3-D+p}{2}) \Gamma^2 (\frac{D-p-1}{2} )}{ \Gamma (D-p-1)} (q^2_{\perp})^{\frac{D-p-5}{2}} 
\nonumber \\
&& \times \left[  \left(2 - \frac{(p+1)(D-p-3)}{D-2} \right) + \left( 4(D-p-2) -2 \right)  \right]  \;,
\label{EDS9}
\end{eqnarray}
where the first round bracket in the second line comes from the dilaton, while the second round bracket comes from the RR contribution. This can be compared to the sum of \eqref{finalampST_RR} and \eqref{finalampFT_Dil}, which agrees with what is written above.

The total eikonal, including both leading and subleading contributions, is defined as $\delta (b,E)=\delta^{(1)}(E,b)+\delta^{(2)}(E,b)$, where the $\delta^{(i)}(E,b)$ have been defined in section \ref{HElimit}. From the various expressions computed in this section we arrive at the following expression,
\begin{eqnarray}
&& \delta (b, E) = \frac{N \kappa_D^2 \tau_p}{4}  E\frac{\Gamma 
( \frac{D-p-4}{2})}{ \pi^{\frac{D-p-2}{2}} b^{D-p-4}} 
+\frac{(N \kappa_D^2 \tau_p)^2 \, E \,\Gamma^2 ( \frac{D-p-3}{2})
\Gamma (D-p - \frac{7}{2})}{16 \,\pi^{D-p - \frac{3}{2}} \Gamma (D-p-4) 
\,\,b^{2D-2p-7}} \nonumber \\
&& + \frac{(N \kappa_D^2 \tau_p)^2 \, E \, \Gamma (D-p - \frac{7}{2})
\Gamma^2 ( \frac{D-p-1}{2})}{16 (3+p-D) \Gamma (D-p-1)  
\,\,\pi^{D-p - \frac{3}{2}}\,\, b^{2D-2p-7}}   \nonumber \\
&& \times  
\Bigg\{ -  \frac{8(D-p-2) (D-p-3)}{D-2} \nonumber \\
&& +  \frac{(p+1) (D-p-3)}{D-2}  + \frac{4 (D-p-2) 
(D-2p -4)}{D-2}\nonumber \\
&& + [ 4(D-p-2) -2] +  [2 - \frac{(p+1)(D-p-3)}{D-2}]    \Bigg\} \;,
\label{chiTOTALEFINALE}
\end{eqnarray}
where $\tau_p$ is the physical D$p$-brane tension, $\tau_p = \frac{T_p}{\kappa_D}$. The first line contains the leading contribution given by the tree diagram with a graviton exchange and the subleading term of one-loop diagram with two graviton exchanges, the third line gives  the contribution of the one-loop seagull diagram and the fourth line gives the contribution of the one-loop diagram with the 3-graviton vertex. Finally the first square bracket in the last line gives the contribution of the one-loop diagram with the RR fields attached to the D$p$-branes, while the last square bracket gives that of the dilaton attached to the D$p$-branes. 

It is easy to show, in this extended theory of gravity, that the subleading contribution contained inside the big curly brackets vanishes. In this case the sum of the leading and subleading eikonal reduces just to the expression in the first line of (\ref{chiTOTALEFINALE}).  This is in agreement with the same result obtained in \cite{D'Appollonio:2010ae} for $D=10$ and the results found in section \ref{HElimit}.

\subsection{Pure Einstein Gravity}
\label{Einstein}

In this section we will consider the case of pure Einstein gravity. Let us start by considering the one-point graviton amplitude where only the tree diagram with the graviton exchange and the
one-loop diagram with the three-graviton vertex contribute. They are given in momentum space by (\ref{gra13}) and (\ref{3gra1}), respectively. Going to position space we find,
\begin{eqnarray}
&&\langle \eta_{\mu \nu} + 2 \kappa_D h_{\mu \nu} \rangle \nonumber \\
&&  = \left[ 1- \frac{D-p-3}{D-2} \left( \frac{R_p}{r}\right)^{D-p-3} + \frac{1}{2}
\left( \frac{D-p-3}{D-2} \right)^2 \left( \frac{R_p}{r}\right)^{2(D-p-3)}+ \dots  \right]\eta_{\parallel \mu \nu}
\nonumber \\
&& + \left[ 1 + \frac{p+1}{D-2}  \left( \frac{R_p}{r}\right)^{D-p-3}  \right. \nonumber \\
&& \left.  - \frac{1}{4} 
\left( \frac{  (D-p-3)^2 (p+1) }{2 (D-2) (D-p-2)} - 2 \left( \frac{p+1}{D-2}  \right)^2 \right) 
\left( \frac{R_p}{r}\right)^{2(D-p-3)}+ \dots  \right] \eta_{\perp \mu \nu} \nonumber \\
&& - \frac{1}{4 (5+p-D)} 
\left( \frac{(D-p-3)^2 (p+1)}{2 (D-2) (D-p-2)} - \frac{2(p+1) (D-p-3)}{D-2} \right) \nonumber \\
&& \times
\left( \eta_{\perp \mu \nu} - 2(D-p-3) \frac{r_\mu r_\nu}{r^2} \right)  \left( \frac{R_p}{r} \right)^{2(D-p-3)} \;, \label{EDS11}
\end{eqnarray}
where in the right-hand-side we have added the contribution of the flat Minkowski metric for $r \rightarrow \infty$. Notice that in the equation above we have now included the term in the third line of (\ref{3gra1}) that was neglected  in  subsection \ref{1point} and the term  in the second line of the same equation that was cancelled by the additional contributions of the dilaton and RR field. It can be checked that the term in the third line of (\ref{3gra1}), that we have neglected, gives the second term in the second to last line of (\ref{EDS11}).

To make contact with existing literature let us consider the case $D=4$ and $p=0$ where,
\begin{eqnarray}
N \tau_0 = \frac{N T_0}{\kappa_4}  \equiv M ~~~;~~~R_p \rightarrow 4 G_N M \;.
\label{EDS12}
\end{eqnarray}
Then (\ref{EDS11}) becomes,
\begin{eqnarray}
\langle \eta_{\mu \nu} + 2 \kappa_4 h_{\mu \nu} \rangle &=&  \left[ 1 - \frac{4MG_N}{2 r} + \frac{1}{8} \left( \frac{4MG_N}{r}\right)^2   + \dots \right]\eta_{00} \nonumber \\
&& + \left[ 1 + \frac{4MG_N}{2 r} + \frac{1}{4}(\frac{1}{4} +1 ) \left( \frac{4MG_N}{r}\right)^2   \right] \eta_{ij} \nonumber \\
&& + \, ( \frac{1}{8} -1) \frac{r_i r_j}{2 r^2}  \left( \frac{4MG_N}{r}\right)^2 \;,
\label{EDS13}
\end{eqnarray}
where the second term in the two round brackets in the last line comes from the term in the third line of (\ref{3gra1}). The subscript $0$ corresponds to the time coordinate, while $i,j$ correspond to the three spatial coordinates.  It is easy to check that the previous metric satisfies the following condition at each order in $G_N$,
\begin{eqnarray}
\partial^\nu h_{\nu \mu} - \frac{1}{2} \partial_\mu h =0 ~~;~~~ h \equiv h_{\mu \nu} \eta^{\mu \nu} \;.
\label{EDS13a}
\end{eqnarray}
If we want the one-point amplitude in the harmonic gauge the term of order $G_N^2$ must satisfy (54) of \cite{BjerrumBohr:2002ks} instead of the equation above. This is obtained by neglecting in (\ref{EDS13}) the second term in the two round brackets. With this gauge choice (\ref{EDS13}) becomes,
\begin{eqnarray}
\langle g_{\mu \nu} \rangle &=& \left( 1 - \frac{2MG_N}{r} + \frac{2 M^2 G_N^2}{r^2} \right) \eta_{00} +
\left( 1 + \frac{2MG_N}{r} + \frac{M^2 G_N^2}{r^2} \right) \eta_{ij}  \nonumber \\
&& \qquad + \frac{r_i r_j}{r^2} \frac{M^2 G_N^2}{r^2} \;.
\label{EDS13b}
\end{eqnarray}

In the final part of this subsection we consider the leading and subleading eikonal in the case of pure Einstein gravity. It can be easily obtained from the one in  (\ref{chiTOTALEFINALE}) by neglecting the last line. We find that,
\begin{eqnarray}
&& \delta (b, E) = \frac{N \kappa_D^2 \tau_p}{4}  E\frac{\Gamma 
( \frac{D-p-4}{2})}{ \pi^{\frac{D-p-2}{2}} b^{D-p-4}} 
+\frac{(N \kappa_D^2 \tau_p)^2 \, E \,\Gamma^2 ( \frac{D-p-3}{2}) 
\Gamma (D-p - \frac{7}{2})}{16 \,\pi^{D-p - \frac{3}{2}} \Gamma (D-p-4) 
\,\,b^{2D-2p-7}} \nonumber \\
&& + \frac{(N \kappa_D^2 \tau_p)^2 \, E \, \Gamma (D-p - \frac{7}{2})
\Gamma^2 ( \frac{D-p-1}{2})}{16 (3+p-D) \Gamma (D-p-1)  
\,\,\pi^{D-p - \frac{3}{2}}\,\, b^{2D-2p-7}}   \nonumber \\
&& \times  
\Bigg\{ - 4 (D-p-2) +  \frac{(p+1) (D-p-3)}{D-2} \Bigg\} \;.
\label{chiEINSTEIN}
\end{eqnarray}
If we look at the case for $D=4$ and $p=0$, we see that the last term in the first line does not contribute and  regularising the first term,
\begin{eqnarray}
\frac{\Gamma (\frac{D-p-4}{2})}{b^{D-p-4}} \Longrightarrow  -2 \log b \;,
\label{EDS14}
\end{eqnarray}
we find that,
\begin{eqnarray}
&&\delta^{(D=4; p=0)} = - 4 G_N ME \log b + \frac{\pi (G_N M)^2 E}{2b} \left(8 - \frac{1}{2} \right) \nonumber \\
&& =  - 4  G_N ME \log b +  \frac{15 \pi (G_N M)^2 E}{4b} \;,
\label{EDS15}
\end{eqnarray}
where the first term in the round bracket comes from the seagull diagram, while the second comes from the one-loop diagram with the 3-graviton vertex. It agrees with the classical part of the eikonal derived in \cite{Bjerrum-Bohr:2016hpa} and with the eikonal derived in \cite{Akhoury:2013yua,Luna:2016idw}. From the eikonal we can derive the deflection angle for a massless particle,
\begin{eqnarray}
\theta = - \frac{1}{E} \frac{\partial}{\partial b} \delta^{(D=4; p=0)} = \frac{4 G_N M}{b} + \frac{15 \pi (G_N M)^2}{4b^2} + \dots \;,
\label{EDS16}
\end{eqnarray}
where the dots refer to terms with higher powers of $b$ in the denominator and we have used the expression for the deflection angle given by \eqref{eq:deflanglinearbmassless}. The first term is the old result from Einstein, while the second term agrees with recent calculations in \cite{Akhoury:2013yua,Bjerrum-Bohr:2016hpa,Luna:2016idw, Bjerrum-Bohr:2018xdl}. The results in this subsection are also considered as a probe-limit of the more general pure gravity scenario we will consider in chapter \ref{chap:graveik}.

Using \eqref{chiEINSTEIN} we can also calculate the deflection angle for $D$ dimensions and $p=0$. We find,
\begin{eqnarray}
\theta = - \frac{1}{E} \frac{\partial}{\partial b} \delta^{(p=0)} = \sqrt{\pi}\frac{\Gamma \left( \frac{D}{2} \right)}{\Gamma \left( \frac{D-1}{2} \right)} \left( \frac{R_s}{b}\right)^{D-3} + \frac{\sqrt{\pi}}{2} \frac{\Gamma \left(D-\frac{1}{2} \right)}{\Gamma \left(D-2 \right)}\left( \frac{R_s}{b}\right)^{2D-6} \;,
\end{eqnarray}
where $R_s$ is the ``Schwarzschild radius" defined in appendix \ref{sec:geodesics}. Comparing this result with \eqref{eq:geogendnull}, where the deflection angle has been calculated from the metric for the $D$-dimensional generalisation of a Schwarzschild black hole, we find perfect agreement. Note that we cannot compare the result for general $p$ because the D-brane coupling used in \cite{Emparan:2009at} is different to the one we are using here.


\section{Discussion}
\label{sec:discussionSUGRA}

In this chapter we have discussed how to extract, from scattering amplitudes, classical quantities such as the classical solution related to the backreaction of a heavy source and the eikonal describing a scattering process in the Regge regime. The general ideas are well known and have been exploited in several previous papers to obtain these quantities in the limit of large distance or impact parameter, see for instance~\cite{Duff:1974xx,Bertolini:2000jy}. Furthermore we have presented a detailed analysis of the first subleading corrections to the limit mentioned above by focusing on type II supergravity in the presence of a stack of parallel D$p$-branes as an example. In the case of the eikonal, these corrections are determined by the subleading energy contributions in the Regge regime and so probe the structure of the gravitational theory in more detail. For instance the leading eikonal receives contributions only from ladder diagrams where gravitons are exchanged, while the subleading eikonal involves diagrams with different topologies and lower spin states. This raises the possibility, at the first subleading order, that the eikonal should be described by an operator instead of a simple phase since inelastic processes become possible. Note that in gravitational theories with a higher derivative modification of the 3-graviton vertex this already happens at the level of leading eikonal~\cite{Camanho:2014apa}.

\chapter{The Eikonal in Kaluza-Klein Gravity}\label{chap:kkeik}

This chapter is based on the paper \cite{KoemansCollado:2019lnh}. Here we discuss the eikonal in the context of Kaluza-Klein gravity with a background manifold, ${\mathbb R}^{1,D-2} \times S^1$.

This chapter is structured as follows. In section \ref{kktheory} we introduce the action for Einstein gravity  coupled to a real massless scale in $D$ dimensions and its compactification to $D-1$ dimensions  on a circle, resulting in a Kaluza-Klein tower of charged massive scalars coupled to the massless graviton, dilaton and $U(1)$ gauge field. Although there are also the massive Kaluza-Klein excitations of states of these aforementioned fields, they contribute to inelastic scattering of the  massive charged scalars in $D-1$ dimensions. We will focus on the eikonal of elastic scattering processes in this chapter. In section \ref{bornapproxkk} we look at 2 $\rightarrow $ 2 Born scattering (single graviton exchange) of scalars in a compactified background, ${\mathbb R}^{1,D-2} \times S^1$  by quantizing the momentum along the circle as usual. We show how the resulting amplitude  can be interpreted in ${\mathbb R}^{1,D-2}$ as the scattering of 2 charged scalar fields with Kaluza-Klein  masses $m_1 $ and $m_2$ via single graviton, photon and dilaton exchange. We also perform  consistency checks on the amplitude to show how we may recover the expected results in the $R \rightarrow \infty $ and $R \rightarrow 0 $ limits where $R$ is the radius of the compactified circle. In appendix \ref{app:kkexp} we show that in impact parameter space the corresponding one-loop contribution to the  2 $\rightarrow $ 2  scattering in ${\mathbb R}^{1,D-2} \times S^1$ is related to the square of the tree-level expression in precisely the way required for the eikonal phase to exponentiate. There is however a technical limitation in our proof which requires us to restrict to the case $D=5$. In section \ref{kinlimits} we look at various limits of the leading order eikonal expression. We also relate the subleading eikonal in the Kaluza-Klein theory in a particular limit to the subleading eikonal from scattering massless states from a stack of $D$p-branes. In section \ref{disc} we briefly discuss our results.

\section{Kaluza-Klein Theory} \label{kktheory}

In this section we briefly consider some aspects of Kaluza-Klein theory \cite{Kaluza:1921tu,Overduin:1998pn} which we will use throughout the chapter. For definiteness we consider the case $D=5$ and thus ${\mathbb R}^{1,4}$ compactified to ${\mathbb R}^{1,3} \times S^1  $. We first need to identify the correct charges of the massive Kaluza-Klein scalars. We start with the $D=5$ gravity action coupled to a real massless scalar field,
\be
S_5 = \int d^5x  \sqrt{-\hat{g} } \, \left( \frac{1}{2 \kappa_5^2} {\hat R}   - \frac{1}{2} \partial_M \Phi \partial_N \Phi {\hat g}^{MN} \right) \;,
\label{eq:5DKKaction}
\ee
with indices $M,N=0, \ldots, 4 $ and $\kappa_5^2 = 8 \pi G_5 $, ${\hat{g} }  = \det ({\hat g}_{MN} ) $. Note that the hat denotes $D=5$ quantities. Taking the standard form for the Kaluza-Klein $D=5$ line element (with coordinate $x^5 $  parametrizing the circle of radius $R$),
 \be
 d {\hat S}^2 = \phi^{-1/3}g_{\mu \nu} dx^\mu dx^\nu + \phi^{2/3}(dx^5 + \lambda A_\mu dx^\mu )^2  \;,
 \label{eq:5DKKmetric}
 \ee
where $g_{\mu \nu}, A_\mu $ and $\phi$ denote the $D=4$ graviton, $U(1)$ gauge field and dilaton respectively. Substituting the metric given by \eqref{eq:5DKKmetric} into \eqref{eq:5DKKaction}, including both the $D=4$ massless modes arising from the $D=5$ metric as well as the infinite tower of massive Kaluza-Klein states arising from $\Phi$, we find that the $D=4$ action \cite{Blau} is given by,
\begin{eqnarray}
S_4 &=&  \int d^4x  \sqrt{- g } \, \biggl(  \frac{1}{2 \kappa_4^2} { R}   -\frac{1}{4}\phi  F_{\mu \nu}F^{\mu \nu} -  \frac{1}{96 \pi G_4} \frac{1}{\phi^2} \partial_\mu \phi \partial^\mu \phi   \nonumber \\
&& - \frac{1}{2} \sum_{n \in {\mathbb Z }} \left( ( D_\mu \Phi_n )( D^\mu \Phi^{*}_n  ) + m_n^2 \phi^2  \Phi^{*}_n  \Phi_n \right) \biggr) \;,
\end{eqnarray}
where  $\kappa_4^ 2 = 8 \pi G_4 $, we note that the $D=5$ and $D=4$ gravitational constants are related by $G_5 = 2 \pi R G_4 $ and the parameter $\lambda $ appearing in the $D=5$ metric ansatz, \eqref{eq:5DKKmetric}, is fixed to be $\lambda = \sqrt{16\pi G_4 }$ in order to obtain the  canonical form of the Maxwell term in the $S_4$ action. It is conventional to redefine the dilaton $\phi = e^{\sigma \sqrt{48\pi G_4}}$  which leads to a canonical kinetic term for $\sigma$. With this substitution we find,
\begin{eqnarray}
\label{KK4D}
S_4 &=&  \int d^4x  \sqrt{- g } \, \biggl(  \frac{1}{2 \kappa_4^2} { R}   -\frac{1}{4} e^{\sigma \sqrt{48\pi G_4}}  F_{\mu \nu}F^{\mu \nu}  -   \frac{1}{2}  \partial_\mu \sigma \partial^\mu \sigma  \nonumber \\
&& - \frac{1}{2} \sum_{n \in {\mathbb Z }} \left( ( D_\mu \Phi_n )( D^\mu \Phi^{*}_n  ) +  e^{-\sigma \sqrt{48\pi G_4}} m_n^2  \Phi^{*}_n  \Phi_n \right) \biggr) \;.
\end{eqnarray}
Expanding the terms in the action about $\sigma = 0 $ we see that the $\Phi_n$ fields are massive, charged complex scalars, with $ D_\mu \Phi_n  = ( \partial_\mu \Phi_n -iQ_n A_\mu \Phi_n ) $ and charges given by $Q_n  =n  \sqrt{16 \pi G_4}/R $.

\section{Scattering on ${\mathbb R}^{1,D-2} \times S^1$ in the Born approximation} \label{bornapproxkk}

In this section we want to consider the case of large $s$ fixed $t$ eikonal scattering but where one of the spatial transverse directions has been compactified on a circle of radius $R$. The calculation is similar to the uncompactified case, discussed in section \ref{sec:treelevelamp}, except that we have to replace an integration over continuous momentum with a sum over discrete quantized momentum along the $S^1$ when Fourier transforming the amplitude into impact parameter space. 

Let us define $q = ({\bf q}', q_s)$ with $q_s = n/R $, $n \in \mathbb{Z}$ being the integer valued momentum number. Correspondingly, we partition the impact parameter $b$ previously defined over $D-2$ transverse directions into its components along the $D-3$ transverse directions and the circle $S^1 $ and so we define $b = ( {\bf b}', b_s ) $. It will also be useful to define the new Mandelstam variable $s'$  pertaining to $D-1$ dimensional spacetime,  so $s' = -(p_1'+p_2')^2 $ where $'$ denotes momenta restricted to $ {\mathbb R}^{1,D-2}$. We thus have the relation $ s = s' - \frac{(n_1 +n_2)^2}{R^2} $.

We note that in order to get the corresponding impact parameter space amplitude in ${\mathbb R}^{1,D-2} \times S^1$ we need to perform the following change,
\begin{equation}
  \label{eq:changeconttoKK}
     \int d^{D-2}q \rightarrow \int d^{D-3}q' \frac{1}{R}  \sum_{n  \in {\mathbb Z } } \;,
\end{equation}
in the expression we have discussed earlier for the leading eikonal, which we reproduce here,
\begin{equation}
  \label{eq:defneikonal}
   \delta^{(1)} (D) =  { \tilde{\cal A}}_1 = \frac{1}{4Ep} \int \frac{d^{D-2}q}{(2\pi )^{D-2}}  e^{i q \cdot b }   {\cal A}_1 \;,
\end{equation}
where $E=E_1+E_2$ and ${\cal A}_1$ is the relevant tree-level amplitude. Using the fact that  $q$ and $q'$ are related via $ q^2 = (q')^2 + (n/R)^2 $, the impact parameter space amplitude for scattering on ${\mathbb R}^{1,D-2} \times S^1$ can be deduced from \eqref{eq:1geamp} and be written as, 
\begin{eqnarray}
  \label{eq:compeikonal}
i\delta^{(1)}(D-1,R) \equiv i { \tilde{\cal A}}_1 (D-1,R)  &= &i\frac{\kappa_D^2 (s' - m_1^2 - m_2^2 - 2 m_1 m_2)}{2} \int  \frac{d^{D-3}q'}{(2\pi )^{D-2}} e^{i q' \cdot b' }  \nonumber \\
&& \qquad \times \frac{1}{R}  \sum_{n  \in {\mathbb Z } } \frac {e^{ib_s n/R}}{(q')^2 + (n/R)^2 } \;,
\end{eqnarray}
where we have defined $m_1^2 = (n_1/R)^2 $ and $m_2^2 = (n_2/R)^2 $ which we recognise as the Kaluza-Klein masses associated with compactification on $S^1$. Note that we have taken the 5-dimensional masses to be 0 as suggested by \eqref{eq:5DKKaction}. We have also defined, $n=n_1+n_3$. The integration over the $q' $ momenta can be carried out explicitly giving,
\begin{eqnarray}
i{ \tilde{\cal A}}_1 (D-1, R)  &=& i\frac{\kappa_D^2 }{2} \frac{s' - m_1^2 -m_2^2 - 2 m_1 m_2  }{(2\pi)^{\frac{D-1}{2}} }\frac{1}{ {(b')}^{\frac{D-5}{2} }}  \frac{1}{R} \nonumber \\
&& \quad \times  \sum_{n  \in {\mathbb Z } } e^{ib_s n/R}  {\biggl( \frac{|n|}{R} \biggr)^{\frac{D-5}{2} } K_{\frac{D-5}{2} } \biggl( \frac{|n|b'}{R} \biggr)   \label{TLC} \;,
}\end{eqnarray}
where $K_n(x) $ is a Bessel K function. In the ultra-relativistic limit $ s' \gg m_1^2, m_2^2  $ we find,
\be
\label{RTLC}
i { \tilde{\cal A}}_1 (D-1, R)  = i\frac{\kappa_D^2 }{2} \frac{s'}{(2\pi)^{\frac{D-1}{2}} }\frac{1}{ {(b')}^{\frac{D-5}{2} }}  \frac{1}{R}  \sum_{n  \in {\mathbb Z } } e^{ib_s n/R}  {\left( \frac{|n|}{R} \right)^{\frac{D-5}{2} } K_{\frac{D-5}{2} } \left( \frac{|n|b'}{R} \right) } \;.
\ee
We can check the consistency of \eqref{RTLC} by considering the two limits $R \rightarrow 0 $  and $R \rightarrow  \infty $ in which we expect to recover the expression found in \eqref{eq:1geimp} for scattering in pure gravity on $\mathbb{R}^{1,D-1}$ with $D-1$ and $D$  spacetime dimensions respectively. Note that we also show the first sign of exponentiation \cite{Kabat:1992tb,Giddings:2010pp,Akhoury:2013yua} at one-loop explicitly in appendix \ref{app:kkexp}.

\subsubsection{$R \to 0$ Limit}

First we consider the limit $R \rightarrow 0 $. In this limit we expect only the  zero momentum, $n=0$, Kaluza-Klein mode to contribute to the sum over t-channel states. This can be seen using the known expansion of the Bessel K function, $ K_\nu (z)  \sim \sqrt{\pi /{2z} } e^{-z} +... $ as $z \rightarrow \infty $ which applies when $n\neq 0 $.  For the  $n=0 $ contribution we have to consider $ K_\nu (z) $ for $z  \rightarrow  0 $. The behaviour in this limit is,
\begin{eqnarray}
\label{Kz0}
K_\nu (z)  &=& \frac{1}{2} \left[ \Gamma(\nu) \left( \frac{z}{2} \right)^{-\nu } \left(1 + \frac{z^2}{4(1+\nu ) } + \ldots \right)  \right. \nonumber \\
&& \quad \left. + \, \Gamma(-\nu) \left( \frac{z}{2} \right)^{\nu } \left(1 + \frac{z^2}{4(1+\nu ) } + \ldots \right) \right] \;.
\end{eqnarray}
Hence we find,
\begin{equation}
\lim_{n \rightarrow 0} \, \left[ \left( \frac{|n|}{R} \right)^{\frac{D-5}{2}} K_{\frac{D-5}{2}} \left( \frac{|n|b'}{R} \right) \right] =  \left( \frac{1}{2} \right)^{\frac{7-D}{2}} \Gamma \left(\frac{D-5}{2} \right) \frac{1}{(b')^{\frac{D-5}{2}}} \;.
\end{equation}   
Using this, the final result  for the amplitude  $ \tilde{\cal A}_1 (D-1,R)  $ as $R \rightarrow 0 $ is,
\begin{equation}
 \label{AR0}
 \lim_{R \rightarrow 0} \, [ i { \tilde{\cal A}}_1 (D-1,R) ] = i s' G_{D-1} \frac{ \Gamma \left( \frac{D-5}{2}  \right) }{\pi^{\frac{D-5}{2}} b'^{D-5} } \;,
 \end{equation} 
where $G_{D-1} $ is the  gravitational constant in the $D-1$ non-compact spacetime dimensions and is related to the one in $D$ dimensions via the familiar Kaluza-Klein relation $ G_D = 2\pi R G_{D-1} $. It is clear that the right hand side of the expression \eqref{AR0} is the same as $\tilde{\cal A}_1 (D-1)$ and the corresponding eikonal phase $ \delta^{(1)}(D-1,R \rightarrow 0) = \delta^{(1)}(D-1)$ discussed in section \ref{sec:eikonal} in the ultra-relativistic limit. So we recover the expected result, namely  the usual expression for high energy tree-level scattering but with $D-1$ non-compact spacetime dimensions.

\subsubsection{$R \to \infty$ Limit}
 
We now consider the opposite limit of $\delta^{(1)}(D-1, R) $ with $R \rightarrow \infty $ where we should recover the expression $\delta^{(1)}(D)$ given in  \eqref{eq:1geimp}. Noting that in this limit the discrete sum over momenta becomes a continuous integral $ \frac{1}{2 \pi R} \sum_{n} \rightarrow \frac{1}{2\pi} {\int}_{-\infty}^{\infty} dq $ the amplitude becomes,
 \begin{equation}
  \label{ }
  i{ \tilde{\cal A}}_1 (D-1, R \rightarrow \infty )  =   i\frac{4 \pi G_D s  }{ (2 \pi)^{\frac{D-1}{2}} (b')^{D-4}  } I \left( \frac{b_s}{b'} \right) \;,
 \end{equation}
where we use the fact that $s' \rightarrow s $ as $R \rightarrow \infty $ and the integral $I(b_s/b' ) $  is given by,
\begin{equation}
I \left( \frac{b_s}{b'} \right) = {\int}_{-\infty}^{\infty} d{\tilde q } \, e^{i {\tilde q }(b_s/b' ) }   \left( |{\tilde q}| \right)^{\frac{D-5}{2}}   K_{\frac{D-5}{2}} \left( |{\tilde q}| \right) \;,
 \end{equation}
 with ${\tilde q} = q \,b' $. The integral $ I(b_s/b' )  $ can be computed using the cosine integral transform formula for the  function $x^\nu K_\nu(x) $,
\begin{equation}
\int _0^\infty dx \, {\rm cos}(xy) x^{\pm \mu } K_\mu (a x) = \sqrt{\pi/4} \, \left(2a\right)^{\pm \mu} \Gamma\left( \pm\mu +\frac{1}{2} \right) ( y^2 +a^2 )^{\mp \mu -\frac{1}{2}  } \;.
\end{equation}
In our case, $\mu = (D-5)/2$, $a=1 $ and $y = b_s/b'$ and so we find,
\begin{equation}
i { \tilde{\cal A}}_1 (D-1, R \rightarrow \infty )  = i \frac{4\pi G_D \, s }{ (2 \pi)^{\frac{D-1}{2}} (b')^{D-4}  } \frac{  {2}^{\frac{D-5}{2}} \sqrt{\pi} \, \Gamma\left( \frac{D-4}{2} \right)} {\left( b_s^2/b'^2 +1 \right)^{\frac{D-4}{2}} } \;.
\end{equation}
Using the relation  between the impact parameters $b,  b' $ and $b_s $,   $ b^2 = b'^2 +b_s^2 $   we can see that,
\begin{equation}
i { \tilde{\cal A}}_1 (D-1, R \rightarrow \infty ) = { i \tilde{\cal A}}_1 (D) \;,
\end{equation}
which is equivalent to \eqref{eq:1geimp} in the ultra-relativistic limit as expected.

\subsubsection{Kaluza-Klein Decomposition}

In the discussion above we have focused on the ultra-relativistic limit $ s' \gg m_1^2, m_2^2  $ for simplicity. We will now consider the more general case so that later on we can consider the comparison of scattering amplitudes and eikonal phases in the Kaluza-Klein theory with those previously studied in chapter \ref{chap:sugraeik} which included massless dilatons elastically scattering off a large stack of $Dp$-branes. 

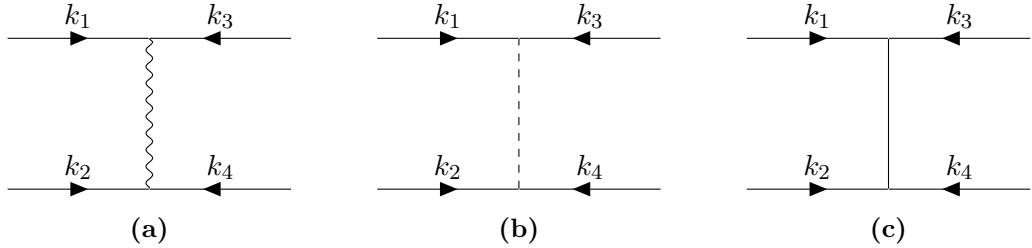
\begin{figure}[h]
  \begin{subfigure}[t]{0.3\textwidth}
    \centering
    \begin{tikzpicture}
	    \begin{feynman}
			\vertex (a) at (-2,-2) {};
			\vertex (b) at ( 2,-2) {};
			\vertex (c) at (-2, 0) {};
			\vertex (d) at ( 2, 0) {};
			\vertex[circle,inner sep=0pt,minimum size=0pt] (e) at (0, 0) {};
			\vertex[circle,inner sep=0pt,minimum size=0pt] (g) at (0, -2) {};
			\diagram* {
			(c) -- [fermion,edge label=$k_1$] (e) -- [anti fermion,edge label=$k_3$] (d),
			(g) -- [boson] (e),
			(a) -- [fermion,edge label=$k_2$] (g) -- [anti fermion,edge label=$k_4$] (b),
			};
	    \end{feynman}
    \end{tikzpicture}
    \caption{}
    \label{fig:3a}
  \end{subfigure}
  \quad
  \begin{subfigure}[t]{0.3\textwidth}
    \centering
    \begin{tikzpicture}
	    \begin{feynman}
			\vertex (a) at (-2,-2) {};
			\vertex (b) at ( 2,-2) {};
			\vertex (c) at (-2, 0) {};
			\vertex (d) at ( 2, 0) {};
			\vertex[circle,inner sep=0pt,minimum size=0pt] (e) at (0, 0) {};
			\vertex[circle,inner sep=0pt,minimum size=0pt] (g) at (0, -2) {};
			\diagram* {
			(c) -- [fermion,edge label=$k_1$] (e) -- [anti fermion,edge label=$k_3$] (d),
			(g) -- [scalar] (e),
			(a) -- [fermion,edge label=$k_2$] (g) -- [anti fermion,edge label=$k_4$] (b),
			};
	    \end{feynman}
    \end{tikzpicture}
    \caption{}
    \label{fig:3b}
  \end{subfigure}
  \quad
  \begin{subfigure}[t]{0.3\textwidth}
    \centering
    \begin{tikzpicture}
	    \begin{feynman}
			\vertex (a) at (-2,-2) {};
			\vertex (b) at ( 2,-2) {};
			\vertex (c) at (-2, 0) {};
			\vertex (d) at ( 2, 0) {};
			\vertex[circle,inner sep=0pt,minimum size=0pt] (e) at (0, 0) {};
			\vertex[circle,inner sep=0pt,minimum size=0pt] (g) at (0, -2) {};
			\diagram* {
			(c) -- [fermion,edge label=$k_1$] (e) -- [anti fermion,edge label=$k_3$] (d),
			(g) -- (e),
			(a) -- [fermion,edge label=$k_2$] (g) -- [anti fermion,edge label=$k_4$] (b),
			};
	    \end{feynman}
    \end{tikzpicture}
    \caption{}
    \label{fig:3c}
  \end{subfigure}
  \caption{The various constituent diagrams found in tree-level scalar scattering with graviton exchange on ${\mathbb R}^{1,D-2} \times S^1$ when decomposed into separate dilaton, gauge field and dimensionally reduced graviton contributions. The solid lines represent scalar states, the dashed line represent gauge fields and the wavy lines represent gravitons. Note that the internal solid line in \ref{fig:3c} represents a massless dilaton state.}
  \label{fig:treegravcompact}
\end{figure}

If we return to  \eqref{eq:compeikonal} and consider the momentum space amplitude from which it is calculated, we have,
\be 
 i {{\cal  A}}_1 (D-1, R ) = -i 2\pi R\kappa_5^2\frac{ (s' - m_1^2 -m_2^2 - 2 m_1 m_2 )^2 }{t' - (n_1 +n_3)^2/R^2 } \;.
\ee
This corresponds to a single $D=5$ graviton exchange between the massive states $\Phi_n $ in the Kaluza-Klein tower arising from compactification on $ {\mathbb R}^{1,D-2} \times S^1$, where we recall that $m_1^2 = n_1^2/R^2$, $m_2^2 = n_2^2/R^2 $. This scattering is  inelastic if $n_1 \neq  -n_3  $  in which case massive $D=4$ states are exchanged. We wish to focus here on elastic scattering so we will chose kinematics such that $n_1 = -n_3 $ (and hence $ n_2 = -n_4 $) but with both $n_1 $ and $n_2 $ non-zero in general. Our amplitude $ {{\cal  A}}_1 (D-1, R ) $ may be then written as,
\be 
 i {{\cal  A}}_1 (D-1, R ) = -2i \frac{\kappa_4^2 }{t'} \left(\frac{1}{2} (s' - m_1^2 -m_2^2 )^2 - 2 m_1m_2 (s' - m_1^2 -m_2^2)  + 2 m_1^2 m_2^2  \right) \;.
\ee
The first term in the round brackets is almost of the form of the diagram involving  a single massless graviton exchange in $D=4$ between two massive scalars, as given in \eqref{eq:1geamp} in $D=4$. It is instructive to rewrite ${{\cal  A}}_1 (D-1, R )$ in the form,
\begin{eqnarray}
i {{\cal  A}}_1 (D-1, R ) &=& -2i \frac{\kappa_4^2 }{t'} \biggl(\frac{1}{2} [ (s' - m_1^2 -m_2^2 )^2 - m_1^2m_2^2 ] - [ 2 m_1m_2 (s' - m_1^2 -m_2^2)]  \nonumber \\
&& \qquad   + 3 m_1^2 m_2^2  \biggr) \;. \label{KKdecomp}
\end{eqnarray}
Now the first term in the square brackets is precisely the  contribution from a single $D=4$ graviton. Based on the fact  that a massless $D=5$ graviton gives rise to a massless graviton, photon and dilaton in $D=4$ (along with their massive Kaluza-Klein  excitations) we would expect the term in the second square brackets to correspond to massless photon exchange and finally that the last term to correspond to massless dilaton exchange. The several Feynman diagrams are shown in figure \ref{fig:treegravcompact}.

Let us check that this is the case. The diagram for single photon exchange via the t-channel is easily computed using $D=4$ scalar electrodynamics. Given two charged massive scalars of masses $m_1, m_2$ and electric charges $q_1, q_2$, the resulting momentum space amplitude, $\mathcal{A}_{\text{phot}}$, is found to be, 
\be
i \mathcal{A}_{\text{phot}} = 2i q_1 q_2 \frac{( s' -m_1^2 -m_2^2)}{t'} \;.
\ee
Now in the Kaluza-Klein case we previously read off the charges from the action and so we can write $q_1 = \sqrt{2} \kappa_4 n_1/R $ and  $q_2 = \sqrt{2}\kappa_4 n_2/R $  giving 
\be
i \mathcal{A}_{\text{phot}} = 4i \kappa_4^2 m_1 m_2 \frac{( s' -m_1^2 -m_2^2)}{t'} \;,
\ee
which is precisely the contribution of the second square brackets in \eqref{KKdecomp}.

Finally we need to show that the final term in \eqref{KKdecomp} corresponds to a single dilaton exchange in $D=4$. This is straightforward as it only depends on the coupling between the massive Kaluza-Klein scalars $\Phi_n$ and the dilaton $\sigma$ which from the action \eqref{KK4D} is found to be $ -i \sqrt{6} \kappa_4 m_n^2 $. The contribution, $\mathcal{A}_{\text{dil}}$ is thus,
\be 
i \mathcal{A}_{\text{dil}} = -6 i \kappa_4^2 \frac{m_1^2 m_2^2}{t'} \;,
\ee
which is exactly the last term in \eqref{KKdecomp}.

\section{Various Kinematic Limits of the Eikonal} \label{kinlimits}

\subsection{The Leading Eikonal in the Ultra-Relativistic Limit}

In the ultra-relativistic case where $s' \gg m_1^2, m_2^2 $ and where we recall that $m_1^2 = n_1^2/R^2, m_2^2 =n_2^2/R^2 $ are the Kaluza-Klein masses, we expect the eikonal to be related to the deflection angle or time delay of a relativistic particle moving in the background geometry corresponding to the Aichelburg-Sexl (A-S) shock-wave metric \cite{Camanho:2014apa}. However since in our case one of the transverse directions is compactified on a circle, we have to reconsider the form of the A-S metric on $ {\mathbb R}^{1,D-2} \times S^1$.

First lets review the non-compact case. In ${\mathbb R}^{1,D-1} $ the form of the A-S metric is, 
\be 
ds^2 = dudv + h(u, x^i)du^2 + \sum_{i=1}^{D-2} \, (dx^i)^2 \;,
\ee 
where $u, v$ are the usual light-cone coordinates and $x^i$, $i=1,\ldots,D-2$ are the flat transverse coordinates. For the non-compact case we will follow the discussion in \cite{Camanho:2014apa} to make contact with the eikonal. The stress-energy that sources the A-S metric is due to a relativistic particle moving in the $v$-direction carrying momentum $-P_u$ ($P_u < 0 $  producing a shock-wave at $u=0$),
\be
T_{uu} = -P_u \delta(u) \delta^{D-2}(x^i) \;.
\ee
The Einstein equations reduce to,
\be 
\nabla_{\perp}^2 h(u, x^i) = -16 \pi G_D |P_u| \delta(u) \delta^{D-2}(x^i) \;.
\ee
The function $ h(u, x^i) $ is clearly related to the Green's function of the Laplace operator on the flat transverse space, so the general solution is (for $D>4 $),
 \be
 h(u, x^i ) =  \frac{4 \Gamma \left( \frac{D-4}{2} \right)}{\pi^{\frac{D-4}{2} } } \frac{G_D |P_u| \delta(u) }{r^{D-4}} \;,
 \ee
with $r^2 = \sum_{i=1}^{D-2} \, (x^i)^2 $. In order to have continuity in $v$ as a second particle  with momentum $p_v$ moves across the shock-wave in the $u$ direction at transverse distance $r =b$, one can remove the singular term $\delta(u)$  in the metric by defining the new coordinate $v_{new}$,
\be
v = v_{new}  + \frac{4 \Gamma \left( \frac{D-4}{2} \right)}{\pi^{\frac{D-4}{2} } } \frac{G_D |P_u|  }{b^{D-4}}\theta(u) \;,
\ee
where $ \theta(u)$ is the usual Heaviside step function. This leads to a corresponding Shapiro time delay as the particle crosses the shock wave given by,
\be
\Delta v =  \frac{4 \Gamma \left( \frac{D-4}{2} \right)}{\pi^{\frac{D-4}{2} } } \frac{G_D |P_u|}{b^{D-4}} \;.
\ee
This result is consistent with the leading eikonal $\delta^{(1)}(D)$ of \eqref{eq:1geimp} with
\be
\delta^{(1)}(D) = -p_v \Delta v|_{r=b} \;,
\ee 
where we identify, $s =  4p_vP_u $.

Now lets consider the case when the shock-wave is due to a particle moving in ${\mathbb R}^{1,D-2} \times S^1 $. The form of the metric becomes,
\be
\label{ASCOMP}
ds^2 = dudv + h(u, x^i, y)du^2 + \sum_{i=1}^{D-3} \, (dx^i)^2 + dy^2 \;,
\ee
where $y$ is the coordinate on $S^1$ with $y \sim y +2\pi R $. Note that 
$(x^i, y) $ are still transverse to the shock-wave. The stress-energy tensor in this case becomes,
\be
T_{uu} = -P_u \delta(u) \delta^{D-3}(x^i) \frac{1}{2\pi R}\sum_{n  \in {\mathbb Z } } e^{in y/R} \;,
\ee
where we have  given the representation of the delta function on $S^1 $ in terms of the quantised momentum modes. The Einstein equations reduce to,

\be
(\nabla_{\perp}^2 +\partial_y^2 )h(u, x^i, y) = -16 \pi G_D |P_u| \delta(u) \delta^{D-3}(x^i)\frac{1}{2\pi R}\sum_{n  \in {\mathbb Z } } e^{in y/R} \;,
\ee
where $\nabla_{\perp}^2   $ is the Laplacian on ${\mathbb R}^{D-3} $. The solution to $ h(u, x^i, y)$ is,
\begin{eqnarray}
h(u,x^i,y) &=&  16\pi |P_u| \, \frac{G_D}{2 \pi R}\,  \sum_{n  \in {\mathbb Z } } e^{in y/R} \int \frac{d^{D-3}q}{{(2\pi)}^{D-3}} \frac{ e^{i q\cdot x} }{(q^2 +n^2/R^2 ) } \delta(u) \cr
\cr
\cr
& = & 8 \frac{|P_u|}{(2\pi)^{\frac{D-5}{2} }} \, \frac{G_D}{2 \pi R} \, \frac{1}{r^{\frac{D-5}{2} } }\,  \sum_{n  \in {\mathbb Z } } e^{iy n/R}  { \left( \frac{|n|}{R} \right)^{\frac{D-5}{2} } K_{\frac{D-5}{2} } \left( \frac{|n|r}{R} \right) \delta(u) } \;, \nonumber \\
\end{eqnarray}
where $K_\nu(x) $ are Bessel K functions and $r^2 = \sum_{i=1}^{D-3} \, (x^i)^2 $.

Considering a second particle moving through this geometry, now separated from the first by the impact vector $b = (b', b_s) $ i.e. the particle approaches at minimum distance $r = b'$ in ${\mathbb R}^{D-3}$ and  $y = b_s $ along the $S^1 $. The corresponding Shapiro time delay is,
 \be
 \Delta v =  8 \frac{|P_u|}{(2\pi)^{\frac{D-5}{2}} } \, \frac{G_D}{2 \pi R} \, \frac{1}{b'^{\frac{D-5}{2} } }\,  \sum_{n  \in {\mathbb Z } } e^{ib_s n/R}  {\left( \frac{|n|}{R} \right)^{\frac{D-5}{2} } K_{\frac{D-5}{2} } \left( \frac{|n|b'}{R} \right)} \;.
 \ee
This agrees with the leading eikonal expression \eqref{RTLC} in the ultra-relativistic limit  $ s' \gg m_1^2, m_2^2 $,
\be
\delta^{(1)}(D-1,R) =  - p_v \Delta v|_{r=b, y = b_s} \;,
\ee
where we have taken $s'= 4p_vP_u $. From the previous analysis of the eikonal $\delta^{(1)}(D-1,R)$ in the limits  $R \rightarrow 0 $ and  $R \rightarrow \infty $ we can see that the shock-wave metric \eqref{ASCOMP} reduces to the usual Aichelburg-Sexl form in $D-1$ and $D$ non-compact dimensions respectively.

\subsection{The Leading Eikonal in the Large Kaluza-Klein Mass Limit}

In this section we will take the large Kaluza-Klein mass limit for one of the scattering particles and study the eikonal for a massless probe. This will allow us to connect this analysis with the deflection angle of a massless probe in the background of a Schwarzschild black hole in the ${\mathbb R}^{1,D-2}$ non-compact sector. In order to do so we will let $n_1=n_3=0$ (such that our probe has zero Kaluza-Klein mass) and $m_2^2 > s' \gg t'$. We will first work in $D=5$. In this limit we find that \eqref{eq:compeikonal} becomes,
\begin{equation}
i\delta^{(1)} = i\tilde{\mathcal{A}}_{1} \approx i \kappa_5^2 E_1 m_2 \int \frac{{\textrm{d}}^{2}q_{\bot}'}{(2\pi)^{3}}e^{iq_{\bot}' \cdot b'} \frac{1}{R} \frac{1}{(q_{\bot}')^2} \;,
\end{equation}
where we have also used that for $m_2 \gg 0$ we have $s' - m_2^2 \sim 2 E_1 m_2$ where $E_1$ is the energy of the massless probe. Using the fact that $\kappa_5^2=8 \pi G_5 = 8 \pi (2 \pi R G_4)$ we have,
\begin{eqnarray}\label{eq:leadeikLKKlimit}
i \delta^{(1)} & \approx & \frac{i 8 \pi G_4}{(2 \pi)^2} (E_1 m_2) \int {\textrm{d}}^{2}q_{\bot}'e^{iq_{\bot}' \cdot b'} \frac{1}{(q_{\bot}')^2} \nonumber \\
& = & - 4 G_4 m_2 E_1 \log{b} \;.
\end{eqnarray}
We can now relate this to the leading order contribution to the deflection angle. Using the fact that at leading order for a massless probe we can relate the eikonal to the leading contribution to the deflection angle using \eqref{eq:deflanglinearbmassless} we then find from \eqref{eq:leadeikLKKlimit} that the deflection angle is given by,
\begin{equation}
\theta^{(1)} = - \frac{1}{E_1} \frac{\partial \delta^{(1)}}{\partial b} = \frac{4 G_4 m_2}{b} \;.
\end{equation}
This is the well known expression for the leading contribution to the deflection angle of a massless probe in the background of a Schwarzschild black hole (see equation \eqref{eq:geo4null}). In fact this leading contribution is found to be universal. We find the same expression for the leading contribution to the deflection angle of a Reissner-Nordstrom black hole \cite{Eiroa:2002mk,Sereno:2003nd} as well as for the EMd black hole described in appendix \ref{angleEMDbh}. We will see in the next section that this EMd black hole is in fact the relevant black hole solution for the case we are considering here. This universality is due to the fact that no matter what source you are scattering from, the high energy limit of the Born amplitude is always the same \cite{D'Appollonio:2010ae,D'Appollonio:2013hja}.

We can also generalise this to arbitrary spacetime dimensions $D$. In this case we find that \eqref{eq:compeikonal} becomes,
\begin{eqnarray}
i\delta^{(1)} & \approx & \frac{i 8 \pi G_{D-1}}{(2 \pi)^{D-3}} (E m_2) \int {\textrm{d}}^{D-3} q_{\bot}'e^{iq_{\bot}' \cdot b'} \frac{1}{(q_{\bot}')^2} \nonumber \\
& = & \frac{i 8 \pi G_{D-1} E m_2}{4 \pi^{\frac{D-3}{2}}} \Gamma{\left(\frac{D-5}{2} \right)} \frac{1}{b^{D-5}} \;.
\end{eqnarray}
We therefore find for the deflection angle,

\begin{eqnarray}
\theta & \approx & \frac{4 G_{D-1} m_2}{ \pi^{\frac{D-5}{2}}} \Gamma{\left(\frac{D-3}{2} \right)} \frac{1}{b^{D-4}} \nonumber \\
& = & \sqrt{\pi} \frac{\Gamma{\left(\frac{D-1}{2} \right)}}{\Gamma{\left(\frac{D-2}{2} \right)}} \left( \frac{R_s}{b} \right)^{D-4} \;,
\end{eqnarray}
where we the $D$-dimensional generalisation of the Schwarzschild radius is given by,

\begin{equation}
8 G_{D-1} m_2 = \frac{D-3}{\Gamma{\left(\frac{D-2}{2} \right)}}  \pi^{\frac{D-4}{2}} R_s^{D-4} \;.
\end{equation}
We find that this equation for the deflection angle is equivalent to the leading term of \eqref{eq:geogendnull} found in appendix \ref{sec:geodesics} (with the above definition of the Schwarzschild radius in this context). Note that here the dimensionality is shifted by 1 since the black hole resides in the $D-1$ uncompact dimensions.

\subsection{The Subleading Eikonal in the Large Kaluza-Klein Mass Limit}

In this section we find the subleading corrections to the eikonal for high energy scattering of a neutral scalar particle off a heavy Kaluza-Klein state extending the analysis on the leading eikonal in the previous section. There we showed the leading eikonal in the scalar-gravity theory compactified on ${\mathbb R}^{1,D-2} \times S^1$ reproduced the expected leading contribution to the deflection angle in the case of a EMd black hole living in $D-1$ dimensions. Which we noted was equivalent to the leading contribution of the Schwarzschild black hole result. However as is well known, massive Kaluza-Klein states also couple to the dilaton and massless gauge fields that are part of the higher-dimensional metric and as we know from \cite{D'Appollonio:2010ae,D'Appollonio:2013hja} as well as the parallel discussion in chapter \ref{chap:sugraeik} these interactions should start contributing at the level of the subleading eikonal.

We know that for general spacetime and world-volume dimensions there is a subleading contribution to the eikonal for a scalar scattering off of a stack of $D$p-branes as we saw in chapter \ref{chap:sugraeik}. However in the case of $D=4, p=0$ this subleading contribution vanishes. Note that in $D=4, p=0$ case the stack of $N$ $D0$-branes carry a net electric charge $N \mu_0$ and the RR field coupling to the $D0$-branes is just an abelian gauge field $C_\mu$. Since the states exchanged in the $D0$-brane scattering case are equivalent to the states exchanged between the scalar probe and heavy Kaluza-Klein state we would like to extend the results found in chapter \ref{chap:sugraeik} to the case we are considering here.

In order to do this we will argue that both the supergravity action and the Kaluza-Klein action, \eqref{KK4D}, are equivalent in the appropriate normalisation. We will map both actions to the field normalisations used in \eqref{eq:horowitzS} so that we can also use the deflection angle results presented in appendix \ref{angleEMDbh}. From \eqref{eq:bulkaction}, with $D=4$, $p=0$ and ignoring the boundary part, we have the supergravity action,
\begin{eqnarray}\label{eq:bulkactionSUGRA}
S_{\text{SUGRA}}= \int d^{4} x \sqrt{-g} \left[ \frac{1}{2 \kappa_4^2} R- \frac{1}{2} 
\partial_{\mu}\phi\, \partial^{\mu}\phi -
\frac{1}{4}e^{-\sqrt{6}\kappa_4 \phi }F^2_{2} \right] \text{ .}
\end{eqnarray}
We can normalise this action in the same as in \eqref{eq:horowitzS} by changing,
\be
\kappa_4 \rightarrow \frac{1}{\sqrt{2}} \;, \quad \phi \rightarrow 2 \phi \;, \quad C^{\mu} \rightarrow 2 C^{\mu} \;,
\ee
where $C^{\mu}$ is the gauge field associated with the field strength $F_2$. We then find that,
\begin{eqnarray}
\label{4dsugra}
S_{\text{SUGRA}} = \int d^{4} x \sqrt{-g} \left[ R- 2 \partial_{\mu}\phi\, \partial^{\mu}\phi - e^{-2\sqrt{3}\phi}F^2_{2} \right]  \;.
\end{eqnarray}
This is now in the same normalisation as \eqref{eq:horowitzS} and we can see that the parameter $\alpha$ takes the value $\sqrt{3}$. Note that in this normalisation the mass and charge of the $D0$-brane become,
\be
M = \frac{N T_0}{\kappa_4} \rightarrow \sqrt{2} N T_0 \qquad Q = N \mu_0 \rightarrow 2 N \mu_0 = 2 \sqrt{2} N T_0 \;,
\ee
where we have used the fact that $\mu_p = \sqrt{2} T_p$. Hence we find a relationship between the mass and charge given by $Q=2M$.

We can now consider the Kaluza-Klein action \eqref{KK4D}. In this case we find that the action after the substitutions,
\be
\kappa_4 \rightarrow \frac{1}{\sqrt{2}} \;, \quad \sigma \rightarrow 2 \sigma \;, \quad C^{\mu} \rightarrow 2 C^{\mu} \;,
\ee
becomes,
\begin{eqnarray}
S_{\text{KK}} &=& \int d^{4} x \sqrt{-g} \biggl[ R- 2 \partial_{\mu}\sigma\, \partial^{\mu}\sigma - e^{-2\sqrt{3}\sigma}F^2_{2} \nonumber \\
&& - \frac{1}{2} \sum_{n \in {\mathbb Z }} \left( ( D_\mu \Phi_n )( D^\mu \Phi^{*}_n  ) +  e^{-2 \sqrt{3} \sigma} m_n^2  \Phi^{*}_n  \Phi_n \right) \biggr] \;.
\end{eqnarray}
We find that the massless sector of this action is equivalent to the action \eqref{eq:horowitzS} with $\alpha=\sqrt{3}$. We can also look at what happens to the mass and charge in this case,
\be
M = \frac{n}{R} \qquad Q = \sqrt{2} \kappa_4 \frac{n}{R} \rightarrow 2 \frac{n}{R} \;,
\ee
and so again we find that $Q=2M$. This analysis suggests that, since the actions can be mapped to each other, we can simply use the results previously derived in the case of a $D0$-brane for the scattering from a heavy Kaluza-Klein mode which we are considering here. This then implies that we have,
\be
{\delta}^{(2)}_{\text{KK}} (R) = 0 \;,
\ee
where ${\delta}^{(2)}_{\text{KK}} (R)$ is the subleading eikonal in the Kaluza-Klein theory with $D=4$ uncompact dimensions. We can now compare this to the deflection angle result derived in appendix \ref{angleEMDbh}. There we found that the subleading contribution to the deflection angle is given by,
\be
\phi^{(2)} = \frac{3}{8} \frac{1}{b^2} \left(\pi M^2 \left(\sqrt{1+\frac{2Q^2}{M^2}} + 9 \right) - 3 \pi Q^2 \right) \;.
\ee
Using the fact that in the normalisation used we have the relation $Q=2M$ we find that,
\be
\phi^{(2)} = 0 \;.
\ee
Which is consistent with the fact that the subleading contribution to the eikonal has been found to be zero.

\section{Discussion} \label{disc}

In this chapter we have investigated eikonal scattering in one of the simplest examples of a Kaluza-Klein theory, namely 5D Einstein gravity coupled to a real, massless 5D free scalar field, compactified to 4D on a circle. Despite this there is a richness to the 4D theory since it contains, in the massless sector, a massless scalar originating from the 5D massless scalar and a graviton, $U(1)$  gauge file and a dilaton, which originate from the 5D graviton. The massive sector contains a Kaluza-Klein tower of massive scalars charged with respect to the gauge field whose mass and charge satisfy the relation $Q= 2M $ in appropriate units. Their masses and charges have the usual interpretation of quantized momentum states moving around the circle. In considering $2 \rightarrow 2$ scattering of these massive 4D scalars there are both elastic and inelastic processes involved. Thus from the 4D point of view it is quite a complex system. It has already been previously shown that for simple scalar gravity theories in non-compact dimensions, contributions with a higher number of gravitons exchanged exponentiate into a phase \cite{Kabat:1992tb,Giddings:2010pp,Akhoury:2013yua}. The proof that the eikonal exponentiates to all orders in the Kaluza-Klein model considered here is something that would be interesting to investigate. One may intuitively expect that the exponentiation holds at all orders since in the limit of infinite or zero compactification radius $R$ we recover the 5D or 4D Einstein-scalar theories, for which we know the eikonal phase exponentiates. But explicitly proving this even at the one-loop level in the Kaluza-Klein model, as we have shown in appendix \ref{app:kkexp}, is rather non-trivial. 

An interesting test of our expression for the eikonal phase in the compactified case were the comparisons with deflection angle calculations in the various corresponding background geometries. We found agreement in the ultra-relativistic limit where the corresponding geometry was the compactified version of the Aichelburg-Sexl shock wave metric. We also considered the heavy Kaluza-Klein mass limit where we found agreement with the deflection angle calculated from a Einstein-Maxwell-dilaton black hole.

\chapter{The Eikonal in Einstein Gravity}\label{chap:graveik}

This chapter is based on the paper \cite{KoemansCollado:2019ggb} where we discuss the eikonal in pure Einstein gravity. In this chapter we study the scattering of two massive scalars in order to be able to describe the dynamics of binary black holes by using the eikonal. We also keep the spacetime dimension general which provides more general results and an easy way to regulate IR singularities. 

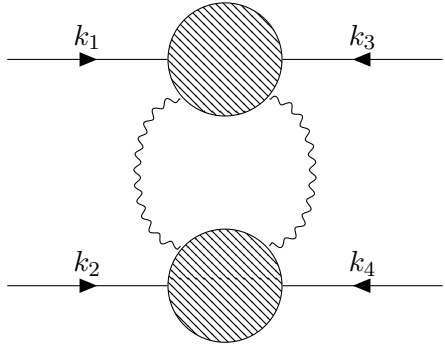
\begin{figure}[h]
  \centering
  \begin{tikzpicture}[scale=1.5]
	\begin{feynman}
      \vertex[blob, minimum size=1.5cm] (m) at (0, 1) {\contour{white}{}};
      \vertex[blob, minimum size=1.5cm] (m2) at (0, -1) {\contour{white}{}};
	  \vertex[circle,inner sep=0pt,minimum size=0pt] (c) at (0, 0) {};      
      
      \vertex (a) at (2,1) {};
      \vertex (b) at (-2,1) {};
      \vertex (a2) at (2,-1) {};
      \vertex (b2) at (-2,-1) {};
	  \vertex (d) at (0.3,0.65) {};
	  \vertex (d2) at (-0.3,0.65) {};  
	  \vertex (e) at (0.3,-0.65) {};
	  \vertex (e2) at (-0.3,-0.65) {};      
      
      \diagram* {
      (a) -- [fermion,edge label=$k_3$, swap] (m),
      (b) -- [fermion,edge label=$k_1$] (m),
      (a2) -- [fermion,edge label=$k_4$, swap] (m2),
      (b2) -- [fermion,edge label=$k_2$] (m2),
      (d) -- [boson, half left, looseness=1.0] (e),
      (d2) -- [boson, half right, looseness=1.0] (e2),
      };
    \end{feynman}
    \end{tikzpicture}
  \caption{A figure illustrating the procedure outlined at the beginning of section \ref{sec:amplitude} and described by equation \eqref{eq:ampmaster} for the one-loop amplitude. The solid lines represent massive scalars and the wavy lines represent gravitons. The shaded blob is described by equation \eqref{eq:amp4pt}.}
  \label{fig:4ptglue}
\end{figure}

The chapter is structured as follows. In section~\ref{sec:amplitude} we introduce the basic objects needed for our analysis, i.e. the tree-level on-shell vertices between two massive scalars and one and two gravitons. The field theory limit of a string expression provides a rather simple $D$-dimensional expression that we use to derive the relevant part of the amplitude with two graviton exchanges, see figure~\ref{fig:4ptglue}. We then extract the box and the triangle contributions that determine the 2PM eikonal phase. In section~\ref{sec:eikonal} we discuss the exponentiation pattern mentioned above and obtain explicit expressions for the 1PM and 2PM $D$-dimensional eikonal. As a check we derive the deflection angle in various probe-limits where it is possible to compare with a geodesic calculation in the metric of an appropriate black hole finding perfect agreement. Section~\ref{sec:discussionGRAV} contains a brief discussion on the possible relevance of our result for the study of the 3PM eikonal.

\section{Massive Scalar Scattering} \label{sec:amplitude}

In this section we focus on the $2 \rightarrow 2$ gravitational scattering process between two massive scalars in $D$ spacetime dimensions with both one and two graviton exchanges. As mentioned in the introduction, we are interested in extracting the classical contributions to this process, so instead of calculating the full amplitude by using the standard Feynman rules, we glue on-shell building blocks that capture just the unitarity cuts needed for reconstructing the classical eikonal. While this approach is by now commonly used in a $D=4$ setup, it is possible to implement it in general $D$ \cite{Bern:2011qt, Bern:2019crd, Huber:2019fea} and here we follow the method established in section \ref{2braneamps}, now including mass terms for the scalars.

For the one graviton exchange (1PM order) amplitude we can use, as an effective vertex, the on-shell three-point amplitude between two identical massive scalars and a graviton. In the standard Feynman vertex\footnote{We use the mostly plus convention for the metric and, as usual, we consider a two times signature to satisfy all on-shell constraints in the case of three-point functions.} $-i \kappa_D \left (k_{1 \mu}k_{2 \nu} + k_{1 \nu}k_{2 \mu} - (k_1 k_2 - m^2) \eta_{\mu \nu} \right )$, we can then drop the last two terms since they are proportional to $q^2$ and use the on-shell amplitude,
\begin{equation}
\begin{tikzpicture}[baseline=(eq)]
    \begin{feynman}[inline=(eq)]
      \vertex[blob, minimum size=1.0cm] (m) at ( 0, 0) {\contour{white}{}};
      \vertex (a) at (0,-2) {};
      \vertex (c) at (-2, 0) {};
      \vertex (d) at ( 2, 0) {};
      \vertex (eq) at (0,-1) {};
      \diagram* {
      (a) -- [boson,edge label=$q$] (m),
      (c) -- [fermion,edge label=$k_1$] (m),
      (d) -- [fermion,edge label=$k_2$, swap] (m),
      };
    \end{feynman}
\end{tikzpicture}
 = A_3^{\mu\nu}(k_1, k_2, q)
=  -i \kappa_D \left (k_{1}^{\mu}k_{2}^{\nu} + k_{1}^{\nu}k_{2}^{\mu} \right ) \;, \label{eq:amp3pt}
\end{equation}
where $\kappa_D = \sqrt{8 \pi G_N}$ and $G_N$ is the $D$-dimensional Newton's gravitational constant.

For two graviton exchange (2PM order) we need the corresponding four-point amplitude as the new ingredient. A particularly compact expression for this amplitude can be obtained by taking the field theory limit of the 2-tachyon 2-graviton amplitude in the Neveu-Schwarz string calculated by using the KLT approach. The result is,
\begin{eqnarray}
\hat{A}^{\alpha\beta;\rho\sigma}_4(k_1, k_2, q_1, q_2) &=& \frac{2\kappa_D^2 (k_2 q_1) (k_1 q_1)}{(q_1q_2) }\left[ \frac{k_2^\rho k_1^\alpha}{k_2 q_1}  + \frac{k_2^\alpha k_1^\rho}{k_1 q_1} +\eta^{\rho \alpha} \right] \nonumber \\
&& \qquad \times \left[ \frac{k_2^\sigma k_1^\beta}{k_2 q_1}  + \frac{k_2^\beta k_1^\alpha}{k_1 q_1} +\eta^{\sigma \beta} \right]\!.
\label{KLT24c}
\end{eqnarray}
By using the on-shell conditions it is possible to verify that~\eqref{eq:amp3pt} is symmetric under the exchange of the two scalars or the two gravitons and that it reproduces the known results for $D\to 4$, see for instance equations~(2.19) and~(2.2) of~\cite{Cachazo:2017jef}. For our purposes it will be convenient to use a different form for the amplitude where we have used momentum conservation and on-shell conditions to express $k_2$ in terms of $k_1$, $q_1$ and $q_2$,
\begin{eqnarray} \label{eq:amp4pt}
&&
\begin{tikzpicture}[baseline=(eq)]
    \begin{feynman}[inline=(eq)]
      \vertex[blob, minimum size=1.0cm] (m) at ( 0, 0) {\contour{white}{}};
      \vertex (a) at (-1,-2) {};
      \vertex (b) at ( 1,-2) {};
      \vertex (c) at (-2, 0) {};
      \vertex (d) at ( 2, 0) {};
      \vertex (eq) at (0,-1) {};
      \diagram* {
      (a) -- [boson,edge label=$q_1$] (m),
      (b) -- [boson,edge label=$q_2$,swap] (m),
      (c) -- [fermion,edge label=$k_1$] (m),
      (d) -- [fermion,edge label=$k_2$, swap] (m),
      };
    \end{feynman}
\end{tikzpicture}
 = A_4^{\alpha\beta;\rho\sigma}(k_1, k_2, q_1, q_2) \nonumber \\
 &&=\frac{-2\kappa_D^2 [(k_1 q_1) + (q_2 q_1)] (k_1 q_1)}{(q_1q_2) } \biggl( \frac{(k_1 + q_2 )^\rho k_1^\alpha}{(k_1 q_1) + (q_2 q_1)}  - \frac{(k_1  +q_1)^\alpha k_1^\rho}{k_1 q_1} +\eta^{\rho \alpha} \biggr) \nonumber \\
&& \qquad \times \biggl( \frac{(k_1 + q_2 )^\sigma k_1^\beta}{(k_1 q_1) + (q_1 q_2)}  - \frac{(k_1 +q_1)^\beta k_1^\alpha}{k_1 q_1} +\eta^{\sigma \beta} \biggr) \;.
\end{eqnarray}
Of course this expression is equivalent to~\eqref{KLT24c} on-shell, but~\eqref{eq:amp4pt} is transverse in the following slightly more general sense: it vanishes whenever the polarization of a graviton takes the form $\epsilon_{\mu\nu}=\zeta_{\mu} q_\nu+\zeta_\nu q_\mu$ just by using the on-shell conditions and  momentum conservation to rewrite products between momenta such as $k_i k_j$ (without the need of using it to rewrite the products between momenta and the arbitrary vectors $\zeta_{\mu}$).

In the next subsection we derive the classical ${\cal O}(G_N)$ contribution by gluing two amplitudes~\eqref{eq:amp3pt} with the de Donder propagator,
\begin{equation}\label{eq:dedpro}
[G(q)]^{\mu \nu; \rho \sigma} = \frac{-i}{2q^2} \left ( \eta^{\mu \rho} \eta^{\nu \sigma} + \eta^{\mu \sigma} \eta^{\nu \rho} - \frac{2}{D-2} \eta^{\mu \nu} \eta^{\rho \sigma} \right )\,.
\end{equation} 
In subsection \ref{sec:oneloopamp} we obtain the ${\cal O}(G_N^2)$ result by gluing the gravitons of two copies of the amplitude~\eqref{eq:amp4pt}. In all the four scalar amplitudes obtained in this section we denote the two incoming particles with momenta $k_1$ and $k_2$ and outgoing momenta $k_3$ and $k_4$. The particles $1$ and $3$ have mass $m_1$, while the particles $2$ and $4$ have mass $m_2$, see for instance figure~\ref{fig:3ptglue}. 

\subsection{One Graviton Exchange} \label{sec:treelevelamp}

Using the gluing procedure outlined above we can calculate the tree-level four-point massive scalar scattering by gluing two amplitudes~\eqref{eq:amp3pt} with a de Donder propagator~\eqref{eq:dedpro} and obtain,
\begin{equation}\label{eq:amptree}
i \mathcal{A}_1 = [G(k_1+k_3)]_{\mu_1 \nu_1; \mu_2 \nu_2}\; A^{\mu_1\nu_1}_3 (k_1, k_3, -k_1-k_3) A_3^{\mu_2\nu_2}(k_2, k_4, k_1+ k_3) \;.
\end{equation}
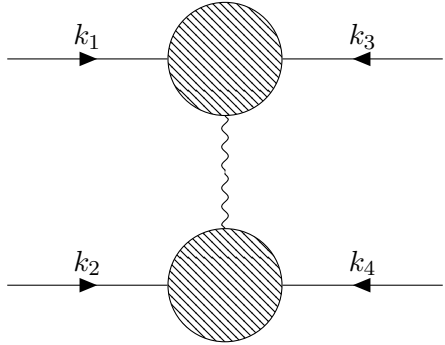
\begin{figure}[h]
  \centering
  \begin{tikzpicture}[scale=1.5]
	\begin{feynman}
      \vertex[blob, minimum size=1.5cm] (m) at (0, 1) {\contour{white}{}};
      \vertex[blob, minimum size=1.5cm] (m2) at (0, -1) {\contour{white}{}};
	  \vertex[circle,inner sep=0pt,minimum size=0pt] (c) at (0, 0) {};      
      
      \vertex (a) at (2,1) {};
      \vertex (b) at (-2,1) {};
      \vertex (a2) at (2,-1) {};
      \vertex (b2) at (-2,-1) {};
      \diagram* {
      (a) -- [fermion,edge label=$k_3$, swap] (m),
      (b) -- [fermion,edge label=$k_1$] (m),
      (a2) -- [fermion,edge label=$k_4$, swap] (m2),
      (b2) -- [fermion,edge label=$k_2$] (m2),
      (c) -- [boson] (m),
      (c) -- [boson] (m2),
      };
    \end{feynman}
    \end{tikzpicture}
  \caption{A figure illustrating the procedure outlined at the beginning of section \ref{sec:amplitude} and described by equation \eqref{eq:amptree} for the tree-level amplitude. The solid lines represent massive scalars and the wavy lines represent gravitons. The shaded blob is described by equation \eqref{eq:amp3pt}.}
  \label{fig:3ptglue}
\end{figure}
We then find,
\begin{equation}
i \mathcal{A}_{1} = \frac{2i \kappa_D^2}{q^2} \left ( \frac{1}{2}(s- m_1^2 - m_2^2)^2 - \frac{2}{D-2} m_1^2 m_2^2 \right )  =  \frac{2 i \kappa_D^2 \gamma(s)}{q^2} \label{eq:1geamp} \;,
\end{equation}
where $q \equiv k_1+k_3$ is the momentum exchanged between the two massive scalars and we have defined the quantity,
\begin{equation}
\gamma(s)=2(k_1 k_2)^2 - \frac{2}{D-2} m_1^2 m_2^2 = \frac{1}{2} (s-m_1^2-m_2^2)^2 - \frac{2}{D-2} m_1^2 m_2^2 \;. \label{eq:1gegms}
\end{equation}
In the high energy limit and after moving into impact parameter space we can see that this contribution grows as $E_i$ (since $G_N M^*$ is constant) and violates perturbative unitarity at large energies, we will come back to this point when discussing the two graviton exchange amplitude. By construction, this result just captures the pole contribution in $t$ of the amplitude, but this is sufficient to extract the classical interaction between two well separated particles. This is more clearly seen by transforming the amplitude to impact parameter space. We can do this by using the equations developed in chapter \ref{chap:background}, specifically equation \eqref{eq:eikips}. Terms in~\eqref{eq:1geamp} that are regular as we take $t\to 0$ yield only delta-function contributions localised at $\mathbf{b}=0$ and so can be neglected. 

We can now use \eqref{eq:eikips} and \eqref{eq:impoformu} to find the impact parameter space expression of the tree-level contribution,
\begin{equation}
i \tilde{\mathcal{A}}_{1} = \frac{i \kappa_D^2 \gamma(s)}{2 E p} \frac{1}{4 \pi^{\frac{D-2}{2}}} \Gamma \left( \frac{D}{2}-2 \right) \frac{1}{ \mathbf{b}^{D-4}} \;. \label{eq:1geimp}
\end{equation}
This result agrees with known results \cite{Kabat:1992tb} and as discussed in more detail in section \ref{sec:eikonal} is related to the result for the first order contribution to the deflection angle in the post-Minkowskian expansion.

\subsection{Two Graviton Exchanges} \label{sec:oneloopamp}

In this subsection we discuss the gluing procedure at one-loop. Schematically we have,
\begin{eqnarray}\label{eq:ampmaster}
i \mathcal{A}_2 &=& \int \frac{d^{D} k}{(2\pi)^{D}} [G(k)]_{\alpha_1 \beta_1 ; \alpha_2 \beta_2}\; [G(k+q)]_{\rho_1 \sigma_1 ; \rho_2 \sigma_2} \; \nonumber \\
&& \qquad \quad \times A_4^{\alpha_1 \beta_1 ; \rho_1 \sigma_1} (k_1, k_3, k, -k-q) A_4^{\alpha_2 \beta_2 ; \rho_2 \sigma_2}(k_2, k_4, -k, k+q) \;,
\end{eqnarray}
where $A_4$ is the four-point amplitude given by \eqref{eq:amp4pt}, we recall that $q \equiv k_1+k_3$ is the momentum exchanged between the two massive scalars, $k$ is the momentum in the loop and $[G]$ represents the graviton propagator~\eqref{eq:dedpro}.

\begin{figure}[h]
  \begin{subfigure}[t]{0.45\textwidth}
    \centering
      \begin{tikzpicture}
	    \begin{feynman}
			\vertex (a) at (-2.5,-2) {};
			\vertex (b) at ( 2.5,-2) {};
			\vertex (c) at (-2.5, 0) {};
			\vertex (d) at ( 2.5, 0) {};
			\vertex[circle,inner sep=0pt,minimum size=0pt] (e) at (-1, 0) {};
			\vertex[circle,inner sep=0pt,minimum size=0pt] (f) at (1, 0) {};
			\vertex[circle,inner sep=0pt,minimum size=0pt] (g) at (-1, -2) {};
			\vertex[circle,inner sep=0pt,minimum size=0pt] (h) at (1, -2) {}; 
			\diagram* {
			(c) -- (e) -- (f) -- (d),
			(g) -- (e),
			(h) -- (f),
			(a) -- (g) -- (h) -- (b),
			};
	    \end{feynman}
  \end{tikzpicture}
    \caption{}
    \label{fig:amp1}
  \end{subfigure}
  \quad
  \begin{subfigure}[t]{0.45\textwidth}
    \centering
    \begin{tikzpicture}
		\begin{feynman}
			\vertex (c) at (-2.5, 0) {};
			\vertex (d) at ( 2.5, 0) {};
			\vertex (x) at (-2.5, -2) {};
			\vertex (y) at ( 2.5, -2) {};
			\vertex[circle,inner sep=0pt,minimum size=0pt] (e) at (0, -1) {};
			\vertex[circle,inner sep=0pt,minimum size=0pt] (m) at (0, 0) {};
			\vertex[circle,inner sep=0pt,minimum size=0pt] (a) at (-1, -2) {};
			\vertex[circle,inner sep=0pt,minimum size=0pt] (b) at (1, -2) {}; 
     
			\diagram* {
			(c) -- (m) -- (d),
			(a) -- (m),
			(b) -- (m),
			(x) -- (a) -- (b) -- (y),
			};
		\end{feynman}
  \end{tikzpicture}
    \caption{}
    \label{fig:amp2}
  \end{subfigure}
  \caption{The two topologies of integrals that contribute to the two graviton exchange amplitude in the classical limit. In \ref{fig:amp1} we have the box topology and in \ref{fig:amp2} we have the triangle topology. The integral structure in \ref{fig:amp2} receives contributions from various Feynman diagrams, including those with a three-point vertex in the bulk. We can ignore other integral structures, such as bubble and tadpoles, since they do not contribute in the classical limit.}
  \label{fig:1}
\end{figure}
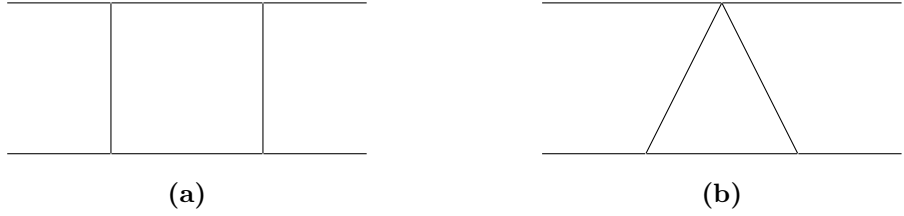

In order to interpret the expression found after attaching the relevant vertices using \eqref{eq:ampmaster} we need to rewrite it in terms of the relevant integral topologies which are schematically shown in figure \ref{fig:1}. In order to do this we define an operation denoted as $\mathcal{S}_{n}[\mathcal{A}_2]$ which searches the full expression, $\mathcal{A}_2$ resulting from \eqref{eq:ampmaster} and yields the integrand with $n$ number of propagators. Starting from the maximum number of propagators which in this case is $n=4$, we have,
\begin{equation}
\mathcal{S}_{4}[\mathcal{A}_2] = a_{\square} = \int \frac{d^{D} k}{(2\pi)^{D}} \frac{1}{k^2} \frac{1}{(q+k)^2} \frac{1}{(k_1 + k)^2 + m_1^2} \frac{1}{(k_2 - k)^2 + m_2^2} \mathcal{N}_{\square} \;,
\end{equation}
where we have set all the momenta in the internal propagators on-shell in $\mathcal{N}_{\square}$ since terms proportional to any propagator would cancel with one of the propagators in the denominator and therefore not contribute to the diagram with the above pole structure. Note that we have identified the pole structure above with the so called scalar box integral topology. An explicit expression for the numerators will be given in the upcoming subsections. 

We now want to search further in order to find the integrand with 3 poles. So now we have,
\begin{equation}
\mathcal{S}_{3}[\mathcal{A}_2 - a_{\square}] = a_{\triangle} = \int \frac{d^{D} k}{(2\pi)^{D}} \frac{1}{k^2} \frac{1}{(q+k)^2} \frac{1}{(k_1 + k)^2 + m_1^2} \mathcal{N}_{\triangle} \;,
\end{equation}
where we are searching the difference between the full expression, $\mathcal{A}_2$, and the part already extracted for the box diagram, $a_{\square}$. We have also set the momenta in the internal propagators on-shell in, $\mathcal{N}_{\triangle}$, for the same reasons described previously. Note that we have identified the pole structure above with the so called triangle integrals. It should be mentioned that one also extracts the crossed box and "inverted" triangle (i.e. the contribution with the opposite massive scalar propagator) by searching for the relevant pole structures.

Once the procedure described above has been completed the classical contributions to each of the expressions above are determined by implementing the scaling limit mentioned in section \ref{sec:eikonalamplitudebg}.

For two graviton exchanges we have two amplitude topologies that contribute; the box and triangle integrals, which are shown in figure \ref{fig:1}. The masses can be of the same order or much smaller than the centre of mass energy and of course the integrals take different forms in these two cases. In section \ref{app:integrals} we focus on the case $s\sim m_i^2$ and evaluate the first terms in the high energy expansion~\eqref{eq:heml} for the box and triangle integrals. In the ultra-relativistic case one recovers the massless results that can be found for instance in~\cite{Ellis:2007qk}.

\subsubsection{Box contribution}\label{sec:boxamplitude}

From the procedure outlined at the start of this subsection we find the following expression for the numerator, $\mathcal{N}_{\square}$, of the box diagram contribution to the two graviton exchange amplitude,
\begin{equation}
\mathcal{N}_{\square} = 4 \kappa_D^4 \gamma^2(s) \;,
\end{equation}
where $\gamma(s)$ has been defined in \eqref{eq:1gegms}. Writing this by including the integration over the loop momenta as well as including the contribution from the crossed box diagram we find,~
\begin{equation}\label{eq:fullboxamp}
i \mathcal{A}_{2} = 4 \kappa_D^4 (\gamma^2(s) \mathcal{I}_4(s,t) + \gamma^2(u) \mathcal{I}_4(u,t)) \;.
\end{equation}
where the integrals $\mathcal{I}_4(s,t)$ and $\mathcal{I}_4(u,t)$ have been computed in detail in section \ref{app:boxintegrals} and are defined as,
\begin{eqnarray}
\mathcal{I}_4(s,t) = \int \frac{d^{D} k}{(2\pi)^{D}} \frac{1}{k^2} \frac{1}{(q+k)^2} \frac{1}{(k_1 + k)^2 + m_1^2} \frac{1}{(k_2 - k)^2 + m_2^2} \;, \\
\mathcal{I}_4(u,t) = \int \frac{d^{D} k}{(2\pi)^{D}} \frac{1}{k^2} \frac{1}{(q+k)^2} \frac{1}{(k_3 + k)^2 + m_1^2} \frac{1}{(k_2 - k)^2 + m_2^2} \;.
\end{eqnarray}
Substituting the results for the integrals we find the leading contribution in the limit described by \eqref{eq:heml},
\begin{equation}
i \mathcal{A}^{(1)}_{2} =  -\frac{\pi^{\frac{D}{2}}}{(2\pi)^D} \frac{\pi}{2} \frac{4 \kappa_D^4 \gamma^2(s)}{\sqrt{(k_1 k_2)^2 - m_1^2 m_2^2}} \frac{\Gamma^2(\frac{D}{2}-2) \Gamma(3-\frac{D}{2})}{\Gamma(D-4)} (q^2)^{\frac{D}{2}-3} \;.
\end{equation}
The details of how to take the limit described by \eqref{eq:heml} when performing the integrals required to yield this result is given in section \ref{app:boxintegrals}. Moving to impact parameter space using \eqref{eq:eikips} and \eqref{eq:impoformu} we find that,
\begin{equation}
i \tilde{\mathcal{A}}^{(1)}_{2} = - \frac{\kappa_D^4 \gamma^2(s)}{(Ep)^2} \frac{1}{128 \pi^{D-2}} \Gamma^2 \left(\frac{D}{2}-2 \right) \frac{1}{\mathbf{b}^{2D-8}} \;.
\label{eq:boxleading}
\end{equation}
In the limit~\eqref{eq:heml}, this contribution grows as $E_i^2$ (since $G_N M^*$ is constant). Comparing with \eqref{eq:1geimp} we easily see that $i \tilde{\mathcal{A}}^{(1)}_{2}=\frac{1}{2} (i \tilde{\mathcal{A}}_{1})^2$, which is the first sign of the eikonal exponentiation as discussed in more detail in section \ref{sec:eikonalapproxbg}; the exponential of the tree-level amplitude will account for the first leading energy contributions of all higher loop amplitudes.

We can also look at the subleading contribution, in the limit described by \eqref{eq:heml}, to the two graviton exchange box diagram (as we will see in section \ref{sec:eikonal} this contributes to the second order of the post-Minkowskian expansion). Using the result for the subleading contribution to the integrals $\mathcal{I}_4(s,t)$ and $\mathcal{I}_4(u,t)$ found in  \eqref{eq:ij4subleading} we have,
\begin{equation}\label{eq:boxsubleadingbq}
i \mathcal{A}^{(2)}_{2} = \frac{i 2 \kappa_D^4 \gamma^2(s) \sqrt{\pi}}{(4\pi)^{\frac{D}{2}}} \frac{m_1+m_2}{(k_1 k_2)^2 - m_1^2 m_2^2} \frac{\Gamma \left(\frac{5-D}{2}\right) \Gamma^2\left(\frac{D-3}{2}\right)}{\Gamma(D-4)} (q^2)^{\frac{D-5}{2}}\;.
\end{equation}
At large energies this result scales as $E_i$ exactly as $\mathcal{A}_{1}$. This contribution should be exponentiated by the first subleading terms in the energy expansion of the higher loop contributions and so provides a new contribution to the eikonal phase. In impact parameter space~\eqref{eq:boxsubleadingbq} becomes,
\begin{equation}
i \tilde{\mathcal{A}}^{(2)}_{2} = \frac{i \kappa_D^4 \gamma^2(s)}{64 \pi^{D-\frac{3}{2}}} \frac{m_1+m_2}{Ep((k_1 k_2)^2 - m_1^2 m_2^2)} \frac{\Gamma \left(\frac{2D-7}{2}\right) \Gamma^2\left(\frac{D-3}{2}\right)}{\Gamma(D-4)} \frac{1}{\mathbf{b}^{2D-7}} \;. \label{eq:boxsubleadingb}
\end{equation}
We have checked that the results in this subsection agree in $D=4$ with equivalent results \cite{Kabat:1992tb, Akhoury:2013yua, Luna:2016idw,Damour:2017zjx,Bjerrum-Bohr:2018xdl}. Let us stress that equation~\eqref{eq:boxsubleadingb} vanishes in the $D\to 4$ limit because of the presence of the factor of $\Gamma(D-4)$ in the denominator. Thus, for $D>4$ there is a contribution to the eikonal from the box integral which becomes trivial in the four dimensional case. In general, this contribution is crucial in order to match, in the probe-limit, with the geodesic calculations as discussed in section~\ref{sec:eikonal} and in chapter \ref{chap:sugraeik} for the massless probe case $m_1 \gg 0, m_2=0$.

The subsubleading contributions to the box diagram are naively expected to be finite in the limit described by \eqref{eq:heml}, but there is actually a log-divergent term in the amplitude, as discussed for the massless case in~\cite{Amati:1990xe,Ciafaloni:2018uwe}, see also~\cite{Bellini:1992eb,Dunbar:1994bn} for an explicit evaluation of the same $2\to 2$ one-loop process with external gravitons. This contribution comes from using \eqref{eq:fullboxamp} and the next order in the expansion of the box integral, which in our case yields,
\begin{eqnarray}\label{eq:boxsubsubleadingbq}
&& i \mathcal{A}^{(3)}_{2} = 4 \kappa_D^4 \gamma^2 (s) \frac{i}{8(4\pi)^{\frac{D}{2}}} \Gamma \left(\frac{4-D}{2}\right)\, \frac{\Gamma^2\!\left(\frac{D-2}{2}\right)}{\Gamma(D-4)}  (q^2)^{\frac{D-4}{2}} \frac{1}{D-4} \nonumber \\
&& \times \left[ \frac{4(5-D)}{(k_1 k_2)^2 - m_1^2 m_2^2} \left( 1 + \frac{2k_1 k_2 \, \text{arcsinh} \left( \sqrt{\frac{\sigma-1}{2}} \right)}{\sqrt{(k_1k_2)^2 - m_1^2 m_2^2}}  \right)  + i \frac{\pi (D-4) (k_1+k_2)^2}{[(k_1k_2)^2 - m_1^2 m_2^2]^{3/2}} \right] \nonumber \\
&& - 4 \kappa_D^4 \psi(s)  \frac{i}{(4\pi)^{\frac{D}{2}}} \frac{\text{arcsinh}\left(\sqrt{\frac{\sigma-1}{2}}\right)}{ \sqrt{(k_1k_2)^2 - m_1^2 m_2^2}}\,\Gamma \left(\frac{6-D}{2}\right)\, \frac{\Gamma^2\!\left(\frac{D-4}{2}\right)}{\Gamma(D-4)} (q^2)^{\frac{D-4}{2}} \;,
\end{eqnarray}
where we have defined $\sigma = \frac{- k_1 k_2}{m_1 m_2}$ and,
\begin{eqnarray}
\psi(s) &=& -(2 k_1 k_2)\left( (2 k_1 k_2)^2 -\frac{4 m_1^2 m_2^2}{D-2} \right) \nonumber \\
&=& \left(s-m_1^2-m_2^2\right) \left( \left(s-m_1^2-m_2^2\right)^2 -\frac{4 m_1^2 m_2^2}{D-2} \right) \;.
\end{eqnarray}
Note that the last term in \eqref{eq:boxsubsubleadingbq} comes from expressing the $\gamma^2(u)$ from the second term in \eqref{eq:fullboxamp} in terms of $\gamma^2(s)$, i.e. we have $\gamma^2(u)=\gamma^2(s)+t\,\psi(s)+\mathcal{O}(t^2)$. We can also write the result above in impact parameter space for which we find,
\begin{eqnarray}\label{eq:boxsubsubleadingbqips}
&& i \tilde{\mathcal{A}}^{(3)}_{2} = \frac{\kappa_D^4 \gamma^2 (s)}{Ep} \frac{i}{128 \pi^{D-1}} \Gamma^2\!\left(\frac{D-2}{2}\right) \frac{1}{(\mathbf{b}^2)^{D-3}} \nonumber \\
&& \times \left[ \frac{4(5-D)}{(k_1 k_2)^2 - m_1^2 m_2^2} \left( 1 + \frac{2k_1 k_2 \, \text{arcsinh} \left( \sqrt{\frac{\sigma-1}{2}} \right)}{\sqrt{(k_1 k_2)^2 - m_1^2 m_2^2}}  \right) + i \frac{\pi (D-4) (k_1+k_2)^2}{[(k_1k_2)^2 - m_1^2 m_2^2]^{3/2}} \right] \nonumber \\
&& + \frac{\kappa_D^4 \psi(s)}{Ep} \frac{i}{8 \pi^{D-1}} \frac{\text{arcsinh}\left(\sqrt{\frac{\sigma-1}{2}}\right)}{ \sqrt{(k_1 k_2)^2 - m_1^2 m_2^2}} \Gamma^2\!\left(\frac{D-2}{2}\right) \frac{1}{(\mathbf{b}^2)^{D-3}} \;.
\end{eqnarray}
By using $\text{arcsinh} \, y=\log(y+\sqrt{y^2+1})$ in equation \eqref{eq:boxsubsubleadingbq} we can see that the second term on the second line and the term on the last line are log-divergent at large energies. It is interesting to highlight the following points. First, the same arcsinh-function arising from this subsubleading contribution also appears in the recent 3PM result~\cite{Bern:2019nnu, Bern:2019crd}. Then these terms violate perturbative unitarity in the $s/m_i^2 \to \infty$ limit and~\cite{Amati:1990xe} conjectured that they should resum to provide a quantum correction to the eikonal phase. This contribution is relevant in the discussion of the Reggeization of the graviton, for a recent discussion see~\cite{Melville:2013qca} and references therein. Finally the contribution~\ref{eq:boxsubsubleadingbqips} provides an additional imaginary part to $\tilde{\cal A}_2$ beside that coming from the leading term~\eqref{eq:boxleading}. In~\cite{Amati:1990xe}, it was shown that this subleading imaginary part vanishes in the $D=4$ massless case. Since the last term in the second line vanishes in $D=4$, here we find through a direct calculation that the same result holds also for the scattering of massive scalars. We will briefly come back to these points in section~\ref{sec:discussionGRAV}.

\subsubsection{Triangle contribution}
\label{sec:triangleamplitude}

Following the procedure outlined at the beginning of this subsection we find that the expression for the numerator, $\mathcal{N}_{\triangle}$, for the triangle-like contributions, with the $m_1$ massive scalar propagator, is given by,
\begin{eqnarray}
\mathcal{N}_{\triangle} &=& \kappa_D^4 \left( \frac{16(D-3) (k \, k_2)^2 m_1^4}{(D-2)q^2} \right. \nonumber \\ 
&& \;\; \left. + \, 4m_1^2 \left[ 2m_1^2 m_2^2 \frac{D^2 - 4D + 2}{(D-2)^2} - 2m_1^2 s  + m_1^4 + (m_2^2-s)^2 \right] \right) \;, 
\label{eq:triintgrand}
\end{eqnarray}
where we have already neglected some terms which are subleading in the limit given by \eqref{eq:heml} (i.e. don't contribute classically at second post-Minkowskian order). As we've done before we can express this in terms of an integral basis in which the expression \eqref{eq:triintgrand} becomes,
\begin{eqnarray}
&&\kappa_D^4 \biggl\{ \frac{16(D-3) k_{2 \mu} k_{2 \nu} m_1^4}{(D-2)q^2} \mathcal{I}_{3}^{\mu \nu}(m_1) \nonumber \\
&& + 4m_1^2 \left[ 2m_1^2 m_2^2 \frac{D^2 - 4D + 2}{(D-2)^2} - 2m_1^2 s  + m_1^4 + (m_2^2-s)^2 \right] \mathcal{I}_{3}(m_1) \biggr\} \nonumber \\
&& \qquad + m_1 \leftrightarrow m_2 \;,
\end{eqnarray}
where we recall that $k$ is the loop momentum and we have now included the contribution coming from the equivalent diagram with the $m_2$ massive scalar propagator. We have also used the definitions for the integrals,
\begin{eqnarray}
\mathcal{I}_{3}^{\mu \nu}(m_i) &=& \int \!\frac{d^Dk}{(2\pi)^D}\, \frac{1}{k^2}  \,\frac{1}{(q+k)^2} \, \frac{1}{(k+k_i)^2 + m_i^2} k^{\mu} k^{\nu} \;, \\
\mathcal{I}_{3}(m_i) &=& \int \!\frac{d^Dk}{(2\pi)^D}\, \frac{1}{k^2} \,\frac{1}{(q+k)^2} \,\frac{1}{(k+k_i)^2+m_i^2} \;.
\end{eqnarray}
Substituting the appropriate results for these integrals in the limit described by \eqref{eq:heml}, which are calculated in section \ref{app:triintegrals}, yields,
\begin{eqnarray}
i\mathcal{A}^{(2)}_2 &=& i \frac{2 \kappa_D^4 \sqrt{\pi}}{(4\pi)^{\frac{D}{2}}} \frac{\Gamma{\left( \frac{5-D}{2} \right)} \Gamma^2{\left( \frac{D-3}{2} \right)}}{\Gamma{\left( D-3 \right)}}(q^2)^{\frac{D-5}{2}} (m_1+m_2) \biggl\{  (s-m_1^2 -m_2^2)^2 \nonumber \\
&& - \frac{4m_1^2 m_2^2 }{(D-2)^2}  - \frac{(D-3) \left( (s-m_1^2 -m_2^2)^2 - 4 m_1^2 m_2^2 \right)  }{4 (D-2)^2} \biggr\} \label{eq:triamp} \;,
\end{eqnarray}
where we have again neglected subleading terms which do not contribute at second post-Minkowskian order. We can write equation \eqref{eq:triamp} in impact parameter space,
\begin{eqnarray}
i \tilde{\mathcal{A}}^{(2)}_{2} &=& i \frac{\kappa_D^4}{64 \pi^{D-\frac{3}{2}} \, Ep} \frac{\Gamma \left(\frac{2D-7}{2}\right) \Gamma^2\left(\frac{D-3}{2}\right)}{\Gamma(D-3)} \frac{m_1+m_2}{\mathbf{b}^{2D-7}} \biggl\{ (s-m_1^2 -m_2^2)^2 \nonumber \\
&& - \frac{4m_1^2 m_2^2 }{(D-2)^2}  - \frac{(D-3) \left( (s-m_1^2 -m_2^2)^2 - 4 m_1^2 m_2^2 \right)  }{4 (D-2)^2} \biggr\} \;. \label{eq:triips}
\end{eqnarray}
The results in this subsection agree with results for $D=4$ found in \cite{Luna:2016idw,Damour:2017zjx,Bjerrum-Bohr:2018xdl}. We have not considered the subleading triangle-like contribution explicitly in this subsection because we have found that it does not contribute to the log-divergent terms we discuss in sections \ref{sec:boxamplitude} and \ref{sec:discussionGRAV}. This should be clear from the results for the various integrals in section \ref{app:triintegrals}. Note also that these subleading contributions do not produce contributions to the real part of $i \mathcal{A}_2$.

\section{The Eikonal and Two-Body Deflection Angles} \label{sec:eikonal}

In this section we discuss explicit expressions for the eikonal and two-body deflection angles using the amplitudes derived in section \ref{sec:amplitude}. We also discuss various probe-limits for both general $D$ and $D=4$ in order to compare with existing results in the literature.

Using the various relations found in section \ref{sec:eikonalapproxbg} and the results from section \ref{sec:amplitude} we can write the leading (1PM) and first subleading eikonals (2PM). Using equation \eqref{eq:1geamp} we find for the leading eikonal,
\begin{eqnarray}
\delta^{(1)}(s,m_i,\mathbf{b}) &=&   \frac{16 \pi G_N \gamma (s)}{4Ep} \frac{\Gamma ( \frac{D-4}{2})}{4 \pi^{\frac{D-2}{2}} \mathbf{b}^{D-4}} \nonumber \\
  &=&   \frac{ \pi G_N \Gamma ( \frac{D-4}{2})}{\pi^{\frac{D-2}{2}} \mathbf{b}^{D-4}}  \frac{(s-m_1^2 -m_2^2)^2 - \frac{4}{D-2} m_1^2 m_2^2}{\sqrt{(s-m_1^2 -m_2^2)^2 - 4 m_1^2 m_2^2 }} \;.
\label{S12}
\end{eqnarray}
We can verify the exponentiation of the eikonal at one-loop level by looking at the leading one-loop contribution, \eqref{eq:boxleading}, which we reproduce below,
\begin{eqnarray}
i \tilde{\mathcal{A}}^{(1)}_{2} = - \frac{\kappa_D^4 \gamma^2(s)}{(Ep)^2} \frac{1}{128 \pi^{D-2}} \Gamma^2 \left(\frac{D}{2}-2 \right) \frac{1}{\mathbf{b}^{2D-8}} = \frac{1}{2} (i \delta^{(1)})^2  \;.
\end{eqnarray}
Notice that this includes the appropriate numerical coefficient as required for the second line in \eqref{eq:sumofleadingcont} to hold.

Summing equations \eqref{eq:boxsubleadingb} and \eqref{eq:triips} we find for the subleading eikonal,
\begin{eqnarray}
&&\delta^{(2)}(s,m_i,\mathbf{b}) = 
\frac{(8 \pi G_N)^2 (m_1+m_2)}{Ep \, \pi^{D - \frac{3}{2}} }\, \frac{ \Gamma ( \frac{2D-7}{2}) \Gamma^2 ( \frac{D-3}{2})}{16 \,\mathbf{b}^{2D-7}} \nonumber \\
&& \times \left\{ \frac{ \gamma^2 (s)}{\Gamma (D-4) \left[ (s-m_1^2 -m_2^2)^2 - 4 m_1^2 m_2^2\right]}  + \frac{1}{4 \Gamma (D-3)}  \right. \nonumber \\
&&  \left. \times \left[ (s-m_1^2 -m_2^2)^2 - \frac{4m_1^2 m_2^2 }{(D-2)^2}  - \frac{(D-3) \left( (s-m_1^2 -m_2^2)^2 - 4 m_1^2 m_2^2 \right)  }{4 (D-2)^2} \right] \right\} \;. \nonumber \\
\label{S16}
\end{eqnarray}

The relation between the eikonal and the scattering angle relevant for discussing the post-Minkowskian expansion as well as comparing with results found using general relativity has been discussed in section \ref{sec:introeikpwd} and is given up to 2PM order by,
\begin{equation}
\theta = -\frac{1}{p}\frac{\partial}{\partial \mathbf{b}} \left( \delta^{(1)} + \delta^{(2)} \right) + \ldots  \label{eq:eikdefangle}
\end{equation}
where as previously stated $p$ is the absolute value of the space-like momentum in the center of mass frame of the two scattering particles.

\subsection{Various Probe Limits in Arbitrary $D$}\label{sec:probeind}

The corresponding deflection angle for the leading eikonal is given by,
\begin{eqnarray}
\theta^{(1)} = - \frac{1}{p} \frac{\partial}{\partial \mathbf{b}} \delta^{(1)} = \frac{4 \pi G_N \Gamma (\frac{D-2}{2}) \sqrt{s}}{\pi^{\frac{D-2}{2}} \mathbf{b}^{D-3}} \,\, \frac{(s-m_1^2 -m_2^2)^2 - \frac{4}{D-2} m_1^2 m_2^2}{(s-m_1^2 -m_2^2)^2 - 4 m_1^2 m_2^2} \;. 
\label{S13}
\end{eqnarray}
In the limit where both masses are zero (the ACV limit~\cite{Amati:1987wq}) the deflection angle can be written as follows,
\begin{eqnarray}
\theta^{(1)}_{ACV} = \frac{ \sqrt{\pi} \Gamma ( \frac{D}{2} )}{ \Gamma ( \frac{D-1}{2})} 
\left( \frac{R_s}{\mathbf{b}}\right)^{D-3}~;~~ 
R_s^{D-3} = \frac{16\pi G_N M^*}{(D-2)\Omega_{D-2}}~~;~~ 
\Omega_{D-2}=
\frac{2 \pi^{\frac{D-1}{2}}}{\Gamma ( \frac{D-1}{2})} \;,
\label{S14}
\end{eqnarray}
where $R_s$ is the effective Schwarzschild radius in $D$ dimensions, $M^*=\sqrt{s}$ or $M^*=m_1,m_2$ depending on which scale is larger, and $\Omega_{D-2}$ is the volume of a $(D-2)$-dimensional sphere. 

In the probe-limit with $m_2=0$ and $m_1=M$ where the mass, $M \gg \sqrt{s-M^2}$, we find that the deflection angle becomes,
\begin{eqnarray}
\theta^{(1)}_{\text{null}} = \frac{4 \pi G_N \Gamma (\frac{D-2}{2}) M}{\pi^{\frac{D-2}{2}} \mathbf{b}^{D-3}} \;,
\label{S15}
\end{eqnarray}
which is equal to the deflection angle that is obtained from the first term of the eikonal in \eqref{chiTOTALEFINALE} for $p=0$ and with the identification $N \tau_0=M$. This is also consistent with \eqref{eq:geogendnull} with the Schwarzschild radius defined as in \eqref{S14} with $M^*=M$.

In order to compare with the more general results for timelike geodesics in a $D$-dimensional Schwarzschild background obtained in appendix \ref{sec:geodesics} we can also take the timelike probe-limit. In this limit we have as before $m_1=M \gg m_2$ where $m_2 = m \neq 0$, so we have $\sqrt{s} \sim M$ and $(s-m_1^2 -m_2^2) \sim 2 E_2 M$. Using this we find,
\begin{equation}
\theta^{(1)}_{\text{timelike}} = \frac{\sqrt{\pi } \Gamma \left(\frac{D}{2}-1\right) \left((D-2) E_2^2-m^2\right)}{2 (E_2^2-m^2) \Gamma \left(\frac{D-1}{2}\right)} \left( \frac{R_s}{\mathbf{b}} \right)^{D-3} \;,
\label{s20}
\end{equation}
where we have used the definition of the Schwarzschild radius given in \eqref{S14}. This agrees with equations \eqref{eq:geoPhi} and \eqref{eq:geophi1} by using the relation, $J \simeq |p| |\mathbf{b}|$.

The subleading contribution to the deflection angle is given by using \eqref{eq:eikdefangle} and \eqref{S16},
\begin{eqnarray}
&& \theta^{(2)} = \frac{(8 \pi G_N)^2 (m_1+m_2)}{Ep^2 \pi^{D - \frac{3}{2}} }\, \frac{ 2\Gamma ( \frac{2D-5}{2}) \Gamma^2 ( \frac{D-3}{2})}{16 \,\mathbf{b}^{2D-6}} \nonumber \\
&& \times \left\{ \frac{ \gamma^2 (s)}{\Gamma (D-4) \left[ (s-m_1^2 -m_2^2)^2 - 4 m_1^2 m_2^2\right]}  + \frac{1}{4 \Gamma (D-3)}  \right. \nonumber \\
&&  \times \left. \left[ (s-m_1^2 -m_2^2)^2 - \frac{4m_1^2 m_2^2 }{(D-2)^2}  - \frac{(D-3) \left( (s-m_1^2 -m_2^2)^2 - 4 m_1^2 m_2^2 \right)  }{4 (D-2)^2} \right] \right\}  \;. \nonumber \\
\label{S17}
\end{eqnarray}
The subleading eikonal and deflection angle do not contribute in the limit when both masses are zero for any value of $D$ as discussed in \cite{Ciafaloni:2018uwe}. In the probe-limit where $m_2=0$ and $m_1 \equiv M \gg E_2$ we find for the subleading eikonal,
\begin{eqnarray}
\delta^{(2)}_{\text{null}} &=&  \frac{  (8\pi G_N M)^2   E_2 \, \Gamma ( \frac{2D-7}{2} ) 
\Gamma^2 ( \frac{D-3}{2})}{16 \pi^{D - \frac{3}{2}} \Gamma (D-4) \mathbf{b}^{2D-7} }  \nonumber \\
&& \quad + \, \frac{  (8\pi G_N M)^2   E_2 \, \Gamma ( \frac{2D-7}{2} ) 
\Gamma^2 ( \frac{D-3}{2})}{16 \pi^{D - \frac{3}{2}} \Gamma (D-3)  \mathbf{b}^{2D-7} } 
\left( 1 - \frac{D-3}{4 (D-2)^2} \right) \;.
\label{S18}
\end{eqnarray}
The term in the first line is equal to the second term in the first line of \eqref{chiTOTALEFINALE} for $p=0$, while the term in the second line is equal to the sum of the terms in the third and fourth line of \eqref{chiTOTALEFINALE} for $p=0$. The corresponding deflection angle is given by,
\begin{eqnarray}
\theta^{(2)}_{\text{null}} &=&  \frac{  (8\pi G_N M)^2 \, \Gamma ( \frac{2D-5}{2} ) 
\Gamma^2 ( \frac{D-3}{2})}{8 \pi^{D - \frac{3}{2}} \Gamma (D-4) \mathbf{b}^{2D-6} }  \nonumber \\
&& + \frac{  (8\pi G_N M)^2 \, \Gamma ( \frac{2D-5}{2} ) 
\Gamma^2 ( \frac{D-3}{2})}{8 \pi^{D - \frac{3}{2}} \Gamma (D-3)  \mathbf{b}^{2D-6} } 
\left( 1 - \frac{D-3}{4 (D-2)^2} \right) \nonumber \\
&=& \frac{\sqrt{\pi } \Gamma \left(D-\frac{1}{2}\right)}{2 \Gamma (D-2)} \left( \frac{R_s}{\mathbf{b}} \right)^{2(D-3)} \;,
\label{S19}
\end{eqnarray}
where we have used the definition of the Schwarzschild radius given in \eqref{S14}. This agrees with equation \eqref{eq:geogendnull}. We can similarly look at the timelike probe-limit described before \eqref{s20}. In this case we find that \eqref{S17} becomes,
\begin{eqnarray}
\theta^{(2)}_{\text{timelike}} &=& \frac{\sqrt{\pi } \Gamma \left(D-\frac{5}{2}\right)}{\Gamma (D-2)} \nonumber \\
&& \quad \times \frac{\left((2 D-5) (2 D-3) E_2^4+6 (5-2 D) E_2^2 m^2+3 m^4\right) }{8 (E_2^2-m^2)^2 } \left( \frac{R_s}{\mathbf{b}} \right)^{2(D-3)} \;. \nonumber \\
\end{eqnarray}
We can easily check, by using the relation $J \simeq |p| |\mathbf{b}|$, that this agrees with equations \eqref{eq:geoPhi} and \eqref{eq:geophi2} as expected.

\subsection{Various Probe Limits in $D=4$}\label{sec:probeind4}

We will now set $D=4$ in the various equations obtained in the previous subsection. We find that the leading eikonal in $D=4$ is equal to,
\begin{eqnarray}
\delta^{(1)} = - 2G_N \frac{2 \gamma (s)}{2Ep} \log \mathbf{b}  = - 2 G_N \frac{(s-m_1^2 -m_2^2)^2 - 2 m_1^2 m_2^2}{\sqrt{(s-m_1^2 -m_2^2)^2 - 4 m_1^2 m_2^2 }} \log \mathbf{b} \;,
\label{S1}
\end{eqnarray}
while the deflection angle is given by,
\begin{eqnarray}
\theta^{(1)} = - \frac{1}{p} \frac{\partial}{\partial \mathbf{b}} \delta^{(1)} = \frac{4 G_N \sqrt{s}}{\mathbf{b}}  \frac{(s-m_1^2 -m_2^2)^2 - 2 m_1^2 m_2^2}{(s-m_1^2 -m_2^2)^2 - 4 m_1^2 m_2^2} \;, 
\label{S2}
\end{eqnarray}
where we recall $E = \sqrt{s}$, $p$ is the absolute value of the three-dimensional momentum in the center of mass frame of the two scattering particles and we have used equations \eqref{eq:1gegms} and \eqref{eq:1geEP}. Note that these results are equivalent to the expressions found in section \ref{sec:leadingeikbg}.

In the limit where both masses are zero (ACV limit) one gets,
\begin{eqnarray}
\delta_{ACV}^{(1)} = - 2 G_N s \,\,\log \mathbf{b} ~~;~~\theta_{ACV}^{(1)}
 = \frac{4G_N \sqrt{s}}{\mathbf{b}}= \frac{2 R_s}{\mathbf{b}}\;,
\label{S4}
\end{eqnarray}
which agrees with results found in \cite{Ciafaloni:2018uwe}. In the probe-limit where $m_2=0$ and $m_1=M$ we find the following eikonal,
\begin{eqnarray}
\delta_{\text{null}}^{(1)} = - 2 G_N (s-M^2) \log \mathbf{b}  \sim - 4 G_N M E_2 \log \mathbf{b} \;,
\label{S5}
\end{eqnarray}
where in the rest frame of the massive particle we have again used that $s-M^2 = 2ME_2$. Notice that equation \eqref{S5}  agrees with the first term of \eqref{EDS15}. For the deflection angle we find,
\begin{eqnarray}
\theta_{\text{null}}^{(1)} = \frac{4 G_N \sqrt{s}}{\mathbf{b}} = \frac{4 G_N M}{\mathbf{b}} \;,
\label{S6}
\end{eqnarray}
when we assume that $M \gg E_2$. This agrees with the well known expression for the leading contribution to the deflection angle of a Schwarzschild black hole reproduced in \eqref{eq:geo4null}. Taking the timelike probe-limit of \eqref{S2} as described in the previous subsection where, $m_1=M \gg m_2$ where $m_2 = m \neq 0$, we find,
\begin{equation}
\theta_{\text{timelike}}^{(1)} = \frac{R_s (2E_2^2 - m^2)}{E_2^2-m^2}\frac{1}{\mathbf{b}} \;,
\end{equation}
which we find agrees with the first contribution to \eqref{eq:geogendtime4} as well as equivalent results in \cite{Damour:2017zjx}.

The subleading eikonal in $D=4$ is found to be,
\begin{eqnarray}
\delta^{(2)} &=& \frac{(8 \pi G_N)^2 (m_1+m_2)}{64 Ep \pi \mathbf{b}} \left[ \frac{15}{16} (s-m_1^2 -m_2^2)^2 - \frac{3}{4} m_1^2 m_2^2 \right] \nonumber \\
&=&  \frac{\pi G_N^2 (m_1+m_2)}{ 2\mathbf{b}  \sqrt{(s-m_1^2 -m_2^2)^2 - 4m_1^2 m_2^2} }
\left[ \frac{15}{4} (s-m_1^2 -m_2^2)^2 - 3 m_1^2 m_2^2 \right] \;.
\label{S7}
\end{eqnarray}
The factor of $m_1+m_2$ in front implies that the subleading eikonal in the massless  limit is vanishing~\cite{Amati:1990xe},
\begin{eqnarray}
\delta^{(2)}_{ACV} =0 ~~;~~~ \theta_{ACV}^{(2)} =0 \;.
\label{S8}
\end{eqnarray}
This also implies that there is no contribution of order $1/\mathbf{b}^2$ to the deflection angle which is consistent with the result found in the previous subsection that this contribution is zero for any number of spacetime dimensions. From equation \eqref{S7} we can compute the deflection angle,
\begin{eqnarray}
\theta^{(2)} =- \frac{1}{p} \frac{\partial}{\partial \mathbf{b}} \delta^{(2)} &=& \frac{\pi G_N^2 (m_1+m_2) \sqrt{s}}{ \mathbf{b}^2 \left[ (s-m_1^2 -m_2^2)^2 - 4m_1^2 m_2^2\right] } \nonumber \\
&& \quad \times \left[ \frac{15}{4} (s-m_1^2 -m_2^2)^2 - 3 m_1^2 m_2^2 \right].
\label{S9}
\end{eqnarray}
In the probe-limit where $m_2 =0$ and $m_1=M$ we find,
\begin{eqnarray}
\delta_{\text{null}}^{(2)} = \frac{ 15 \pi G_N^2 M}{8\mathbf{b}} (s-M^2) \sim \frac{15 \pi (G_N M)^2E_2}{4\mathbf{b}} \;,
\label{S10}
\end{eqnarray}
which is equal to the second term of \eqref{EDS15}. For the deflection angle we instead get,
\begin{eqnarray}
\theta_{\text{null}}^{(2)} =  \frac{15 \pi (G_N M)^2}{4\mathbf{b}^2}\;,
\label{S11}
\end{eqnarray}
which we find agrees with the subleading contribution found in \eqref{eq:geo4null}. We can also take the timelike probe-limit of \eqref{S9} for which we find,
\begin{equation}
\theta_{\text{timelike}}^{(2)} = \frac{3 \pi R_s^2 (5E^2_2 - m^2)}{16 (E_2^2- m^2)}\frac{1}{\mathbf{b}^2} \;.
\end{equation}
This agrees with the second contribution to \eqref{eq:geogendtime4} as well as equivalent results in \cite{Damour:2017zjx}.


\section{Discussion}
\label{sec:discussionGRAV}

In this chapter we have studied the classical gravitational interaction between two massive scalars in $D$-dimensions up to 2PM order. As usual the spacetime dimension can be used as an infrared regulator and physical observables, such as the deflection angle discussed in section~\ref{sec:eikonal} have a smooth $D\to 4$ limit. The structure of the $D$-dimensional result is in some aspects richer than the one found in $D=4$. For instance the box integral provides not only the contribution necessary to exponentiate the leading energy behaviour of the tree-level diagram, but also a new genuine contribution to the subleading classical eikonal, see~\eqref{eq:boxsubleadingb}.

The box integral also provides a subsubleading contribution~\eqref{eq:boxsubsubleadingbq} that for $D\not=4$ has a new imaginary part, while its real part has a structure which also appears in the ${\cal O}(G_N^3)$ amplitude presented in~\cite{Bern:2019nnu, Bern:2019crd}. In the ultra-relativistic limit $s\gg m_i^2$, this contribution is log-divergent and, if it does exponentiate as suggested in~\cite{Amati:1990xe}, it would provide a new {\em quantum} contribution, $\delta^{(2)}_{q}$, to the eikonal. For instance, from~\eqref{eq:boxsubsubleadingbq} in $D=4$ one would obtain\footnote{The triangle contributions discussed in section~\ref{sec:triangleamplitude} do not yield any log-divergent term.}
\begin{equation}
\delta^{(2)}_{q} \simeq  \frac{12 G_N^2 s}{\pi b^2} \log \frac{s}{m_1m_2} = 12 \frac{G_N s}{\hbar} \frac{\lambda_P^2}{\pi b^2} \log \frac{s}{m_1m_2}~,
\label{d2qua}
\end{equation}
where we have taken the limit $s\gg m_i^2$ in order to compare with equation~(5.18) of\footnote{That equation should have an extra factor of $\lambda_P$ and $\delta^{(n)}_{\rm here}= 2\delta^{(n-1)}_{\rm there}$.}~\cite{Amati:1990xe} and $\lambda_P$ is the Planck length. By restoring the factors of $\hbar$ we can see from~\eqref{S1} that $\delta^{(1)}/\hbar$ is dimensionless and is therefore the combination that is exponentiated. On the other hand, $\delta_q^{(2)}$ is dimensionless without the need for any factor of $\hbar$ (see the first expression in~\eqref{d2qua} or equivalently it can be written in terms of $\lambda_P^2$ if we extract a factor of $1/\hbar$), which highlights its quantum nature.

\chapter{The Two-Body Hamiltonian in Einstein Gravity}\label{chap:potential}

In this chapter we will discuss how to derive the Hamiltonian and associated potential for a binary system in Einstein gravity up to 2PM order using the results we have derived in chapter \ref{chap:graveik}. We will give a brief overview of the derivation of the deflection angle using the Hamiltonian and proceed to derive the $D$-dimensional Hamiltonian. We will also briefly discuss our result in the context of the PN expansion.

\section{Two-Body Deflection Angle from the Hamiltonian}

The generic two-body Hamiltonian can be written as,
\begin{equation}
H(\bs r,\bs p)= \sqrt{m_1^2+\bs p^2}+\sqrt{m_2^2+\bs p^2} + V(\bs r, \bs p) \;, \label{eq:3PMHam}
\end{equation}
where in general we can write the potential as,
\begin{equation}\label{eq:undetpot}
V(\bs r, \bs p) = \sum_{n=1}^{\infty} c_n(\bs p^2) G_N^n \left( \frac{1}{|\bs r|} \right)^{n(D-3)} \;,
\end{equation}
with undetermined coefficients $c_n(\bs p^2)$ that do not depend on $\bs r$. We also know that the momentum can be decomposed in terms of its radial and angular components as,
\begin{equation}
\mathbf{p}^2=p_r^2+\frac{J^2}{r^2}. \label{eq:boldptoprJ}
\end{equation}
Note that we will use the symbols $|\bs r|$ and $r=|\bs r|$ interchangeably in this chapter. Hamilton's equations are given by,
\begin{equation}\label{eq:hamiltonseqs}
\dot{p} = - \frac{\partial H}{\partial q}~~;~~\dot{q} = \frac{\partial H}{\partial p} \;,
\end{equation}
where here we have $q=\{r, \theta\}$ and $p=\{p_r,J\}$. Using \eqref{eq:3PMHam} with \eqref{eq:hamiltonseqs} we find the four equations of motion,
\begin{eqnarray}
\dot{r} = \frac{\partial H}{\partial p_r} &=& p_r \left( \frac{1}{ \sqrt{m_1^2+\bs p^2}} + \frac{1}{ \sqrt{m_2^2+\bs p^2}} + 2 \frac{\partial V}{\partial (p^2)} \right) \;, \\
\dot{\theta} = \frac{\partial H}{\partial J} &=& \frac{J}{r^2} \left( \frac{1}{ \sqrt{m_1^2+\bs p^2}} + \frac{1}{ \sqrt{m_2^2+\bs p^2}} + 2 \frac{\partial V}{\partial (p^2)} \right) \;, \\
\dot{p}_r = - \frac{\partial H}{\partial r} &=&  \frac{J^2}{r^3} \left( \frac{1}{ \sqrt{m_1^2+\bs p^2}} + \frac{1}{ \sqrt{m_2^2+\bs p^2}} \right) - \frac{\partial V}{\partial r} \;, \\
\dot{J} = - \frac{\partial H}{\partial \theta} &=& 0 \;.
\end{eqnarray}
Using the first two of these equations we can write down the equation of motion for the angle as a function of $r=|\mathbf{r}|$,
\begin{equation}
\frac{\partial \theta}{\partial r} = \frac{J}{r^2 p_r} \;.
\end{equation}
Integrating this equation we can then write the full deflection angle as,
\begin{equation}
\theta + \pi = 2 \int_{r_0}^{\infty} dr \frac{J}{r^2 \sqrt{\bs p^2(r,E) - \frac{J^2}{r^2}}} \;,\label{eq:chifrompr}
\end{equation}
where $r_0$ is the minimum separation distance given by,
\begin{equation}\label{eq:prtor0cond}
\frac{\partial r}{\partial \theta} \Bigr|_{r=r_0} = 0 \implies p_r(r_0) = 0 \implies \mathbf{p}^2(r_0,E) - \frac{J^2}{r_0^2} = 0\;.
\end{equation}
To find a perturbative solution to \eqref{eq:chifrompr} we observe that the momentum variable $\bs p$ can be expanded in a post-Minkowskian expansion \cite{Bern:2019crd, Kalin:2019rwq, Bjerrum-Bohr:2019kec} and so we can write,
\begin{equation}\label{psquared}
\bs p^2(r,E)= p_0^2(E) + \sum_{n=1}^{\infty} \frac{f_n(E) G_N^n}{r^{n(D-3)}} \;,
\end{equation}
where we have introduced a new set of undetermined coefficients $p_0$ and $\{f_n\}$ and we are using $r=|\bs r|$. We will see later that these coefficients are related to the deflection angle (and therefore also the eikonal). Using \eqref{psquared} and \eqref{eq:boldptoprJ} in \eqref{eq:chifrompr} we can write the deflection angle as,
\begin{equation}
\theta + \pi = 2 \int_{r_0}^{\infty} dr \frac{J}{r^2} \left( p_0^2(E) + \sum_{n}  \frac{f_n(E) G_N^n}{r^{n(D-3)}} - \frac{J^2}{r^2} \right)^{-\frac{1}{2}} \;.
\end{equation}
We can now remove the $J$ dependence of the expression above by using the fact that $r_0$ is defined by \eqref{eq:prtor0cond}. Doing so we find,
\begin{equation}
\theta + \pi = 2 \int_{r_0}^{\infty} dr \frac{r_0}{r^2} \left(\frac{p_0^2(E) + \sum_{n} \frac{f_n(E) G_N^n}{r^{n(D-3)}}} {p_0^2(E) + \sum_{n} \frac{f_n(E) G_N^n}{r_0^{n(D-3)}}} -\frac{r_0^2}{r^2}  \right)^{-\frac{1}{2}} \;.
\end{equation}
Performing the substitution, $u=r_0/r$, as we have done many times in appendix \ref{app:geodesics} for similar integrals, we find,
\begin{equation}
\theta + \pi = 2 \int_{0}^{1} du \left(\frac{p_0^2(E) + \sum_{n} f_n(E) G_N^n \left( \frac{u}{r_0} \right)^{n(D-3)}}{p_0^2(E) + \sum_{n} f_n(E) G_N^n \left( \frac{1}{r_0} \right)^{n(D-3)}}-u^2 \right)^{-\frac{1}{2}} \;.
\end{equation}
Although this integral can be expressed in a closed form as found in \cite{Bjerrum-Bohr:2019kec} it is instructive here to instead perform a perturbative expansion in $G_N$. Expanding the integrand above as a power series in $G_N$ we find that,
\begin{eqnarray}
&& \theta + \pi \approx 2 \int_{0}^{1} du \Biggl[ \frac{1}{\sqrt{1-u^2}} - G_N \frac{f_1 (u^{D-3}-1)}{2 p_0^2(1-u^2)^\frac{3}{2} } \left( \frac{1}{r_0} \right)^{D-3}  \nonumber \\
&& + G_N^2 \left( \frac{3 f_1^2 \left(u^{D-3}-1\right)^2}{8 p_0^4 \left(1-u^2\right)^{\frac{5}{2}}} - \frac{\left(u^{D-3}-1\right) \left(-f_1^2+f_2 p_0^2 u^{D-3}+f_2 p_0^2\right)}{2 p_0^4 \left(1-u^2\right)^{\frac{3}{2}}} \right)  \left( \frac{1}{r_0} \right)^{2(D-3)}  \Biggr] \;, \nonumber \\ 
\end{eqnarray}
where we have only kept terms up to 2PM order. The integrals over $u$ can be readily computed and one finds,
\begin{eqnarray}\label{eq:genr0Danglepot}
&& \theta \approx G_N \left( \frac{\sqrt{\pi} f_1 \Gamma \left(\frac{D-2}{2}\right)}{p_0^2 \Gamma \left(\frac{D-3}{2}\right)} \right) \left( \frac{1}{r_0} \right)^{D-3}  + G_N^2 \frac{\sqrt{\pi }}{2 p_0^4} \left(\frac{\Gamma \left(D-\frac{5}{2}\right) \left((D-4) f_1^2+2 f_2 p_0^2\right)}{\Gamma (D-3)}   \right. \nonumber \\
&& \, \left. -  \frac{(D-3) f_1^2 \Gamma \left(\frac{D-2}{2}\right)}{\Gamma \left(\frac{D-3}{2}\right)} \right)  \left( \frac{1}{r_0} \right)^{2(D-3)} \;.
\end{eqnarray}
The expression above is expressed in terms of $r_0$, however we are interested in an expression for the angle in terms of the angular momentum, $J$. This will allow us to make direct comparisons with our previous results in chapter \ref{chap:graveik} calculated using the eikonal. To find the relation between $r_0$ and $J$ we can perturbatively solve the equation $p_r(r_0)=0$ as indicated by \eqref{eq:prtor0cond}. Up to 2PM we have,
\begin{equation}
p_0^2(E) + \sum_{n=1}^{2} f_n(E) G_N^n \left( \frac{1}{r_0} \right)^{n(D-3)} - \frac{J^2}{r_0^2} = 0 \;,
\end{equation}
and we find that,
\begin{equation}\label{eq:r0toJforgenDpot}
r_0 \approx J \left( \frac{1}{p_0} - G_N \frac{f_1 p_0^{D-6}}{2 J^{D-3}} \right) \;.
\end{equation}
Finally we can substitute \eqref{eq:r0toJforgenDpot} into \eqref{eq:genr0Danglepot} to express the deflection angle in terms of $J$,
\begin{eqnarray}\label{eq:2PMDdimfiangle}
\theta_J & \approx & \frac{\sqrt{\pi} f_1 \Gamma \left(\frac{D}{2}-1\right)}{\Gamma \left(\frac{D-3}{2}\right)} \left( \frac{G_N \, p_0^{D-5}}{J^{D-3}} \right) \nonumber \\
&& \quad + \frac{\sqrt{\pi } \Gamma \left(D-\frac{5}{2}\right) \left((D-4) f_1^2+2 f_2 p_0^2\right)}{2 \Gamma (D-3)}  \left( \frac{G_N \, p_0^{D-5}}{J^{D-3}} \right)^2 \;.
\end{eqnarray}
We have found that this expression agrees with known results \cite{Bern:2019crd, Antonelli:2019ytb, Kalin:2019rwq, Bjerrum-Bohr:2019kec} in $D=4$.

\section{The Two-Body Potential}

In this section we will explicitly derive the 2PM two-body Hamiltonian for a system of binary black holes in $D$-dimensions. In order to do so we first need to establish the relationship between the various sets of undetermined coefficients defined in the previous section. From \eqref{eq:undetpot} we have a set $\{ c_n \}$ of undetermined coefficients which we would like to relate to the coefficients, $\{ f_n \}$ in \eqref{psquared}.

To relate these two sets of coefficients we find an expression for $\bs p^2$ by perturbatively solving,
\begin{equation}
\sqrt{m_1^2+\bs p^2}+\sqrt{m_2^2+\bs p^2} + \sum_{n=1}^{2} c_n(\bs p^2) \left( \frac{G_N}{|\bs r|} \right)^{n(D-3)} = E \;, \label{eq:2PMHamERel}
\end{equation}
where we have truncated the potential at 2PM order.

At order $\mathcal{O}(G_N^0)$ we have,
\begin{eqnarray}
&& \sqrt{m_1^2+ p_0^2}+\sqrt{m_2^2+ p_0^2} = E \nonumber \\
&& \implies p_0^2 = \frac{E^4-2 E^2 \left(m_1^2+m_2^2\right)+\left(m_1^2-m_2^2\right)^2}{4 E^2} = \frac{ (k_1 k_2)^2 - m_1^2 m_2^2}{E^2}  \;, \label{eq:p2zero}
\end{eqnarray}
where $k_1 k_2$ is the usual product of incoming momenta of the two massive scalars as described in section \ref{eq:convandkin}. Notice that this expression for $p_0$ is equivalent to the expression we use for the absolute value of the space-like momentum in the center of mass frame of the two scattering particles which we have usually denoted as $p$ and can be found in equation \eqref{eq:1geEP}.

We can now move on to the first order in $G_N$. Since the $\mathcal{O}(G_N)$ contributions to the first two terms in \eqref{eq:2PMHamERel} will reproduce the value, $E$, due to the result in \eqref{eq:p2zero} we can collect all the terms at order $G_N$ and find a solution for $f_1$ by asserting that the collection of terms at order $G_N$ yields $0$. Substituting \eqref{psquared} into \eqref{eq:2PMHamERel} up to $\mathcal{O}(G_N)$ and doing so we find,
\begin{eqnarray}
&&\frac{G_N}{r^{D-3}} \left(c_1+\frac{f_1 E}{\left(E^2+m_1^2-m_2^2\right)}+\frac{f_1 E}{\left(E^2-m_1^2+m_2^2\right)}\right) = 0  \nonumber \\
&& \quad \implies c_1 = - \frac{E}{2 E_1 E_2} f_1 \;,  \label{eq:c1}
\end{eqnarray}
where we have used the expressions for $E_1$ and $E_2$ which are given by,
\begin{equation}
E_1 = \frac{E^2+m_1^2-m_2^2}{2E}~~;~~E_2 = \frac{E^2-m_1^2+m_2^2}{2E} \;.
\end{equation}
We now perform an equivalent step to order, $\mathcal{O}(G_N^2)$, by keeping all the terms up to $\mathcal{O}(G_N^2)$ in \eqref{eq:2PMHamERel}. In this case we find an expression for $c_2$ in terms of $f_1$ and $f_2$,
\begin{eqnarray}\label{eq:c2}
&& c_2 = \frac{E}{8 E_1^2 E_2^2 \left(E^4-\left(m_1^2-m_2^2\right)^2\right)^2} \left(-16 E_1^2 E_2^2 E^2 f_2 \left(E^4-\left(m_1^2-m_2^2\right)^2\right) \right. \nonumber \\
&& \left. +  16 E_1 E_2 E f_1 \left(E^8+E^4 \left(-4 m_1^4+6 m_1^2 m_2^2-4 m_2^4\right)+4 E^2 \left(m_1^2-m_2^2\right)^2 \left(m_1^2+m_2^2\right) \right.  \right. \nonumber \\
&& \left. \left. - \left(m_1^2-m_2^2\right)^2 \left(m_1^4+m_2^4\right)\right)+f_1^2 \left(E^8+2 E^4 \left(m_1^2-m_2^2\right)^2-3 \left(m_1^2-m_2^2\right)^4\right)\right) \;. \nonumber \\
\end{eqnarray}
Using these results we can now fully relate, up to 2PM order, the undetermined coefficients of the potential, $\{ c_i \}$, to the coefficients for the expansion of the momentum, $\{f_i\}$ which enter into the expression for the deflection angle in \eqref{eq:2PMDdimfiangle}. Note that the results in this section are independent of the spacetime dimension $D$.

\subsection{Deriving the Potential from the Deflection Angle}

In this section we will use the results we have developed in the previous sections with the results from chapter \ref{chap:graveik} to write down the 2PM $D$-dimensional Hamiltonian for a binary system in Einstein gravity.

From \eqref{eq:2PMDdimfiangle} we know that the 1PM contribution to the deflection angle can be written in terms of $f_1$ and $p_0$ as,
\begin{equation}
\theta_J^{\text{1PM}} = \frac{\sqrt{\pi} f_1 \Gamma \left(\frac{D}{2}-1\right)}{\Gamma \left(\frac{D-3}{2}\right)} \left( \frac{G_N \, p_0^{D-5}}{J^{D-3}} \right) \;.
\end{equation}
We can then use the expression for the angle derived at 1PM using the eikonal, \eqref{S13}, as well as the expression for $p_0$, \eqref{eq:p2zero}, to find the equation for $f_1$,
\begin{equation}
f_1 = \frac{2 \Gamma\left( \frac{D-3}{2} \right)}{\pi^{\frac{D-3}{2}} E} \gamma(s) \;,
\end{equation}
where we have used the fact that $J=p_0 b$ as well as the definition of $\gamma(s)$ given in \eqref{eq:1gegms}. Using the corresponding expression for $c_1$ found in \eqref{eq:c1} we also find,
\begin{equation}\label{eq:c1final}
c_1 = - \frac{\Gamma\left( \frac{D-3}{2} \right)}{\pi^{\frac{D-3}{2}} E_1 E_2} \gamma(s) \;.
\end{equation}
Similarly the expression for the 2PM contribution to the deflection angle in terms of the $\{f_i\}$ and $p_0$ is given by,
\begin{equation}
\theta_J^{\text{2PM}} = \frac{\sqrt{\pi } \Gamma \left(D-\frac{5}{2}\right) \left((D-4) f_1^2+2 f_2 p_0^2\right)}{2 \Gamma (D-3)}  \left( \frac{G_N \, p_0^{D-5}}{J^{D-3}} \right)^2 \;.
\end{equation}
Comparing the expression above with the result for the 2PM contribution to the angle derived in chapter \ref{chap:graveik}, given by \eqref{S17}, we find for $f_2$,
\begin{eqnarray}
f_2 &=& \frac{2}{\pi^{D-3} E} \Gamma^2 \left(\frac{D-3}{2}\right) \left((m_1+m_2) \Lambda(s) -(D-4)\frac{(E-m_1-m_2)}{E^2 p^2} \gamma^2(s)  \right) \;, \nonumber \\
\end{eqnarray}
where to simplify the result we have defined $\Lambda(s)$ as,
\begin{equation}\label{eq:2geLs}
\Lambda(s) = (s-m_1^2 -m_2^2)^2 - \frac{4m_1^2 m_2^2 }{(D-2)^2}  - \frac{(D-3) \left( (s-m_1^2 -m_2^2)^2 - 4 m_1^2 m_2^2 \right)  }{4 (D-2)^2} \;.
\end{equation}
Using the expression for $c_2$ derived in \eqref{eq:c2} with the expression above we find that we can write $c_2$ as,
\begin{eqnarray}\label{eq:c2final}
c_2 &=& \frac{\pi ^{-D} E \Gamma \left(\frac{D-3}{2}\right)}{2 E_1^2 E_2^2 \left(E^4-\left(m_1^2-m_2^2\right)^2\right)^2} \left(-\frac{8 \pi ^3 E_1^2 E_2^2 \Gamma \left(\frac{D-3}{2}\right)}{E p^2} \left(E^4-\left(m_1^2-m_2^2\right)^2\right) \right. \nonumber \\
&& \times \left. \left(E^2 \Lambda(s) p^2 (m_1+m_2)-\gamma^2(s) (D-4) (E-m_1-m_2)\right) \right. \nonumber \\
&& \left. + 8 \gamma(s) \pi ^{\frac{D+3}{2}} E_1 E_2 \left(E^8+E^4 \left(-4 m_1^4+6 m_1^2 m_2^2-4 m_2^4\right)  \right. \right.\nonumber \\
&& \left. \left. +4 E^2 \left(m_1^2-m_2^2\right)^2 \left(m_1^2+m_2^2\right)-\left(m_1^2-m_2^2\right)^2 \left(m_1^4+m_2^4\right)\right) \right. \nonumber \\
&& \left. +\frac{\pi ^3 \gamma^2(s) \Gamma \left(\frac{D-3}{2}\right) \left(E^8+2 E^4 \left(m_1^2-m_2^2\right)^2-3 \left(m_1^2-m_2^2\right)^4\right)}{E^2}\right) \;.
\end{eqnarray}
We now have all the necessary results to write down the full 2PM two-body Hamiltonian in Einstein gravity in arbitrary $D$-dimensions. Taking \eqref{eq:3PMHam}, \eqref{eq:undetpot}, \eqref{eq:c1final} and \eqref{eq:c2final} we find,
\begin{eqnarray}
&& H(\bs r,\bs p)= \sqrt{m_1^2+\bs p^2}+\sqrt{m_2^2+\bs p^2} - \frac{G_N}{|\bs r|^{D-3}} \left[ \frac{\Gamma\left( \frac{D-3}{2} \right)}{\pi^{\frac{D-3}{2}}} \frac{\gamma(s)}{E^2 \zeta} \right] \nonumber \\
&& + \frac{G_N^2}{|\bs r|^{2D-6}} \Biggl[ \frac{\Gamma \left(\frac{D-3}{2}\right)}{4 \pi^D \gamma ^5 \zeta ^3 m^5} \Biggl\{\gamma(s) \pi ^{\frac{D+3}{2}} m^2 \left(4 (2 \zeta -1) \mu ^2 \right.  \nonumber \\
&& \left.  +m^2 \left(\gamma ^4+\gamma ^2 (4-16 \zeta )+4 \zeta -5\right) + 4 \mu  m \left(\gamma ^2 (8 \zeta -2)-4 \zeta +5\right)\right)   \nonumber \\
&&   - \frac{2 \pi ^3 \Gamma \left(\frac{D-3}{2}\right)}{\mu ^2 \left(\sigma ^2-1\right)} \left\{ \gamma^2(s)[ (3 \zeta -1) \mu ^2 \left(\sigma ^2-1\right) + 2 \gamma ^3  (D-4) \zeta ^2 m^2 (\gamma - 1)] \right.  \nonumber \\
&& \left.  +2 \Lambda(s) \gamma ^3 \zeta ^2 \mu ^2 m^4 \left(\sigma ^2-1\right)\right\}   \Biggr\} \Biggr] \;,
\end{eqnarray}
where we have defined the symbols,
\begin{eqnarray}
&& \sigma = \frac{- k_1k_2}{m_1 m_2}~;~\zeta = \frac{E_1 E_2}{E^2}~;~ m=m_1+m_2 \nonumber \\
&& \qquad \gamma = \frac{E}{m}~;~\mu = \frac{m_1 m_2}{m}~;~\nu = \frac{\mu}{m} \;.
\end{eqnarray}
To compare our results here with known results in the literature we can set the spacetime dimension to the usual $D=4$. Also using the explicit expressions for $\gamma(s)$, \eqref{eq:1gegms}, and $\Lambda(s)$, \eqref{eq:2geLs}, we find,
\begin{eqnarray}\label{eq:full2PMHamD4}
&& H(\bs r,\bs p)= \sqrt{m_1^2+\bs p^2}+\sqrt{m_2^2+\bs p^2} + \frac{G_N}{|\bs r|} \left[ \frac{\mu^2}{\zeta \gamma^2}\left( 1 - 2 \sigma^2  \right) \right] \nonumber \\
&& + \frac{G_N^2}{|\bs r|^{2}} \left[ \frac{\nu ^2 m^3}{\gamma ^2 \zeta } \left(\frac{3}{4} \left(1-5 \sigma ^2\right) -\frac{4 \nu  \sigma \left(1-2 \sigma ^2\right)}{\gamma  \zeta } - \frac{\nu^2 (1-\zeta )  \left(1-2 \sigma ^2\right)^2 }{2 \gamma ^3 \zeta ^2}\right) \right] \;, \nonumber \\
\end{eqnarray}
where we have rearranged the 2PM contribution to make comparison with the literature easier. We can quickly see this agrees with equivalent results in \cite{Bern:2019nnu,Antonelli:2019ytb,Bern:2019crd,Cristofoli:2019neg}.

\subsection{PN Expansion}

In this subsection we will set $D=4$ and perform the PN expansion of the 2PM potential that we have calculated in the previous subsection. We will also explicitly compare the static (for simplicity) component with results in the literature and collect all the terms from our 2PM result that are relevant at 5PN.

From the 2PM potential found in \eqref{eq:full2PMHamD4}, we can perform an expansion in $v_{\infty}^2$ by using the relation we found in \eqref{eq:EinPNexp}, which we rewrite here,
\begin{equation}
E^2 = 2m_1m_2 \sqrt{1+v_{\infty}^2} + m_1^2 + m_2^2 \;.
\end{equation}
Performing the expansion using the relation above we find,
\begin{eqnarray}\label{eq:PMtoPNexpanpot}
&& V(\bs r) = \frac{G_N m_1 m_2}{|\bs r|} \Biggl[ -1 - \frac{\left(3 m_1^2+8 m_1 m_2+3 m_2^2\right) v_{\infty}^2}{2 \left(m_1+m_2\right)^2 }  \nonumber \\
&& + \frac{\left(5 m_1^2+2 m_1 m_2+5 m_2^2\right) v_{\infty}^4}{8 \left(m_1+m_2\right)^2 } + \ldots \Biggr] + \frac{G_N^2 m_1 m_2}{|\bs r|^2} \Biggl[ \frac{ \left(m_1^2+3 m_1 m_2+m_2^2\right)}{2 \left(m_1+m_2\right)} \nonumber \\
&& + \frac{\left(10 m_1^4+67 m_1 m_2^3+117 m_1^2 m_2^2+67 m_1^3 m_2+10 m_2^4\right) v_{\infty}^2}{4 \left(m_1+m_2\right){}^3}\nonumber \\
&& + \frac{\left(27 m_1^6+55 m_1 m_2^5-137 m_1^2 m_2^4-333 m_1^3 m_2^3-137 m_1^4 m_2^2+55 m_1^5 m_2+27 m_2^6\right) v_{\infty}^4}{16 \left(m_1+m_2\right){}^5}\nonumber \\
&& \qquad + \ldots \Biggr] \;,
\end{eqnarray}
where the ellipses denote contributions at higher orders in the velocity.

\subsubsection{Static Potential}

We can easily observe that the first term in \eqref{eq:PMtoPNexpanpot} is the Newton potential, as expected at 0PN. If we collect the static terms (i.e. the terms that are non-zero when we set $v_{\infty}=0$) we find that the static potential up to 2PM accuracy is given by,
\begin{eqnarray}\label{eq:PNexpStaticPot}
V_{\text{static}} =-\frac{G_N m_1 m_2}{|\bs r|} + \frac{G_N^2 m_1 m_2 \left(m_1^2+3 m_1 m_2 +m_2^2\right)}{2 \left(m_1+m_2\right) |\bs r|^2} \;.
\end{eqnarray}
Although the first term trivially reproduces the Newtonian result the second term does not reproduce the results in \cite{Foffa:2019hrb}. However as explained in \cite{Bern:2019nnu} the PM potential is equivalent, in the PN expansion, to results in the literature only after a canonical transformation. We can therefore try to identify the explicit canonical transformation required in order to make the above result match with the literature. 

We can use an ansatz for the transformation of the form,
\begin{equation}
(r,p) \rightarrow (R,P) = \left( \left(1+A \frac{G_N}{r} \right) r + B \frac{G_N}{r} p, p \right) \;,
\end{equation}
where $A, B$ are undetermined constants and we have not transformed the momenta since it does not appear in \eqref{eq:PNexpStaticPot}. For this transformation the Poisson bracket yields,
\begin{equation}
\{R,P\}_{r,p} = 1 - \frac{G_N p}{r^2} B = 1 \;,
\end{equation} 
where we have set this equal to $1$ as required for a canonical transformation. Using this we find that we require $B=0$. Substituting the remaining part of the transformation into our static potential and expanding in $G_N$ we find,
\begin{equation}
V_{\text{static}} =-\frac{G_N m_1 m_2}{|\bs r|} + \frac{G_N^2}{|\bs r|^2} \left(\frac{m_1 m_2 \left(m_1^2+3 m_1 m_2 + m_2^2\right)}{2 \left(m_1+m_2\right)} +  m_1 m_2 A \right) \;.
\end{equation}
From this we find the value for $A$ required to connect to the results found in \cite{Foffa:2019hrb} is,
\begin{equation}
A = - \frac{m_1 m_2}{2(m_1+m_2)} \;.
\end{equation}
We then have,
\begin{equation}
V_{\text{static}} =-\frac{G_N m_1 m_2}{|\bs r|} + \frac{G_N^2 m_1 m_2 \left(m_1+m_2\right)}{2|\bs r|^2} \;.
\end{equation}
This reproduces the relevant results in \cite{Foffa:2019hrb}. We have  therefore produced an explicit check that the static components of our 2PM potential match known results in the PN expansion.

\subsubsection{Collecting Terms Relevant for the 5PN Potential}

In this subsection we will collect terms coming from the 2PM potential that are relevant at 5PN order. In principle one could use this result in combination with the static component of the 5PN potential \cite{Foffa:2019hrb} and the relevant results coming from the 3PM component of the potential \cite{Bern:2019nnu,Bern:2019crd} to reconstruct most of the potential at 5PN accuracy.

From the discussion in section \ref{sec:pmexpansionbg} we can quickly identify which orders in velocity we need to collect from the 1PM and 2PM contributions to the potential in order to find the contributions relevant to the 5PN potential. For 1PM we clearly need the $\mathcal{O}(v_{\infty}^{10})$ terms in the velocity expansion and equivalently for 2PM we need the $\mathcal{O}(v_{\infty}^8)$ terms. So we find,
\begin{eqnarray}
&& \tilde{V}_{\text{5PN}} = - \frac{G_N m_1 m_2 }{|\bs r|} \frac{ v_{\infty}^{10}}{256 \left(m_1+m_2\right)^6} \biggl(77 m_1^6+206 m_1 m_2^5+199 m_1^2 m_2^4+156 m_1^3 m_2^3 \nonumber \\
&& + 199 m_1^4 m_2^2+206 m_1^5 m_2+77 m_2^6 \biggr) - \frac{G_N^2 m_1 m_2}{|\bs r|^2} \frac{v_{\infty}^8}{256 \left(m_1+m_2\right)^9}  \biggl(425 m_1^{10}  \nonumber \\
&& + 1879 m_1 m_2^9+1617 m_1^2 m_2^8 -5685 m_1^3 m_2^7-17737 m_1^4 m_2^6-23781 m_1^5 m_2^5\nonumber \\
&& -17737 m_1^6 m_2^4-5685 m_1^7 m_2^3+1617 m_1^8 m_2^2+1879 m_1^9 m_2+425 m_2^{10}\biggr) \;,
\end{eqnarray}
where the result is given in the PM gauge.

\chapter{Conclusions and Outlook}\label{chap:conclusion}

In this thesis we have considered binary systems in various theories of gravity with the ultimate goal of computing quantities relevant for describing the conservative dynamics of binary black holes. Computing these quantities directly using general relativity has proven to be a difficult task \cite{bertotti1956gravitational, Kerr1959, BERTOTTI1960169, Portilla:1979xx, Westpfahl:1979gu, Portilla:1980uz}. For this reason, in this thesis we have developed techniques using quantum field theory to simplify this problem. Using the eikonal approximation the calculation translates into calculating only a subset of Feynman diagrams from all the Feynman diagrams present at each order in $G_N$. In addition to this we also need to be able to identify the classical contributions for which the eikonal approximation technique provides a relatively simple algorithm as described towards the end of chapter \ref{chap:background}. Chapters \ref{chap:sugraeik} and \ref{chap:kkeik} were developed to understand the eikonal approximation techniques in more detail whilst in chapters \ref{chap:graveik} and \ref{chap:potential} we applied those techniques in the context of binary black holes in Einstein gravity. In this final chapter we will discuss and summarise the various results in each chapter and give some outlook on future avenues and extensions that could be researched using the material developed in this thesis.

In chapter \ref{chap:sugraeik} we studied the possibility that the eikonal could be described by an operator instead of a simple phase in the context of type II supergravity focusing on the scattering of massless perturbative states from a stack of D$p$-branes. The two main points of our analysis for the subleading eikonal in chapter \ref{chap:sugraeik} are that the relevant information is encoded in the onshell three and four-point vertices (see section~\ref{2braneamps}) and that its derivation requires us to disentangle cross terms between leading and subleading energy contributions (see section~\ref{HElimit}). In the scenario under study, the inelastic contributions which grow with energy in the one-loop amplitudes, are completely accounted for by the cross terms mentioned above and thus the final expression for the first subleading eikonal in supergravity is described fully by the elastic processes and is given by,
\begin{equation}
\delta^{(2)}(E,b) = \frac{(N T_p \kappa_D)^2 E}{16 \pi^{D-p-\frac{3}{2}}} \frac{\Gamma^2 \left(\frac{D-p-3}{2} \right) \Gamma \left(\frac{2D-2p-7}{2} \right)}{\Gamma \left(D-p-4 \right)} \frac{1}{b^{2D-2p-7}} \label{eq:sugrasubeik}   \;,
\end{equation}
which agrees with previous results in~\cite{D'Appollonio:2010ae}. We do not know a general argument proving that this is always the case and so we think that it would be interesting to check this pattern both in more complicated theories and at further subleading orders (higher orders in the PM expansion). For instance, an analysis of the eikonal operator in string theory beyond the leading order~\cite{Amati:1987wq,Amati:1987uf,Amati:1988tn} is missing. Of course we could use the full four-point string amplitudes in our derivation of section~\ref{2braneamps} simply by reinstating the $\alpha'$ dependence that in maximally supersymmetric theories appears just in the overall combination of gamma functions,
\begin{equation}
  \label{eq:gammafst}
   \frac{\Gamma ( 1 -\frac{\alpha' s}{4} ) \Gamma ( 1- \frac{\alpha't}{4} )
\Gamma ( 1 -\frac{\alpha'u}{4} ) }{\Gamma ( 1  + \frac{\alpha's}{4}) \Gamma ( 1 +\frac{\alpha't}{4} ) \Gamma ( 1 +\frac{\alpha'u}{4} )}\;.
\end{equation}
For instance in the first contribution to the dilaton to dilaton scattering we analysed, this amounts to including the factor defined above in the vertex~\eqref{STampRR}. However this is not sufficient to reconstruct the full string eikonal as, of course, we need to include also the contributions due to the exchanges of the leading and subleading Regge trajectories between the D$p$-branes and the perturbative states. It seems possible to generalise, along these lines, the analysis of chapter \ref{chap:sugraeik} to the full string setup by using the formalism of the Reggeon vertex~\cite{Ademollo:1989ag,Ademollo:1990sd,Brower:2006ea, D'Appollonio:2013hja}. 

On the more conceptual side, it would be interesting to check if in more general cases, such as those with two arbitrary masses $m_1,m_2$ at higher orders in the PM expansion or in cases with non-zero angular momentum, the subleading contributions to the eikonal are still universal or if there are inelastic effects that induce differences between the various states as we know happens when the 3-graviton vertex is modified~\cite{Camanho:2014apa}.

In chapter \ref{chap:kkeik} we studied the scattering of massless states in Kaluza-Klein gravity with non-zero Kaluza-Klein mass. Recent work \cite{Khalil:2018aaj} has investigated the particular case of binary Einstein-Maxwell-dilaton black holes in the post-Newtonian approximation. We believe that our result for the leading eikonal in Kaluza-Klein theories,
\begin{eqnarray}
\delta^{(1)} (D-1, R)  &=& \frac{\kappa_D^2 }{2} \frac{s' - m_1^2 -m_2^2 - 2 m_1 m_2  }{(2\pi)^{\frac{D-1}{2}} }\frac{1}{ {(b')}^{\frac{D-5}{2} }}  \frac{1}{R} \nonumber \\
&& \qquad \times \sum_{n  \in {\mathbb Z } } e^{ib_s n/R}  {\biggl( \frac{|n|}{R} \biggr)^{\frac{D-5}{2} } K_{\frac{D-5}{2} } \biggl( \frac{|n|b'}{R} \biggr) \;,
}\end{eqnarray}
could be relevant in the study of such binary systems in the first post-Minkowskian approximation. The massless sector of the Kaluza-Klein action given by \eqref{KK4D} supported the well known Einstein-Maxwell-dilaton black hole solutions whose extremal $Q=2M$ limit gave deflection angles consistent with the eikonal calculation in the limit where one of the 2 massive Kaluza-Klein scalars is taken to be very heavy and the other is taken to be massless. 

The techniques developed in the chapters mentioned above were then systematically applied in the more general case of two massive scalars with arbitrary masses $m_1,m_2$ in chapter \ref{chap:graveik}. Doing so allowed us to compute the $D$-dimensional two-body deflection angles up to 2PM, $\mathcal{O}(G_N^2)$, via the eikonal approximation method.

An interesting feature of the 2PM eikonal phase for massive scalars that we obtained in chapter \ref{chap:graveik} is that its double-massless limit smoothly reproduces the massless result up to 2PM order. This is also valid for the quantum contribution mentioned in section \ref{sec:discussionGRAV}. Using our results we can see that in the ultra-relativistic limit we have,
\begin{equation}
\theta^{(2)}_{\text{ACV}} = 0~~~;~~~~\delta^{(2)}_{q, \text{ACV}} =  \frac{12 G_N^2 s}{\pi b^2} \log (s)\;,
\end{equation}
which reproduces the known 2PM results in the ultra-relativistic limit found in~\cite{Amati:1990xe}. 

The same property does not seem to hold in the 3PM case for the result by Bern et al \cite{Bern:2019nnu,Bern:2019crd} and it would be very interesting to understand the origin of this mismatch. Specifically there are certain $\log(m_i)$ divergences in their result. In \cite{Bern:2019nnu,Bern:2019crd} the authors argue that these $\log(m_i)$ divergences present when taking the $m_i \to 0$ limit are due to the hierarchy of scales when performing the integrals necessary to compute their result. This however does not seem to resolve the divergence present in the ultra-relativistic limit where we find a $\log(s)$ divergence,
\begin{eqnarray}
&&\lim_{\substack{s \rightarrow \infty}} (\ldots) \sinh^{-1}\left(\frac{1}{2} \sqrt{\frac{s-(m_1+m_2)^2}{m_1 m_2}}\right) \sim - \frac{16 G_N^3 s^{3/2}}{b^3} \log \left(\frac{s}{m_1 m_2} \right) \;. \label{eq:sinhlim1}
\end{eqnarray}
The prefactor here is of the right order in $M^*$ to contribute\footnote{Specifically it is of the order $M^{*3}$ as required at 3PM. Recall from the discussion in section \ref{sec:probeind} that in the ultra-relativistic (or double-massless) limit $M^*=\sqrt{s}$.}. This suggests that this quantity really does contribute to the angle in the ultra-relativistic limit when we take the impact parameter to be large.

There has recently been a new 3PM result by Damour \cite{Damour:2019lcq} which differs from the 3PM result by Bern et al. Performing the same check on the $\text{arcsinh}(\ldots)$ component of the result by Damour we find,
\begin{equation}
\lim_{\substack{s \rightarrow \infty}} (\ldots) \sinh^{-1}\left(\frac{1}{2} \sqrt{\frac{s-(m_1+m_2)^2}{m_1 m_2}}\right) \sim - \frac{768 G_N^3 m_1^2 m_2^2}{s^{1/2} b^3} \log \left(\frac{s}{m_1 m_2} \right) \to 0 \;, \label{eq:sinhlim2}
\end{equation}
where in the last step we have taken $s \to \infty$ explicitly. We quickly see that in this case the problematic $\log(s)$ divergence in the ultra-relativistic limit is not present due to the factor of $s^{-1/2}$. Although scaling arguments similar to the above suggest that the massless probe limit is smoothly reproduced in both cases, we can also observe that the double massless limit of both of these results does not reproduce the equivalent 3PM contribution calculated in \cite{Amati:1990xe, Amati:1992zb} as is also recently noted in a paper by Bern et al \cite{Bern:2020gjj}.
\begin{figure}[h]
  \centering
  \includegraphics[scale=0.55]{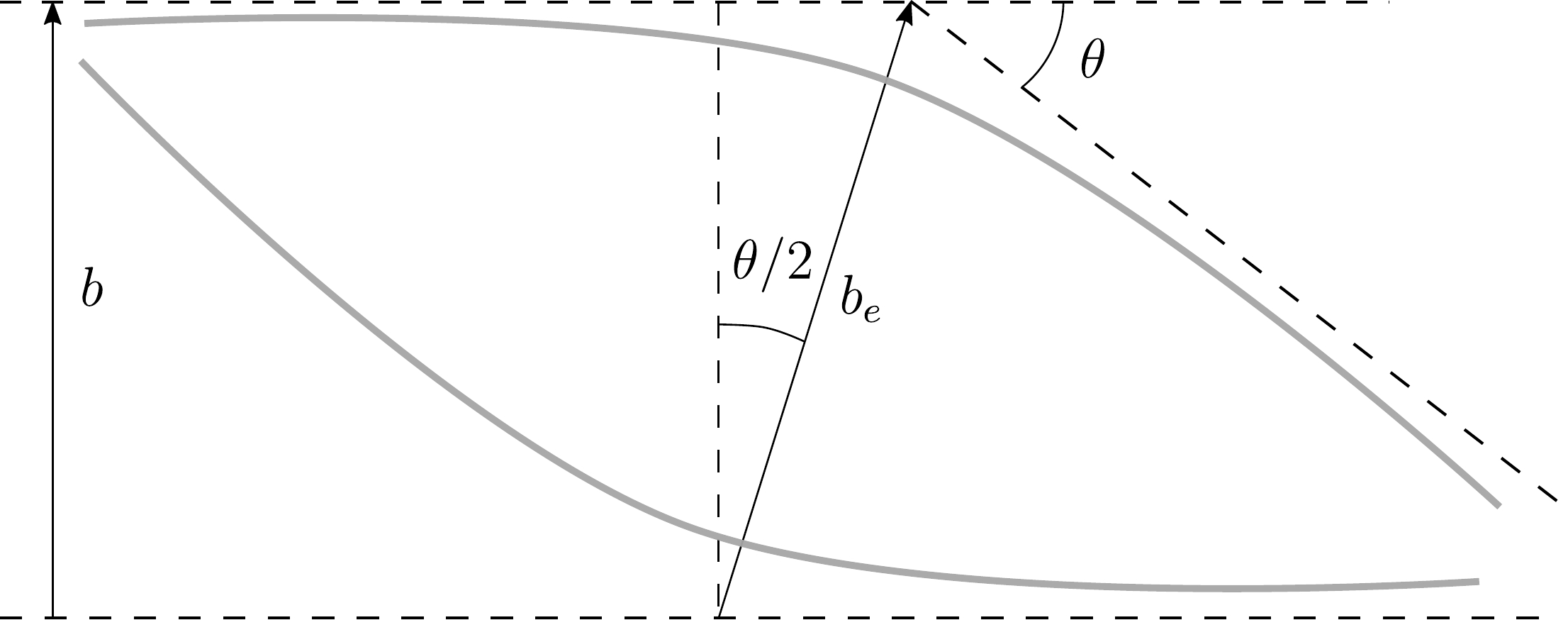}
  \caption{A figure illustrating the two-body scattering process with the variables required to go beyond 2PM level when discussing the deflection angle.}
  \label{fig:scatangles}
\end{figure}

We believe that the results discussed in this thesis can be applied to higher orders in the PM expansion beyond the 2PM level considered explicitly here. This could provide an alternative method with which to check various known results at the 3PM level and beyond. Including the relevant higher order eikonal corrections beyond 2PM, one could use the techniques discussed in section \ref{sec:introeikpwd} by considering a modification to \eqref{eq:fullleadingsaddleaJexprLINEAR} where we do not linearise in $\phi$. This is schematically given by,
\begin{eqnarray}
&&\frac{a_J(s,m_i)}{4Ep} \approx \frac{J}{2p^2} \left( \frac{J}{p} \right)^{-i \delta_1} \frac{2\pi}{\sqrt{-\det(i J S''(x_0,\phi_0))}} \nonumber \\
&& \qquad \times \exp \left[i J \left( -\tilde{\delta}_1 \log \left(x_0 \right) + \sum_{k=1}^{\infty} \frac{\tilde{\delta}_{k+1}}{x_0^k} +2 x_0 \sin \phi_0 - 2 \phi_0 \right) \right] \;,
\end{eqnarray}
where $(x_0, \phi_0)$ is the appropriate saddle point calculated perturbatively up to the desired PM order. The schematic above along with the expression relating the partial waves to the deflection angle given by \eqref{eq:anglefromaJ} has been has been checked perturbatively up to $\mathcal{O}(G^5)$ and found to agree with the more common expression \cite{Bjerrum-Bohr:2018xdl, Bern:2020gjj} for the deflection angle derived using kinematic arguments and given by,
\begin{eqnarray}\label{eq:angfromeikinbhigherPM}
\sin \frac{\theta}{2} &=& - \frac{1}{2 p} \frac{\partial}{\partial b_e}  \delta(s, b_e) \nonumber \\
\implies \tan \frac{\theta}{2} &=& \frac{1}{2p} \left( \frac{\delta_1}{b} + \sum_{k=1}^{\infty} \frac{k \delta_{k+1}}{b^k} \cos^{k-1} \frac{\theta}{2} \right)
\end{eqnarray}
where we have used $b = b_e \cos \frac{\theta}{2}$ as recently discussed in \cite{Bern:2020gjj} and illustrated in figure \ref{fig:scatangles} in order to be able to consistently use these equations beyond the 2PM level.

Another interesting development would be to generalise the analytic bootstrap approach of~\cite{Amati:1990xe,Amati:1992zb,Ciafaloni:2018uwe} beyond the massless $D=4$ case. In that approach the quantum part of the eikonal $\delta^{(2)}_q$ plays an important role in the derivation of the subsequent {\em classical} PM order and we expect that a similar pattern is valid also beyond the setup of~\cite{Amati:1990xe,Amati:1992zb,Ciafaloni:2018uwe}. This approach has the potential to provide an independent derivation of the 3PM eikonal phase both in the massless higher dimensional case and in the physically interesting case of the massive scattering in $D=4$. 

Chapter \ref{chap:potential} discusses how to derive the potential from the dynamical information calculated using our eikonal approach. Starting from the two-body deflection angle we were able to reverse-engineer and find the two-body Hamiltonian required to reproduce the two-body deflection angle result. This is an important step in our approach because although all the dynamical information is neatly contained in the two-body deflection angle, the Hamiltonian form is useful for experimentalists and others which use these results to aid the creation of more accurate gravitational wave templates.

There are many more interesting and exciting avenues of research aside from the ones mentioned in more detail above. These include studying cases with non-zero angular momentum in order to study the case of binary spinning black holes \cite{Bini:2017xzy,Vines:2017hyw, Vines:2018gqi, Guevara:2018wpp, Guevara:2019fsj, Maybee:2019jus, Damgaard:2019lfh}, computing higher order results in both the PM and PN expansion \cite{Foffa:2019hrb, Bern:2019nnu, Bern:2019crd, Bini:2019nra, Damour:2019lcq}, studying these processes in modified theories of gravity \cite{Camanho:2014apa,DAppollonio:2015fly, Brandhuber:2019qpg, Huber:2019ugz}, as well as including fine structure to better understand the dynamics of binary neutron stars \cite{Bini:2020flp}. Finally there have been recent developments in connecting the amplitude results directly to the two-body deflection angle both bypassing the Hamiltonian approach as well as the eikonal approach \cite{Kalin:2019rwq, Bjerrum-Bohr:2019kec, Kalin:2019inp} and these methods could be studied in more detail.

\appendix

\chapter{Geodesics in Black Hole Backgrounds}\label{app:geodesics}

In this appendix we will be describing geodesics in a variety of black hole backgrounds which are relevant to the main text.  

We can write the action for a massive relativistic particle of mass $m$ in a background $g_{\mu \nu}$ as,
\begin{equation}
S = \frac{1}{2} \int d \tau \left( e(\tau)^{-1} \dot{x}^2 -m^2 e(\tau) \right) \label{eq:geo1} \;,
\end{equation}
where $\dot{x}^2 = g_{\mu \nu} \frac{d x^{\mu}}{d \tau} \frac{d x^{\nu}}{d \tau}$ and $e(\tau)$ is an auxiliary field. The equation of motion for $e(\tau)$ is then easily obtained,
\begin{equation}
e^2 = - \frac{\dot{x}^2}{m^2} \label{eq:geo2} \;.
\end{equation}
Using this equation we will be able to calculate the deflection angle for null-like or time-like probes for the different black hole backgrounds we will consider.

\section{Geodesics in a $D$-dimensional Schwarzschild \\ Background} \label{sec:geodesics}

The metric for a $D$-dimensional Schwarzschild black hole is given by,
\begin{equation}
g_{\mu \nu} dx^{\mu} dx^{\nu} = - \left(1-\left(\frac{R_s}{r}\right)^n \right) dt^2 + \left(1-\left(\frac{R_s}{r}\right)^n \right)^{-1} dr^2 + r^2 d \Omega_{n+1}^2 \label{eq:geo3} \;,
\end{equation}
where $n=D-3$ and $R_s$ is the Schwarzschild radius given by,
\begin{equation}
R_s^{D-3} = \frac{16\pi G_N M}{(D-2)\Omega_{D-2}}~~;~~ 
\Omega_{D-2}=
\frac{2 \pi^{\frac{D-1}{2}}}{\Gamma ( \frac{D-1}{2})} \;.
\end{equation}
For simplicity we will use the spherical symmetry of the Schwarzschild solution and work in the equatorial plane and so we set all the angles, $\theta_i= \pi/2$, such that we are left with only one angle, $\phi$, in the impact plane. Inserting \eqref{eq:geo3} into \eqref{eq:geo2} and using the reparameterization invariance to set $e=1$ we find,
\begin{equation}
-m^2 = - \left(1-\left(\frac{R_s}{r}\right)^n \right) \left( \frac{dt}{d \tau} \right)^2 + \left(1-\left(\frac{R_s}{r}\right)^n \right)^{-1} \left( \frac{dr}{d \tau} \right)^2 + r^2 \left(\frac{d \phi}{d \tau} \right)^2 \label{eq:geo4} \;.
\end{equation}
Since the metric is independent of both $t$ and $\phi$ we have two constants of the motion which are given by,
\begin{eqnarray}
- E &=& - \left(1 - \left(\frac{R_s}{r}\right)^n \right)  \frac{dt}{d\tau} \;, \\
J &=& r^2 \frac{d\phi}{d\tau} \;,
\end{eqnarray}
where $E$ and $J$ parametrize the energy and total angular momentum of the system respectively. Notice that the symbol $E$ in this appendix corresponds to $E_2$ in sections \ref{sec:probeind} and \ref{sec:probeind4} where we consider various probe-limits to the 1PM and 2PM two-body deflection angles. Substituting these quantities into \eqref{eq:geo4} we can find an expression for $dr/d\tau$ and using the chain rule we can find the corresponding expression for $d\phi/dr$. The deflection angle is then given by,
\begin{equation}
\Phi = 2 \int_{r_0}^{\infty} dr \left(\frac{d\phi}{dr} \right) - \pi \;, 
\end{equation}
where $r_0$ is the point of closest approach. Inserting the relevant quantities into the expression above and expanding in powers of $(R_s/r_0)^n$ we notice that we can express the result as a series,
\begin{equation}
\Phi = 2 \sum_{j=1}^{\infty} \int_{0}^{1} du A_j(u) \left(\frac{R_s}{r_0}\right)^{jn} \;,
\end{equation}
where we have used the substitution $u=r_0/r$ and explicitly evaluated the integrals up to third order and found the following results,
\begingroup 
\allowdisplaybreaks
\begin{align}
& \int_{0}^{1} du A_1(u) = \frac{\sqrt{\pi} \, \Gamma \left(\frac{n+1}{2}\right) \left(E^2 (n+1)-m^2\right)}{4 \left(E^2-m^2\right) \Gamma \left(\frac{n}{2}+1\right)} \;, \\
\nonumber \\
& \int_{0}^{1} du A_2(u) = \frac{\sqrt{\pi}}{16 \left(E^2-m^2\right)^2} \Biggl(\frac{\Gamma \left(n+\frac{1}{2}\right)}{\Gamma (n+1)} (E^4 (4 (n+2) n+3) \nonumber \\*
& - 6 E^2 m^2 (2 n+1)+3 m^4 ) - \frac{4 E^2 \Gamma \left(\frac{n+1}{2}\right) \left(E^2 (n+1)-m^2\right)}{\Gamma \left(\frac{n}{2}\right)}\Biggr)  \;, \\
\nonumber \\
& \int_{0}^{1} du A_3(u) = \frac{\sqrt{\pi }}{32 \left(m^2-E^2\right)^3} \Biggl(-\frac{2 E^2 \Gamma \left(\frac{n+1}{2}\right) \left(E^2 (n-2)+4 m^2\right) \left(E^2 (n+1)-m^2\right)}{\Gamma \left(\frac{n}{2}\right)} \nonumber \\*
& + \frac{2 E^2 \Gamma \left(n+\frac{1}{2}\right) \left(E^4 (4 (n+2) n+3)-6 E^2 m^2 (2 n+1)+3 m^4\right)}{\Gamma (n)} \nonumber \\*
& +\frac{\Gamma \left(\frac{3 n}{2}+\frac{1}{2}\right)}{\Gamma \left(\frac{3 n}{2}+1\right)} \bigl[E^6 (-(n+1)) (3 n+1) (3 n+5)+15 E^4 m^2 (n+1) (3 n+1)\nonumber \\*
& -15 E^2 m^4 (3 n+1)+5 m^6\bigr] \Biggr) \;.
\end{align}
\endgroup
These results in addition to the relation up to third order between the point of closest approach and the angular momentum can be used to express the result for the deflection angle up to third order in $(R_s/J)^n$. The relation between $r_0$ and $J$ is found by evaluating,
\begin{equation}
\frac{dr}{d \tau} \biggr\rvert_{r=r_0} = E^2-m^2 \left(1 - \frac{R_s}{r_0}\right)^n - \frac{J^2}{r_0^2} \left(1 - \frac{R_s}{r_0}\right)^n = 0 
\end{equation}
and up to third order in $J(R_s/J)^n$ is found to be,
\begin{eqnarray}
&& r_0 \approx J \biggl(\frac{1}{\sqrt{E^2-m^2}} - \frac{1}{2} E^2 \left(E^2-m^2\right)^{\frac{n-3}{2}} \left( \frac{R_s}{J} \right)^{n} \nonumber \\
&& - \frac{1}{8} E^2 \left(E^2 (2 n+1)-4 m^2\right) \left(E^2-m^2\right)^{n-\frac{5}{2}} \left(\frac{R_s}{J}\right)^{2 n}\biggr) \;.
\end{eqnarray}
Putting together the relevant quantities and expanding the resulting expression as a power series in $(R_s/J)^n$ we find that the deflection angle is given by,
\begin{eqnarray}
&& \Phi = \sum_{j=1}^{\infty} \Phi_j \left( \frac{R_s}{J} \right)^{j(D-3)} \label{eq:geoPhi} \;,
\end{eqnarray}
where we have substituted for $n=D-3$ and we have for the first 3 terms,
\begin{align}
&\Phi_1 = \frac{\sqrt{\pi } \Gamma \left(\frac{D}{2}-1\right) (E^2-m^2)^{\frac{D-5}{2}} \left[(D-2) E^2-m^2\right]}{2 \Gamma \left(\frac{D-1}{2}\right)} \;, \label{eq:geophi1} \\
\nonumber \\
& \Phi_2 = \frac{\sqrt{\pi } \Gamma \left(D-\frac{5}{2}\right) (E^2-m^2)^{D-5}}{8 \Gamma (D-2)} \nonumber \\
& \times \left[(2 D-5) (2 D-3) E^4+6 (5-2 D) E^2 m^2+3 m^4\right] \;, \label{eq:geophi2} \\
\nonumber \\
& \Phi_3 = \frac{\sqrt{\pi } \Gamma \left(\frac{3 D}{2}-4\right) (E^2-m^2)^{\frac{3 (D-5)}{2}}}{16 \Gamma \left(\frac{3 D}{2}-\frac{7}{2}\right)} \nonumber \\
& \times \bigl[(3 D-8) (3 D-4) (D-2) E^6-15 (D-2) (3 D-8) E^4 m^2 \nonumber \\
& +15 (3 D-8) E^2 m^4-5 m^6\bigr] \label{eq:geophi3} \;.
\end{align}
There are a few limiting cases of the results above that we can use to compare with known results. The result for a null geodesic in general $D$ is given by the $m=0$ case of the equations above. We find,
\begin{eqnarray}
\Phi &=& \frac{\sqrt{\pi } \Gamma \left(\frac{D}{2}\right)}{\Gamma \left(\frac{D-1}{2}\right)}\left( \frac{R_s}{b} \right)^{D-3} + \frac{\sqrt{\pi } \Gamma \left(D-\frac{1}{2}\right)}{2 \Gamma (D-2)} \left( \frac{R_s}{b} \right)^{2(D-3)} \nonumber \\
&&  + \frac{\sqrt{\pi } \Gamma \left(\frac{3 D}{2}-1\right)}{6 \Gamma \left(\frac{3 D}{2}-\frac{7}{2}\right)} \left( \frac{R_s}{b} \right)^{3(D-3)} + \ldots \;, \label{eq:geogendnull}
\end{eqnarray}
where we have also used that when $m=0$ we have $J = Eb$. Comparing this expression with the results found in appendix D of \cite{Collado:2018isu} we find agreement for the first two terms (note the third term was not calculated in the aforementioned reference). We can also look at the $D=4$ timelike geodesic ($m \neq 0$) case where we find,
\begin{eqnarray}
\Phi &=& \frac{\left(2 E^2-m^2\right)}{\sqrt{E^2-m^2}} \left( \frac{R_s}{J} \right) + \frac{3\pi}{16} \left(5 E^2-m^2\right) \left( \frac{R_s}{J} \right)^2 \nonumber \\
&&  +\frac{\left(64 E^6-120 E^4 m^2+60 E^2 m^4-5 m^6\right)}{12 (E^2-m^2)^{3/2}} \left( \frac{R_s}{J} \right)^3 + \ldots \;, \label{eq:geogendtime4}
\end{eqnarray}
which has been checked and agrees with equivalent results in \cite{Damour:2017zjx,Bern:2019nnu, Antonelli:2019ytb}. Finally by taking the $m=0$ case of the expression above we can also look at the case of a null geodesic in $D=4$. This yields,
\begin{equation}
\Phi = \frac{2R_s}{b} + \frac{15\pi}{16} \left( \frac{R_s}{b} \right)^{2} + \frac{16}{3}\left( \frac{R_s}{b} \right)^{3}  + \ldots  \;. \label{eq:geo4null}
\end{equation}
We find that this agrees with well known results. 

Note that to relate these results to those found when studying the supergravity eikonal in chapter \ref{chap:sugraeik} we need to use the relation between the Schwarzschild radius and the various constants of the stack of ($p=0$) D-branes given by,
\begin{equation}
R_s^{D-3} = \frac{N \kappa_D T_{p=0}}{\Omega_{D-2}(D-2)} \;.
\end{equation}

\section{Null-Like Geodesics in a Einstein-Maxwell-dilaton \\ Black Hole Background} \label{angleEMDbh}

We can consider black hole solutions in a theory with a graviton, dilaton and photon with an action given by \cite{Horne:1992zy},
\be
\label{eq:horowitzS}
S = \int d^4 x \sqrt{-g} \left(R - 2(\nabla \Phi)^2 - e^{-2 \alpha \Phi}F^2 \right) \;,
\ee
where units have been taken such that $\kappa=1/\sqrt{2}$ and the spacetime dimension has been taken to be $D=4$. The solution describing static, spherically-symmetric black holes in this theory is given by,
\begin{eqnarray}
\label{genericmetric}
ds^2 &=& -\left( 1 - \frac{r_{+}}{r} \right)\left( 1 - \frac{r_{-}}{r} \right)^{\gamma} dt^2 + \left( 1 - \frac{r_{+}}{r} \right)^{-1}\left( 1 - \frac{r_{-}}{r} \right)^{-\gamma} dr^2 \nonumber \\
&& \qquad + r^2 \left( 1 - \frac{r_{-}}{r} \right)^{1-\gamma} d \Omega^2  \\
e^{2 \phi} &=&  \left( 1- \frac{r_-}{r} \right)^{\sqrt{1 -\gamma^2}} \\
F_{tr}  &=& \frac{Q}{r^2}  \;,
\end{eqnarray}
where $\gamma={\frac{1-\alpha^2}{1+\alpha^2}}$ and we define the various symbols introduced, which are related to the mass and charge of the black hole, via,
\be
2M = r_{+} + \gamma r_{-} \qquad 2 Q^2 = (1+\gamma)r_{+}r_{-} \;.
\ee
Inverting the equations above we find that we can write,
\be
r_{+} = M + M \sqrt{1-\frac{2 \gamma}{1+\gamma}\left(\frac{Q}{M} \right)^2 } \qquad r_{-} = \frac{M}{\gamma} - \frac{M}{\gamma} \sqrt{1-\frac{2 \gamma}{1+\gamma}\left(\frac{Q}{M} \right)^2 } \;.
\ee
We will consider the deflection angle for general parameter $\gamma$ below, however it is worth noting at this stage that the above solution takes an interesting form in the specific case when $\alpha = \sqrt{3}$ and we take the extremal limit  $r_- = r_+ $,  or correspondingly  $Q = 2M$. The metric then becomes,
\be
ds^2 = -\left( 1 + \frac{r_+}{r} \right)^{\frac{1}{2}} dt^2 + \left( 1 + \frac{r_+}{r} \right)^{-\frac{1}{2}} dr^2 + r^2\left( 1 + \frac{r_+}{r} \right)^{\frac{3}{2} } d\Omega^2 \;.
\ee

As we have done in the case of the Schwarzschild black hole in the previous subsection we can use the spherical symmetry and take $\theta$ to be in the equatorial plane, $\theta=\pi/2$. We can then use \eqref{eq:geo2} to write the equation for a massless probe in the metric, \eqref{genericmetric}, that we are considering here as,
\begin{eqnarray}
0 &=& - \left( 1 - \frac{r_{+}}{r} \right)\left( 1 - \frac{r_{-}}{r} \right)^{\gamma} \left( \frac{dt}{d \tau} \right)^2 +  \left( 1 - \frac{r_{+}}{r} \right)^{-1}\left( 1 - \frac{r_{-}}{r} \right)^{-\gamma} \left( \frac{dr}{d \tau} \right)^2 \nonumber \\
&& \quad + r^2 \left( 1 - \frac{r_{-}}{r} \right)^{1-\gamma} \left(\frac{d \phi}{d \tau} \right)^2  \label{eq:KKBHgeoeq} \;.
\end{eqnarray}
As in this previous subsection, the metric we are considering here is independent of both $t$ and $\phi$ and so we have two constants of the motion which are given by,
\begin{eqnarray}
- E &=& - \left( 1 - \frac{r_{+}}{r} \right)\left( 1 - \frac{r_{-}}{r} \right)^{\gamma}  \frac{dt}{d\tau} \;, \\
J &=&  r^2 \left( 1 - \frac{r_{-}}{r} \right)^{1-\gamma} \frac{d\phi}{d\tau} \;.
\end{eqnarray}
Substituting for all the relevant quantities in \eqref{eq:KKBHgeoeq} and solving for $d\phi/dr$ we find that the deflection angle is given by,
\be
\Phi + \pi = 2 \int_{r_0}^{\infty} dr \frac{1}{r^2 \left( 1 - \frac{r_{-}}{r} \right)^{1-\gamma}} \left[ \frac{\left( 1 - \frac{r_{+}}{r_0} \right) \left( 1 - \frac{r_{-}}{r_0} \right)^{\gamma}}{r_0^2 \left( 1 - \frac{r_{-}}{r_0} \right)^{1-\gamma}} - \frac{\left( 1 - \frac{r_{+}}{r} \right) \left( 1 - \frac{r_{-}}{r} \right)^{\gamma}}{r^2 \left( 1 - \frac{r_{-}}{r} \right)^{1-\gamma}} \right]^{-\frac{1}{2}} \;.
\ee
Using the substitution $u=r_0/r$ we find,
\begin{eqnarray}
\Phi + \pi &=& 2 \int_{0}^{1} du \left[\left( 1 - \frac{r_{+}}{r_0} \right) \left( 1 - \frac{r_{-} u}{r_0} \right)\left( \frac{r_0 - r_{-}}{r_0 - r_{-}u} \right)^{2\gamma-1} \right. \nonumber \\
&& \quad \left. - u^2 \left( 1 - \frac{r_{+} u}{r_0} \right)\left( 1 - \frac{r_{-}u}{r_0} \right)     \right]^{-\frac{1}{2}} \;.
\end{eqnarray}
We want to solve this integral as an expansion in $1/r_0$ as we have done in the previous subsection, so expanding the integrand in powers of $1/r_0$ up to third order we have,
\begin{eqnarray}
\Phi + \pi &=& 2 \left( \int_{0}^{1} du \frac{1}{\sqrt{1-u^2}} + \int_{0}^{1} du A_1(u) \left(\frac{1}{r_0}\right) + \int_{0}^{1} du A_2(u) \left(\frac{1}{r_0}\right)^2 \right. \nonumber \\
&& \left. \quad  + \int_{0}^{1} du A_3(u) \left(\frac{1}{r_0}\right)^3 +  \ldots \right) \;,
\end{eqnarray}
where we have explicitly solved for the relevant integrals and found,
\begin{eqnarray}
\int_{0}^{1} du A_1(u) &=& 2 M \;, \\
\int_{0}^{1} du A_2(u) &=& \sqrt{\frac{(\gamma+1)M^2 - 2\gamma Q^2}{(\gamma+1)M^2}} M^2 \left(\frac{\pi}{16 \gamma^2} - \frac{1}{\gamma} - \frac{\pi}{16} + 1 \right) \nonumber \\
&& + \frac{1}{16 \gamma^2} [(\pi (31\gamma^2-1)+16(1-3\gamma)\gamma) M^2 + \pi (1-7\gamma)\gamma Q^2 ] \;, \\
\int_{0}^{1} du A_3(u) &=& \frac{M}{48 \gamma^3}  (\gamma -1) \sqrt{\frac{(\gamma +1) M^2-2 \gamma  Q^2}{(\gamma +1) M^2}} \left(2 M^2 \left(3 \pi  \left(17 \gamma ^2+\gamma -1\right) \right. \right. \nonumber \\\
&& \left. \left.  + 4 \gamma  (1-23 \gamma ) \right) +\gamma Q^2  (32 \gamma +\pi  (3-21 \gamma )) \right) \nonumber \\
&& + \frac{M}{48 \gamma^3 (\gamma +1)}\left(3 \gamma  (\pi Q^2  (\gamma +1) ((23 \gamma -14) \gamma +3)  \right. \nonumber \\\
&& \left.  - 8 \gamma  (\gamma  (19 \gamma +10)-1)) - 2M^2 (\gamma +1) (3 \pi  ((\gamma  (47 \gamma -16)-2) \gamma +1)  \right. \nonumber \\\
&& \left.  - 4 \gamma  ((145 \gamma -24) \gamma +1)) \right) \;.
\end{eqnarray}
Note that we have rewritten the results for the integrals in terms of the black hole mass and charge. We would like to further write this result in terms of the impact parameter $b$ and so we must identify the relation between $r_0$ and $b$ as we have also done in the previous subsection. We find that up to third order,
\begin{eqnarray}
&& r_0 \approx b + \frac{2 \gamma ^2 M \sqrt{1-\frac{2 \gamma  Q^2}{(\gamma +1) M^2}}-2 \gamma  M \sqrt{1-\frac{2 \gamma  Q^2}{(\gamma +1) M^2}}-6 \gamma ^2 M+2 \gamma  M}{4 \gamma ^2} \nonumber \\
&& + \frac{-7 \gamma ^2 M^2+\gamma ^2 M^2 \sqrt{1-\frac{2 \gamma  Q^2}{(\gamma +1) M^2}}-M^2 \sqrt{1-\frac{2 \gamma  Q^2}{(\gamma +1) M^2}}+M^2+3 \gamma ^2 Q^2-\gamma  Q^2}{4 b \gamma ^2} \;. \nonumber \\
\end{eqnarray}
Substituting this and expanding as appropriate we find,
\begin{eqnarray}
&& \Phi = \frac{4 M}{b} + \frac{1}{b^2} \frac{\pi}{8 \gamma ^2}   \left(M^2 \left(\gamma ^2 \left(-\left(\sqrt{\frac{(\gamma +1) M^2-2 \gamma  Q^2}{(\gamma +1) M^2}}-31\right)\right) \right. \right.  \nonumber \\
&& \left. \left. + \sqrt{\frac{(\gamma +1) M^2-2 \gamma  Q^2}{(\gamma +1) M^2}}-1\right)+\gamma  (1-7 \gamma ) Q^2\right)   \nonumber \\
&& - \frac{1}{b^3} \frac{4}{3 \gamma ^2}  \left(M \left(-34 \gamma ^2 M^2+2 \gamma ^2 M^2 \sqrt{1-\frac{2 \gamma  Q^2}{(\gamma +1) M^2}}-\gamma ^2 Q^2 \sqrt{1-\frac{2 \gamma  Q^2}{(\gamma +1) M^2}} \right. \right. \nonumber \\
&& \left. \left. - 2 M^2 \sqrt{1-\frac{2 \gamma  Q^2}{(\gamma +1) M^2}}+\gamma  Q^2 \sqrt{1-\frac{2 \gamma  Q^2}{(\gamma +1) M^2}}+2 M^2+15 \gamma ^2 Q^2-3 \gamma  Q^2\right)\right) \nonumber \\
&& \quad + \ldots \;.
\end{eqnarray}

We can check this result by computing the $\gamma=1$ case which corresponds to a Reissner–Nordstrom black hole. Choosing $\gamma=1$ above we find,
\be
\Phi \approx \frac{4M}{b} + \frac{15 \pi}{4} \frac{M^2}{b^2} - \frac{3 \pi}{4} \frac{Q^2}{b^2} + \frac{128 M^3}{3 b^3}-\frac{16 M Q^2}{b^3} \;,
\ee
as expected for a Reissner–Nordstrom black hole \cite{Eiroa:2002mk,Sereno:2003nd}.

We can also compute the relevant deflection angle in the Kaluza-Klein case by setting  $\gamma=-1/2$ (equivalently $\alpha=\sqrt{3}$). In this case we find that the deflection angle is given by,
\begin{eqnarray}
&& \Phi \approx \frac{4M}{b} + \frac{3}{8} \frac{1}{b^2} \left(\pi M^2 \left(\sqrt{1+\frac{2Q^2}{M^2}} + 9 \right) - 3 \pi Q^2 \right) \nonumber \\
&& \quad + \frac{1}{b^3} \left(4 M Q^2 \left(\sqrt{\frac{2 Q^2}{M^2}+1}-7\right)+M^3 \left(8 \sqrt{\frac{2 Q^2}{M^2}+1}+\frac{104}{3}\right)\right) \;.
\end{eqnarray}

\section{Null-Like Geodesics in a Kerr Background}\label{sec:deflangleequat}

The metric for a Kerr black hole is given by,
\begin{eqnarray}
ds^2 &=& - \left(1 - \frac{R_s r}{\Sigma} \right) dt^2 + \frac{\Sigma}{\Delta}dr^2 + \Sigma d\theta^2  + \left(r^2 + a^2 +   \frac{R_s r a^2}{\Sigma} \right) \sin^2 \theta d\phi^2 \nonumber \\
&& \qquad -  \frac{2 R_s r a \sin^2\theta}{\Sigma} dt d\phi \;, \label{eq:kerrmetric}
\end{eqnarray}
where we have defined,
\begin{eqnarray}
R_s &=& 2 G_N M \\
a &=& \frac{J}{M} \\
\Sigma &=& r^2 + a^2 \cos^2 \theta \\
\Delta &=& r^2 - R_s r + a^2 \;.
\end{eqnarray}
We will be calculating null-like geodesics in the equatorial plane for computational simplicity. Note however that unlike the previous two subsections there is no spherical symmetry and we can therefore not generalise these results for general $\theta$. Taking \eqref{eq:kerrmetric} with $\theta=\pi/2$ yields,
\begin{equation}
ds^2 = - \left(1 - \frac{R_s}{r} \right) dt^2 + \frac{r^2}{\Delta}dr^2 + \left(r^2 + a^2 + \frac{R_s a^2}{r} \right) d\phi^2 -  \frac{2 R_s a }{r} dt d\phi \;.
\end{equation}
Since the metric is independent of $t$ and $\phi$ we have at least two constants of the motion. These are,
\begin{eqnarray}
- E &=& - \left(1 - \frac{R_s}{r} \right) \frac{dt}{d\tau} - \frac{R_s a }{r} \frac{d\phi}{d\tau} \;, \\
J &=& \left(r^2 + a^2 + \frac{R_s a^2}{r} \right) \frac{d\phi}{d\tau} - - \frac{ R_s a }{r} \frac{dt}{d\tau} \;.
\end{eqnarray}
As before we use \eqref{eq:geo2} to calculate the null-like geodesics in the metric given by \eqref{eq:kerrmetric} and in this case we find,
\begin{equation}
0 =  - \left(1 - \frac{R_s}{r} \right) \left( \frac{dt}{d \tau} \right)^2 + \frac{r^2}{\Delta} \left( \frac{dr}{d \tau} \right)^2 + \left(r^2 + a^2 + \frac{R_s a^2}{r} \right) \left(\frac{d \phi}{d \tau} \right)^2 - \frac{2 R_s a }{r} \frac{dt}{d\tau} \frac{d\phi}{d\tau}  \;.
\end{equation}
Substituting for the constants of the motion and solving for $d\phi/dr$ we get,
\begin{equation}
\frac{d\phi}{dr} = \frac{b(r-R_s)+a R_s}{a^2 + r(r-R_s)} \left( (a^2 - b^2 + r^2) + \frac{R_s}{r}(a-b)^2) \right)^{-\frac{1}{2}} \;, \label{eq:KerrDef1}
\end{equation}
where we have used that $J=Eb$ for null-like geodesics. Note that this equation has been checked for $a=0$ in which case it reduces to the Schwarzschild version of this equation as expected. The point of closest approach, $r_0$, is related to the impact parameter by,
\begin{equation}
\frac{dr}{d\tau} \bigr|_{r=r_0} \Rightarrow \quad b = \frac{\sqrt{r_0^2(a^2 + r_0^2 - R_sr_0)} - aR_s}{r-R_s} \;. \label{eq:KerrDef2}
\end{equation}
Substituting this into \eqref{eq:KerrDef1} we find,
\begin{eqnarray}
&&\frac{d\phi}{dr} = \frac{a(r-r_0)R_s + r_0(R_s-r)\sqrt{a^2 +r_0(r_0-R_s)}}{(R_s-r_0)(a^2 + r(r-R_s))} \left[ \frac{r-r_0}{r(r_0-R_s)^2} \times \right. \nonumber \\
&& \left. \left( r^2(r_0-R_s)^2 +  r r_0 (r_0-R_s)^2 + R_s (-2a^2r_0+r_0^2 (-r_0+R_s)) \right. \right. \nonumber \\
&& \left. \left. + 2a R_s r_0 \sqrt{a^2 + r_0^2 -r_0 R_s} \right) \right]^{-\frac{1}{2}} \;. \label{eq:KerrDef3}
\end{eqnarray}
As before we can use the substitution $u=r_0/r$ and expand in powers of $1/r_0$. We can then write the deflection angle up to third order as,
\begin{eqnarray}
\Phi + \pi &=& 2 \left( \int_{0}^{1} du \frac{1}{\sqrt{1-u^2}} + \int_{0}^{1} du A_1(u) \left(\frac{1}{r_0}\right) + \int_{0}^{1} du A_2(u) \left(\frac{1}{r_0}\right)^2 \right. \nonumber \\
&& \left. \quad  + \int_{0}^{1} du A_3(u) \left(\frac{1}{r_0}\right)^3 + \ldots \right) \;,
\end{eqnarray}
where we have found,
\begin{eqnarray}
\int_{0}^{1} du A_1(u) &=& R_s \;, \\
\int_{0}^{1} du A_2(u) &=& \frac{1}{32} R_s \left((15 \pi -16) R_s-32 a\right) \;, \\
\int_{0}^{1} du A_3(u) &=&  \frac{1}{96} R_s \left(48 a^2+24 (8-5 \pi ) a R_s+(244-45 \pi ) R_s^2\right)  \;,
\end{eqnarray}
We can then write the angle as,
\begin{eqnarray}
\Phi &=& \frac{2R_s}{r_0} +  \frac{1}{r_0^2} \left( -a R_s + \frac{(15\pi - 16)}{32}R_s^2  \right) \nonumber \\
&& \quad + \frac{1}{r_0^3} \left( \frac{a^2 R_s}{2} + \frac{(8-5 \pi) a R_s^2}{4} + \frac{(244-45\pi)R_s^3}{96} \right) + \ldots \;. \label{eq:KerrDef5}
\end{eqnarray}
We now need to find the relation between $r_0$ and the impact parameter $b$ in order to be able to express the equation above in terms of $b$. We find that up to third order,
\begin{equation}
r_0 \approx b - \frac{R_s}{2} + \frac{1}{b} \left( -\frac{a^2}{2}+a R_s-\frac{3 R_s^2}{8} \right) \;.
\end{equation}
Now substituting this into \eqref{eq:KerrDef5} we can express the deflection angle in terms of impact parameter $b$ after an appropriate expansion. Doing so we find,
\begin{equation}
\Phi = \frac{2R_s}{b} +  \frac{1}{b^2} \left( -2 a R_s + \frac{15\pi}{16} R_s^2 \right) + \frac{1}{b^3} \left( 2a^2 R_s	- \frac{5 \pi}{2} a R_s^2 + \frac{16}{3}R_s^3 \right) + \ldots \;. \label{eq:KerrDef4}
\end{equation}
We see that the expression above agrees with known results \cite{Iyer:2009hq}.

\chapter{Auxiliary Integrals} \label{app:aux}

In this appendix we will give results for various sub-integrals which appear in chapter \ref{chap:integrals}. In section \ref{app:boxintegrals} we find integrals such as,
\begin{eqnarray}
\mathcal{\hat{I}}^{(a)}_{t_2} &=& \int_{-\infty}^{\infty} dt_2 \, dt_4 |t_2|^{2m} t_2^{2n} \exp \left[  -\frac{(t_2\; t_4)}{T_0} \left(\begin{matrix} m_1^2 & \tilde{k}^2 \\ \tilde{k}^2 & m_2^2 \end{matrix} \right) \left(\begin{matrix} t_2 \\ t_4 \end{matrix} \right)  \right] \nonumber \\
&=& \frac{\sqrt{\pi T_0}}{m_2} \left(\frac{ m_2^2 T_0}{m_1^2 m_2^2 - \tilde{k}^4}\right)^{m+n+\frac{1}{2}} \Gamma \left(m+n+\frac{1}{2} \right) \;,
\end{eqnarray}
\begin{eqnarray}
\mathcal{\hat{I}}^{(b)}_{t_2} &=& \int_{-\infty}^{\infty} dt_2 \, dt_4 |t_2|^{2m} t_4^{2n} \exp \left[  -\frac{(t_2\; t_4)}{T_0} \left(\begin{matrix} m_1^2 & \tilde{k}^2 \\ \tilde{k}^2 & m_2^2 \end{matrix} \right) \left(\begin{matrix} t_2 \\ t_4 \end{matrix} \right)  \right] \nonumber \\
&=& \frac{T_0^{m+n+1}}{m_1^{2m+1} m_2^{2n+1}} \Gamma \left(m+\frac{1}{2} \right) \Gamma \left(n+\frac{1}{2} \right) \nonumber \\
&& \qquad \times \, _2F_1\left(m+\frac{1}{2},n+\frac{1}{2};\frac{1}{2};\frac{\tilde{k}^4}{m_1^2 m_2^2}\right) \;,
\end{eqnarray}
\begin{eqnarray}
\mathcal{\hat{I}}^{(c)}_{t_2} &=& \int_{-\infty}^{\infty} dt_2 \, dt_4 |t_2|^{2m} t_2 t_4 \exp \left[  -\frac{(t_2\; t_4)}{T_0} \left(\begin{matrix} m_1^2 & \tilde{k}^2 \\ \tilde{k}^2 & m_2^2 \end{matrix} \right) \left(\begin{matrix} t_2 \\ t_4 \end{matrix} \right)  \right] \nonumber \\
&=&  -\frac{\tilde{k}^2 \sqrt{\pi T_0}}{m_2^3} \left(\frac{ m_2^2 T_0}{m_1^2 m_2^2 - \tilde{k}^4}\right)^{m+\frac{3}{2}} \Gamma \left(m+\frac{3}{2} \right) \;,
\end{eqnarray}
and similarly we find,
\begin{eqnarray}
\mathcal{\hat{I}}^{(a)}_{t_4} &=& \int_{-\infty}^{\infty} dt_2 \, dt_4 |t_4|^{2m} t_4^{2n} \exp \left[  -\frac{(t_2\; t_4)}{T_0} \left(\begin{matrix} m_1^2 & \tilde{k}^2 \\ \tilde{k}^2 & m_2^2 \end{matrix} \right) \left(\begin{matrix} t_2 \\ t_4 \end{matrix} \right)  \right] \nonumber \\
&=& \frac{\sqrt{\pi T_0}}{m_1} \left(\frac{m_1^2 T_0}{m_1^2 m_2^2 - \tilde{k}^4}\right)^{m+n+\frac{1}{2}} \Gamma \left(m+n+\frac{1}{2} \right) \;,
\end{eqnarray}
\begin{eqnarray}
\mathcal{\hat{I}}^{(b)}_{t_4} &=& \int_{-\infty}^{\infty} dt_2 \, dt_4 |t_4|^{2m} t_2^{2n} \exp \left[  -\frac{(t_2\; t_4)}{T_0} \left(\begin{matrix} m_1^2 & \tilde{k}^2 \\ \tilde{k}^2 & m_2^2 \end{matrix} \right) \left(\begin{matrix} t_2 \\ t_4 \end{matrix} \right)  \right] \nonumber \\
&=& \frac{T_0^{m+n+1}}{m_1^{2n+1} m_2^{2m+1}} \Gamma \left(m+\frac{1}{2} \right) \Gamma \left(n+\frac{1}{2} \right) \nonumber \\
&& \qquad \times \, _2F_1\left(m+\frac{1}{2},n+\frac{1}{2};\frac{1}{2};\frac{\tilde{k}^4}{m_1^2 m_2^2}\right) \;,
\end{eqnarray}
\begin{eqnarray}
\mathcal{\hat{I}}^{(c)}_{t_4} &=& \int_{-\infty}^{\infty} dt_2 \, dt_4 |t_2|^{2m} t_2 t_4 \exp \left[  -\frac{(t_2\; t_4)}{T_0} \left(\begin{matrix} m_1^2 & \tilde{k}^2 \\ \tilde{k}^2 & m_2^2 \end{matrix} \right) \left(\begin{matrix} t_2 \\ t_4 \end{matrix} \right)  \right] \nonumber \\
&=&  -\frac{\tilde{k}^2 \sqrt{\pi T_0}}{m_1^3} \left(\frac{ m_1^2 T_0}{m_1^2 m_2^2 - \tilde{k}^4}\right)^{m+\frac{3}{2}} \Gamma \left(m+\frac{3}{2} \right) \;.
\end{eqnarray}
Note that we have assumed that $n,m$ are even for all the expressions above, the integrals yield zero otherwise. For two absolute values we can have the integral,
\begin{eqnarray}
\mathcal{\hat{I}}_{t_2, t_4} = \int_{-\infty}^{\infty} dt_2 \, dt_4 |t_2 t_4| \exp \left[  -\frac{(t_2\; t_4)}{T_0} \left(\begin{matrix} m_1^2 & \tilde{k}^2 \\ \tilde{k}^2 & m_2^2 \end{matrix} \right) \left(\begin{matrix} t_2 \\ t_4 \end{matrix} \right)  \right] \;,
\end{eqnarray}
which appears at subsubleading order in \eqref{eq:xtraconti4ltoi4ssl}. We can perform an integral of the form above by expanding,
\begin{equation}
e^{-2 \frac{t_2 t_4}{T_0} \tilde{k}^2} = \sum_{n=0}^{\infty} \frac{1}{(2n)!} \left[ \left( 2 \frac{t_2 t_4}{T_0} \tilde{k}^2 \right)^2 \right]^n + \text{odd terms} \;,
\end{equation}
where we have focused only on even terms as they are the only ones that can contribute to the integral. We therefore have,
\begin{eqnarray}
\mathcal{\hat{I}}_{t_2, t_4} &=& \int_{0}^{\infty} dt_2^2 \, dt_4^2 \sum_{n=0}^{\infty} \frac{1}{(2n)!} \left( 4 \frac{t_2^2 t_4^2}{T_0^2} \tilde{k}^4 \right)^n e^{-m_1^2 \frac{t_2^2}{T_0}} e^{-m_2^2 \frac{t_4^2}{T_0}} \nonumber \\
&=& \int_{0}^{\infty} dx_2 \, dx_4 e^{-x_2} e^{-x_4} \frac{T_0^2}{m_1^2 m_2^2} \sum_{n=0}^{\infty} \frac{x_2^n x_4^n}{(2n)!} \left( 4 \frac{\tilde{k}^4}{m_1^2 m_2^2} \right)^n  \nonumber \\
&=& \frac{T_0^2}{m_1^2 m_2^2} \sum_{n=0}^{\infty} \frac{(n!)^2}{(2n)!} \left( 4 \frac{\tilde{k}^4}{m_1^2 m_2^2}\right)^n \nonumber \\
&=& \frac{T_0^2}{m_1^2 m_2^2 - \tilde{k}^4} + \frac{T_0^2 \tilde{k}^2}{(m_1^2 m_2^2 - \tilde{k}^4)^{3/2}} \arctan \left[\frac{\tilde{k}^2}{\sqrt{m_1^2 m_2^2 - \tilde{k}^4}} \right] \;.
\end{eqnarray}
We also need the following integrals which can be solved in a similar way,
\begin{eqnarray}
\mathcal{\hat{I}}^{(a)}_{t_2, t_4} &=& \int_{-\infty}^{\infty} dt_2 \, dt_4 |t_2 t_4| t_2^2 \exp \left[  -\frac{(t_2\; t_4)}{T_0} \left(\begin{matrix} m_1^2 & \tilde{k}^2 \\ \tilde{k}^2 & m_2^2 \end{matrix} \right) \left(\begin{matrix} t_2 \\ t_4 \end{matrix} \right)  \right] \nonumber \\
&=& \frac{T_0^3 (\tilde{k}^4 +4 m_1^2 m_2^2)}{2 m_1^2 \left(m_1^2 m_2^2-\tilde{k}^4\right)^2} - \frac{3 T_0^3 \tilde{k}^2 m_1^2 m_2^2}{2 m_1^2 \left(m_1^2 m_2^2-\tilde{k}^4\right)^{3/2}} \arctan \left[ \frac{\tilde{k}^2}{\sqrt{m_1^2 m_2^2-\tilde{k}^4}} \right]   \;, \nonumber \\
\end{eqnarray}
\begin{eqnarray}
\mathcal{\hat{I}}^{(b)}_{t_2, t_4} &=& \int_{-\infty}^{\infty} dt_2 \, dt_4 |t_2 t_4| t_4^2 \exp \left[  -\frac{(t_2\; t_4)}{T_0} \left(\begin{matrix} m_1^2 & \tilde{k}^2 \\ \tilde{k}^2 & m_2^2 \end{matrix} \right) \left(\begin{matrix} t_2 \\ t_4 \end{matrix} \right)  \right] \nonumber \\
&=& \frac{T_0^3 (\tilde{k}^4 +4 m_1^2 m_2^2)}{2 m_2^2 \left(m_1^2 m_2^2-\tilde{k}^4\right)^2} - \frac{3 T_0^3 \tilde{k}^2 m_1^2 m_2^2}{2 m_2^2 \left(m_1^2 m_2^2-\tilde{k}^4\right)^{3/2}} \arctan \left[ \frac{\tilde{k}^2}{\sqrt{m_1^2 m_2^2-\tilde{k}^4}} \right]   \;, \nonumber \\
\end{eqnarray}
\begin{eqnarray}
\mathcal{\hat{I}}^{(c)}_{t_2, t_4} &=& \int_{-\infty}^{\infty} dt_2 \, dt_4 |t_2 t_4| t_2 t_4 \exp \left[  -\frac{(t_2\; t_4)}{T_0} \left(\begin{matrix} m_1^2 & \tilde{k}^2 \\ \tilde{k}^2 & m_2^2 \end{matrix} \right) \left(\begin{matrix} t_2 \\ t_4 \end{matrix} \right)  \right] \nonumber \\
&=& -\frac{3 T_0^3 \tilde{k}^2}{2 \left(m_1^2 m_2^2-\tilde{k}^4\right)^2} - \frac{T_0^3( 2\tilde{k}^2 + m_1^2 m_2^2)}{2\left(m_1^2 m_2^2-\tilde{k}^4\right)^{5/2}} \arctan \left[ \frac{\tilde{k}^2}{\sqrt{m_1^2 m_2^2-\tilde{k}^4}} \right] \;. \nonumber \\
\end{eqnarray}
The only non-trivial sub-integral needed for the triangle integrals in section \ref{app:triintegrals} is given by,
\begin{eqnarray}
\mathcal{\hat{I}}_{t_3} &=& \int_{0}^{\infty} dt_3 \exp \left[- \frac{m_i^2 t_3^2}{T_0} \right] |t_3| t_3^n \nonumber \\
&=& \frac{1}{2} m_i^{-n-2} T_0^{\frac{n}{2}+1} \Gamma \left(\frac{n}{2}+1\right) \;.
\end{eqnarray}

\chapter{Exponentiation of Tree Level Amplitude on ${\mathbb R}^{1,3} \times S^1$} \label{app:kkexp}

In this appendix we will show that by calculating the leading contribution to the one-loop amplitude on ${\mathbb R}^{1,3} \times S^1$ we can show the first signs of exponentiation of the tree-level amplitude in Kaluza-Klein theory as discussed in chapter \ref{chap:kkeik}. We will keep the spacetime dimensions general but will have to restrict to $D=5$ later in the calculation. We will compute the one-loop amplitude in the ultra-relativistic limit but the results should easily extend to the case where both Kaluza-Klein masses are kept large.

The method we use is the standard one of introducing Schwinger parameters $t_i$ corresponding to the 4 internal propagators of the box diagram as we have done in chapter \ref{chap:integrals}. Starting from the expression for the one-loop amplitude in normal flat space calculated in section \ref{sec:oneloopamp}, using \eqref{eq:changeconttoKK} and making a change of variables from $t_i $ to $T, t_2,t_3, t_4$ where $T = \sum_i t_i $ we can express the one-loop amplitude in the ultra-relativistic limit as,
\begin{eqnarray}
i { {\cal A}}_2 (D-1, R  )  &=&  4i \kappa^4 (s'^2)^2 \frac{1 }{(2\pi)^{D-1}} \int _0^\infty dT\, {\displaystyle  \prod_{i=2}^4 }\int _0^\infty dt_i \nonumber \\
&& \qquad \times \int  \frac{ d^{D-1} q'}{(2\pi)^{D-1}} \frac{1}{2\pi R} \,  \sum_{n  \in {\mathbb Z } } e^{-T\left( q + A/2T \right)^2 } \;,
\end{eqnarray}
where we use the notation $  \left( q + A/2T \right)^2 = \left( q' + A'/2T \right)^2 + \left( n/R+ A_s/2T \right)^2   $  with    $q = (q', n/R) $ being  the loop momentum on  ${\mathbb R}^{1,D-2} \times S^1  $
and $A = (A', A_s) $ with 
\begin{eqnarray}
A' &=& (- 2p_3't_2 - 2p_1't_2 -2 p_1't_3 +2 p_2' t_4) \cr 
\cr
A_s &=& \frac{1}{R} (- 2n_3 t_2 - 2n_1 t_2 -2 n_1 t_3 +2 n_2 t_4) \;.
\end{eqnarray}
The integral over the $q'$ momenta is just a Gaussian integral and can be easily performed,
\begin{eqnarray}
i { {\cal A}}_2 (D-1, R  )  &=&  \frac{4i \kappa^4 (s'^2)^2 }{2 \pi R} \frac{\pi ^{\frac{D-1}{2} }}{(2\pi)^{D-1}} \int _0^\infty dT  \,  T^{-{\frac{D-1}{2}}}{\displaystyle  \prod_{i=2}^4 }\int _0^\infty dt_i   \nonumber \\
&& \qquad \times  \sum_{n  \in {\mathbb Z } } e^{-T\left( n/R + A_s/2T \right)^2} e^{\frac{f}{T} +t_2 t}  \;,
\end{eqnarray}
where we have defined, $f = t(-t_2^2 -t_2 t_3 -t_2t_4 ) + t_3t_4s$. Changing to new variables $ \alpha_i \equiv t_i/T $,
\begin{eqnarray}
i{ {\cal A}}_2 (D-1, R  )  &=& \frac{4i \kappa^4 (s'^2)^2 }{2 \pi R} \frac{\pi ^{\frac{D-1}{2} }}{(2\pi)^{D-1}} \int _0^\infty dT  \,  T^{3-{\frac{D-1}{2}}}{\displaystyle  \prod_{i=2}^4 }\int _0^\infty d\alpha_i   \nonumber \\
&& \qquad \times  \sum_{n  \in {\mathbb Z } } e^{-T\left( n/R + {\tilde A}_s  \right)^2} e^{T {\tilde f} } \;, \label{A1loop}
\end{eqnarray}
with  $ {\tilde A}_s = (- n_3 \alpha_2 - n_1 \alpha_2 - n_1 \alpha_3 + n_2 \alpha_4)/R  $  and $ {\tilde f} =  t(-\alpha_2^2  - \alpha_2 \alpha_3 -\alpha_2 \alpha_4 )  +\alpha_3 \alpha_4 s + \alpha_2 t$. If we combine the $n$ independent terms that multiply $T$ in the exponential term in \eqref{A1loop}, we find after some algebra that this can be expressed as, 
  \begin{eqnarray}
 &-&\alpha_2^2 t'  -\alpha_2 \alpha_3 \left( t' - (n_3+n_1)^2/R^2 +n_1(n_1+n_3)/R^2 \right) - \alpha_2 \alpha_4 \left(t'  -(n_3+n_1)^2/R^2 \right. \cr
 \cr
 &-& \left. n_2(n_3+n_1)/R^2 \right)  -\alpha_3^2  n_1^2/R^2  - \alpha_4^2 n_2^2/R^2  - \alpha_3 \alpha_4 \left(  n_1^2/R^2 + n_2^2/R^2  \right) \cr
 \cr
 &+&  \alpha_3 \alpha_4 s' + \alpha_4 t'- \alpha_2 (n_3+n_1)^2 /R^2 \;.
  \end{eqnarray}
The original high energy limit $s \rightarrow \infty $ with $t$ kept fixed, can be understood in the compactified case as the limit $s' \rightarrow \infty$ with $t'$ fixed and the Kaluza-Klein momenta $n_i/R$, $i = 1\ldots4 $ fixed. Now in the $s' \rightarrow \infty$ limit, the only non-vanishing contributions in the $\alpha_3, \alpha_4 $ integrals come from the regions $\alpha_3, \alpha_4  \rightarrow 0 $ so as to keep the term in the exponential from diverging. These integrals may then be solved after a wick rotation in  ${\mathbb R}^{1,D-2}$ which yield Gaussian integrals resulting in a factor $  \frac{\pi}{2 T \sqrt{-{E'_e}^4} }$ where $E'_e $ is the corresponding energy in Euclidean space. Combining these results we find that the amplitude then reduces to, 
 \begin{eqnarray}
i{  { \cal A}}_2 (D-1, R  )  &= &\frac{4i \kappa^4 (s'^2)^2 }{2 \pi R} \frac{\pi ^{\frac{D+1}{2} }}{(2\pi)^{D-1}}  \frac{1}{2 T \sqrt{-{E'_e}^4} }  \int _0^\infty dT 
T^{2-{\frac{D-1}{2}}} \int _0^\infty d\alpha_2 \sum_{n  \in {\mathbb Z } }\cr
\cr
 && \times \, e^{-T\left( |t'|\alpha_2 (1-\alpha_2 )    +  \alpha_2 (n_3+n_1)^2/R^2  + n^2/R^2  + 2  \alpha_2 n(n_3+n_1) /R^2 \right)    } \;.
 \end{eqnarray}
We now wish to transform the amplitude to impact parameter space to check that the result is equivalent to the square of the tree-level diagram in impact parameter space, as we anticipate from the exponentiation of the eikonal phase \cite{Kabat:1992tb,Giddings:2010pp,Akhoury:2013yua}. Given that on ${\mathbb R}^{1,D-2} \times S^1  $,  the transverse momentum is given by $q  = ({q'}, (n_1+n_3)/R )$, with $|t'| = {q'}^{2} $, we find that the impact space expression for the one-loop amplitude ${ \tilde  { \cal A}}_2 (D-1, R  ) $ is given by, 
\begin{equation}
i { \tilde  { \cal A}}_2 (D-1, R  ) = \frac{1 }{2s'}\int  \frac{d^{D-3}q'}{(2\pi )^{D-3}} e^{i q' \cdot b' }  \frac{1}{2 \pi R}  \sum_{n_1+n_3  \in {\mathbb Z } } e^{ib_s (n_1+n_3)/R}    i{ \cal A}_2 (D-1, R  ) \;,
\end{equation}
which we can see by combining \eqref{eq:defneikonal} with \eqref{eq:changeconttoKK}. The integration over the angular variables in the momentum integral $\int d^{D-3}q'$ can be carried out yielding a Bessel function of the first kind,
\begin{eqnarray}
&& i { \tilde  { \cal A}}_2 (D-1, R  )  = -\frac{4 \kappa^4 (s'^2)^2 }{(2 s')^2} \frac{\pi^{\frac{D+1}{2} }}{(2\pi)^{D-1}} \frac{1}{ (2\pi)^{\frac{D-3}{2}}} (b')^{\frac{5-D}{2}} \nonumber \\
&& \times \frac{1}{(2 \pi R)^2} \int _0^\infty dT  \,  T^{\frac{5-D}{2}}  \int_0^1 d \alpha_2 \sum_{m, n \in {\mathbb Z } } e^{im b_s/R} e^{-in b_s/R} \nonumber \\
&& \times \left(  \int_0^\infty dq'    e^{-T q'^2 \alpha_2(1-\alpha_2)}  {( q' )}^{\frac{D-3}{2}}   J_{\frac{D-5}{2} }(q' b')  \right) e^{-T \left( \alpha_2 \frac{m^2}{R^2} +(1-\alpha_2 )\frac{n^2}{R^2} \right)}  \;.
\end{eqnarray}
The integration over the remaining momentum magnitude, $ q' $, can be performed using the result for the integral \cite{gradshteyn1996table},
\begin{equation}
\int_0^\infty  e^{-\alpha x^2 } x^{\nu +1} J_\nu( \beta x) dx = \frac{\beta^\nu }{(2 \alpha )^{\nu +1 } } \, e^{ - \beta^2/4\alpha} \;,
\end{equation}
where in our case we have $ \alpha  = T \alpha_2(1-\alpha_2 ) , \,  \nu = \frac{D-5}{2},  \, \beta = b'$. We then find,
\begin{eqnarray}
i { \tilde  { \cal A}}_2 (D-1, R)  &=&  -\frac{\pi^\frac{D+1}{2} }{ {(2\pi)}^\frac{3D-5}{2}}  \frac{1}{(2s')^2}  \frac{4 \kappa^4 (s'^2)^2 }{(2 \pi R)^2}  \sum_{m, n \in {\mathbb Z }} e^{im b_s/R} e^{-in b_s/R}  \int_0^1  d\alpha_2 \int_0^\infty d {\hat T}  \cr \nonumber
  \cr 
  \cr 
  && \quad \times \, {({\hat T})}^{\frac{5-D}{2}} \frac{1}{(2 {\hat T} )^\frac{D-3}{2}  } \frac{1}{\left( \alpha_2 (1- \alpha_2) \right)^{\frac{7-D}{2}}}  \,  e^{-b'^2/4 {\hat T} } \, e^{ -\frac{\hat T}{R^2} \left( \frac{m^2}{1-\alpha_2 }  + \frac{n^2}{ \alpha_2 } \right) } \;.
\end{eqnarray}
To perform the integration over $\alpha_2$ it is convenient to make the change of variables, $u = \frac{\alpha_2}{ 1- \alpha_2}$, leading to,
\be
I_1 = \int_0^1  d\alpha_2  \frac{ e^{ -\frac{\hat T}{R^2} \left( \frac{m^2}{1-\alpha_2 }  + \frac{n^2}{\alpha_2} \right) } }{\left( \alpha_2 (1- \alpha_2 ) \right)^{\frac{7-D}{2} }  } = e^{ -\frac{\hat T}{R^2} (m^2 +n^2) } \int_0^\infty du \frac{(1+u)^{5-D}}{u^{\frac{7-D}{2}}} 
e^{ -\frac{\hat T}{R^2} \left( m^2 u + \frac{n^2}{u} \right) } \;.
\ee
From \cite{gradshteyn1996table} we find the following integral identity,
\be
\int_0^\infty x^{\nu - 1} e^{- \beta/4x -\gamma x } \, dx = {\left( \frac{\beta}{\gamma} \right)}^{\frac{\nu}{2}} \, K_\nu (\sqrt{\beta \gamma} \,  ) \;.
\ee
Although the remaining integration over $u$ that we need to perform is not quite in this form for arbitrary $D$ it is for $D=5$. So taking $D=5$ we find, 
\be 
I_1 = K_0\left(\sqrt{\frac{2 n m {\hat T}}{R^2} } \, \right) \;.
\ee
Putting all of the above together we find that the one-loop amplitude for $D=5$ becomes,
\begin{eqnarray}
i { \tilde  { \cal A}}_2 (D-1, R  )|_{D=5}   &=&  -\frac{\pi^3 }{ {(2\pi)}^5}  \frac{1}{(2s')^2}  \frac{4 \kappa^4 (s'^2)^2 }{(2 \pi R)^2}  \sum_{m, n \in {\mathbb Z }}  e^{im b_s/R -in b_s/R}  \cr
\cr
  &&\times \int_0^\infty  \frac{d {\hat T} }{2 {\hat T}} \,e^{-b'^2/4 {\hat T} -\frac{\hat T}{R^2} (m^2 + n^2 ) } K_0\left(\sqrt{\frac{2 n m {\hat T}}{R^2} }\right) \;.
\end{eqnarray}
An interesting integral identity involving Bessel K-functions is the so called 'duplicating' relation \cite{gradshteyn1996table},
\be
\label{Kdup}
\int_0^\infty e^{-\frac{x}{2} - \frac{1}{2x} (z^2 +\omega^2 ) }\, K_\nu \left( \frac{z\omega}{x}\right)\, \frac{dx}{x} = 2 K_\nu(z) K_\nu (\omega ) \;.
\ee 
To make use of this identity we need to perform the change of variables 
$ {\tilde T} = 1/{\hat T} $ and then $T' = \frac{{b'}^2}{2} {\tilde T}$. Performing these substitutions we then have,
\begin{eqnarray}
i{ \tilde  { \cal A}}_2 (D-1, R  )|_{D=5} &=&  -\frac{\pi^3 }{ {(2\pi)}^5}   \frac{ \kappa^4 s'^2 }{(2 \pi R)^2} \sum_{m, n \in {\mathbb Z }}  e^{im b_s/R -in b_s/R} \nonumber \\
&& \times  \int_0^\infty  \frac{d { T'} }{{T'}} \,e^{- \frac{T'}{2}  -\frac{{b'}^2}{2 T' R^2} (m^2 + n^2 ) } K_0\left(\frac{n m{b'}^2}{T'R^2}\right) \;.
\end{eqnarray}
Now using the identity given in \eqref{Kdup} with $z^2 = \frac{m^2 {b'}^2}{R^2} $ and  $\omega^2 = \frac{n^2 {b'}^2}{R^2} $  (and $z = +\sqrt{z^2} = \frac{|m| b'}{R}, \omega = +\sqrt{\omega^2} = \frac{|n| b'}{R} $ ), we arrive at the final form of ${ \tilde  { \cal A}}_2 (D-1, R  )|_{D=5}$,
\begin{eqnarray}
i{ \tilde  { \cal A}}_2 (D-1, R  )|_{D=5} & = &-\frac{\pi^3 }{ {(2\pi)}^5}   \frac{ \kappa^4 s'^2 }{(2 \pi R)^2}  \sum_{m, n \in {\mathbb Z }}  e^{im b_s/R -in b_s/R}\, K_0\left(\frac{|m| b'}{R} \right) K_0\left(\frac{|n| b'}{R} \right) \cr 
\cr
\cr
 & = & -\frac{1}{2}\left( \frac{s' \kappa^2}{4 \pi}\sum_{m \in {\mathbb Z }} \frac{ e^{im b_s/R} }{2 \pi R} K_0\left(\frac{|m| b'}{R} \right)\right)^2 \cr
 \cr
 \cr 
 &=&  \frac{1}{2}\left(  i { \tilde{\cal A}}_1 (D-1,R)|_{D=5}   \right)^2 \;,
\end{eqnarray}
where we have used the ultra relativistic form of the tree-level scattering amplitude \eqref{RTLC}, for which  $s'$ is much greater than the Kaluza-Klein masses of the scattering particles. This result corroborates our intuition that the eikonal exponentiates into a phase as suggested earlier. Note that although we have taken the ultra-relativistic limit for simplicity this result should extend beyond the ultra-relativistic case.

\addcontentsline{toc}{chapter}{Bibliography}

\providecommand{\href}[2]{#2}\begingroup\raggedright\endgroup

\end{document}